# 第一章  引论

## 1.1  概述

铁磁流体由铁磁性质的固相微粒和不导磁的液态基础载体组成。但它基本上仍然是一类液体，具有液体的一切共性，并且遵守液体运动的三个基本定律，即质量守恒定律，动量守恒定律，能量守恒定律。然而不同于普通流体的是，在三个守恒定律的数学形态上，除连续性方程外，在运动方程和能量方程中，都包含有磁作用项或蕴涵磁作用因素。最重要的诸如磁作用对粘性的影响，在运动方程中的磁力力矩，能量方程中的磁内能、磁化功等。

外磁场对铁磁流体的控制，不需要通过附加的机械装置，就可以改变铁磁流体的运动状态，这样超距的控制，在所有流体中是独一无二的特点。铁磁流体内固相微粒的尺寸比一般工业两相内的固相颗粒要小好几个数量级。这种纳米级尺寸的固相微粒，在铁磁流体中两相运动的滞后是完全可以忽略的。但是，一旦施加外磁场，则固相微粒因受到磁场力和磁场力矩的作用，而与基载液体在速度上不再同步，即平移速度和转动速度的滞后就将出现，滞后引起的粘性阻力和粘性阻力矩就将磁场力和磁场力矩传递到基载液体上。所以说没有两相流效应，就谈不上外磁场对铁磁流体的控制。

要使两相流效应足以对整个基载液产生控制作用，其条件是固相微粒必须为细小。这样占铁磁流体总体积不超过 10%的固相物质能够提供巨大的两相接触界面，而且尺寸细小，微粒就有热运动。热运动是保持固相微粒均匀分布于铁磁流体中的根本前提。目前的铁磁流体固相微粒的直径为 8nm~12nm，大约相当于 100 个单分子大，可以认为是大分子，而且微粒在铁磁流体中是稀疏相。这样，统计物理学和分子运动论就成为研究固相微粒的行为的有力工具。使用统计物理学和分子运动论分析铁磁流体得出的结果使人满意。最显著的例子就是 Langevin 函数给出的铁磁流体的磁化强度。

和普通两相流一样，铁磁流体也存在连续性问题。在铁磁流体中，任何一点处能遇到的，要么是固相微粒，要么是基载液体，不可能两相同处于一点。铁磁性的固相微粒分散在基载液中，微粒所在点处，物理性质与周围液体完全不同，并且是陡峭地突然改变，是一种矩形脉冲式的变化。将这些分散的脉冲物理性质化为连续的物理性质，本书的办法是在每个固相微粒所在点的邻域内利用 Dirac 的 $\delta$ 函数取该邻域内物理性质之平均值。尽管各个邻域相对于流场很小，但它们是无间断地互相接连而充满整个流场，以致邻域上的物理性质之平均值在流场中也成为连续的、平滑的。在此基础上列出的动力学方程只在平均值的意义上成立。但它们仍能给出有用的结果。

## 1.2  铁磁流体
### 1.2.1  铁磁流体的组成成分

铁磁流体的主要成分是液态的基础载体和固相的铁磁性微粒。

能够作为基础载体的液态物质很多。诸如水、酒精、煤油、酯类、醚类、硅油类、矿物油类、水银及其合金，等等。液态的基础载体简称基载液，它是铁磁流体能够具有流体性质的根本因素。基载液主要是按照铁磁流体的用途来选择。对于大量使用而且顾及成本低廉，则最好选择可以回收后循环使用的液体，例如矿物油类；对于特殊用途，像在高温低压下能够较长期保持不干涸的铁磁流体，则硅油是合适的选择对象；对于医疗临床使用，水作为基载液是最容易接受的。就体积而言，基载液占铁磁流体的绝大部分，通常为 90%，甚至更多。

固相磁性微粒均采用铁磁性材料，如铁、钴、镍、磁铁矿（$Fe_3O_4$）、二氧化铬（$CrO_2$）、钡铁氧体、锶铁氧体，以及一些稀土金属的合金，等等。铁磁流体的名称并不表示这类流体具有铁磁性质，只是因为它们含有铁磁性的固相微粒。由于固相微粒按体积只占铁磁流体的 5%~10%，加上固相微粒热运动的影响，铁磁流体的磁化率仅达到 2~3，远低于铁磁物质，但比顺磁物质高很多，所以它是一类超顺磁体。

在铁磁流体中，基载液是不导磁的，只有固相物质具有磁化性能。外磁场如何能够通过体积分量

不到 10%的固相来控制整个铁磁流体？达到这一点有两个基本要求：一是液固两相有足够大的接触面积，二是液、固两相必须是均匀地混合在一起。其办法就是，使固相物质成为极细小的微粒状，其直径只有 $10^{-8} \sim 10^{-9}$m 的量级。设体积为 $V_f$ 的铁磁流体中含有 $N$ 个微粒，则固相的体积 $V_p$ 是

$$V_p = N V_{p1}$$

$V_{p1}$ 是一个固相微粒的体积。上式两边除以 $V_f$，则有

$$\frac{V_p}{V_f} = \frac{N}{V_f} \frac{1}{6} \pi d_p^3$$

式中 $V_p/V_f = \phi_p$ 是固相物质在铁磁流体中的体积分量；$N/V_f = n_v$ 是单位体积铁磁流体中所含的固相微粒的个数，即体积数密度；$d_p$ 是单个微粒的直径。上式可以写成

$$n_v = \frac{6\phi_p}{\pi d_p^3} \tag{1.1}$$

取 $\phi_p = (0.05 \sim 0.10)$，$d_p = (5 \sim 10)10^{-9}$m，代入式（1.1）得

$$n_v = (10^{23} \sim 10^{24})颗/\text{m}^3 = (10^{17} \sim 10^{18})颗/\text{cm}^3$$

这样巨大的体积数密度提供的两相之接触界面也是极其巨大的。设 $S_p$ 是单位体积铁磁流体内固相微粒的总表面积，$S_{p1}$ 是一个微粒的表面积，则有 $S_p = n_v S_{p1}$，用式（1.1）和 $S_{p1} = \pi d_p^2$ 代入，就得

$$S_p = \frac{6\phi_p}{d_p} \tag{1.2}$$

取 $\phi_p = 0.06$，$d_p = 8 \times 10^{-9}$m，于是由式（1.2）算出

$$S_p = 45 \times 10^6 \, \text{m}^2/\text{m}^3 = 45 \, \text{m}^2/\text{cm}^3$$

这个结果说明，在 1 毫升的铁磁流体中，两相之间界面面积竟然有 45 m² 之大。这个数字并非特殊，而是相当典型的。按表面粘附条件，外磁场通过控制固相微粒来控制整个铁磁流体是不难理解的。

通常粒度小到 $10^{-6}$m 的微粒在液体中就有 Brown 运动。所以粒度为 $10^{-9}$m 的固相微粒在铁磁流体中具有很激烈的 Brown 运动。微粒的 Brown 运动是其周围的基载液分子碰撞的结果。基载液分子碰撞固相微粒的方向是随机的。碰撞的频率不下于每秒 $10^{21}$ 次。所以固相微粒的运动必然具有高度的随机性质，并且在每个方向都有相同的几率。从而保证固相微粒在铁磁流体中的均匀分布。

但是，Brown 运动同时也造成了固相微粒之间互相碰撞的机会。由于微粒之间相互作用的磁性力和 van der Waals 力，碰撞之后可能不再分开，逐渐结合成更大的聚集体。聚集体大到一定程度之后，就丧失 Brown 运动的能力，沉淀现象随之发生。此时铁磁流体的磁性能变坏以至于最终消失。为了克服聚集问题，在铁磁流体中加入了第三种重要成分，即分散剂。分散剂是一类长链大分子，它们有一

个极性端，极性端后面的尾部是非极性的。分散剂链分子的长度与固相微粒的直径有相同的量级。极性端吸附于固相微粒的表面上，其尾部在基载液分子的碰撞下作自由的摆动，这样的摆动也是一种热运动，具有热运动的能量。这种摆动的能量构成了防止固微粒聚集的能垒[1]。可以作为分散剂的物质包括油酸、亚油酸、卵磷酯、聚醚酸、聚磷酸衍生物、聚胺类，等等。分散剂的选择，要考虑和基载液体及固相微粒的材料性质相匹配。尤其是分散剂不能大量地溶解于基载液中，以至失掉其应有的功能。

### 1.2.2 铁磁流体的胶体稳定性

依照气体分子运动论，每一个分子热运动的能量等于 $Ck_0T$，其中 $k_0$ 是每个气体分子的气体常数，它是一个普适的物理常数，$k_0 = 1.38 \times 10^{-23} N \cdot m/K$，$T$ 是绝对温度，$C$ 是对应平均动量的因数，平均动量有三种，即最可几动量、算术平均动量、均方根动量。由能量均分定理，气体分子具有的广义动量平方或广义坐标平方形式的热运动能量（诸如平动动能 $(m_1V)^2/(2m_1)$、转动动能 $(J_1\omega)^2/(2J_1)$、振动动能 $(\mu u)^2/(2\mu)$、弹性位移势能 $(kx)^2/(2k)$ 等等），其能量都等于 $Ck_0T$，所以下面的式（1.3）是普适的：

$$E_{T1} = Ck_0T \tag{1.3}$$

上式左方下标 "$T$" 表示热运动，下标 "$1$" 表示一个分子。铁磁流体中的固相微粒的尺寸大约相当于100个单分子，所以可以看作是"大分子"，借助气体分子运动论分析固相微粒的行为，结果令人满意。于是气体分子运动论遂成为研究铁磁流体的有力工具之一。

1.在重力场中铁磁流体的胶体稳定性

铁磁流体的固相微粒在重力场中的势能是固相微粒的重量与它所排开的同体积的铁磁流体重量之差和高度的乘积。一个固相微粒的重量是 $\rho_{NP}V_{p1}g$，同样体积的铁磁流体的重量是 $\rho_f V_{p1}g$，则一个微粒的重力场势能是

$$(\rho_{NP} - \rho_f)V_{p1}gh = E_{g1}$$

式中 $h$ 是固相微粒所处的高度，固相微粒的体积 $V_{p1} = \pi d_p^3/6$，如果固相微粒的热运动能 $E_{T1}$ 大于势能 $E_{g1}$，则固相微粒在铁磁流体中仍然有 Brown 运动而不致沉淀，即

$$Ck_0T \geq (\rho_{NP} - \rho_f)gh \cdot \frac{\pi}{6}d_p^3$$

由此得出

$$d_p \leq \left[\frac{6Ck_0T}{\pi(\rho_{NP} - \rho_f)gh}\right]^{1/3} \tag{1.4}$$

取一种矿物油+Fe$_3$O$_4$ 的铁磁流体，固相微粒的 $\rho_{NP} = 5240 kg/m^3$，铁磁流体的 $\rho_f = 1113 kg/m^3$，重力加速度 $g = 9.81 m/s^2$，$h = 0.1m$，$C = 1$，$T = 298K$，代入式（1.4）算出

$$d_p \leq 12.2 \times 10^{-9} \, \text{m}$$

若固相微粒的直径不大于 12nm，则在铁磁流体中的固相微粒不会因重力作用而沉淀。这就是说铁磁流体具有合乎要求的贮存性能，它能够长期地保持胶体状态。

2. 在外磁场中铁磁流体的稳定性

一颗固相微粒在外磁场 $\overrightarrow{H}$ 中的磁势能 $E_{m1}$ 是它的磁矩 $\overrightarrow{m_{p1}}$ 与 $\overrightarrow{H}$ 的点积，即

$$E_{m1} = -\mu_0 \overrightarrow{m_{p1}} \cdot \overrightarrow{H} = -\mu_0 m_{p1} H \cos\theta$$

式中 $\theta$ 是两矢量 $\overrightarrow{m_{p1}}$ 和 $\overrightarrow{H}$ 之间的夹角。$E_{m1}$ 的最大数值是当 $\cos\theta = 1$，此时两矢量 $\overrightarrow{m_{p1}}$ 和 $\overrightarrow{H}$ 是共线的，固相微粒受到的磁场力（即 Kelvin 力）也是最大的，若固相微粒的热运动能 $Ck_0T$ 大于值 $\mu_0 m_{p1} H$，即

$$Ck_0T \geq \mu_0 m_{p1} H$$

则铁磁流体在外磁场中能够稳定地保持其胶体状态。

用 $m_{p1} = M_p V_{p1} = M_p \pi d_p^3 / 6$ 代入上面的不等式，就得

$$d_p \leq \left[ \frac{6Ck_0T}{\pi\mu_0 M_p H} \right]^{1/3} \tag{1.5}$$

设微粒材料是 $Fe_3O_4$，其磁化强度 $M_p = 4.46 \times 10^5 \, \text{A/m}$，外磁场强度取为 $4 \times 10^5 \, \text{A/m}$，温度为常温 $T = 298\text{K}$，$\mu_0 = 4\pi \times 10^{-7} \, N/A^2$，取 $C = 1$，代入式（1.5）中，得到

$$d_p \leq 3.27 \times 10^{-9} \, \text{m}$$

实际使用的微粒直径都大于上述计算的结果，而 $H = 4 \times 10^5 \, \text{A/m}$ 仅是中等的磁场强度。于是就提出分散剂的必要性。

3.铁磁流体内微粒磁性集结而引起的胶体稳定性问题

铁磁流体内的固相微粒尺寸小于其材料的磁晶单畴临界尺寸，所以它们属于磁单畴或亚磁单畴的微粒。每个微粒都是一个磁偶极子，它在自己的周围形成磁场。若另一个磁偶极子靠近它，而在它的磁场内具有了磁势能，这就是偶极子对之间的磁势能。

设偶极子对由偶极子（即磁性固相微粒）$\overrightarrow{m_{p1}}$ 和 $\overrightarrow{m_{p2}}$ 形成，它们之间的磁吸引势能是[1]

$$E_{dd} = -\frac{\mu_o}{4\pi |\vec{r}|^3} m_{p1} m_{p2} [3\cos\theta\cos(\theta - \varphi) - \cos\varphi] \tag{1.6}$$

式中 $|\vec{r}|$ 是两偶极子中心距离矢量的模，$|\vec{r}| = d_s + a_{p1} + a_{p2}$，$d_s$ 是两偶极子表面之间的距离，$a_{p1}$ 和 $a_{p2}$ 分别是偶极子 $\overrightarrow{m_{p1}}$ 和 $\overrightarrow{m_{p2}}$ 的半径，$\theta$ 是两偶极子中心联线与矢量 $\overrightarrow{m_{p1}}$ 或 $\overrightarrow{m_{p2}}$ 的夹角，$\varphi$ 是两矢量 $\overrightarrow{m_{p1}}$ 和 $\overline{m_{p2}}$ 的夹角。当矢量 $\overrightarrow{m_{p1}}$ 和 $\overrightarrow{m_{p2}}$ 共线并且两偶极子表面相接触时，偶极子对之间的磁吸引势能数值最大。

即在式（1.6）中取 $d_s = 0$，$\theta = \varphi = 0$，于是有

$$E_{dd} = -\frac{\mu_0}{4\pi(a_{p1} + a_{p2})^3} m_{p1} m_{p2} \tag{1.7}$$

倘若取两微粒的材料相同，$M_{p1} = M_{p2} = M_p$，并且粒度相同，$a_{p1} = a_{p2} = d_p/2$，同时用 $m_p = M_p \pi d_p^3/6$ 代入，得到

$$E_{dd} = -\frac{\pi}{144} \mu_0 M_p^2 d_p^3 \tag{1.8}$$

两偶极子接触以后，能依靠热运动将两者分开的条件是 $Ck_0 T \geq E_{dd}$，于是解出

$$d_p \leq \left(\frac{144 Ck_0 T}{\pi \mu_0 M_p^2}\right)^{1/3} \tag{1.9}$$

对于材料为 Fe₃O₄ 的微粒，在常温下的临界尺寸是

$$d_p \leq 0.908 \times 10^{-8} \text{m}$$

通常铁磁流体的固相微粒取在 8nm 到 12nm 之间，但这都是名义上的平均尺寸，实际微粒是大小不等的。上述 9.08nm 这个临界条件基本上就是实际应用的平均值，或者接近实际应用的平均值。所以，在铁磁流体中必定存在很多大于 9.08nm 的微粒。这种尺寸的微粒一旦碰撞，就集结在一起不再分离，形成更大的粒度。其后在不断地碰撞中不断地"长大"，最终失去 Brown 运动的能力，沉淀随之发生。在此过程中，铁磁流体的磁化性能逐减弱，最终丧失磁性。这是又一个需要分散剂的原因。

## 1.3 铁磁流体的磁松弛过程
### 1.3.1 概述

所谓松弛过程，就是当铁磁流体所处的外部条件（主要指外磁场）发生改变时，铁磁流体内部从原来的状态改变到与外部条件相适应的状态，这就是松弛。磁松弛就是铁磁流体的磁化强度 $\overline{M}$ 随外磁场强度 $\overline{H}$ 改变的过程。铁磁流体的磁化和退磁都属于磁松弛。铁磁流体磁化的驱动力是外磁场，外磁场的作用是使铁磁流体内固相微粒的磁矩矢量在适应的程度上转向外磁场矢量 $\overline{H}$ 的方向，从而显示出磁性。铁磁流体退磁的驱动力是温度 $T$，确切地说就是热运动能 $k_0 T$。热运动的纷乱无规的随机性质使铁磁流体内大量的固相微粒之磁矩朝向各个方向的几率相同，从而铁磁流体不表现出宏观上的磁性。$k_0 T$ 越大，乱得越厉害，变乱的进程也更迅速，所以铁磁流体的磁化强度是外磁场和温度两者的作用达到平衡的结果。

松弛既是一个过程，当然完成这样的过程需要一定的时间。铁磁流体磁松弛有两种机制：一种是由微粒的磁矩旋转来实现，这称为内禀性过程；另一种是由微粒自身的旋转来完成，这种过程称为非

内禀性。内禀性过程所需的时间是 Neel 扩散时间 $t_N$，非内禀性过程所需的时间是 Brown 扩散时间 $t_B$。

$t_N$ 和 $t_B$ 是铁磁流体两个重要的物理参数。

### 1.3.2　内禀性磁松弛过程

铁磁流体中所使用的固相微粒均是铁磁性材料。铁磁性材料磁化有两个特点：一是这类材料具有磁畴结构，其磁松弛过程是以磁畴为单位旋转的。每个磁畴的体积很小，大约只有 $10^{-12}\,\mathrm{m^3} \sim 10^{-10}\,\mathrm{m^3}$，但相对于材料的原子来说，却是非常巨大，每个磁畴内含有几十亿个原子。所以，以磁畴为单位旋转的磁效应远比以原子为单位旋转的磁效应大得多。故而铁磁性材料的磁化率要比顺磁性材料大好几个数量级。

另一个特点是磁化各向异性，即存在容易磁化方向和不容易磁化方向。铁磁物质和任何导磁物质一样，其磁化必需外磁场对其作功，这种功就是磁化功 $w_m$，即

$$\delta w_m = \mu_0 H dM \tag{1.10}$$

需要磁化功最低的方向，就是铁磁材料的容易磁化的方向。在图 1-1 中绘有 $A$ 和 $B$ 两条示意的磁化曲线。按照式（1.10）所表述的磁化功就是磁化曲线和纵坐标轴 $M$ 之间所包围的面积。图 1-1 中曲线 $A$ 与 $M$ 轴之间所包围的面积比曲线 $B$ 的小，所以曲线 $A$ 所代表的方向是容易磁化的方向，而曲线 $B$ 则不是。两条曲线之间的阴影面积就是磁化功之差，称作为磁晶各向异性能。这个各向异性能是磁化方向与容易磁化方向之间的夹角 $\theta_K$ 的函数。其函数形式是 $\sin\theta_K$ 偶次方的级数形式。级数的前 $n$ 项为：

$$E_K = K_1 \sin^2\theta_K + K_2 \sin^4\theta_K + \cdots + K_n \sin^{2n}\theta_K$$

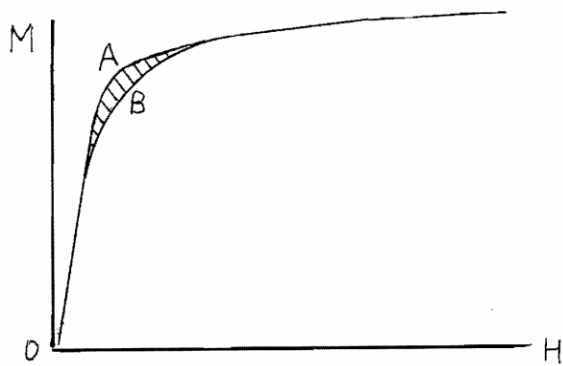

图 1-1　一种铁磁性材料沿不同方向的起始磁化曲线

式中，$E_K$ 是单位体积内的磁晶各向异性能，$K_1$，$K_2$，…，$K_n$ 均是实验得到的系数，并且均具有体积能的单位 $N\cdot m/m^3$。在一级近似下，只取级数的第一项：

$$E_K = K_1 \sin^2\theta_K$$

对于一个固相磁性微粒，各向异性能为

$$E_{K1} = V_{p1} K_1 \sin^2\theta_K = \frac{1}{6}\pi d_p^3 K_1 \sin^2\theta_K$$

当 $\theta_K = \pm\pi/2$ 时能量 $E_K$ 达到最大峰值，即

$$(E_K)_{\max} = K_1, \quad (E_{K1})_{\max} = V_{p1}K_1 = \frac{1}{6}\pi d_p^3 K_1 \tag{1.11}$$

在固态铁磁性物质内，原子运动的主要形式是旋转振动。振动的驱动能量是热能，所以大量原子随机热振动能量的统计平均值就是 $k_0T$。这种随机热振动频率的分布遵守统计物理学中的 Maxwell-Boltzmann 分布律[①]，即

$$df = \frac{f}{z}\exp\left(-\frac{E_{K1}}{k_0T}\right)d\theta_K$$

式中，$f$ 是统计平均振动频率，$z$ 是配分函数。如果在振动中达到角度 $\theta_K$ 的频率是 $f_\theta$，则在区间 $[\theta_K, \theta_K + d\theta_K]$ 内的振动次数是

$$df = f_\theta d\theta_K$$

代入 M-B 分布式中，就得

$$f_\theta = \frac{f}{z}\exp\left(-\frac{E_{K1}}{k_0T}\right)$$

当 $\theta_K = 0$ 或 $\pi$ 时，有 $E_{K1} = 0$，此时的振动频率记为 $f_0$，由上面的 M-B 分布式可以得到 $f_0 = f/z$，于是 M-B 分布就成为

$$f_\theta = f_0 \exp\left(-\frac{E_{K1}}{k_0T}\right) \tag{1.12}$$

对于一般的磁性固体材料，这个最低能位的原子振动频率 $f_0$ 的值，数量级为 $10^{12} \sim 10^{13}\mathrm{s}^{-1}$，McNab 等人得出在煤油+Fe$_3$O$_4$ 的铁磁流体中，固相微粒的 $f_0$ 之值约为 $10^9\mathrm{s}^{-1}$。

当 $\theta_K = \pm\pi/2$ 时，$E_K$ 达到最大值。一个固相微粒在此能位最高峰时的振动频率是

$$f_{\pm\pi/2} = f_0 \exp\left(-\frac{K_1 V_{p1}}{k_0T}\right) \tag{1.13}$$

只有在原子旋转振动的能量高于 $K_1 V_{p1}$ 时，才有可能实现磁矩的转向，所以 $f_{\pm\pi/2}$ 就是可能实现磁矩转向的频率。$f_{\pm\pi/2}$ 的量纲是 $\mathrm{s}^{-1}$，故其倒数就是时间，称为 Neel 松弛时间或 Neel 扩散时间，它就是微粒中的磁畴由原来的方向转为外磁场方向所需的时间，即实现固相微粒的磁矩转向的时间

---

[①] 请参阅参考文献[1]第 3 章或参考文献[3]的第 4 章。亦可在统计物理学的书籍中找到。

$$t_N = \frac{1}{f_{\pm\pi/2}} = \frac{1}{f_0}\exp\left(\frac{K_1 V_{p1}}{k_0 T}\right) \tag{1.14}$$

由式（1.13）可以看出，磁晶各向异性能的最高能位峰值 $K_1 V_{p1}$ 越高，原子可能越过这个能位的频率 $f_{\pm\pi/2}$ 越低，从而扩散时间 $t_N$ 越长。系数 $K_1$ 是常数，由式（1.14）可见，固相微粒的体积 $V_{p1}$ 越大，$t_N$ 以指数关系增长。

铁磁流体的固相微粒的平均尺寸是磁单畴的或亚磁单畴的，所以固相微粒的磁畴旋转也就是微粒的磁矩旋转。如果微粒内部的磁畴旋转速度远快于微粒本体在基载液中的旋转速度，从而磁松弛取决于磁畴旋转，这样铁磁流体称为内禀性铁磁流体。

### 1.3.3 非内禀性磁松弛过程

在外磁场作用下，悬浮于基载液中的固相微粒的磁矩转向外磁场的机制有两种。一是固体微粒内部的磁畴相对于微粒本体旋转而趋向外磁场方向；一是固相微粒本体旋转而拖带其磁畴转向外磁场。磁畴相对于微粒本体的旋转是相对运动，而微粒本体的旋转是牵连速度。所以磁矩的转速应是两者之和。虽然两种速度同时存在，但是小体积的微粒，其磁畴旋转远快于微粒本体的旋转，故其磁松弛过程取决于磁畴之旋转，这就是前面所讨论的内禀性磁松弛过程。大体积的微粒，其本体在磁力矩和其他外力矩的作用下，旋转速度远快于内部磁畴的旋转，磁松弛过程取决于微粒本体的旋转。这种磁化过程就是非内禀性的。

就非内禀性的磁化过程来讲，问题就集中于固相微粒本体在基载液中的旋转。此时驱动旋转的力矩是磁力矩 $\overrightarrow{L_{m1}}$

$$\overrightarrow{L_{m1}} = \mu_0 \overrightarrow{m_{p1}} \times \overrightarrow{H} \tag{1.15a}$$

式中矢量 $\overrightarrow{m_{p1}}$ 是一个固相微粒的磁矩。若外磁场强度矢量 $\overrightarrow{H}$ 和 $\overrightarrow{m_{p1}}$ 之间的夹角是 $\varphi$，则上式可以写成

$$L_{m1} = \mu_0 m_{p1} H \sin\varphi \tag{1.15b}$$

当微粒旋转时，必然受到基载液对其施加的粘性阻力矩。可以想像粘性阻力矩的大小一定和微粒之旋转速度 $d\varphi/dt$ 成正比，即

$$L_\tau \propto d\varphi/dt$$

或写成等式

$$\overrightarrow{L_\tau} = -C_\tau \frac{d\vec{\varphi}}{dt}$$

式中 $C_\tau$ 是阻力矩系数，负号表示 $L_\tau$ 是阻力矩。$L_\tau$ 的单位为 $N \cdot m$，旋转角速度 $d\varphi/dt$ 的单位为 $s^{-1}$，故 $C_\tau$ 的单位是 $N \cdot m \cdot s$。单个微粒的旋转运动方程是

$$J_1 \frac{d^2\vec{\varphi}}{dt^2} = \overrightarrow{L_m} + \overrightarrow{L_\tau} + \overrightarrow{L_B}$$

式中 $\overline{L_B}$ 称为 Brown 力矩。它因微粒相互擦撞产生。设上面方程的各项共线，则可以写成

$$J_t \frac{d^2\varphi}{dt^2} = -\mu_0 m_{p1} H \sin\varphi - C_\tau \frac{d\varphi}{dt} + L_B$$

上式右方第一项磁力矩前面加上负号是因为磁力矩总是使 $\varphi$ 角减小。微粒在基载液中的旋转运动是一种极低 Reynolds 数的运动，故运动方程左边的惯性项可以略去，而后整个式子通乘以 $\sin\varphi/(\mu_0 m_{p1} H)$，就得

$$\frac{C_\tau}{\mu_0 m_{p1} H} \sin\varphi \frac{d\varphi}{dt} = -\sin^2\varphi + \frac{(\sin\varphi) L_B}{\mu_0 m_{p1} H}$$

将上式左方的系数改写

$$\frac{C_\tau}{\mu_0 m_{p1} H} = \frac{C_\tau}{2k_0 T} \cdot \frac{2k_0 T}{\mu_0 m_{p1} H}$$

注意 $C_\tau$ 的单位是 $N \cdot m \cdot s$，而 $k_0 T$ 的单位是 $N \cdot m$，所以两者的比值是时间单位 s，于是将此比值定义为 $t_B$，同时用 $\alpha$ 表示右方的第二个比值，所以有

$$t_B = \frac{C_\tau}{2k_0 T}, \qquad \alpha = \frac{\mu_0 m_{p1} H}{k_0 T} \tag{1.16}$$

$t_B$ 称为 Brown 松弛时间或 Brown 扩散时间。将式（1.16）代入运动方程，并加以改写得

$$-\frac{2}{\alpha} t_B \frac{d\cos\varphi}{dt} = -(1 - \cos^2\varphi) + \frac{L_B \sin\varphi}{\mu_0 m_{p1} H}$$

将上式两边取统计平均，方程右方第二项分母 $\mu_0 m_{p1} H$ 是常数，而分子 $L_B$ 和 $\sin\varphi$ 之间是统计独立的，故

$$\left( \frac{L_B \sin\varphi}{\mu_0 m_{p1} H} \right)_e = \frac{(L_B)_e (\sin\varphi)_e}{\mu_0 m_{p1} H}$$

下标"$e$"表示数学期望，即统计平均。$L_B$ 是热运动产生的 Brown 力矩，这种力矩无论方向或大小都是随机的，杂乱无序的，所以 $L_B$ 的统计平均值是零。于是，对于大量微粒的统计平均值之旋转运动方程成为

$$\frac{2}{\alpha} t_B \frac{d(\cos\varphi)_e}{dt} = (1 - \cos^2\varphi)_e$$

方程的右边是统计平均值。在统计物理学中，一个物理量 $A$ 的统计平均值 $A_e$ 的计算式是

$$A_e = \frac{\int_\Omega A \exp\left(-\frac{E}{k_0 T}\right) d\Omega}{\int_\Omega \exp\left(-\frac{E}{k_0 T}\right) d\Omega} \tag{1.17}$$

式（1.17）右方的分母称作为配分函数，$E$ 是总能量，$\Omega$ 称为相宇或相空间。在当前所讨论的问题中，$A$ 就是 $1 - \cos^2\varphi$，$E$ 是微粒在磁场中的磁势能 $-\mu_0\overrightarrow{m_{p1}} \cdot \overrightarrow{H} = -\mu_0 m_{p1} H \cos\varphi$，在铁磁流体大量固相微粒的磁矩从四面八方转向外磁场方向，它们的磁矩矢量在空间扫描出一个半径为 $m_{p1}$ 的圆球表面，这个圆球面就是相空间 $\Omega$。于是，运动方程成为

$$\frac{2}{\alpha} t_B \frac{d(\cos\varphi)_e}{dt} = \frac{\int_0^\pi [(1 - \cos^2\varphi) \exp(\alpha\cos\varphi)](2\pi m_{p1} \sin\varphi) m_{p1} d\varphi}{\int_0^\pi [\exp(\alpha\cos\varphi)](2\pi m_{p1} \sin\varphi) m_{p1} d\varphi}$$

注意右方积分中 $\alpha\cos\varphi = \mu_0 m_{p1} H \cos\varphi/(k_0 T) = \mu_0\overrightarrow{m_{p1}} \cdot \overrightarrow{H}/(k_0 T) = -E/(k_0 T)$，将右方分子分母约去常数 $2\pi m_{p1}^2$ 之后，改写成便于积分的形式

$$\frac{2}{a} t_B \frac{d(\cos\varphi)_e}{dt} = \frac{\int_0^\pi (1 - \cos^2\varphi) \exp(\alpha\cos\varphi) d(-\cos\varphi)}{\int_0^\pi \exp(\alpha\cos\varphi) d(-\cos\varphi)} =$$

$$1 - \frac{\int_0^\pi (\alpha\cos\varphi)^2 \exp(\alpha\cos\varphi) d(\alpha\cos\varphi)}{\alpha^2 \int_0^\pi \exp(\alpha\cos\varphi) d(\alpha\cos\varphi)} =$$

$$1 - \frac{\left\{[(\alpha\cos\varphi)^2 - 2(\alpha\cos\varphi) + 2] \exp(\alpha\cos\varphi)\right\}_0^\pi}{\alpha^2 [\exp(\alpha\cos\varphi)]_0^\pi} =$$

$$1 - \frac{(\alpha^2 + 2\alpha + 2)e^{-\alpha} - (\alpha^2 - 2\alpha + 2)e^\alpha}{\alpha^2(e^{-\alpha} - e^\alpha)}$$

于是得出统计平均的旋转运动方程为

$$\frac{d}{dt}(\cos\varphi)_e = \left(\coth\alpha - \frac{1}{\alpha}\right)\frac{1}{t_B} \tag{1.18a}$$

就一个固相微粒而言，其磁矩 $\overrightarrow{m_{p1}}$ 在外磁场 $\overrightarrow{H}$ 方向的投影是它的有效磁矩 $m_p$，即

$$m_{p1} \cos\varphi = m_p$$

对于大量的固相微粒，就需要按统计平均值考虑，因为 $m_{p1}$ 是常数，故

$$m_{p1}(\cos\varphi)_e = (m_p)_e$$

设在体积为 $V_f$ 的铁磁流体中含有 $N$ 个固相微粒，则上式可改写成

$$\frac{Nm_{p1}}{V_f}(\cos\varphi)_e = \frac{N}{V_f}(m_p)_e$$

注意 $m_{p1} = V_{p1}M_p$， $N(m_p)_e = V_f M$，其中 $M_p$ 是固相微粒材料最大磁化强度，$M$ 是铁磁流体的磁化强度，

$NV_{p1}/V_f = V_p/V_f = \phi_p$， $\phi_p$ 是铁磁流体中固相物质的体积分量。于是上式成为

$$(\cos\varphi)_e = \frac{M}{\phi_p M_p} \tag{1.19}$$

将式（1.19）代入式（1.18a），由于 $\phi_p$ 和 $M_p$ 是常数遂得

$$\frac{dM}{dt} = \phi_p M_p \left(\coth\alpha - \frac{1}{\alpha}\right)\frac{1}{t_B} \tag{1.18b}$$

式（1.18b）就是铁磁流体非内禀性的磁化运动方程。上式积分，设初始条件是 $t = 0$， $M(0) = 0$，并且设 $\alpha$ 与 $t$ 无关，就有

$$M(t) = \phi_p M_p \left(\coth\alpha - \frac{1}{\alpha}\right)\frac{t}{t_B} \tag{1.20}$$

当 $t = t_B$ 时，就得到和外界条件（主要是外磁场强度 $H$ 和温度 $T$）相适应 的平衡磁化强度

$$M = \phi_p M_p \left(\coth\alpha - \frac{1}{\alpha}\right) \tag{1.21}$$

式（1.21）称为 Langevin 方程，而 $\coth\alpha - (1/\alpha)$ 称为 Langevin 函数，记为 $L(\alpha)$，即

$$L(\alpha) = \coth\alpha - \frac{1}{\alpha} \tag{1.22}$$

### 1.4  铁磁流体在外磁场中所受的力、力矩和所具的磁势能

在铁磁流体中取出一个六面微元控制体作为分析的单元。这个单元相对于铁磁流体流场而言非常微小，以致于在其中参数的变化可以认为是线性变化，即采用 Taylor 展开的前两项，这当然带来很大的简便。相对于铁磁流体的固相微粒而言，这个微元又足够大，它其中包含了大量的固相微粒，从而其参数变化符合连续性和光滑可微的要求。

#### 1.4.1  铁磁流体在外磁场中所受到的磁力

磁力是铁磁流体所特有的一种彻体力。实际上外磁场只对铁磁流体内的铁磁性固相微粒起作用，而基载液是不导磁的，对于外磁场的存在没有任何响应。在铁磁流体的混合流分析中，是将固相和液相的胶体混合物作为一个整体来对待，也就是将固相微粒磁化的效应摊派到整个铁磁流体混合物上，结果就是式（1.21）所表示的那样。

铁磁流体磁化的基本机理是每个单畴或亚单畴的固相微粒体的磁矩方向，都在某种程度上趋近于外磁场的方向。所谓的某种程度是指微粒的磁势能与微粒的热运动能相平衡的结果。对于非均匀磁场，由于各点磁场强度不同，对于微粒磁矩的约束能力也不同，结果各处的固相微粒磁矩方向趋向外磁场的程度不一致，其表现就是铁磁流体的非均匀磁化。在图 1-2 的（b）图上可以看到分子电流是变化的。

这些电流都是所在平面上的平均值，电流的下标是其电流环的法线方向。电流环法线的正方向按右手螺旋规则确定，如图（c）所示。

在图（c）中，平面表面 $add'a'$ 的分子电流之平均值是

$$\frac{1}{4}\left[\left(i_z' + \frac{\partial i_z'}{\partial x}\frac{dx}{2} + \frac{\partial i_z'}{\partial z}\frac{dz}{2}\right) + \left(i_z' + \frac{\partial i_z'}{\partial y}\frac{dy}{2} + \frac{\partial i_z'}{\partial z}\frac{dz}{2}\right) + \right.$$

$$\left.\left(i_z' - \frac{\partial i_z'}{\partial x}\frac{dx}{2} + \frac{\partial i_z'}{\partial z}\frac{dz}{2}\right) + \left(i_z' - \frac{\partial i_z'}{\partial y}\frac{dy}{2} + \frac{\partial i_z'}{\partial z}\frac{dz}{2}\right)\right] = i_z' + \frac{\partial i_z'}{\partial z}\frac{dz}{2}$$

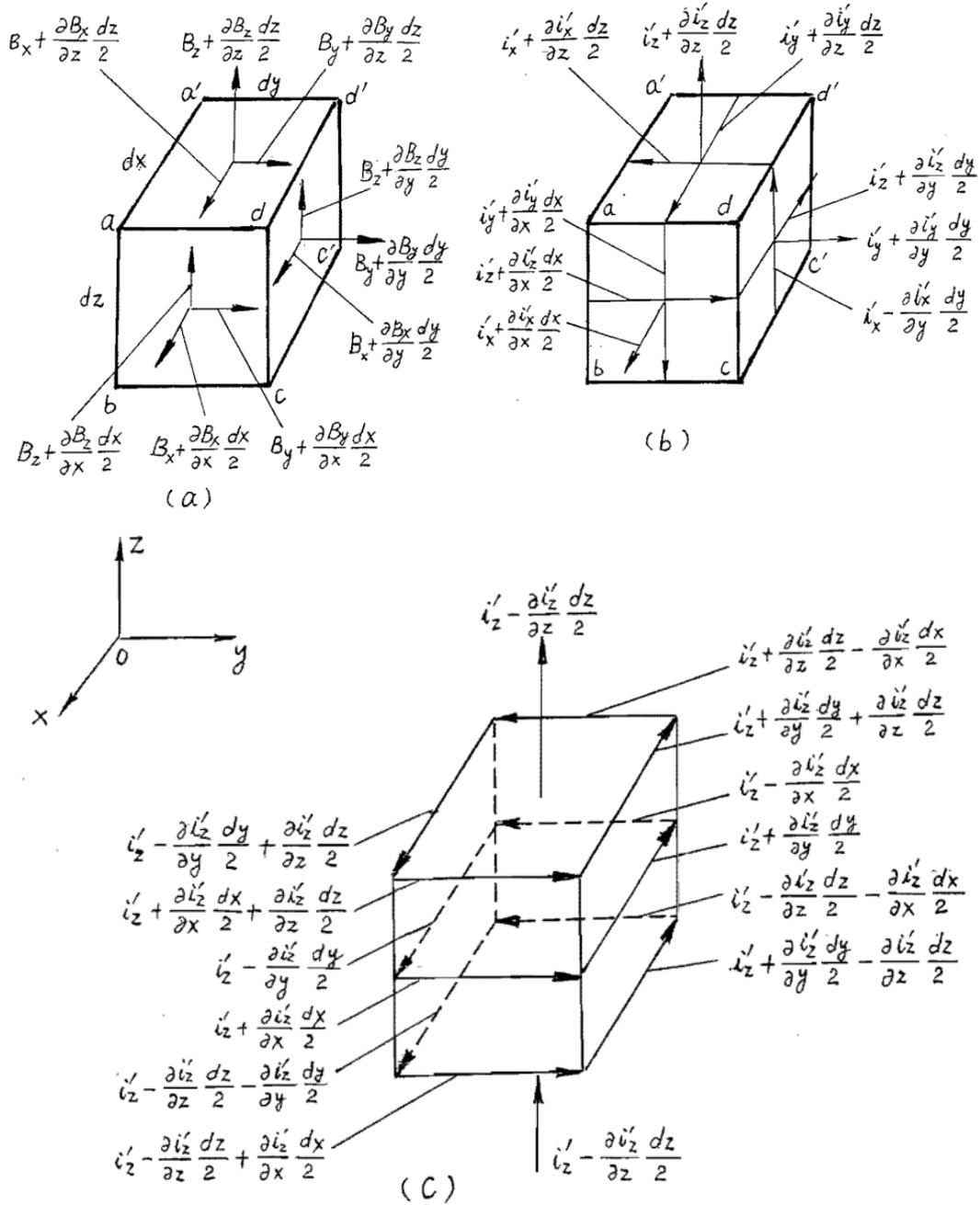

图 1-2　铁磁流体的微元控制体上的外磁场和分子电流环

同样的办法可以得出下平面 $bcc'b'$ 的分子电流的平均值是

$$i'_z - \frac{\partial i'_z}{\partial z}\frac{dz}{2}$$

其它四个平面表面上分子电流的平均值依次是 $i'_x + \frac{\partial i'_x}{\partial x}\frac{dx}{2}$，$i'_x - \frac{\partial i'_x}{\partial x}\frac{dx}{2}$，$i'_y + \frac{\partial i'_y}{\partial y}\frac{dy}{2}$，$i'_y - \frac{\partial i'_y}{\partial y}\frac{dy}{2}$。所有

这些平均值都用各电流环平面的法向矢量表示，如图 1-2 的（b）和（c）。

设在微元控制体中心处的外磁场强度是 $\overrightarrow{B_0}$，$\overrightarrow{B_0}$ 常称为磁化场。矢量 $\overrightarrow{B_0}$ 在 $x$，$y$，$z$ 三个方的分量是

$B_x$，$B_y$，$B_z$，则

$$\overrightarrow{B_0} = \vec{i}B_x + \vec{j}B_y + \vec{k}B_z$$

因为外磁场不均匀，所以在微元控制体的六个表面上，磁化场分量都有增量，这就是图 1-2 的（a）图
所示的那样。在磁化场强度较高的表面上，固相微粒的磁矩方向与磁化场方向一致的程度较高，若以
平面分子电流环作为固相微粒的电磁学模型，则在微元控制体表面上存在未被抵消的分子电流。并且
磁化场强度较高的表面上，分子电流强度也较高，磁化场强度较低的表面上，分子电流强度也较低，
如图 1-2 的（b）图所示。

若有传导电流穿过微元控制体，并且传导电流的面密度是 $\vec{j_c}$，则微元控制体内的磁化场 $\overrightarrow{B_0}$ 就是有

旋的，即

$$\nabla \times \overrightarrow{B_0} = \mu_0 \vec{j_c}$$

上式两边展开写成

$$\left(\vec{i}\frac{\partial}{\partial x} + \vec{j}\frac{\partial}{\partial y} + \vec{k}\frac{\partial}{\partial z}\right) \times (\vec{i}B_x + \vec{j}B_y + \vec{k}B_z) = \mu_0(\vec{i}j_x + \vec{j}j_y + \vec{k}j_z)$$

左边相乘以后，按方向拆开就得

$$\left.\begin{array}{l} \mu_0 j_x = \dfrac{\partial B_z}{\partial y} - \dfrac{\partial B_y}{\partial z} \\[2mm] \mu_0 j_y = \dfrac{\partial B_x}{\partial z} - \dfrac{\partial B_z}{\partial x} \\[2mm] \mu_0 j_z = \dfrac{\partial B_y}{\partial x} - \dfrac{\partial B_x}{\partial y} \end{array}\right\} \tag{1.23}$$

由图 1-2 的（b）图可以看到，在磁化场不均匀的情况下，微元六面体各表面上的束缚分子电流不
相等，此现象说明有未抵消的束缚分子电流穿过微元六面体。设这种未被抵消的束缚分子电流的面密
度为 $\vec{j'}$，于是六面体中的铁磁流体磁化强度 $\overrightarrow{M}$ 是有旋的，即

$$\nabla \times \overrightarrow{M} = \vec{j'}$$

注意到 $\overrightarrow{M} = \vec{i'} = \overrightarrow{n^0}i'$，$\overrightarrow{n^0}$ 是分子电流环平面的法向单位矢，依照 $i'$ 的方向用右手螺旋定则确定 $\overrightarrow{n^0}$ 之正向。

将上式两边展开，而后按方向拆成下面的三个式子

$$j'_x = \frac{\partial M_z}{\partial y} - \frac{\partial M_y}{\partial z} = \frac{\partial i'_z}{\partial y} - \frac{\partial i'_y}{\partial z}$$
$$j'_y = \frac{\partial M_x}{\partial z} - \frac{\partial M_z}{\partial x} = \frac{\partial i'_x}{\partial z} - \frac{\partial i'_z}{\partial x} \Bigg\} \qquad (1.24)$$
$$j'_z = \frac{\partial M_y}{\partial x} - \frac{\partial M_x}{\partial y} = \frac{\partial i'_y}{\partial x} - \frac{\partial i'_x}{\partial y}$$

电流段在外磁场中所受到的力，遵守 Ampere 定律，即

$$d\vec{F} = I d\vec{l} \times \vec{B}$$

由图 1-2 的（a）图和（b）图可见，微元六面体各表面上的束缚电流段与所在表面上的磁化场分量之间只有平行和垂直的关系，所以两者的矢性积只能是零或代数积。对照图 1-2 的（a）图和（b）图，在表 1-1 中列出了微元六面体各表面上的法向磁力和切向磁力。每一个力的第一下标表示该力所在平面的法线方向，第二下标表示该力本身的方向。于是两下标相同的力是法向力，而两下标不同的力就是切向力。

铁磁流体在非均匀外磁场中所受的力 $\vec{f_m}$

将表 1-1 中的各项乘积展开，舍去二阶小量，然后直列相加，就得三个方向的磁力之分量：

$$\sum dF_{nx} = \left( i'_y \frac{\partial B_y}{\partial x} + B_y \frac{\partial i'_y}{\partial x} + i'_z \frac{\partial B_z}{\partial x} + B_z \frac{\partial i'_z}{\partial x} - i'_x \frac{\partial B_y}{\partial y} - B_y \frac{\partial i'_x}{\partial y} - i'_x \frac{\partial B_z}{\partial z} - B_z \frac{\partial i'_x}{\partial z} \right) dxdydz$$

$$\sum dF_{ny} = \left( i'_z \frac{\partial B_z}{\partial y} + B_z \frac{\partial i'_z}{\partial y} + i'_x \frac{\partial B_x}{\partial y} + B_x \frac{\partial i'_x}{\partial y} - i'_y \frac{\partial B_z}{\partial z} - B_z \frac{\partial i'_y}{\partial z} - i'_y \frac{\partial B_x}{\partial x} - B_x \frac{\partial i'_y}{\partial x} \right) dxdydz$$

$$\sum dF_{nz} = \left( i'_x \frac{\partial B_x}{\partial z} + B_x \frac{\partial i'_x}{\partial z} + i'_y \frac{\partial B_y}{\partial z} + B_y \frac{\partial i'_y}{\partial z} - i'_z \frac{\partial B_x}{\partial x} - B_x \frac{\partial i'_z}{\partial x} - i'_z \frac{\partial B_y}{\partial y} - B_y \frac{\partial i'_z}{\partial y} \right) dxdydz$$

将磁感应强度的 Gauss 散度定理，即 $\frac{\partial B_x}{\partial x} + \frac{\partial B_y}{\partial y} + \frac{\partial B_z}{\partial z} = 0$ 代入上面三式，得

$$f_{mx} = \frac{\sum dF_{nx}}{dxdydz} = i'_x \frac{\partial B_x}{\partial x} + i'_y \frac{\partial B_y}{\partial x} + i'_z \frac{\partial B_z}{\partial x} + B_y \left( \frac{\partial i'_y}{\partial x} - \frac{\partial i'_x}{\partial y} \right) - B_z \left( \frac{\partial i'_x}{\partial z} - \frac{\partial i'_z}{\partial x} \right)$$

$$f_{my} = \frac{\sum dF_{ny}}{dxdydz} = i'_x \frac{\partial B_x}{\partial y} + i'_y \frac{\partial B_y}{\partial y} + i'_z \frac{\partial B_z}{\partial y} + B_z \left( \frac{\partial i'_z}{\partial y} - \frac{\partial i'_y}{\partial z} \right) - B_x \left( \frac{\partial i'_y}{\partial x} - \frac{\partial i'_x}{\partial y} \right)$$

$$f_{mz} = \frac{\sum dF_{nz}}{dxdydz} = i'_x \frac{\partial B_x}{\partial z} + i'_y \frac{\partial B_y}{\partial z} + i'_z \frac{\partial B_z}{\partial z} + B_x \left( \frac{\partial i'_x}{\partial z} - \frac{\partial i'_z}{\partial x} \right) - B_y \left( \frac{\partial i'_z}{\partial y} - \frac{\partial i'_y}{\partial z} \right)$$

表 1-1　微元六面体表面上的法向磁力和切向磁力

| 微元体的表面 | 表面外法线 $n$ | $dF_{nx}$ | $dF_{ny}$ | $dF_{nz}$ |
|---|---|---|---|---|
| $abcd$ | $x$ | $(i'_y + \frac{\partial i'_y}{\partial x}\frac{dx}{2})dydz(B_y + \frac{\partial B_y}{\partial x}\frac{dx}{2})$ <br> $(i'_z + \frac{\partial i'_z}{\partial x}\frac{dx}{2})dzdy(B_z + \frac{\partial B_z}{\partial x}\frac{dx}{2})$ | $-(i'_y + \frac{\partial i'_y}{\partial x}\frac{dx}{2})dydz(B_x + \frac{\partial B_x}{\partial x}\frac{dx}{2})$ | $-(i'_z + \frac{\partial i'_z}{\partial x}\frac{dx}{2})dzdy(B_x + \frac{\partial B_x}{\partial x}\frac{dx}{2})$ |
| $a'b'c'd'$ | $-x$ | $-(i'_y - \frac{\partial i'_y}{\partial x}\frac{dx}{2})dydz(B_y - \frac{\partial B_y}{\partial x}\frac{dx}{2})$ <br> $-(i'_z - \frac{\partial i'_z}{\partial x}\frac{dx}{2})dzdy(B_z - \frac{\partial B_z}{\partial x}\frac{dx}{2})$ | $(i'_y - \frac{\partial i'_y}{\partial x}\frac{dx}{2})dydz(B_x - \frac{\partial B_x}{\partial x}\frac{dx}{2})$ | $(i'_z - \frac{\partial i'_z}{\partial x}\frac{dx}{2})dydz(B_x - \frac{\partial B_x}{\partial x}\frac{dx}{2})$ |
| $dcc'd'$ | $y$ | $-(i'_x - \frac{\partial i'_x}{\partial y}\frac{dy}{2})dxdz(B_y + \frac{\partial B_y}{\partial y}\frac{dy}{2})$ | $(i'_z + \frac{\partial i'_z}{\partial y}\frac{dy}{2})dzdx(B_z + \frac{\partial B_z}{\partial y}\frac{dy}{2})$ <br> $(i'_x + \frac{\partial i'_x}{\partial y}\frac{dy}{2})dxdz(B_x + \frac{\partial B_x}{\partial y}\frac{dy}{2})$ | $-(i'_z + \frac{\partial i'_z}{\partial y}\frac{dy}{2})dzdx(B_y + \frac{\partial B_y}{\partial y}\frac{dy}{2})$ |
| $abb'a'$ | $-y$ | $(i'_x - \frac{\partial i'_x}{\partial y}\frac{dy}{2})dxdz(B_y - \frac{\partial B_y}{\partial y}\frac{dy}{2})$ | $-(i'_z - \frac{\partial i'_z}{\partial y}\frac{dy}{2})dzdx(B_z - \frac{\partial B_z}{\partial y}\frac{dy}{2})$ <br> $-(i'_x - \frac{\partial i'_x}{\partial y}\frac{dy}{2})dxdz(B_x - \frac{\partial B_x}{\partial y}\frac{dy}{2})$ | $(i'_z - \frac{\partial i'_z}{\partial y}\frac{dy}{2})dzdx(B_y - \frac{\partial B_y}{\partial y}\frac{dy}{2})$ |
| $add'a'$ | $z$ | $-(i'_x + \frac{\partial i'_x}{\partial z}\frac{dz}{2})dxdy(B_z + \frac{\partial B_z}{\partial z}\frac{dz}{2})$ | $-(i'_y + \frac{\partial i'_y}{\partial z}\frac{dz}{2})dydx(B_z + \frac{\partial B_z}{\partial z}\frac{dz}{2})$ | $(i'_x + \frac{\partial i'_x}{\partial z}\frac{dz}{2})dxdy(B_x + \frac{\partial B_x}{\partial z}\frac{dz}{2})$ <br> $(i'_y + \frac{\partial i'_y}{\partial z}\frac{dz}{2})dydz(B_y + \frac{\partial B_y}{\partial z}\frac{dz}{2})$ |
| $bcc'b'$ | $-z$ | $(i'_x - \frac{\partial i'_x}{\partial z}\frac{dz}{2})dxdy(B_z - \frac{\partial B_z}{\partial z}\frac{dz}{2})$ | $(i'_y - \frac{\partial i'_y}{\partial z}\frac{dz}{2})dydx(B_z - \frac{\partial B_z}{\partial z}\frac{dz}{2})$ | $-(i'_x - \frac{\partial i'_x}{\partial z}\frac{dz}{2})dxdy(B_x - \frac{\partial B_x}{\partial z}\frac{dz}{2})$ <br> $-(i'_y - \frac{\partial i'_y}{\partial z}\frac{dz}{2})dydx(B_y - \frac{\partial B_y}{\partial z}\frac{dz}{2})$ |

将式（1.23）和式（1.24）代入 $f_{mx}$ 式的左边，就有

$$f_{mx} = i'_x \frac{\partial B_x}{\partial x} + i'_y \left( \frac{\partial B_x}{\partial y} + \mu_0 j_z \right) + i'_z \left( \frac{\partial B_x}{\partial z} - \mu_0 j_y \right) + B_y j'_z - B_z j'_y$$

上式可以改写成

$$f_{mx} = i'_x \frac{\partial B_x}{\partial x} + i'_y \frac{\partial B_x}{\partial y} + i'_z \frac{\partial B_x}{\partial z} + \mu_0 (i'_y j_z - i'_z j_y) + B_y j'_z - B_z j'_y \tag{1.25}$$

同样可得

$$f_{my} = i'_x \frac{\partial B_y}{\partial x} + i'_y \frac{\partial B_y}{\partial y} + i'_z \frac{\partial B_y}{\partial z} + \mu_0 (i'_z j_x - i'_x j_z) + B_z j'_x - B_x j'_z \tag{1.26}$$

$$f_{mz} = i'_x \frac{\partial B_z}{\partial x} + i'_y \frac{\partial B_z}{\partial y} + i'_z \frac{\partial B_z}{\partial z} + \mu_0 (i'_x j_y - i'_y j_x) + B_x j'_y - B_y j'_x \tag{1.27}$$

将上面三个式子合并写成矢量形式

$$\vec{f_m} = \vec{i'} \cdot \nabla \vec{B_0} + \vec{i'} \times \mu_0 \vec{j_c} + \vec{B_0} \times \vec{j'} \tag{1.28}$$

由于 $\vec{i'}$ 是单位高度上的分子电流环并且 $\vec{i'} = \vec{M}$，故式（1.28）又可写成

$$\vec{f_m} = \vec{M} \cdot \nabla \vec{B_0} + \vec{M} \times (\nabla \times \vec{B_0}) + \vec{B_0} \times (\nabla \times \vec{M}) \tag{1.29}$$

铁磁流体的磁化强度 $\vec{M}$ 是其内部所含的固相微粒的磁矩在磁化场 $\vec{B_0}$ 方向的投影的统计平均值，即式（1.19）所表达的 $M = \phi_p M_p (\cos \varphi)_e$，而 $(\cos \varphi)_e$ 是大量微粒的磁矩 $\vec{m_p}$ 与磁化场 $\vec{B_0}$ 间夹角 $\varphi$ 之统计平均余弦。故在静止状态下铁磁流体的磁化强度矢量总是平行于磁化场的方向。于是式（1.29）右方第一项可藉助矢量恒等式改写

$$\vec{M} \cdot \nabla \vec{B_0} = \frac{M}{B_0} \vec{B_0} \cdot \nabla \vec{B_0} = \frac{M}{B_0} \left[ \frac{1}{2} \nabla (\vec{B_0} \cdot \vec{B_0}) - \vec{B_0} \times (\nabla \times \vec{B_0}) \right] =$$
$$\frac{M}{B_0} \left[ B_0 \nabla B_0 - \vec{B_0} \times (\nabla \times \vec{B_0}) \right] = M \nabla B_0 - \vec{M} \times (\nabla \times \vec{B_0})$$

将上面的结果代入到式（1.29）之右方就得

$$\vec{f_m} = M \nabla B_0 + \vec{B_0} \times (\nabla \times \vec{M}) \tag{1.30}$$

若铁磁流体的磁化曲线可以近似成图 1-3 所示的由两直线段组成折线，在 $H > H_s$ 的区间内 $M = M_s = \text{const}$，即铁磁流体已经饱和磁化，当然就是一种均匀磁化，就有 $\nabla \times \vec{M} = 0$。

在区间 $[0, H_s]$ 内，铁磁流体的磁化率 $\chi_m = \text{const}$，即

$$\vec{M} = \chi_m \vec{H} = \mu_0 \chi_m \vec{B_0}$$

从而

$$\nabla \times \vec{M} = \mu_0 \chi_m \nabla \times \vec{B_0}$$

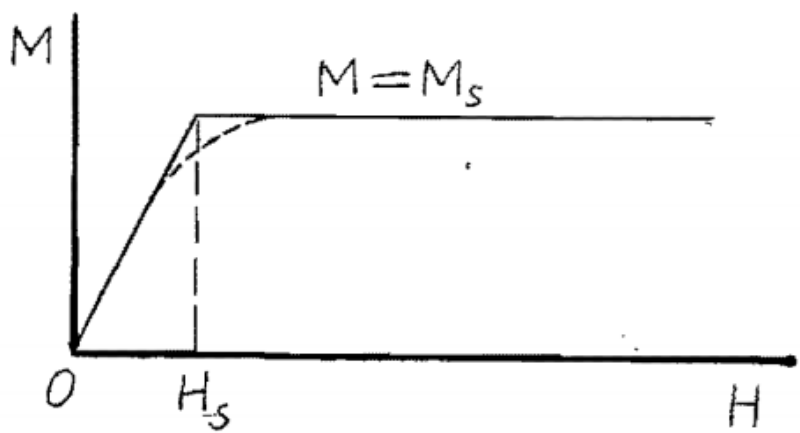

图 1-3　铁磁流体的近似磁化曲线

当没有传导电流穿过时，$\vec{B_0}$ 虽然不均匀，但却无旋，即 $\nabla \times \vec{B_0} = 0$，从而 $\nabla \times \vec{M} = 0$。所以，式（1.30）给出

$$\vec{f_m} = M \nabla B_0 \tag{1.31}$$

在铁磁流体发生运动时，矢量 $\vec{M}$ 和 $\vec{B_0}$ 不一定平行。若铁磁流体的磁化曲线仍可近似成图 1-3 的折线，并且不存在传导电流，则磁力是

$$\vec{f_m} = \vec{f_k} = \vec{M} \cdot \nabla \vec{B_0} \tag{1.32}$$

式（1.31）和式（1.32）是磁场力最常用的形式，称作 Kelvin 力，通常写成 $\vec{f_k}$。

### 1.4.2　铁磁流体在外磁场中所受到的磁力矩

由图 1-2，在微元六面体的每个表面上都有磁力的法向分量和切向分量。法向分量全都通过微元体的几何中心，所以法向分力不产生使微元体旋转的力矩，只有切向分力才造成力矩。力矩矢量 $\vec{L}$ 是力臂矢 $\vec{r}$ 和力矢 $\vec{F}$ 的矢性积，即

$$\vec{L} = \vec{r} \times \vec{F}$$

在微元六面体表面上的磁力之切向分量与坐标轴间只有平行和垂直两种关系。力臂矢的正方向取所有各表面外法线之正方向，即由微元六面体的几何中心指向表面上的切向力，于是

$$dL = \sum |\vec{r} \times \vec{F}| = \sum |\vec{r}| \cdot |\vec{F}| \sin(\vec{r^0}, \vec{F^0})$$

式中，$\vec{r^0}$ 是微元六面体各表面外法线的单位矢，$\vec{F^0}$ 是表面切向力的单位矢。角度 $(\vec{r^0}, \vec{F^0})$ 是

由矢量 $\vec{r}$ 之正方向逆时针转向矢量 $\vec{F}$ 的正方向。所以角度 $(\vec{r^0}, \vec{F^0})$ 只能是 $\pi/2$ 和 $3\pi/2$ 两种情

况。由表 1-1 列出的表面切向力造成 $x$ 轴方向的磁力矩是

$$dL_x = \frac{dz}{2}\left|\left(i'_y + \frac{\partial i'_y}{\partial z}\frac{dz}{2}\right)dx\,dy\left(B_z + \frac{\partial B_z}{\partial z}\frac{dz}{2}\right)\right|\sin\frac{\pi}{2} +$$

$$\frac{dz}{2}\left|\left(i'_y - \frac{\partial i'_y}{\partial z}\frac{dz}{2}\right)dx\,dy\left(B_z - \frac{\partial B_z}{\partial z}\frac{dz}{2}\right)\right|\sin\frac{\pi}{2} +$$

$$\frac{dy}{2}\left|\left(i'_z + \frac{\partial i'_z}{\partial y}\frac{dy}{2}\right)dz\,dx\left(B_y + \frac{\partial B_y}{\partial y}\frac{dy}{2}\right)\right|\sin\frac{3\pi}{2} +$$

$$\frac{dy}{2}\left|\left(i'_z - \frac{\partial i'_z}{\partial y}\frac{dy}{2}\right)dz\,dx\left(B_y - \frac{\partial B_y}{\partial y}\frac{dy}{2}\right)\right|\sin\frac{3\pi}{2}$$

将上式展开并略去二阶小量就得

$$dL_x = (i'_y B_z - i'_z B_y)\,dx\,dy\,dz$$

单位体积的铁磁流体受到的磁力矩在 $x$ 方向的分量是

$$L_x = \frac{dL_x}{dx\,dy\,dz} = i'_y B_z - i'_z B_y$$

同样可得

$$L_y = \frac{dL_y}{dx\,dy\,dz} = i'_z B_x - i'_x B_z$$

$$L_z = \frac{dL_z}{dx\,dy\,dz} = i'_x B_y - i'_y B_x$$

将上面三式合并，写成单位体积铁磁流体，在外磁场作用下所受到磁力矩的矢量关系式

$$\vec{L_m} = \vec{i'} \times \vec{B_0} = \vec{M} \times \vec{B_0} \tag{1.33}$$

对于单纯固相微粒而言，其磁化强度是 $\vec{M_p}$，故单位体积的固相微粒之磁力矩为

$$\vec{L_m} = \vec{M_p} \times \vec{B_0} \tag{1.34a}$$

对于一个固相微粒，受到的磁力矩是

$$\vec{L_{m,1}} = V_{p1}\vec{M_p} \times \vec{B_0} = \vec{m_{p1}} \times \vec{B_0} \tag{1.34b}$$

式中，$V_{p1}$ 是一个固相微粒的体积，$\vec{m_{p1}}$ 是一个固相微粒的磁矩。

由式（1.33）可见：①磁力矩与外磁场是否均匀无关，只取决于当地的外磁场强度；②磁力矩与有无传导电流和剩余束缚电流的存在无关，即外磁场是否有旋和铁磁流体的磁化强

度是否有旋，对磁力矩没有影响。

### 1.4.3 铁磁流体在外磁场中的磁势能

由式（1.29）铁磁流体的单位体积磁力即磁力密度 $\overrightarrow{f_m}$ 是

$$\overrightarrow{f_m} = \overrightarrow{M} \cdot \nabla \overrightarrow{B_0} + \overrightarrow{M} \times (\nabla \times \overrightarrow{B_0}) + \overrightarrow{B_0} \times (\nabla \times \overrightarrow{M})$$

对于均匀磁化的铁磁流体，$\overrightarrow{M} = \text{const}$ ，故

$$\overrightarrow{B_0} \cdot \nabla \overrightarrow{M} = 0$$

将上面两式相加，并且考虑到矢量恒等式，就有

$$\overrightarrow{f_m} = \overrightarrow{M} \cdot \nabla \overrightarrow{B_0} + \overrightarrow{B_0} \cdot \nabla \overrightarrow{M} + \overrightarrow{M} \times (\nabla \times \overrightarrow{B_0}) + \overrightarrow{B_0} \times (\nabla \times \overrightarrow{M}) = \nabla(\overrightarrow{M} \cdot \overrightarrow{B}) \tag{1.35}$$

式（1.35）表示均匀磁化的铁磁流体在外磁场中所受到的磁力是具势的，其势函数就是 $\overrightarrow{M} \cdot \overrightarrow{B_0}$ ，所以磁势能为

$$e_p = -\overrightarrow{M} \cdot \overrightarrow{B_0} \tag{1.36}$$

由式（1.19）和式（1.21）可见铁磁流体的磁化强度是

$$\overrightarrow{M} = \phi_p \overline{M_p}\left(\coth\alpha - \frac{1}{\alpha}\right) = \phi_p \overline{M_p}(\cos\varphi)_e \tag{1.37}$$

式中 $(\cos\varphi)_e$ 是各个固相微粒在外磁场中，磁矩与外磁场之间夹角 $\varphi$ 的余弦统计平均值。设存在一个统计平均角 $\varphi_e$ ，它是

$$\varphi_e = \arccos\left(\coth\alpha - \frac{1}{a}\right) = \arccos[(\cos\varphi)_e] \tag{1.38}$$

则有

$$\cos\varphi_e = (\cos\varphi)_e \tag{1.39}$$

从而式（1.37）写成

$$\overrightarrow{M} = \phi_p \overline{M_p}\cos\varphi_e \tag{1.40}$$

因为矢量 $\overrightarrow{M}$ 是大量固相微粒的磁矩在外磁场 $\overrightarrow{B_0}$ 方向投影的平均值，所以当铁磁流体处于不运动的静止状态时，矢量 $\overrightarrow{M}$ 总与 $\overrightarrow{B_0}$ 共线，从而式（1.36）的磁势能成为

$$e_p = -MB_0 \tag{1.41}$$

用式（1.40）代入式（1.41），有

$$e_p = -(\phi_p M_p \cos\varphi_e)B_0 = -\phi_p \overline{M_p} \cdot \overrightarrow{B_0} \tag{1.42}$$

将式（1.41）微分，得

$$d(-e_p) = M \, dB_0 + B_0 \, dM \tag{1.43}$$

上式右方第一项 $M \, dB_0$ 是磁力所作之功，即移动功，第二项 $B_0 \, dM$ 是磁力矩所作之功，即磁化功。为了更清楚，使用式（1.42）微分。注意 $\phi_p$ 和 $M_p$ 是常数，得

$$d(-e_p) = \phi_p M_p (\cos\varphi_e \, dB_0 + B_0 \, d\cos\varphi_e) \tag{1.44}$$

或写成

$$d(-e_p) = \phi_p M_p \left( \cos\varphi_e \frac{dB_0}{dl} dl + B_0 \frac{d\cos\varphi_e}{d\varphi_e} d\varphi_e \right)$$

右方括号内的第一项

$$\frac{dB_0}{dl} = \frac{\partial B_0}{\partial x}\frac{dx}{dl} + \frac{\partial B_0}{\partial y}\frac{dy}{dl} + \frac{\partial B_0}{\partial z}\frac{dz}{dl}$$

式中的 $dl$ 是

$$\vec{l^0} dl = \vec{i} \, dx + \vec{j} \, dy + \vec{k} \, dz$$

两边点乘以 $\vec{i}$，则得

$$\vec{i} \cdot \vec{l^0} dl = dx$$

此即

$$\frac{dx}{dl} = \vec{l^0} \cdot \vec{i}$$

同样可得

$$\frac{dy}{dl} = \vec{l^0} \cdot \vec{j}, \qquad \frac{dz}{dl} = \vec{l^0} \cdot \vec{k}$$

于是

$$\frac{dB_0}{dl} = \left( \vec{i}\frac{\partial B_0}{\partial x} + \vec{j}\frac{\partial B_0}{\partial y} + \vec{k}\frac{\partial B_0}{\partial z} \right) \cdot \vec{l^0} = (\nabla B_0) \cdot \vec{l^0}$$

故

$$dB_0 = \frac{dB_0}{dl} dl = (\nabla B_0) \cdot \vec{l^0} dl = (\nabla B_0) \cdot \vec{dl} \tag{1.45}$$

又

$$d\cos\varphi_e = -\sin\varphi_e \, d\varphi_e$$

将此结果连同式（1.45）代入到式（1.44）的右方，就得

$$d(-e_p) = \phi_p M_p \cos\varphi_e \nabla B_0 \cdot \vec{dl} - \phi_p M_p B_0 \sin\varphi_e \, d\varphi_e$$

右方第一项注意到式（1.40），第二项写成矢性积的形式，得

$$d(-e_p) = (M\nabla B_0) \cdot d\vec{l} - (\phi_p \overrightarrow{M_p} \times \overrightarrow{B_0}) \cdot \overrightarrow{\varphi_e^0} d\varphi_e$$
$$= (M\nabla B_0) \cdot d\vec{l} - (\phi_p \overrightarrow{M_p} \times \overrightarrow{B_0}) \cdot d\overrightarrow{\varphi_e}$$

（1.46）

因为磁力矩总是使 $\varphi_e$ 减小，故 $d\varphi_e$ 取负值。式（1.46）右方第一项括号内是 Kelvin 力，所以它是磁场力所作之功，即移动功。右方第二项显然是磁力矩所作之功，即磁化功。式（1.46）右方两项和式（1.43）右方两项对应相等，因为由式（1.45）知 $M\,dB_0 = M\nabla B_0 \cdot d\vec{l}$，又由于

$\phi_p M_p$ 是常数，

$$B_0 dM = B_0 d(\phi_p M_p \cos\varphi_e) = -\phi_p M_p B_0 \sin\varphi_e d\varphi_e = -(\phi_p \overrightarrow{M_p} \times \overrightarrow{B_0}) \cdot \overrightarrow{\varphi_e^0} d\varphi_e$$

由此可见式（1.43）和式（1.46）完全相通。

## 1.5 在外磁场中铁磁流体的热力学定律和热力学参数

### 1.5.1. 热力学第一定律应用于铁磁流体

热力学第一定律是一个普适的能量守恒定律。铁磁流体不例外地遵守这个定律。热力学第一定律的数学形式是

$$dE = \delta Q + \delta W$$

式中 $E$ 是内能，$\delta Q$ 是外部加入到系统中的热量，$\delta W$ 是外部对系统所作之功。对于单位体积而言，上式写成

$$de = \delta q + \delta w$$

(1.47)

式（1.47）的物理意义是，外部对系统所加之热和所作之功，均转化为系统内能的增量。

### 1.5.2 铁磁流体的内能

和普通非磁物质不同，铁磁流体的内能至少包括热能和磁能两类。

1.铁磁流体的磁内能

在外磁场中，外磁场对铁磁流体所作的磁功就将转化成为铁磁流体的磁内能。由式（1.21）和式（1.22）所给出的 Langevin 方程可知铁磁流体的磁化强度是

$$M(T,H) = \phi_p M_p \left( \coth\alpha - \frac{1}{a} \right) = \phi_p M_p \left( \coth\frac{\mu_0 V_{p1} M_p H}{k_0 T} - \frac{k_0 T}{\mu_0 V_{p1} M_p H} \right)$$

式中，铁磁流体中固相物质的分量 $\phi_p$，固相物质的磁化强度 $M_p$，以及固相微粒的体积 $V_{p1}$ 均是常数。所以有

$$dM = \frac{\partial M}{\partial T} dT + \frac{\partial M}{\partial H} dH$$

代入到式（1.43）左方第二项中，就得磁化功是

$$\delta w_m = \mu_0 H dM = \mu_0 H \left( \frac{\partial M}{\partial T} dT + \frac{\partial M}{\partial H} dH \right)$$

(1.48)

加入到铁磁流体中的热量，一部分使铁磁流体的温度升高，另一部分改变铁磁流体的磁化强度 $M$，而 $M$ 是温度 $T$ 和外磁场强度 $H$ 的函数，故热量的函数关系为

$$q = q(T,H)$$

从而有

$$\delta q = \frac{\partial q}{\partial T} dT + \frac{\partial q}{\partial H} dH$$

或写成

$$\left.\begin{array}{l} \delta q = c_v(T,H)\,dT + g(T,H)\,dH \\[2mm] c_v(T,H) = \dfrac{\partial q(T,H)}{\partial T}, \qquad g(T,H) = \dfrac{\partial q(T,H)}{\partial H} \end{array}\right\} \tag{1.49}$$

式中，$c_v(T,H)$ 是温度比热容，$g(T,H)$ 是磁比热容。

将式（1.48）和式（1.49）代入热力学第一定律式（1.47）中，就得

$$de = \left(c_v + \mu_0 H \frac{\partial M}{\partial T}\right)dT + \left(g + \mu_0 H \frac{\partial M}{\partial H}\right)dH \tag{1.50}$$

由于内能 $e$ 是与过程无关的热力学参数，所以式（1.50）的右方必定是恰当微分，即

$$\frac{\partial}{\partial H}\left(c_v + \mu_0 H \frac{\partial M}{\partial T}\right) = \frac{\partial}{\partial T}\left(g + \mu_0 H \frac{\partial M}{\partial H}\right)$$

注意 $H$ 和 $T$ 是各自独立的变量，所以上式两边展开之后得出

$$\frac{\partial g}{\partial T} = \frac{\partial c_v}{\partial H} + \mu_0 \frac{\partial M}{\partial T} \tag{1.51}$$

由式（1.49）可以给出铁磁流体的熵是

$$ds = \frac{\delta q}{T} = \frac{1}{T}(c_v dT + g dH) \tag{1.52}$$

由于熵 $s$ 也是与过程无关的热力学参数，所以式（1.52）的右方也必定是恰当微分，即

$$\frac{\partial}{\partial H}\left(\frac{1}{T}c_v\right) = \frac{\partial}{\partial T}\left(\frac{1}{T}g\right)$$

此式给出

$$\frac{\partial c_v}{\partial H} = -\frac{g}{T} + \frac{\partial g}{\partial T} \tag{1.53}$$

联立式（1.51）和式（1.53），得到

$$g(T,H) = \mu_0 T \frac{\partial M}{\partial T}, \qquad \frac{\partial}{\partial H}c_v(T,H) = \mu_0 T \frac{\partial^2 M}{\partial T^2} \tag{1.54}$$

### 1.5.3 铁磁流体在外磁场中的热力学函数

1.内能

由式（1.50）和式（1.54）得

$$de = \left(c_v + \mu_0 H \frac{\partial M}{\partial T}\right)dT + \left(\mu_0 T \frac{\partial M}{\partial T} + \mu_0 H \frac{\partial M}{\partial H}\right)dH \tag{1.55}$$

2.热量

由式（1.49）和式（1.54）得

$$\delta q = c_v(T,H)\,dT + \mu_0 T \frac{\partial M}{\partial T} dH \tag{1.56}$$

3.磁化功

由式（1.48）给出

$$\delta w_m = \mu_0 H \frac{\partial M}{\partial T} dT + \mu_0 H \frac{\partial M}{\partial H} dH \tag{1.57}$$

4.熵

由式（1.52）和式（1.54）得到

$$ds = \frac{c_v}{T} dT + \mu_0 \frac{\partial M}{\partial T} dH \qquad (1.58)$$

5.温度比热容

由式（1.49），$c_v = c_v(T, H)$，则有

$$dc_v = \frac{\partial c_v}{\partial T} dT + \frac{\partial c_v}{\partial H} dH$$

用式（1.54）代入，得

$$dc_v = \frac{\partial c_v}{\partial T} dT + \mu_0 T \frac{\partial^2 M}{\partial T^2} dH \qquad (1.59)$$

6.磁比热容

由式（1.49）和式（1.54）得

$$\left. \begin{aligned} g(T, H) &= \mu_0 T \frac{\partial M}{\partial T} \\ dg &= \left( \mu_0 \frac{\partial M}{\partial T} + \mu_0 T \frac{\partial^2 M}{\partial T^2} \right) dT + \mu_0 T \frac{\partial^2 M}{\partial T \partial H} dH \end{aligned} \right\} \qquad (1.60)$$

在实用中，铁磁流体的 $c_v(T, H)$、$g(T, H)$、$M(T, H)$ 均可用实验得到。在缺乏实验数据时，可以用理论方法估计。$M(T, H)$ 由 Langevin 公式即可得到，而式（1.54）可算出 $g(T, H)$ 和 $c_v(T, H)$ 作为温度 $T$ 和外磁场强度 $H$ 的函数式。

# 第二章　流体力学基础

## 2.1　概述

铁磁流体是一种胶体混合物。除了具有磁性这一特质之外，其基本性质仍然属于液体，所以其运动服从流体力学的规律。

流体力学是连续介质力学的一种。其实流体并不是真正的连续介质。在自然界中，无论是固体、液体、气体都是由单个的分子或原子组成，这些分子或原子之间的空隙远大于它们本身。所以从微观上来看，自然界中没有实在的连续物质。但是，物质内分子或原子的数密度极其大，即使是气态物质，在标准状况下，数密度为 $2.7 \times 10^{19}$ 个/$cm^3$，所有的分子都在不停顿地做热运动，纷扰无序的热运动在各个方向都有相同的几率，故分子运动论所得出的物理性质，全是统计平均值，这些统计平均的物理量具有充分的连续性质。所以，物质自身也就可以假定为连续介质。

铁磁流体在微观上更不能算作连续介质，固相成分是以微粒形式分散存在于铁磁流体中。但其数密度达到 $10^{17} \sim 10^{18}$ 颗/$cm^3$，而且微粒的平均粒度相当于大分子，热运动的随机性质使铁磁流体物理性质的统计平均值在宏观上同样具有充分的连续性。所以将铁磁流体胶体混合物看作为连续介质，于是流体力学成为分析铁磁流体运动的基础。

铁磁流体内的固相微粒，总是理想化地假设为尺寸相同的小圆球体。这些悬浮在基载液体中的小圆球体的运动不外乎两种形式，即旋转和平移。所以铁磁流体内固液两相间的力学关联就是粘性力和粘性力矩。

铁磁流体的组成中，固相的体积分量占铁磁流体总体积的一小部分，通常在 10% 以下。所以固相微粒之间的距离远大于它们本身的直径。在这样的情况下，微粒彼此间互相影响很小，就是所谓的稀疏相，因此，可以采用圆球在无界流中运动的模型来近似地分析固相微粒在基载液中受到的粘性力和粘性力矩。

从物理学来看，流体的运动并没有什么根本性的特殊之处，它仍然遵守物理上的三个基本定律，即质量守恒定律、动量守恒定律和能量守恒定律。这三个定律的数学表达形式就构成决定流体运动的基本方程，它们是连续性方程、运动方程和能量方程。

这些方程的导出，采用微元控制体的方法。微元控制体相对于流场来说是极其微小的，以致于在其中一切物理量的变化都是线性的微量改变。但相对分子或原子而言，无论是它们本身的尺寸或它们之间的距离，微元体是极为巨大的，在其内含有足够大量的分子或原子。使热运动统计结果给出的物理量有一般性，连续而且光滑。

## 2.2　流体的运动和变形

### 2.2.1　概述

考察流体在流动中的行为，基本的方法就是在流体中取出一个微元控制体来分析。这样的微元控制体常称为流体微团。流体的行为不外乎两方面：一方面是如同刚体一般的运动，包括平动和转动；另一方面是变形，包括伸缩变形和剪切变形。刚体力学只讨论运动而不包括变形，弹性力学只讨论变形（和相应的应力），而不涉及运动。流体力学则兼论运动和变形，这是流体力学的一个特点，同时也说明流体力学与刚体力学、弹性力学有共同的相通之处。

### 2.2.2　流体微团的运动

1.平动运动

设流体微团保持其形状和方位不变而沿空间曲线 $S$ 移动，$S$ 的矢径是 $\vec{r}$ ，则微团的平移速度为

$$\vec{U} = \frac{d\vec{r}}{dt} \tag{2.1}$$

在直角坐标系中

$$d\vec{r} = \vec{i}\,dx + \vec{j}\,dy + \vec{k}\,dz \tag{2.2}$$

于是有

$$\vec{U} = \vec{i}\frac{dx}{dt} + \vec{j}\frac{dy}{dt} + \vec{k}\frac{dz}{dt} = \vec{i}\,u + \vec{j}\,v + \vec{k}\,w$$

式中

$$u = \frac{dx}{dt}, \qquad v = \frac{dy}{dt}, \qquad w = \frac{dz}{dt} \tag{2.3}$$

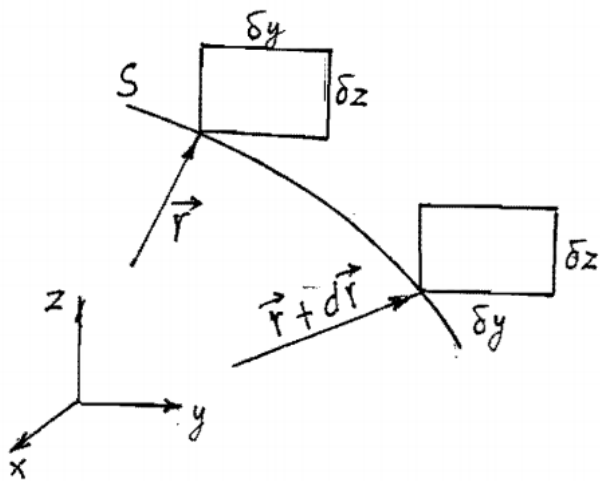

图 2-1    流体微团的平动运动

2.旋转运动

设流体微团保持其形状和空间位置不变，但具有使方位改变的旋转运动，如图 2-2 所示。这是一种刚体的旋转方式，在 $yOz$ 平面上，旋转轴是 $x$ 轴。在转角 $d\phi_x$ 很小的情况下，有

$$d\phi_x \approx \sin(d\phi_x)$$

同时因为转动中，微团的形状和大小均保持不变，故其横边和竖边的转角相等，都是 $d\phi_x$ ，于是由图 2-2 写出

$$d\phi_x = \frac{1}{\delta y}\left[\left(w + \frac{\partial w}{\partial y}\delta y\right) - w\right]dt = \frac{\partial w}{\partial y}dt$$

和

$$d\phi_x = \frac{1}{\delta z}\left[\left(v - \frac{\partial v}{\partial z}\delta z\right) - v\right]dt = -\frac{\partial v}{\partial z}dt$$

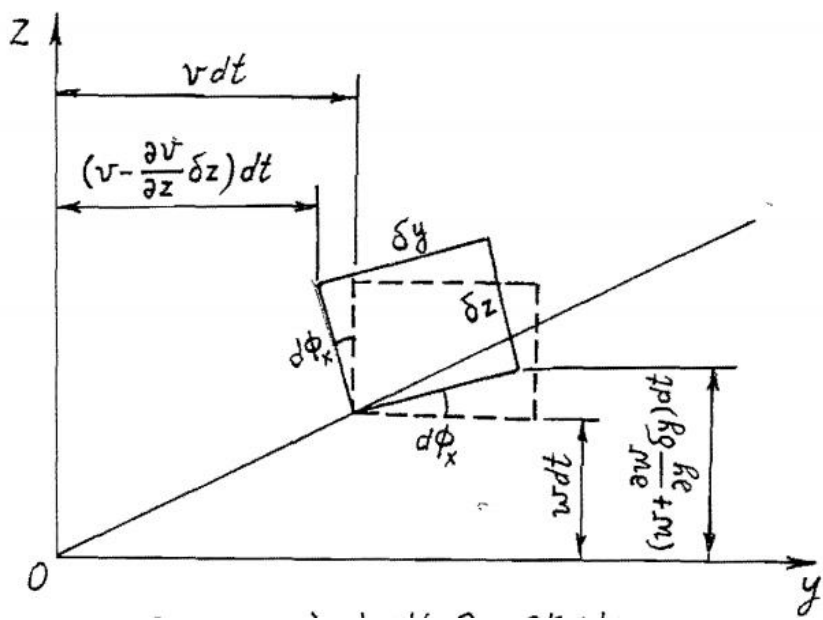

图 2-2  流体微团的旋转运动

将以上两式相加，并且注意到转角对时间的微分就是角速度。就有

$$\omega_x = \frac{d\phi_x}{dt} = \frac{1}{2}\left(\frac{\partial w}{\partial y} - \frac{\partial v}{\partial z}\right)$$

同样可有

$$\omega_y = \frac{d\phi_y}{dt} = \frac{1}{2}\left(\frac{\partial u}{\partial z} - \frac{\partial w}{\partial x}\right) \qquad (2.4)$$

$$\omega_z = \frac{d\phi_z}{dt} = \frac{1}{2}\left(\frac{\partial v}{\partial x} - \frac{\partial u}{\partial y}\right)$$

式（2.4）可合并写成

$$\vec{\omega} = \frac{1}{2}\nabla \times \vec{U} \qquad (2.5)$$

此外定义涡量 $\overline{\Omega}$

$$\vec{\Omega} = \nabla \times \vec{U} = 2\vec{\omega} \qquad (2.6)$$

### 2.2.3  流体微团的变形

1.伸缩变形

若流体在运动中具有速度梯度时，就可能造成流体微团的尺寸和形状的变化。如图 2-3 所示，由于速度梯度的存在，流体微团各点的速度不一致，在 $dt$ 时间内，由位置 $O$ 移动到 $O'$，而边长由 $\delta y$ 和 $\delta z$ 分别变成 $\delta y'$ 和 $\delta z'$。从图中可见

$$\delta y' + v\, dt = \delta y + \left(v + \frac{\partial v}{\partial y}\delta y\right)dt, \qquad \delta z' + w\, dt = \delta z + \left(w + \frac{\partial w}{\partial z}\delta z\right)dt$$

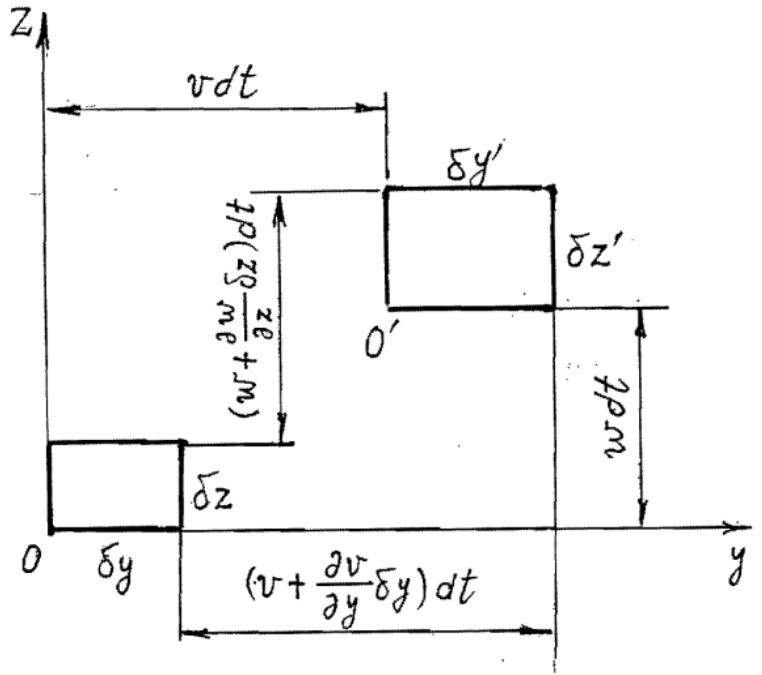

图 2-3　流体微团的伸缩变形

由此得出

$$\Delta y = \delta y' - \delta y = \frac{\partial v}{\partial y}\delta y\, dt, \qquad \Delta z = \delta z' - \delta z = \frac{\partial w}{\partial z}\delta z\, dt$$

定义单位时间内、单位长度的伸缩称为伸缩率 $\varepsilon$，即

$$\left.\begin{array}{l} \varepsilon_{yy} = \dfrac{\Delta y}{\delta y}\dfrac{1}{dt} = \dfrac{\partial v}{\partial y}, \qquad \varepsilon_{zz} = \dfrac{\Delta z}{\delta z}\dfrac{1}{dt} = \dfrac{\partial w}{\partial z} \\[2mm] \varepsilon_{xx} = \dfrac{\Delta x}{\delta x}\dfrac{1}{dt} = \dfrac{\partial u}{\partial x} \end{array}\right\} \tag{2.7}$$

同样有

式中 $\varepsilon$ 的第一下标表示因伸缩而位移的表面之法线方向，第二下标表示位移方向。对于流体微团的体积变化率，可以写出

$$\frac{\delta V_0}{V_0 dt} = \frac{\delta x' \delta y' \delta z' - \delta x \delta y \delta z}{(\delta x \delta y \delta z)dt}$$

$$= \frac{\left(\delta x + \dfrac{\partial u}{\partial x}\delta x\, dt\right)\left(\delta y + \dfrac{\partial v}{\partial y}\delta y\, dt\right)\left(\delta z + \dfrac{\partial w}{\partial z}\delta z\, dt\right) - \delta x \delta y \delta z}{(\delta x \delta y \delta z)dt}$$

分子括弧乘开，略去高阶小项，即得单位时间内体积的相对变化为

$$\frac{\delta V_0}{V_0 dt} = \frac{\partial u}{\partial x} + \frac{\partial v}{\partial y} + \frac{\partial w}{\partial z} = \nabla \cdot \vec{U} \tag{2.8}$$

2.剪切变形

由式（2.8）明显地看到，流体微团的体积变化是由于平行表面的速度不均匀所造成。而剪切变成

菱形，则是垂直于表面的速度不均匀的结果。图 2-4（a）中绘出剪切变形的情况。

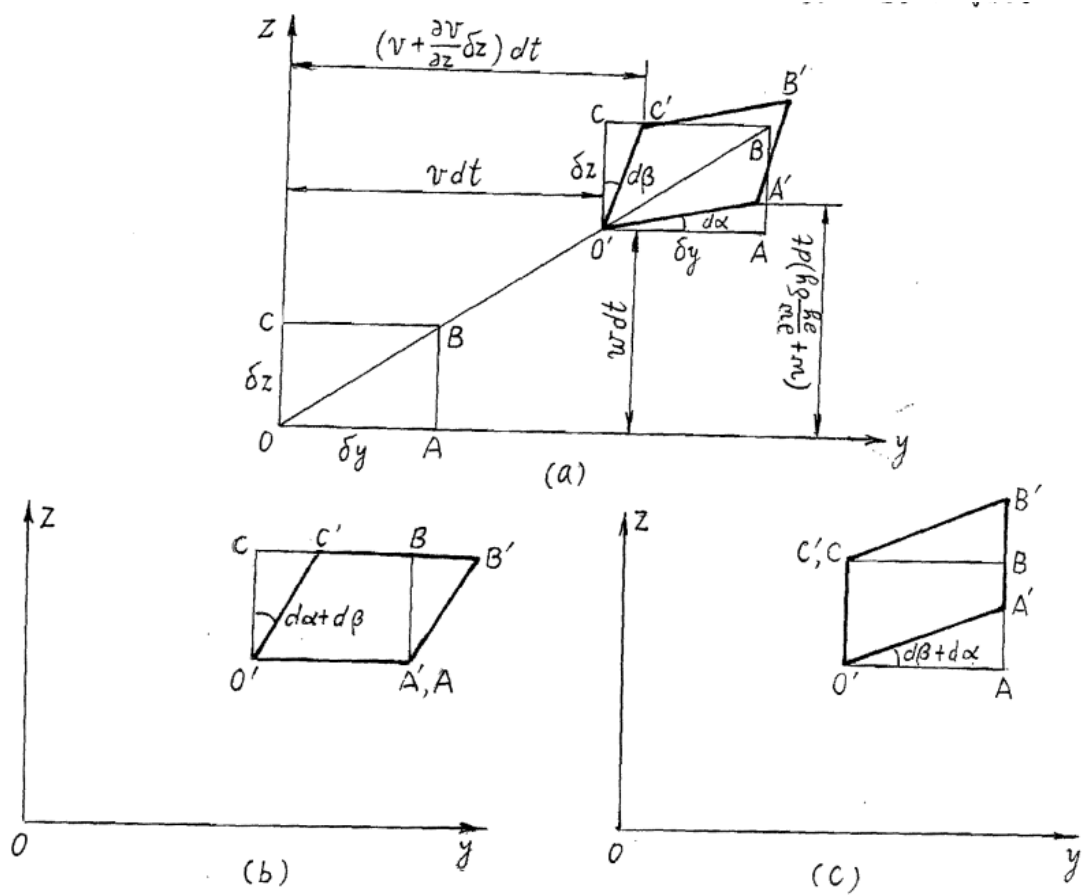

图 2-4　流体微团的剪切变形

在图（a）中，$A$ 点的 $z$ 向速度与 $O$ 点的不相等，在 $dt$ 时间以后形成角度 $d\alpha$；在 $c$ 点的 $y$ 向速度与 $O$ 点的不相等，在 $dt$ 时间之后形成角度 $d\beta$。在 $dt$ 时间中形成的 $d\alpha$ 和 $d\beta$ 均很小，故有

$$d\alpha \approx \sin(d\alpha) = \frac{1}{\delta y}\left[\left(w + \frac{\partial w}{\partial y}\delta y\right)dt - wdt\right]$$

$$d\beta \approx \sin(d\beta) = \frac{1}{\delta z}\left[\left(v + \frac{\partial v}{\partial z}\delta z\right)dt - vdt\right]$$

上两式给出角度的变化率是

$$\frac{d\alpha}{dt} = \frac{\partial w}{\partial y}, \qquad \frac{d\beta}{dt} = \frac{\partial v}{\partial z} \tag{2.9}$$

若将图（a）中的 $O'A'B'C'$ 保持形状不变，顺时针旋转使 $O'A'$ 与 $O'A$ 重合，则如图（b）所示，此时 $O'C'$ 与 $O'C$ 的夹角是 $d\alpha + d\beta$，可以认为由原来的矩形流体微团 $OABC$ 变成菱形的 $O'A'B'C'$ 是由平行于 $y$ 轴的 $CB$ 向右方位错造成。这样的变形率用 $\varepsilon_{zy}$ 表示。下标 $z$ 表示位错平面 $CB$ 的法线方向，$y$ 表示位错的移动方向。定义剪切变形率是角度变化率的平均值，则有

$$\varepsilon_{zy} = \frac{1}{2}\left(\frac{d\alpha}{dt} + \frac{d\beta}{dt}\right) \tag{2.10a}$$

同样，若保持 $O'A'B'C'$ 的形状不变，逆时针旋转使 $O'C'$ 与 $O'C$ 重合，如图（c）所示，此时 $O'A'$ 与 $O'A$

的夹角是 $d\beta + d\alpha$，这样的变形可以认为是平行于 $z$ 轴的 $AB$ 向上方位错造成的，用 $\varepsilon_{yz}$ 表示变形率，

按照剪切变形率是角度变化率的定义，就有

$$\varepsilon_{yz} = \frac{1}{2}\left(\frac{d\beta}{dt} + \frac{d\alpha}{dt}\right) \tag{2.10b}$$

比较以上两式，立即可见 $\varepsilon_{yz} = \varepsilon_{zy}$，然后将式（2.9）代入右方，就得

$$\left.\begin{aligned}
\varepsilon_{yz} = \varepsilon_{zy} = \frac{1}{2}\left(\frac{\partial v}{\partial z} + \frac{\partial w}{\partial y}\right) \\
\varepsilon_{zx} = \varepsilon_{xz} = \frac{1}{2}\left(\frac{\partial w}{\partial x} + \frac{\partial u}{\partial z}\right) \\
\varepsilon_{xy} = \varepsilon_{yx} = \frac{1}{2}\left(\frac{\partial u}{\partial y} + \frac{\partial v}{\partial x}\right)
\end{aligned}\right\} \tag{2.11}$$

同样可以得

式（2.11）可以合并写成

$$\varepsilon_{ij} = \varepsilon_{ji} = \frac{1}{2}\left(\frac{\partial u_i}{\partial x_j} + \frac{\partial u_j}{\partial x_i}\right) \tag{2.12}$$

注意式（2.12）不仅包含式（2.11），而且也包含式（2.7）在内。当 $i \neq j$ 时就是剪切变形，当 $i = j$ 时就是伸缩变形。

## 2.3 流体微团上的应力

### 2.3.1 概述

流体微团处于运动之中时，其参数的改变有两种情况，一种是和运动无关的变化，即在原地随时间改变，另一种是由于运动改变了位置而引起的变化。流体微团相对于流场固然很微小，但它毕竟占有空间，所以在它上面的流体参数并非不变。不过由于它的尺寸微小，参数的变化可以线性化，就是将作为位置函数的参数作 Taylor 展开弃去二阶及其以上的小项，图 2-5 中微元六面体表面上的应力分布就是这样的。

流体微团的变形当然是力作用的结果。因此，应力和变形必定相关联。变形率 $\varepsilon_{ij}$ 的下标 $i, j = 1, 2,$

3。所以 $i, j$ 的排列总共有 9 种，即 $3^2$ 种方式，这表明 $\varepsilon_{ij}$ 是一种二阶张量。每一种变形率都有相应的应

力作用，由此应力 $\tau_{ij}$ 也同样是二阶张量，应力和变形率的一般函数关系可表示为

$$\tau_{ij} = f(\varepsilon_{ij})$$

当下标 $i = j$ 时，即 $\varepsilon_{ii}$ 或 $\varepsilon_{jj}$ 是正应变，相应的 $\tau_{ii}$ 或 $\tau_{jj}$ 是正应力，$i \neq j$ 时 $\varepsilon_{ij}$ 是剪切变形，$\tau_{ij}$ 是剪切应力。

若 $\tau_{ij}$ 和 $\varepsilon_{ij}$ 只是简单的比例关系，即

$$\tau_{ij} = 2\eta\varepsilon_{ij} \tag{2.13}$$

比例系数 $\eta$ 称为流体的动力粘性系数。当 $\eta$ 与 $\varepsilon_{ij}$ 无任何函数关系时，这样的流体就称为牛顿流体。用式（2.12）代入式（2.13），得

$$\tau_{ij} = \tau_{ji} = \eta\left(\frac{\partial u_i}{\partial x_j} + \frac{\partial u_j}{\partial x_i}\right) \tag{2.14}$$

若流动是一维流，则有

$$\tau = \eta\frac{\partial u}{\partial y} \tag{2.15}$$

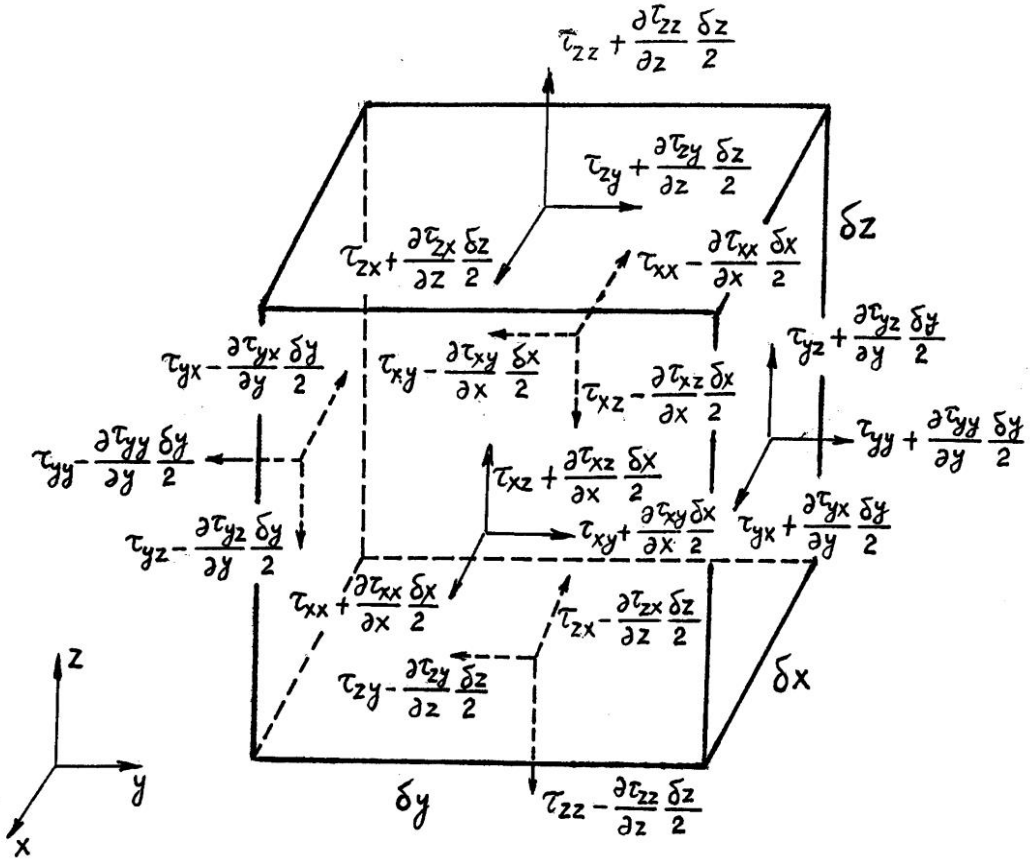

图 2-5　流体微元六面体上的表面应力

由图 2-5 可以看到，在矩形六面微元体上，所有下标位置互换的应力，即 $\tau_{ij}$ 和 $\tau_{ji}$ 在方向上是互相垂直的，所以式（2.14）给出的只是 $\tau_{ij}$ 和 $\tau_{ji}$ 在数值上相等的关系。

2.3.2　流体微团表面应力的合力

在图 2-5 中，微元体上 $x$ 方向的表面力之合成是

$$df_{xx} = \left[\left(\tau_{xx} + \frac{\partial \tau_{xx}}{\partial x}\frac{\delta x}{2}\right) - \left(\tau_{xx} - \frac{\partial \tau_{xx}}{\partial x}\frac{\delta x}{2}\right)\right]\delta y \delta z +$$

$$\left[\left(\tau_{yx} + \frac{\partial \tau_{yx}}{\partial y}\frac{\delta y}{2}\right) - \left(\tau_{yx} - \frac{\partial \tau_{yx}}{\partial y}\frac{\delta y}{2}\right)\right]\delta z \delta x +$$

$$\left[\left(\tau_{zx} + \frac{\partial \tau_{zx}}{\partial z}\frac{\delta z}{2}\right) - \left(\tau_{zx} - \frac{\partial \tau_{zx}}{\partial z}\frac{\delta z}{2}\right)\right]\delta x \delta y =$$

$$\left(\frac{\partial \tau_{xx}}{\partial x} + \frac{\partial \tau_{yx}}{\partial y} + \frac{\partial \tau_{zx}}{\partial z}\right)\delta x \delta y \delta z$$

上式中 $\delta x \delta y \delta z = \delta V_0$，是微元六面体的体积，于是单位体积流体在 $x$ 方向表面应力的合力大小是

$$f_{sx} = \frac{df_{xx}}{\delta x \delta y \delta z} = \frac{\partial \tau_{xx}}{\partial x} + \frac{\partial \tau_{yx}}{\partial y} + \frac{\partial \tau_{zx}}{\partial z} \tag{2.16}$$

将式（2.16）右方的应力张量元素都写成并矢的形式，而后取散度，就有

$$\nabla \cdot (\vec{i}\,\vec{i}\,\tau_{xx} + \vec{j}\,\vec{i}\,\tau_{yx} + \vec{k}\,\vec{i}\,\tau_{zx}) = \left(\vec{i}\frac{\partial}{\partial x} + \vec{j}\frac{\partial}{\partial y} + \vec{k}\frac{\partial}{\partial z}\right)\cdot(\vec{i}\,\vec{i}\,\tau_{xx} + \vec{j}\,\vec{i}\,\tau_{yx} + \vec{k}\,\vec{i}\,\tau_{zx})$$

按照矢量从左边与张量作内积的规律，结果就有

$$\nabla \cdot (\vec{i}\,\vec{i}\,\tau_{xx} + \vec{j}\,\vec{i}\,\tau_{yx} + \vec{k}\,\vec{i}\,\tau_{zx}) = \left(\frac{\partial \tau_{xx}}{\partial x} + \frac{\partial \tau_{yx}}{\partial y} + \frac{\partial \tau_{zx}}{\partial z}\right)\vec{i} = \overrightarrow{f_{sx}} \tag{2.17a}$$

同样，在 $y$ 方面和 $z$ 方向分别是

$$\nabla \cdot (\vec{i}\,\vec{j}\,\tau_{xy} + \vec{j}\,\vec{j}\,\tau_{yy} + \vec{k}\,\vec{j}\,\tau_{zy}) = \left(\frac{\partial \tau_{xy}}{\partial x} + \frac{\partial \tau_{yy}}{\partial y} + \frac{\partial \tau_{zy}}{\partial z}\right)\vec{j} = \overrightarrow{f_{sy}} \tag{2.17b}$$

$$\nabla \cdot (\vec{i}\,\vec{k}\,\tau_{xz} + \vec{j}\,\vec{k}\,\tau_{yz} + \vec{k}\,\vec{k}\,\tau_{zz}) = \left(\frac{\partial \tau_{xz}}{\partial x} + \frac{\partial \tau_{yz}}{\partial y} + \frac{\partial \tau_{zz}}{\partial z}\right)\vec{k} = \overrightarrow{f_{sz}} \tag{2.17c}$$

将式（2.17a）～（2.17c）相加即得

$$\nabla \cdot \tau = \overrightarrow{f_s} \tag{2.18}$$

式（2.18）表明，流体的表面力等于其应力张量之散度。

2.3.3　运动中流体参数对时间的全导数

　　在运动中，流体的参数既随时间改变，又因位置迁移而变化，所以

$$q = q(x, y, z; t)$$

因为流动，流体微团的位置坐标随时间改变，即

$$x = x(t), \qquad y = y(t), \qquad z = z(t)$$

将 $q$ 对 $t$ 取导数，由多变量函数的微分得

$$\frac{dq}{dt} = \frac{\partial q}{\partial t}\frac{dt}{dt} + \frac{\partial q}{\partial x}\frac{dx}{dt} + \frac{\partial q}{\partial y}\frac{dy}{dt} + \frac{\partial q}{\partial z}\frac{dz}{dt}$$

因为 $\dfrac{dx}{dt}=u$， $\dfrac{dy}{dt}=v$， $\dfrac{dz}{dt}=w$， 于是有

$$\frac{dq}{dt}=\frac{\partial q}{\partial t}+u\frac{\partial q}{\partial x}+v\frac{\partial q}{\partial y}+w\frac{\partial q}{\partial z} \tag{2.19}$$

上式亦可写成

$$\frac{dq}{dt}=\frac{\partial q}{\partial t}+\vec{U}\cdot\nabla q \tag{2.20}$$

式中 $\vec{U}=\vec{i}u+\vec{j}v+\vec{k}w$。

在式（2.19）中，右方第一项称为当地导数或定点导数，其余三项称为迁移导数或随体导数。如果过程是稳定的，则有

$$\frac{\partial q}{\partial t}=0 \tag{2.21}$$

## 2.4 流体的输运特性
### 2.4.1 概述
所谓输运就是一种传递过程或转移过程。

流体运动中的输运，在流速不高的层流中，输运主要是流体分子的行为，可以在一定程度上依照分子运动论得到简单的解释。在流速很高的情况下，湍流现象就将发生。湍流是流体微团的脉动运动。这种脉动运动是随机的、不规则的，类似于分子的热运动。湍流中的输运过程，是以流体微团为单位的交换行为。由于流体微团含有大量的单个分子，所以湍流输运效果远远大于层流。这个现象大致可以类比于顺磁体和铁磁体的磁化。通常顺磁物质是由其原子中电子轨道运动所决定的分子磁矩转向外磁场而磁化，而铁磁物质却是以磁畴转向外磁场而磁化。每一磁畴内含有的原子数量达到几十亿的量级。所以顺磁物质的磁化率约为 $10^{-6}\sim10^{-5}$，而铁磁物质的磁化率可达 $10^{4}\sim10^{5}$。由此可以想像与流体微团形式进行的湍流输运和以分子形式进行的层流输运的差距之大。在实用中湍流输运可以完全忽略分子的贡献。于是湍流输运只取决于流动状况。本节只讨论层流输运，它和流动状况无关，只取决于分子的热运动。所以层流的输运系数是流体自身的属性。

### 2.4.2 动量传递
牛顿粘性定律式（2.15）

$$\tau=\eta\frac{\partial u}{\partial y}$$

式中粘性系数 $\eta$ 的单位是 $N\cdot s/m^2$，可见其物理意义是单位面积上传递的冲量。当流体作剪切运动时，即出现速度梯度时，低速层的流体分子与高速层的流体分子之间，通过碰撞而交换动量，一方面阻滞了高速层的流动，另一方面又拖曳低速层提高速度。这样的过程在宏观上表现的就是粘性现象。所以 $\eta$ 通常称作粘性系数，尽管它实质上是一种动量的传递系数。

### 2.4.3 能量传递
就分子运动论的观念而言，热量就是分子运动的动能。Boltzmann 常数 $k_0$ 与 Kelvin 绝对温度 $T$ 的乘积 $k_0T$ 就是热运动能。所以温度是分子运动动能的宏观表现，分子运动动能的交换就是热交换。热交

换的驱动力是温度梯度，高温区 $k_bT$ 值高，即分子运动的动能大。相反在低温区，分子运动的动能较小，通过分子间的碰撞，高能分子将动能传递给低能分子，这种过程被归结为以温度梯度描述的 Fourier 定律：

$$q_{Hx} = -K_{Hx} \frac{\partial T}{\partial x}$$ (2.22a)

对于各向同性物质，其三维情况可写成

$$\overrightarrow{q_H} = -K_H \nabla T$$ (2.22b)

式中 $\overrightarrow{q_H}$ 是热流矢量，它是单位时间内通过单位面积等温面的热通量，$J/(m^2 \cdot s)$ 或者 $N \cdot m/(m^2 \cdot s)$。$K_H$ 称为导热系数或热传导系数，它是在等温面法线的单位长度上，温差为 1K 时，单位时间内通过等温面的热通量，$J/(m \cdot s \cdot K)$ 或 $N \cdot m/(m \cdot s \cdot K)$，温度 $T$ 的单位是 Kelvin 绝对温度 K。

式（2.22a）和式（2.22b）右边的负号表示热流的方向与温度梯度的方向相反，即热从高温处传向低温处的事实。

### 2.4.4 质量传递

流体内的质量传递就是分子本身的交换。在均匀状态下，分子运动的统计情况为各向同性，即一个区域与相邻区域的分子交换互相平衡，净效应为零，即没有质量迁移。在存在密度梯度的不均匀状态下，高密度区扩散到低密度区的分子数量超过从低密度区扩散到高密度区的分子数量，因而净效应表现为由高密度区朝低密度区的"单向"扩散。这种现象归结为 Fick 定律，即

$$q_{mx} = -D_x \frac{\partial \rho}{\partial x}$$ (2.23a)

对于各向同性的三维情况

$$\overrightarrow{q_m} = -D \nabla \rho$$ (2.23b)

式中 $q_m$ 是质量流量，它是单位时间内通过单位面积等密度面的净质量，$kg/(m^2 \cdot s)$，$D$ 称为扩散系数，它是沿等密度面法线的单位长度上，在单位时间内通过的净体积，单位是 $m^3/(m \cdot s)$，$\rho$ 是密度 $kg/m^3$。

对混合物质，讨论其中第 $i$ 种成分在混合物内部扩散时，采用该成分的分密度 $\rho_i$，设混合物质量是 $m$，它是各成分的质量之和，即

$$m = \sum m_i$$

设混合物的体积是 $V$，则有

$$\frac{m}{V} = \sum \frac{m_i}{V}$$

上式即混合物的密度 $\rho$ 是其各成分的分密度 $\rho_i$ 之和

$$\rho = \sum \rho_i$$ (2.24)

于是成分 $i$ 的扩散方程是

$$\overrightarrow{q_{mi}} = -D_i \nabla \rho_i \qquad (2.25)$$

常常使用一种量纲归一化的相对密度，即浓度 $c_i$

$$c_i = \rho_i / \rho \qquad (2.26)$$

由于扩散在混合物的内部进行，所以混合物的密度 $\rho$ 是常数。将式（2.25）两边同除以 $\rho$，并且注意到

$$\frac{\overrightarrow{q_{mi}}}{\rho} = \frac{\rho_i}{\rho} \frac{\overrightarrow{q_{mi}}}{\rho_i} = c_i \overrightarrow{q_{vi}}$$

于是式（2.25）成为

$$c_i \overrightarrow{q_{vi}} = -D_i \nabla c_i \quad 或 \quad \overrightarrow{q_{vi}} = -D_i \nabla (\ln c_i) \qquad (2.27)$$

式中 $\overrightarrow{q_{vi}} = \overrightarrow{q_{mi}} / \rho_i$，它的单位是 m³/(m²·s)。由此可见，体积流量就是速度。所以 $\overrightarrow{q_{vi}}$ 也常称为扩散速度。

## 2.5  流体的热力学特性
### 2.5.1  热力学第一定律

热力学第一定律是普遍的能量守恒定律。它表示外界加入到系统中热量 $\delta Q$ 等于系统的内能增量 $dE$ 与系统对外界作功 $\delta W$ 之和，即

$$\delta Q = dE + \delta W \qquad (2.28a)$$

式中系统对外界所作之功 $\delta W$ 就是膨胀功，所以 $\delta W = p\,dV$，于是

$$\delta Q = dE + p\,dV \qquad (2.28b)$$

对于单位体积流体，式（2.28b）两边同除以体积 $V$，得

$$\delta q = de + \rho p\,dv = de + p\,d(\ln v) \qquad (2.29a)$$

对于单位质量流体，式（2.28b）两边同除以质量 $m$，得

$$\delta q' = de' + p\,dv \qquad (2.29b)$$

上两式中 $v$ 是流体的比容，即其密度 $\rho$ 的倒数。

对于普通的不可压缩流体，其膨胀功 $\delta W = 0$，外界对系统加入的热量全都转为内能。所以对不可压缩流体常使用式（2.29a），而对可压缩流体，使用式（2.29b）更为方便。

### 2.5.2  热力学状态参数

流体的热力学状态有两种，即平衡状态和不平衡状态。不平衡状态是不稳定的暂态，或者认为是一种过渡状态。当系统的外部条件发生变化以后，系统就将从原来的平衡状态转变到与外部相适应的新的平衡状态。这个转变过程的全部经历就称为热力学过程或路径。平衡状态通常指热平衡、力平衡、相平衡、化学平衡等。但是，所有的平衡并不一定是死寂的、凝固的，交换活动仍然存在，但交换的净效应必须是零。所以平衡常常是动平衡。

描述流体性质的参数可以分成两类：一类是和流体热力学过程有关的参数，另一类是与热力学过程无关的参数。后者称为状态参数。热力学过程均是积累的过程，与这种累积变化过程无关的状态参数，在数学上应具全微分的性质。因为具全微分性质的函数，其积分结果与积分路径不相关，只取决于积分限。积分限在热力学上就是起始和终了的状态：

$$\int_1^2 df = f_2 - f_1$$

流体最基本的热力学状态参数是温度 $T$，压力 $p$ 和密度 $\rho$（或比容 $v$）。这三个参数组成的方程就是状态方程：

$$pv = RT \qquad (2.30)$$

此外在热力学中还有三个重要的导出参数，即热内能 $E$，熵 $S$ 和热焓 $H$。它们也全是状态参数。

1. 热内能

热内能就是流体内部所含有的热量。任何物质所含的热量均以温度为指标来度量即

$$E = C_v T \qquad (2.31a)$$

式中 $C_v$ 称为定容热容。将式（2.31a）两边用体积 $V$ 除，就得单位体积的热内能 $e$

$$e = c_v T, \qquad c_v = C_v / V \qquad (2.31b)$$

式中 $c_v$ 称为体积定容比热容，它是单位体积的流体在保持容积不变的情况下，温度升高 1K 时所需的热量，其单位是 $J/(m^3 \cdot K)$ 或 $N \cdot m/(m^3 \cdot K)$。

若将式（2.31a）用质量 $m$ 除，就得单位质量的热内能 $e'$

$$e' = c_v' T, \qquad c_v' = C_v / m \qquad (2.31c)$$

式中 $c_v'$ 称为质量定容比热容，单位是 $J/(kg \cdot K)$ 或 $N \cdot m/(kg \cdot K)$。

由

$$C_v = V c_v = m c_v'$$

所以得

$$c_v = \rho c_v' \qquad (2.32)$$

$\rho$ 是流体的密度。

2. 热焓

热焓的概念在可压缩的流体中十分有用。它既包括了热内能，又包括了压力功，即

$$H = E + pV \qquad (2.33a)$$

由于主要应用于可压缩流体，所以这里只给出单位质量的热焓。将式（2.33a）用质量 $m$ 除，就得到

$$h' = e' + pv \qquad (2.33b)$$

由 $e' = c_v' T$ 和理想气体状态方程 $pv = RT$，则上式成为

$$h' = (c_v' + R)T$$

定义一个新的比热容 $c_p'$，它是

$$c_p' = c_v' + R \qquad (2.34)$$

$c'_p$ 称为质量定压比热容，它是在压力不改变下，单位质量的气体，温度升高 1K 时所需加入的热量。

所以单位是 J/(kg·K) 或 $N·m/(kg·K)$。由于体积在压力不变下可以因加热而膨胀，式（2.34）中 $R$ 就

体现这一点，所以 $c'_p$ 总是大于 $c'_v$。热焓 $h'$ 就是

$$h' = c'_p T \tag{2.35}$$

$c'_p$ 不只是在定压过程中用来计算热焓，在任何过程中计算热焓都应使用 $c'_p$，同样，在任何过程中计算

热内能均应使用 $c'_v$ 或 $c_v$，不只限于定容过程。

对于不可压缩流体 $c'_v$ 和 $c'_p$ 是相等的，所以其热内能和热焓相同。

3.熵

从外界向一个系统加热，以使系统达到一定的温度。若保持流体的体积不变，即定容过程，此时加入的热量全部转变为热内能，即全部用来提高温度。若保持压力不变，即定压过程。则加入的热量只有一部分成为热内能，另一部分转换为膨胀功。所以，欲达到同样的温度，就要加入更多的热量。由此可见热量 $\delta Q$ 和热力学过程有关，它不是状态参数。但是，热量 $\delta Q$ 与系统温度 $T$ 的商却被发现是一个与热力学过程无关的状态参数。所以取名为熵，即

$$dS = \frac{\delta Q}{T} \tag{2.36a}$$

用流体的体积 $V$ 除式（2.36a），就得体积比熵为

$$ds = \delta q/T, \qquad ds = dS/V, \qquad \delta q = \delta Q/V \tag{2.36b}$$

用流体的质量 $m$ 除式（2.36a），就得质量比熵为

$$ds' = \delta q'/T, \qquad ds' = dS/m, \qquad \delta q' = \delta Q/m \tag{2.36c}$$

由上面两式可知

$$ds = \rho ds' \tag{2.37}$$

用 $T$ 除式（2.29a），左边就是 $ds$，即

$$ds = \frac{de}{T} + \frac{\rho p}{T} dv = c_v \frac{dT}{T} + R \frac{dv}{v^2}$$

积分上式得

$$\Delta s = s_2 - s_1 = \left[ c_v \ln T - \frac{R}{v} \right]_1^2 \tag{2.38a}$$

用 $T$ 除式（2.29b），就得

$$ds' = \frac{de'}{T} + \frac{p}{T} dv = c'_v \frac{dT}{T} + R \frac{dv}{v}$$

积分就有

$$\Delta s' = s'_2 - s'_1 = \left[ c'_v \ln T + R \ln v \right]_1^2 \tag{2.38b}$$

由式（2.38a）和（2.38b）可见，无论是体积比熵还是质量比熵，都与热力学过程不相关，这就表明熵是一种状态参数。

    4.流体状态参数之间的关系

    将式（2.33b）两边微分，并且注意到式（2.29b），给出

$$dh' = de' + p\,dv + v\,dp = \delta q' + v\,dp$$

将式（2.36c）代入上式右边第一项，得到

$$dh' = T\,ds' + v\,dp \qquad (2.39)$$

式（2.39）表示 $h'$ 是 $s'$ 和 $p$ 的函数，而 $h'$ 是状态参数，故有全微分

$$dh' = \frac{\partial h'}{\partial s'}ds' + \frac{\partial h'}{\partial p}dp$$

上两式比较，即得

$$T = \frac{\partial h'}{\partial s'}\bigg|_{p}, \qquad v = \frac{\partial h'}{\partial p}\bigg|_{s'} \qquad (2.40a)$$

将式（2.29b）改写成

$$de' = T\,ds' - p\,dv$$

上式表示 $e'$ 是 $s'$ 和 $v$ 的函数，而 $e'$ 是状态参数，故有全微分

$$de' = \frac{\partial e'}{\partial S'}ds' + \frac{\partial e'}{\partial v}dv$$

比较上面两式，即得

$$T = \frac{\partial e'}{\partial s'}\bigg|_{v}, \qquad p = -\frac{\partial e'}{\partial v}\bigg|_{s'} \qquad (2.40b)$$

## 2.6   流体的连续性方程——质量守恒方程

    设微元六面体的体积为 $dV = dxdydz$，其中包含的流体质量为 $dm = \rho dxdydz$，在流场中任取一体积 $V$，其中的流体质量为

$$m = \int_{V} \rho\,dx\,dy\,dz$$

两边同时对时间 $t$ 取导数

$$\frac{dm}{dt} = \int_{V}\frac{d\rho}{dt}dx\,dy\,dz + \rho\left[\frac{d(dx)}{dt}dy\,dz + \frac{d(dy)}{dt}dz\,dx + \frac{d(dz)}{dt}dx\,dy\right] =$$

$$\int_{V}\frac{d\rho}{dt}dx\,dy\,dz + \rho\left[\frac{d}{dx}\left(\frac{dx}{dt}\right) + \frac{d}{dy}\left(\frac{dy}{dt}\right) + \frac{d}{dz}\left(\frac{dz}{dt}\right)\right]dx\,dy\,dz$$

由 $dx/dt = u(x, y, z; t)$；$dy/dt = v(x, y, z; t)$；$dz/dt = w(x, y, z; t)$，故有

$$\frac{d}{dx}\left(\frac{dx}{dt}\right) = \frac{du}{dx} = \frac{\partial u}{\partial x}, \qquad \frac{d}{dx}\left(\frac{dx}{dt}\right) = \frac{du}{dx} = \frac{\partial u}{\partial x}, \qquad \frac{d}{dz}\left(\frac{dz}{dt}\right) = \frac{\partial w}{\partial z}$$

于是就有

$$\frac{dm}{dt} = \int_{V}\left[\frac{d\rho}{dt} + \rho\left(\frac{\partial u}{\partial x} + \frac{\partial v}{\partial y} + \frac{\partial w}{\partial z}\right)\right]dx\,dy\,dz$$

设流场是无源或无汇的场，则 $dm/dt = 0$，于是得到

$$\frac{d\rho}{dt} + \rho\left(\frac{\partial u}{\partial x} + \frac{\partial v}{\partial y} + \frac{\partial w}{\partial z}\right) = 0$$

或写成

$$\frac{d\rho}{dt} + \rho\nabla \cdot \vec{U} = 0 \tag{2.41a}$$

在式（2.20）中将 $q$ 取为密度 $\rho$，有

$$\frac{d\rho}{dt} = \frac{\partial\rho}{\partial t} + \vec{U} \cdot \nabla\rho \tag{2.41b}$$

将式（2.41b）代入式（2.41a）之中，得到

$$\frac{\partial\rho}{\partial t} + \nabla \cdot (\rho\vec{U}) = 0 \tag{2.42a}$$

式（2.41a）和式（2.42a）就是常用的连续性方程一般形式。

对于稳定状况下的可压缩流，$\partial\rho/\partial t = 0$，式（2.42a）成为

$$\nabla \cdot (\rho\vec{U}) = 0 \tag{2.42b}$$

对于不可压缩流体，$\rho = \mathrm{const}$，则式（2.42a）成为

$$\nabla \cdot \vec{U} = 0$$

此即不可压缩流体的连续方程是其速度的散度等于零。无论流动是否稳定，都是如此。

2.7　流体的运动方程——动量守恒方程

运动方程就是牛顿第二定律式

$$m\vec{a} = \vec{F}$$

两边除以流体的体积 $V$ 就得单位体积的方程

$$\rho\vec{a} = \vec{f}$$

加速度 $\vec{a} = d\vec{U}/dt$，作用于单位体积流体上的力有彻体力 $\vec{f_b}$ 和表面力 $\vec{f_s}$，即

$$\rho\frac{d\vec{U}}{dt} = \vec{f_b} + \vec{f_s} \tag{2.43}$$

以下逐项分析式（2.43）：

①惯性项 $\rho\, d\vec{U}/dt$

将式（2.20）中的 $q$ 认定为速度 $\vec{U}$，则有

$$\rho\frac{d\vec{U}}{dt} = \rho\frac{\partial\vec{U}}{\partial t} + \rho\vec{U} \cdot \nabla\vec{U}$$

②彻体力 $\vec{f_b}$

$$\vec{f_b} = \rho \vec{g}$$

式中 $\vec{g}$ 是重力加速度。

③表面力 $\vec{f_s}$

将式（2.18）写成张量的形式

$$\vec{f_s} = \nabla \cdot \tau = \frac{\partial \tau_{ij}}{\partial x_j} \tag{2.44}$$

按牛顿粘性定律式（2.14）

$$\tau_{ij} = \eta \left( \frac{\partial u_i}{\partial x_j} + \frac{\partial u_j}{\partial x_i} \right)$$

当 $i = j$ 时为正应力，在直角坐标系中有

$$\tau_{xx} = 2\eta \frac{\partial u}{\partial x}, \qquad \tau_{yy} = 2\eta \frac{\partial v}{\partial y}, \qquad \tau_{zz} = 2\eta \frac{\partial w}{\partial z}$$

在式 $\tau_{xx}$ 之右方同时加减 $\frac{2}{3}\eta \left( \frac{\partial u}{\partial x} + \frac{\partial v}{\partial y} + \frac{\partial w}{\partial z} \right)$，得

$$\tau_{xx} = 2\eta \frac{\partial u}{\partial x} + \frac{1}{3} \left[ 2\eta \left( \frac{\partial u}{\partial x} + \frac{\partial v}{\partial y} + \frac{\partial w}{\partial z} \right) \right] - \frac{2}{3}\eta \left( \frac{\partial u}{\partial x} + \frac{\partial v}{\partial y} + \frac{\partial w}{\partial z} \right)$$

上式可改写成

$$\tau_{xx} = 2\eta \frac{\partial u}{\partial x} + \frac{1}{3}(\tau_{xx} + \tau_{yy} + \tau_{zz}) - \frac{2}{3}\eta \nabla \cdot \vec{U}$$

定义作用于微元六面体表面上的压力 $p$ 是三个正应力的平均值，并且沿表面外法线的负方向，即

$$p = -\frac{1}{3}(\tau_{xx} + \tau_{yy} + \tau_{zz}) \tag{2.45}$$

于是有

$$\left. \begin{array}{l} \tau_{xx} = -p + 2\eta \dfrac{\partial u}{\partial x} - \dfrac{2}{3}\eta \nabla \cdot \vec{U} \\[2mm] \tau_{yy} = -p + 2\eta \dfrac{\partial v}{\partial y} - \dfrac{2}{3}\eta \nabla \cdot \vec{U} \\[2mm] \tau_{zz} = -p + 2\eta \dfrac{\partial w}{\partial z} - \dfrac{2}{3}\eta \nabla \cdot \vec{U} \end{array} \right\} \tag{2.46}$$

将式（2.46）与（2.14）合并写成

$$\tau_{ij} = -p\delta_{ij} + \eta \left( \frac{\partial u_i}{\partial x_j} + \frac{\partial u_j}{\partial x_i} \right) - \frac{2}{3}\eta (\nabla \cdot \vec{U})\delta_{ij} \tag{2.47}$$

式中 $\delta_{ij}$ 是 Kronecker 符号

$$\delta_{ij} = \begin{cases} 1, & i = j \\ 0, & i \neq j \end{cases} \tag{2.48}$$

将式（2.47）代入式（2.44）中，得表面力 $\overline{f_s}$ 的关系式为

$$\overline{f_s} = -\frac{\partial(p\delta_{ij})}{\partial x_j} + \frac{\partial}{\partial x_j}\left[\eta\left(\frac{\partial u_i}{\partial x_j} + \frac{\partial u_j}{\partial x_i}\right)\right] - \frac{2}{3}\frac{\partial}{\partial x_j}\left[\eta(\nabla\cdot\overline{U})\delta_{ij}\right]$$

注意到式（2.48）则有

$$\frac{\partial(p\delta_{ij})}{\partial x_j} = \frac{\partial p}{\partial x_j} = \nabla p, \qquad \frac{\partial}{\partial x_j}\left[\eta(\nabla\cdot\overline{U})\delta_{ij}\right] = \nabla(\eta\nabla\cdot\overline{U})$$

于是得

$$\overline{f_s} = -\nabla p + \frac{\partial}{\partial x_j}\left[\eta\left(\frac{\partial u_i}{\partial x_j} + \frac{\partial u_j}{\partial x_i}\right)\right] - \frac{2}{3}\nabla(\eta\nabla\cdot\overline{U}) \tag{2.49}$$

将以上得出的惯性力和表面力的关系式代入方程式（2.43）之中，有

$$\rho\frac{\partial\overline{U}}{\partial t} + \rho\overline{U}\cdot\nabla\overline{U} = \overline{f_b} - \nabla p + \frac{\partial}{\partial x_j}\left[\eta\left(\frac{\partial u_i}{\partial x_j} + \frac{\partial u_j}{\partial x_i}\right)\right] - \frac{2}{3}\nabla(\eta\nabla\cdot\overline{U}) \tag{2.50}$$

方程式（2.50）就是单位体积流体的运动方程，或称 Navier-Stokes 方程。它的几种特殊情况如下：

1.稳态流动

在式（2.50）中取

$$\frac{\partial\overline{U}}{\partial t} = 0$$

2.常粘度的可压缩流体

在式（2.50）中取 $\eta = \text{const}$，就得

$$\rho\frac{\partial\overline{U}}{\partial t} + \rho\overline{U}\cdot\nabla\overline{U} = \overline{f_b} - \nabla p + \eta\left(\frac{\partial^2 u_i}{\partial x_j^2} + \frac{\partial^2 u_i}{\partial x_i\partial x_j}\right) - \frac{2}{3}\eta\nabla\nabla\cdot\overline{U} \tag{2.51a}$$

或由 $\dfrac{\partial^2 u_i}{\partial x_j^2} = \nabla^2\overline{U}$，$\dfrac{\partial^2 u_j}{\partial x_i\partial x_j} = \nabla\nabla\cdot\overline{U}$ 代入式（2.51a），给出

$$\rho\frac{\partial\overline{U}}{\partial t} + \rho\overline{U}\cdot\nabla\overline{U} = \overline{f_b} - \nabla p + \eta\nabla^2\overline{U} + \frac{1}{3}\eta\nabla\nabla\cdot\overline{U} \tag{2.51b}$$

3.不可压缩流体

对于不可压缩流体，$\rho = \text{const}$，并且有 $\nabla\cdot\overline{U} = 0$，则式（2.50）成为

$$\rho\frac{\partial\overline{U}}{\partial t} + \rho\overline{U}\cdot\nabla\overline{U} = \overline{f_b} - \nabla p + \frac{\partial}{\partial x_j}\left[\eta\left(\frac{\partial u_i}{\partial x_j} + \frac{\partial u_j}{\partial x_i}\right)\right] \tag{2.52a}$$

4.常粘度的不可压缩流体

式（2.51b）中取 $\nabla \cdot \vec{U} = 0$，得

$$\rho \frac{\partial \vec{U}}{\partial t} + \rho \vec{U} \cdot \nabla \vec{U} = \vec{f_b} - \nabla p + \eta \nabla^2 \vec{U} \qquad (2.52b)$$

## 2.8 流体的涡量方程和 Bernoulli 方程

### 2.8.1 涡量方程

考虑最常用的常粘度情况。将运动方程式（2.51b）改写。在式（2.51b），即

$$\rho \frac{\partial \vec{U}}{\partial t} + \rho \vec{U} \cdot \nabla \vec{U} = \vec{f_b} - \nabla p + \eta \nabla^2 \vec{U} + \frac{1}{3} \eta \nabla \cdot \vec{U}$$

中有些项可以利用矢量恒等式加以改形：

① $\vec{U} \cdot \nabla \vec{U}$

由 $\nabla(\vec{U} \cdot \vec{U}) = 2\vec{U} \cdot \nabla \vec{U} + 2\vec{U} \times (\nabla \times \nabla \vec{U})$，移项得

$$\vec{U} \cdot \nabla \vec{U} = \frac{1}{2} \nabla(\vec{U} \cdot \vec{U}) - \vec{U} \times (\nabla \times \vec{U}) = \frac{1}{2} \nabla U^2 - \vec{U} \times \Omega$$

② $\vec{f_b}$

彻体力 $\vec{f_b}$ 通常是具势的，所以

$$\vec{f_b} = -f_{bj} \vec{e_j^0}$$

其中 $f_{bj}$ 是一常数，负号表示彻体力的方向恒指原点。$\vec{f_b}$ 的势函数是 $\vec{f_b} \cdot \vec{X}$，其梯度为

$$\nabla(\vec{f_b} \cdot \vec{X}) = \nabla(-f_{bj} \vec{e_j^0} \cdot x_k \vec{e_k^0}) = -\nabla(f_{bj} x_k \delta_{jk}) = -\nabla(f_{bj} x_j)$$

由于 $-\nabla(f_{bj} x_j) = -f_{bj} \vec{e_i^0}(\partial x_j / \partial x_i)$，只有当 $i = j$ 时，$\partial x_j / \partial x_i = 1$，当 $i \neq j$ 时，总是有 $\partial x_j / \partial x_i = 0$，所以

$$\nabla(\vec{f_b} \cdot \vec{X}) = -\nabla(f_{bj} x_j) = -f_{bj} \vec{e_j^0} = \vec{f_b}$$

③ $\nabla^2 \vec{U}$

由 $\nabla(\nabla \cdot \vec{U}) = \nabla^2 \vec{U} + \nabla \times (\nabla \times \vec{U}) = \nabla^2 \vec{U} + \nabla \times \vec{\Omega}$，移项就有

$$\nabla^2 \vec{U} = \nabla(\nabla \cdot \vec{U}) - \nabla \times \vec{\Omega}$$

将以上①、②和③的结果代入方程式（2.51b）中，得

$$\rho \frac{\partial \vec{U}}{\partial t} + \rho \left( \frac{1}{2} \nabla U^2 - \vec{U} \times \vec{\Omega} \right) = -\nabla(f_{bj} x_j) - \nabla p + \eta \nabla(\nabla \cdot \vec{U}) - \eta \nabla \times \vec{\Omega} + \frac{1}{3} \eta \nabla(\nabla \cdot \vec{U})$$

上式整理之后，就得

$$\rho \frac{\partial \vec{U}}{\partial t} + \nabla \left[ p + f_{bj}x_j - \frac{4}{3}(\nabla \cdot \vec{U}) \right] + \frac{1}{2}\rho \nabla U^2 = \rho \vec{U} \times \vec{\Omega} - \eta \nabla \times \vec{\Omega} \tag{2.53}$$

式（2.53）仍然是单位体积常粘度流体的运动方程，只是形式上与（2.51b）有所不同而已。

将式（2.53）两边取旋度，利用矢量恒等式得出：

① $\nabla \times \left( \rho \frac{\partial \vec{U}}{\partial t} \right) = \rho \nabla \times \frac{\partial \vec{U}}{\partial t} + \nabla \rho \times \frac{\partial \vec{U}}{\partial t} = \rho \frac{\partial \vec{\Omega}}{\partial t} + \nabla \rho \times \frac{\partial \vec{U}}{\partial t}$

② $\nabla \times \nabla \left[ p + f_{bj}x_j - \frac{4}{3}\eta(\nabla \cdot \vec{U}) \right] = 0$

③ $\nabla \times \frac{1}{2}\rho \nabla U^2 = \frac{1}{2}\rho \nabla \times (\nabla U^2) + \frac{1}{2}\nabla \rho \times (\nabla U^2) = \frac{1}{2}\nabla \rho \times (\nabla U^2)$

④ $\nabla \times (\rho \vec{U} \times \vec{\Omega}) = \nabla \rho \times (\vec{U} \times \vec{\Omega}) + \rho \nabla \times (\vec{U} \times \vec{\Omega}) =$

$$\nabla \rho \times (\vec{U} \times \vec{\Omega}) + \rho(\vec{\Omega} \cdot \nabla \vec{U} - \vec{U} \cdot \nabla \vec{\Omega} + \vec{U}\nabla \cdot \vec{\Omega} - \vec{\Omega}\nabla \cdot \vec{U})$$

式中由于 $\vec{\Omega} = \nabla \times \vec{U}$，故 $\nabla \cdot \vec{\Omega} = \nabla \cdot (\nabla \times \vec{U}) = 0$，于是有

$$\nabla \times (\rho \vec{U} \times \vec{\Omega}) = \nabla \rho \times (\vec{U} \times \vec{\Omega}) + \rho(\vec{\Omega} \cdot \nabla \vec{U} - \vec{U} \cdot \nabla \vec{\Omega} - \vec{\Omega}\nabla \cdot \vec{U})$$

⑤ $\nabla \times (\nabla \times \vec{\Omega}) = \nabla(\nabla \cdot \vec{\Omega}) - \nabla \cdot \nabla \vec{\Omega} = -\nabla^2 \vec{\Omega}$

先将式（2.53）两边取旋度，而后代入上述①~⑤的结果，就得单位体积常粘度流体的涡量方程之一般形式

$$\rho \frac{\partial \vec{\Omega}}{\partial t} + \nabla \rho \times \frac{\partial \vec{U}}{\partial t} + \nabla \rho \times \left( \frac{1}{2}\nabla U^2 \right) = \nabla \rho \times (\vec{U} \times \vec{\Omega}) + \rho(\vec{\Omega} \cdot \nabla \vec{U} - \vec{U} \cdot \nabla \vec{\Omega} - \vec{\Omega}\nabla \cdot \vec{U}) + \eta \nabla^2 \vec{\Omega} \tag{2.54}$$

对于不可压缩流体，$\rho = \text{const}$，$\nabla \rho = 0$，$\nabla \cdot \vec{U} = 0$，则式（2.54）成为单位体积的常粘度不可压缩流的涡量方程

$$\rho \frac{\partial \vec{\Omega}}{\partial t} = \rho \vec{\Omega} \cdot \nabla \vec{U} - \rho \vec{U} \cdot \nabla \vec{\Omega} + \eta \nabla^2 \vec{\Omega} \tag{2.55a}$$

上式亦可写成

$$\rho \frac{d\vec{\Omega}}{dt} = \rho \vec{\Omega} \cdot \nabla \vec{U} + \eta \nabla^2 \vec{\Omega} \tag{2.55b}$$

## 2.8.2 Bernoulli 方程

设：①流体是不可压缩流体，$\rho = \text{const}$，②流动是稳态的，$\partial \vec{U}/\partial t = 0$，③势流，即无旋 $\vec{\Omega} = 0$。于是运动方程（2.53）成为

$$\nabla \left( p + f_{bj}x_j + \frac{1}{2}\rho U^2 \right) = 0，\text{或 } p + f_{bj}x_j + \frac{1}{2}\rho U^2 = C \tag{2.56a}$$

若彻体力仅有重力，则 $f_{bj} = \rho g$ ， $x_j = h$ ，故有

$$p + \frac{1}{2}\rho U^2 + \rho gh = C \qquad (2.56b)$$

方程式（2.56a）和（2.56b）即 Bernoulli 方程。它是水力学的基本方程。

## 2.9 流体的能量守恒方程

能量守恒方程的最基本形式就是热力学第一定律，即

$$dE = \delta Q + \delta W$$

式中 $E$ 是流体的总内能，包括热内能和动能，$Q$ 是外界加入到流体中的热量，$W$ 是外界对流体所作之功。

设在流场中任取一体积为 $V$ 的流体，包围 $V$ 的表面积是 $S$。在体积 $V$ 内，包含的总内能是积分

$$E = \int_V \left( e + \frac{1}{2}\rho U^2 \right) dV = \int_V \rho \left( e' + \frac{1}{2}U^2 \right) dV$$

外界加入到体积 $V$ 中的流体的热量是通过表面积 $S$ 传输的。若热流矢量是 $\overrightarrow{q_H}$ ，则在时间 $dt$ 之内穿过 $S$ 的热量 $\delta Q$ 是

$$\delta Q = \left( \int_S \overrightarrow{q_H} \cdot d\vec{S} \right) dt$$

外界对体积 $V$ 内的流体所作之功，包括彻体力 $\overrightarrow{f_b}$ 所作之功 $W_b$ 和表面力 $\tau$ 所作之功 $W_s$ 。彻体力之功作用于全部体积 $V$ 内，故在时间 $dt$ 中的彻体力的功是

$$\delta W_b = \left[ \int_V (\overrightarrow{f_b} \cdot \vec{U}) dV \right] dt$$

表面力之功作用于表面 $S$ 上，在时间 $dt$ 内表面力的功是

$$\delta W_s = \left[ \int_S (\tau \cdot \vec{U}) \cdot d\vec{S} \right] dt$$

将以上三个式子代入热力学第一定律式中，就有

$$d\int_V \rho \left( e' + \frac{1}{2}U^2 \right) dV = \left[ \int_S \overrightarrow{q_H} \cdot d\vec{S} + \int_V (\overrightarrow{f_b} \cdot \vec{U}) dV + \int_S (\tau \cdot \vec{U}) \cdot d\vec{S} \right] dt$$

使用散度定理，将上式右方对面积 $S$ 的积分改换成对体积 $V$ 的积分：

$$\frac{d}{dt} \int_V \rho \left( e' + \frac{1}{2}U^2 \right) dV = \int_V \nabla \cdot \overrightarrow{q_H} dV + \int_V (\overrightarrow{f_b} \cdot \vec{U}) dV + \int_V \nabla \cdot (\tau \cdot \vec{U}) dV \qquad (2.58)$$

上式左边的 $d/dt$ 可写为积分之内，从而有

$$\frac{d}{dt} \left[ \rho \left( e' + \frac{1}{2}U^2 \right) dV \right] = \left[ \frac{d}{dt} \left( e' + \frac{1}{2}U^2 \right) \right] \rho dV + \left( e' + \frac{1}{2}U^2 \right) \frac{d(\rho dV)}{dt}$$

在流场内没有源或汇，故 $\rho dV = dm = \text{const}$ ，于是 $d(\rho dV)/dt = 0$ ，则

$$\frac{d}{dt} \int_V \rho \left( e' + \frac{1}{2}U^2 \right) dV = \int_V \rho \frac{d}{dt} \left( e' + \frac{1}{2}U^2 \right) dV \qquad (2.59)$$

在方程（2.58）中，各项的被积函数均有连续性质，体积 $V$ 在流场中是任取的，故有

$$\rho \frac{d}{dt}\left(e' + \frac{1}{2}U^2\right) = \nabla \cdot \overrightarrow{q_H} + \overrightarrow{f_b} \cdot \overrightarrow{U} + \nabla \cdot (\tau \cdot \overrightarrow{U}) \tag{2.60}$$

下面给出方程式（2.60）中的各项

①由式（2.20）

$$\rho \frac{d}{dt}\left(e' + \frac{1}{2}U^2\right) = \rho \frac{de'}{dt} + \rho \frac{d}{dt}\left(\frac{U^2}{2}\right) = \rho \frac{\partial e'}{\partial t} + \rho \overrightarrow{U} \cdot \nabla e' + \rho \frac{d}{dt}\left(\frac{U^2}{2}\right)$$

②由 Fourie 定律

$$\nabla \cdot \overrightarrow{q_H} = \nabla \cdot (K_H \nabla T)$$

因为 $\overrightarrow{q_H}$ 是外界加入到流体中的热量，故上式右方取正号。

③ $\nabla \cdot (\tau \cdot \overrightarrow{U})$

注意 $\tau$ 是二阶张量，用张量形式写出[注]

$$\nabla \cdot (\tau \cdot \overrightarrow{U}) = \overrightarrow{e_j^0} \frac{\partial}{\partial x_j} \cdot (\overrightarrow{e_k^0} \overrightarrow{e_i^0} \tau_{ki} \cdot \overrightarrow{e_l^0} u_l) = \overrightarrow{e_j^0} \frac{\partial}{\partial x_j} \cdot (\overrightarrow{e_k^0} \delta_{il} \tau_{ki} u_l) =$$

$$\overrightarrow{e_j^0} \frac{\partial}{\partial x_j} \cdot (\overrightarrow{e_k^0} \tau_{ki} u_i) = \frac{\partial}{\partial x_j}(\overrightarrow{e_j^0} \cdot \overrightarrow{e_k^0} \tau_{ki} u_i) = \frac{\partial}{\partial x_j}(\delta_{jk} \tau_{ki} u_i) = \frac{\partial}{\partial x_j}(\tau_{ji} u_i)$$

于是有

$$\nabla \cdot (\tau \cdot \overrightarrow{U}) = u_i \frac{\partial \tau_{ji}}{\partial x_j} + \tau_{ji} \frac{\partial u_i}{\partial x_j} = \overrightarrow{U} \cdot (\nabla \cdot \tau) + (\tau \cdot \nabla) \cdot \overrightarrow{U} \tag{2.61}$$

由式（2.18）知 $\nabla \cdot \tau = \overrightarrow{f_s}$，以及由式（2.43）知 $\overrightarrow{f_s} = \rho d\overrightarrow{U}/dt - \overrightarrow{f_b}$，于是

$$\nabla \cdot (\tau \cdot \overrightarrow{U}) = \overrightarrow{U} \cdot \rho \frac{d\overrightarrow{U}}{dt} - \overrightarrow{U} \cdot \overrightarrow{f_b} + \tau_{ij} \frac{\partial u_i}{\partial x_j}$$

将①、②、③的结果代入方程（2.60）中：

$$\rho \frac{\partial e'}{\partial t} + \rho \overrightarrow{U} \cdot \nabla e' + \rho \frac{d}{dt}\left(\frac{U^2}{2}\right) = \overrightarrow{f_b} \cdot \overrightarrow{U} + \nabla \cdot (K_H \nabla T) + \rho \frac{d}{dt}\left(\frac{U^2}{2}\right) - \overrightarrow{U} \cdot \overrightarrow{f_b} + \tau_{ij} \frac{\partial u_i}{\partial x_j}$$

消去相同项，并且 $e' = c_V' T$，上式就成为

$$\rho \frac{\partial (c_V' T)}{\partial t} + \rho \overrightarrow{U} \cdot \nabla (c_V' T) = \nabla \cdot (K_H T) + \tau_{ij} \frac{\partial u_i}{\partial x_j}$$

用式（2.47）代入上式右方的最末项，得

---

[注] 此处用的是直角坐标系。

$$\tau_{ij} \frac{\partial u_i}{\partial x_j} = -p\delta_{ij}\frac{\partial u_i}{\partial x_j} + \eta\left(\frac{\partial u_i}{\partial x_j} + \frac{\partial u_j}{\partial x_i}\right)\frac{\partial u_i}{\partial x_j} - \frac{2}{3}\eta(\nabla \cdot \vec{U})\delta_{ij}\frac{\partial u_i}{\partial x_j} =$$

$$-p\nabla \cdot \vec{U} + \eta\left(\frac{\partial u_i}{\partial x_j} + \frac{\partial u_j}{\partial x_i}\right)\frac{\partial u_i}{\partial x_j} - \frac{2}{3}\eta(\nabla \cdot \vec{U})^2$$

引用符号

$$\Phi = \eta\frac{\partial u_i}{\partial x_j}\left(\frac{\partial u_i}{\partial x_j} + \frac{\partial u_j}{\partial x_i}\right) - \frac{2}{3}\eta(\nabla \cdot \vec{U})^2 \tag{2.62}$$

于是流体的能量方程是

$$\rho\frac{\partial(c_v'T)}{\partial t} + \rho\vec{U} \cdot \nabla(c_v'T) = -p\nabla \cdot \vec{U} + \nabla \cdot (K_H\nabla T) + \Phi \tag{2.63}$$

若 $c_v'$ 是常数，则由 $c_v = \rho c_v'$ ，方程（2.63）就成为

$$c_v\frac{\partial T}{\partial t} + c_v\vec{U} \cdot \nabla T = -p\nabla \cdot \vec{U} + \nabla \cdot (K_H\nabla T) + \Phi \tag{2.64}$$

2.10  直角坐标系、圆柱坐标系、圆球坐标系中的流体力学方程

2.10.1  直角坐标系中的不可压缩流体方程组

1.连续性方程

$$\frac{\partial u}{\partial x} + \frac{\partial v}{\partial y} + \frac{\partial w}{\partial z} = 0 \tag{2.65}$$

2.运动方程

$$\rho\left(\frac{\partial u}{\partial t} + u\frac{\partial u}{\partial x} + v\frac{\partial u}{\partial y} + w\frac{\partial u}{\partial z}\right) = f_x - \frac{\partial p}{\partial x} + \eta\left(\frac{\partial^2 u}{\partial x^2} + \frac{\partial^2 u}{\partial y^2} + \frac{\partial^2 u}{\partial z^2}\right) \tag{2.66a}$$

$$\rho\left(\frac{\partial v}{\partial t} + u\frac{\partial v}{\partial x} + v\frac{\partial v}{\partial y} + w\frac{\partial v}{\partial z}\right) = f_y - \frac{\partial p}{\partial y} + \eta\left(\frac{\partial^2 v}{\partial x^2} + \frac{\partial^2 v}{\partial y^2} + \frac{\partial^2 v}{\partial z^2}\right) \tag{2.66b}$$

$$\rho\left(\frac{\partial w}{\partial t} + u\frac{\partial w}{\partial x} + v\frac{\partial w}{\partial y} + w\frac{\partial w}{\partial z}\right) = f_z - \frac{\partial p}{\partial x} + \eta\left(\frac{\partial^2 w}{\partial x^2} + \frac{\partial^2 w}{\partial y^2} + \frac{\partial^2 w}{\partial z^2}\right) \tag{2.66c}$$

3.能量方程

$$\rho\left(\frac{\partial e'}{\partial t} + u\frac{\partial e'}{\partial x} + v\frac{\partial e'}{\partial y} + w\frac{\partial e'}{\partial z}\right) = \frac{\partial}{\partial x}\left(K_H\frac{\partial T}{\partial x}\right) + \frac{\partial}{\partial y}\left(K_H\frac{\partial T}{\partial y}\right) + \frac{\partial}{\partial z}\left(K_H\frac{\partial T}{\partial z}\right) +$$

$$\eta\left[2\left(\frac{\partial u}{\partial x}\right)^2 + 2\left(\frac{\partial v}{\partial y}\right)^2 + 2\left(\frac{\partial w}{\partial z}\right)^2 + \left(\frac{\partial u}{\partial y} + \frac{\partial v}{\partial x}\right)^2 + \left(\frac{\partial v}{\partial z} + \frac{\partial w}{\partial y}\right)^2 + \left(\frac{\partial w}{\partial x} + \frac{\partial u}{\partial z}\right)^2\right] \tag{2.67}$$

4.正应力和切应力

$$\tau_{xx} = -p + 2\eta \frac{\partial u}{\partial x} \qquad \tau_{xy} = \tau_{yx} = \eta \left( \frac{\partial u}{\partial y} + \frac{\partial v}{\partial x} \right) \left.\begin{array}{c} \\ \\ \\ \\ \\ \end{array}\right\}$$

$$\tau_{yy} = -p + 2\eta \frac{\partial v}{\partial y} \qquad \tau_{yz} = \tau_{zy} = \eta \left( \frac{\partial v}{\partial z} + \frac{\partial w}{\partial y} \right) \tag{2.69}$$

$$\tau_{zz} = -p + 2\eta \frac{\partial w}{\partial z} \qquad \tau_{zx} = \tau_{xz} = \eta \left( \frac{\partial w}{\partial x} + \frac{\partial u}{\partial z} \right)$$

### 2.10.2 圆柱坐标系中的不可压缩流体方程组

1.连续性方程

$$\frac{\partial v_r}{\partial r} + \frac{\partial v_\theta}{r \partial \theta} + \frac{\partial v_z}{\partial z} + \frac{v_r}{r} = 0 \tag{2.70}$$

2.运动方程

$$\rho \left( \frac{\partial v_r}{\partial t} + v_r \frac{\partial v_r}{\partial r} + v_\theta \frac{\partial v_r}{r \partial \theta} + v_z \frac{\partial v_r}{\partial z} - \frac{v_\theta^2}{r} \right) =$$
$$f_r - \frac{\partial p}{\partial r} + \eta \left( \frac{\partial^2 v_r}{\partial r^2} + \frac{1}{r} \frac{\partial v_r}{\partial r} + \frac{\partial^2 v_r}{r^2 \partial \theta^2} + \frac{\partial^2 v_r}{\partial z^2} - \frac{2}{r} \frac{\partial v_\theta}{r \partial \theta} - \frac{v_r}{r^2} \right) \tag{2.71a}$$

$$\rho \left( \frac{\partial v_\theta}{\partial t} + v_r \frac{\partial v_\theta}{\partial r} + v_\theta \frac{\partial v_\theta}{r \partial \theta} + v_z \frac{\partial v_\theta}{\partial z} - \frac{v_r v_\theta}{r} \right) =$$
$$f_\theta - \frac{\partial p}{r \partial \theta} + \eta \left( \frac{\partial^2 v_\theta}{\partial r^2} + \frac{1}{r} \frac{\partial v_\theta}{\partial r} + \frac{\partial^2 v_\theta}{r^2 \partial \theta^2} + \frac{\partial^2 v_\theta}{\partial z^2} + \frac{2}{r} \frac{\partial v_r}{r \partial \theta} - \frac{v_\theta}{r^2} \right) \tag{2.71b}$$

$$\rho \left( \frac{\partial v_z}{\partial t} + v_r \frac{\partial v_z}{\partial r} + v_\theta \frac{\partial v_z}{r \partial \theta} + v_z \frac{\partial v_z}{\partial z} \right) = f_z - \frac{\partial p}{\partial z} + \eta \left( \frac{\partial^2 v_z}{\partial r^2} + \frac{1}{r} \frac{\partial v_z}{\partial r} + \frac{\partial^2 v_z}{r^2 \partial \theta^2} + \frac{\partial^2 v_z}{\partial z^2} \right) \tag{2.71c}$$

3.能量方程

$$\rho \left( \frac{\partial e'}{\partial t} + v_r \frac{\partial e'}{\partial r} + v_\theta \frac{\partial e'}{r \partial \theta} + v_z \frac{\partial e'}{\partial z} \right) = \frac{\partial}{\partial r} \left( K_H \frac{\partial T}{\partial r} \right) + \frac{K_H}{r} \frac{\partial T}{\partial r} + \frac{\partial}{r \partial \theta} \left( K_H \frac{\partial T}{r \partial \theta} \right) +$$
$$\frac{\partial}{\partial z} \left( K_H \frac{\partial T}{\partial z} \right) + \eta \left[ 2 \left( \frac{\partial v_r}{\partial r} \right)^2 + 2 \left( \frac{\partial v_\theta}{r \partial \theta} + \frac{v_r}{r} \right)^2 + 2 \left( \frac{\partial v_z}{\partial z} \right)^2 + \right. \tag{2.72}$$
$$\left. \left( \frac{\partial v_\theta}{\partial r} + \frac{\partial v_r}{r \partial \theta} - \frac{v_\theta}{r} \right)^2 + \left( \frac{\partial v_z}{r \partial \theta} + \frac{\partial v_\theta}{\partial z} \right)^2 + \left( \frac{\partial v_z}{\partial r} + \frac{\partial v_r}{\partial z} \right)^2 \right]$$

4.正应力和切应力

$$\tau_{rr} = -p + 2\eta \frac{\partial v_r}{\partial r} \qquad \tau_{r\theta} = \tau_{\theta r} = \eta \left( \frac{\partial v_r}{r \partial \theta} + \frac{\partial v_\theta}{\partial r} - \frac{v_\theta}{r} \right) \left.\begin{array}{c} \\ \\ \\ \\ \\ \end{array}\right\}$$

$$\tau_{\theta\theta} = -p + 2\eta \left( \frac{\partial v_\theta}{r \partial \theta} + \frac{v_r}{r} \right) \qquad \tau_{\theta z} = \tau_{z\theta} = \eta \left( \frac{\partial v_\theta}{\partial z} + \frac{\partial v_z}{r \partial \theta} \right) \tag{2.73}$$

$$\tau_{zz} = -p + 2\eta \frac{\partial v_z}{\partial z} \qquad \tau_{zr} = \tau_{rz} = \eta \left( \frac{\partial v_z}{\partial r} + \frac{\partial v_r}{\partial z} \right)$$

## 2.10.3　圆球坐标系中的不可压缩流体方程组

### 1.连续性方程

$$\frac{\partial v_r}{\partial r} + 2\frac{v_r}{r} + \frac{\partial v_\theta}{r\partial\theta} + \frac{v_\theta}{r}\cot\theta + \frac{1}{\sin\theta}\frac{\partial v_\varphi}{r\partial\varphi} = 0 \tag{2.74}$$

### 2.运动方程

$$\rho\left(\frac{\partial v_r}{\partial t} + v_r\frac{\partial v_r}{\partial r} + v_\theta\frac{\partial v_r}{r\partial\theta} + \frac{v_\varphi}{\sin\theta}\frac{\partial v_r}{r\partial\varphi} - \frac{v_\theta^2 + v_\varphi^2}{r}\right) =$$

$$f_r - \frac{\partial p}{\partial r} + \eta\left(\frac{\partial^2 v_r}{\partial r^2} + \frac{2}{r}\frac{\partial v_r}{\partial r} + \frac{\partial^2 v_r}{r^2\partial\theta^2} + \frac{\cot\theta}{r}\frac{\partial v_r}{r\partial\theta} + \frac{1}{\sin^2\theta}\frac{\partial^2 v_r}{r^2\partial\varphi^2} - \right. \tag{2.75a}$$

$$\left. \frac{2}{r}\frac{\partial v_\theta}{r\partial\theta} - \frac{2}{r\sin\theta}\frac{\partial v_\varphi}{r\partial\varphi} - \frac{2v_r}{r^2} - \frac{2v_\theta}{r^2}\cot\theta\right)$$

$$\rho\left(\frac{\partial v_\theta}{\partial t} + v_r\frac{\partial v_\theta}{\partial r} + v_\theta\frac{\partial v_\theta}{r\partial\theta} + \frac{v_\varphi}{\sin\theta}\frac{\partial v_\theta}{r\partial\varphi} + \frac{v_r v_\theta}{r} - \frac{v_\varphi^2}{r}\cot\theta\right) =$$

$$f_\theta - \frac{\partial p}{r\partial\theta} + \eta\left(\frac{\partial^2 v_\theta}{\partial r^2} + \frac{2}{r}\frac{\partial v_\theta}{\partial r} + \frac{\partial^2 v_\theta}{r^2\partial\theta^2} + \frac{\cot\theta}{r}\frac{\partial v_\theta}{r\partial\theta} + \frac{1}{\sin^2\theta}\frac{\partial^2 v_\theta}{r^2\partial\varphi^2} + \right. \tag{2.75b}$$

$$\left. \frac{2}{r}\frac{\partial v_r}{r\partial\theta} - \frac{2\cot\theta}{r\sin\theta}\frac{\partial v_\varphi}{r\partial\varphi} - \frac{v_\theta}{r^2\sin^2\theta}\right)$$

$$\rho\left(\frac{\partial v_\varphi}{\partial t} + v_r\frac{\partial v_\varphi}{\partial r} + v_\theta\frac{\partial v_\varphi}{r\partial\theta} + \frac{v_\varphi}{\sin\theta}\frac{\partial v_\varphi}{r\partial\varphi} + \frac{v_r v_\varphi}{r}\cot\theta\right) =$$

$$f_\varphi - \frac{1}{\sin\theta}\frac{\partial p}{r\partial\varphi} + \eta\left(\frac{\partial^2 v_\varphi}{\partial r^2} + \frac{2}{r}\frac{\partial v_\varphi}{\partial r} + \frac{\partial^2 v_\varphi}{r^2\partial\theta^2} + \frac{\cot\theta}{r}\frac{\partial v_\varphi}{r\partial\theta} + \right. \tag{2.75c}$$

$$\left. \frac{1}{\sin^2\theta}\frac{\partial^2 v_\varphi}{r^2\partial\varphi^2} + \frac{2}{r\sin\theta}\frac{\partial v_r}{r\partial\varphi} - \frac{2\cot\theta}{r\sin\theta}\frac{\partial v_\theta}{r\partial\varphi} - \frac{v_\varphi}{r^2\sin^2\theta}\right)$$

### 3.能量方程

$$\rho\left(\frac{\partial e'}{\partial t} + v_r\frac{\partial e'}{\partial r} + v_\theta\frac{\partial e'}{r\partial\theta} + \frac{v_\varphi}{\sin\theta}\frac{\partial e'}{r\partial\varphi}\right) = \frac{\partial}{\partial r}\left(K_H\frac{\partial T}{\partial r}\right) + \frac{1}{\sin\theta}\frac{\partial}{r\partial\theta}\left(K_H\frac{\partial T}{r\partial\theta}\sin\theta\right) +$$

$$\frac{1}{\sin^2\theta}\frac{\partial}{r\partial\varphi}\left(K_H\frac{\partial T}{r\partial\varphi}\right) + \eta\left[2\left(\frac{\partial v_r}{\partial r}\right)^2 + 2\left(\frac{\partial v_\theta}{r\partial\theta} + \frac{v_r}{r}\right)^2 + \right.$$

$$2\left(\frac{1}{\sin\theta}\frac{\partial v_\varphi}{r\partial\varphi} + \frac{v_r}{r} + \frac{v_\theta}{r}\cot\theta\right)^2 + \left(\frac{\partial v_r}{r\partial\theta} + \frac{\partial v_\theta}{\partial r} - \frac{v_\theta}{r}\right)^2 + \tag{2.76}$$

$$\left. \left(\frac{1}{\sin\theta}\frac{\partial v_\theta}{r\partial\varphi} + \frac{\partial v_\varphi}{r\partial\theta} - \frac{v_\varphi}{r}\cot\theta\right)^2 + \left(\frac{\partial v_\varphi}{\partial r} + \frac{1}{\sin\theta}\frac{\partial v_r}{r\partial\varphi} - \frac{v_\varphi}{r}\right)^2\right]$$

### 4.正应力和切应力

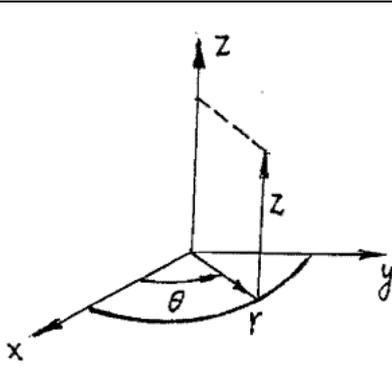

图 2-6 圆柱坐标系

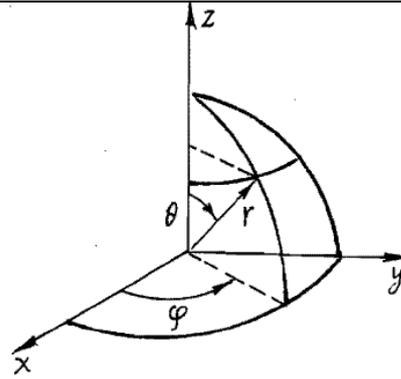

图 2-7 圆球坐标系

$$\left.\begin{aligned}
\tau_{rr} &= -p + 2\eta \frac{\partial v_r}{\partial r} \\
\tau_{\theta\theta} &= -p + 2\eta \left( \frac{\partial v_\theta}{r\partial \theta} + \frac{v_r}{r} \right) \\
\tau_{\varphi\varphi} &= -p + 2\eta \left( \frac{1}{\sin\theta} \frac{\partial v_\varphi}{r\partial \varphi} + \frac{v_r}{r} + \frac{v_\theta}{r} \cot\theta \right) \\
\tau_{r\theta} = \tau_{\theta r} &= \eta \left( \frac{\partial v_r}{r\partial \theta} + \frac{\partial v_\theta}{\partial r} - \frac{v_\theta}{r} \right) \\
\tau_{\theta\varphi} = \tau_{\varphi\theta} &= \eta \left( \frac{1}{r\sin\theta} \frac{\partial v_\theta}{r\partial \varphi} + \frac{\partial v_\varphi}{r\partial \theta} - \frac{v_\varphi}{r} \cot\theta \right) \\
\tau_{\varphi r} = \tau_{r\varphi} &= \eta \left( \frac{\partial v_\varphi}{\partial r} + \frac{1}{\sin\theta} \frac{\partial v_r}{r\partial \varphi} - \frac{v_\varphi}{r} \right)
\end{aligned}\right\} \tag{2.77}$$

## 2.11 流体力学方程的定解条件

分析流场的工作就是求解流体力学方程组。求解这些微分方程得到的只是一般解，包含一些积分常数。确定全部积分常数的值，就能唯一地确定具体问题所需要的解。

定解条件并不是按照数学上的考虑提出的，而是按照实际问题的物理概念给定。所以确定定解条件是建立物理模型的基本内容之一。在大多数情况下，定解条件很难精确表达实际问题的物理状况，原因是要么不能精准了解边界上的物理过程，要么繁难到无法处理。必须采取简化的方法。虽然简化的定解条件决定的结果，仅是近似的解，但在本质上是能反映实际情况的。

定解条件一般包括初始条件和边界条件。初始条件是对时间而言的，与时间相关联的状态，当然不是稳态，而是过渡性质的暂态。边界条件是对空间而言的，它规定流场空间分布的唯一性。

1. 初始条件

选定一个时刻作为初始时间 $t_0$，在此时刻流场参数的具体分布是已知的，即

$$\left.\begin{aligned}
p\big|_{t_0} &= p(x_i; t_0) \\
T\big|_{t_0} &= T(x_i; t_0) \\
\vec{U}\big|_{t_0} &= \vec{U}(x_i; t_0)
\end{aligned}\right\} \tag{2.78}$$

对于与时间无关的稳定状态即定常流动，不需要初始条件。在很多场合中，初始条件的后效影响是逐渐衰减的，直到它的影响小到可以忽略，从而达到与时间无关的所谓定常状态。

2.边界条件

（1）在边界上或在一定位置上，参数的值是已知的，

$$\left. \begin{array}{l} p\big|_{x_i=x_{i0}} = p(x_{i0};t) \\ T\big|_{x_i=x_{i0}} = T(x_{i0};t) \\ \vec{U}\big|_{x_i=x_{i0}} = \vec{U}(x_{i0};t) \end{array} \right\}$$ (2.79)

（2）在边界上或在一定位置上，参数的偏导数是已知的

$$\left. \begin{array}{l} \dfrac{\partial p}{\partial x_i}\bigg|_{x_i=x_{i0}} = c_m \\[2mm] \dfrac{\partial T}{\partial x_i}\bigg|_{x_i=x_{i0}} = c_H \\[2mm] \dfrac{\partial u_j}{\partial x_i}\bigg|_{x_i=x_{i0}} = c_\tau \end{array} \right\}$$ (2.80)

流体参数的偏导数，往往是输运过程的驱动力。上述三种偏导数代表的物理状况是：$\partial p/\partial x_i$ 是流体在 $x_i$ 方向的压力流量的驱动力，$\partial T/\partial x_i$ 代表在 $x_i$ 方向的热传导，$\partial u_j/\partial x_i$ 代表垂直于 $x_i$ 的表面上的粘性剪应力。式（2.80）中，$c_m$、$c_H$、$c_\tau$ 的取值按实际的物理状况确定。例如对于绝热表面，则有 $\partial T/\partial x_i = 0$。

（3）交界面上的连续条件

交界面可以是两种介质的接触面，也可以是一种介质因流动通道几何形状或尺寸大小的突变，以及存在局部的源或汇等而出现的不连续现象。在这类间断面上，流体力学方程不能成立，只能按界面处理。

① 界面上参数值的连续性

在两种介质的交界面上，介质的物性参数，如密度 $\rho$，导热系数 $K_H$，比热容 $c_V$ 等都是突变的。但是不属于物性范畴的力学参数却是连续的。此处所谓的连续就是在交界面上参数的值只有一个，即

$$\left. \begin{array}{l} p_1\big|_{x_i=x_{i0}-0} = p_2\big|_{x_i=x_{i0}+0} \\ T_1\big|_{x_i=x_{i0}-0} = T_2\big|_{x_i=x_{i0}+0} \\ \overline{U_1}\big|_{x_i=x_{i0}-0} = \overline{U_2}\big|_{x_i=x_{i0}+0} \end{array} \right\}$$ (2.81)

上面三式无论界面是两种介质的或是同一介质的，也无论跨过界面时参数变化是否光滑可微，它们都成立。式中的速度条件就是粘性流体的粘附条件，它表示在交界面上既没有相对滑移又无分离。对于理想流，不考虑粘性而在交界面上可以出现相对滑移，以致这一条件成为不必要。

②跨越界面的通量守恒

界面是没有厚度的，它不能积存任何物理量，所以从一侧进入边界的通量是多少，则必然从另一侧出来多少。一般所说的通量指的是质量、热量、动量等，即

$$\left. \begin{array}{l} q_{m1}\big|_{x_{i0}-0} = q_{m2}\big|_{x_{i0}+0} \\ q_{H1}\big|_{x_{i0}-0} = q_{H2}\big|_{x_{i0}+0} \\ \tau_1\big|_{x_{i0}-0} = \tau_2\big|_{x_{i0}+0} \end{array} \right\}$$ (2.82)

可以举出式（2.82）的一些具体例子。

扩散质量流

$$\left. D_1 \frac{\partial c_1}{\partial n}\right|_{x_{i0}-0} = D_2 \left.\frac{\partial c_2}{\partial n}\right|_{x_{i0}+0}$$

压力质量流

$$\rho_1 u_{n1}\big|_{x_{i0}-0} = \rho_2 u_{n2}\big|_{x_{i0}+0}$$

热传导

$$\left. K_{H1}\frac{\partial T_1}{\partial n}\right|_{x_{i0}-0} = K_{H2}\left.\frac{\partial T_2}{\partial n}\right|_{x_{i0}+0}$$ (2.83)

高速流体与固体壁面换热

$$\left. K_H \frac{\partial T_1}{\partial n}\right|_{x_{i0}-0} = h\left(T_\infty - T\big|_{x_{i0}+0}\right)$$

粘性剪切力

$$\left. \eta_1 \frac{\partial \overrightarrow{U_{\tau 1}}}{\partial n}\right|_{x_{i0}-0} = \eta_2 \left.\frac{\partial \overrightarrow{U_{\tau 2}}}{\partial n}\right|_{x_{i0}+0}$$

上述各式中，$n$ 是交界面的法线方向，$\tau$ 是其切线方向。$h$ 称为对流传热系数。$T_\infty$ 是流体的特征温度。

### 2.12 低 Re 数下，圆球在无界粘性流体中的运动

#### 2.12.1 概述

圆球体在流体中的运动有两种基本形式，一种是平移运动，一种是旋转运动。对于铁磁流体而言，其中的固相微粒总是被假定成大小一致的圆球体。由于固相微粒的尺寸极其微小，在一般工业两相流中可以不必考虑这种细微的固体与液相之间的滞后。但是对于铁磁流体，无论是平移速度滞后产生的粘性力，还是旋转速度滞后产生的粘性力矩，都是至关重要的。因为外磁场施加于铁磁流体时，不能磁化的基载液体没有任何磁响应的能力，外磁场只能对磁性固相微粒施加力和力矩。这种力和力矩引起固相微粒与基载液之间在运动上的滞后，伴随而生的就是两者互相作用的粘性力和粘性力矩。对于一个微粒，这种粘性力和粘性力矩很小，但是由于固相微粒的数密度极其巨大，故而粘性力和粘性力矩的总效应非常可观。外磁场施加在固相微粒上的磁力和磁力矩就是通过这种粘性力和粘性力矩传递到整个基载液体上，从而实现对铁磁流体的控制。没有粘性作用，也就谈不上外磁场对铁磁流体的控制，铁磁流体的根本特点就会完全丧失。

设固相微粒的直径 $d_p = 10^{-8}\,\text{m}$，基载液之粘度系数 $\eta_c$ 和密度 $\rho_c$ 分别为 $\eta_c = 4 \times 10^{-3}\,\text{kg(m·s)}^{-1}$，

$\rho_c = 850\,\text{kg/m}^3$，微粒运动速度 $v = 10\,\text{m/s}$，则 Re 数之值为

$$\text{Re} = \frac{\rho_c v d_p}{\eta_c} = 0.021$$

在这样小的 Re 下，显然可以在运动方程中略去惯性项。此外，固相微粒在铁磁流体中所占的体积分量

通常为 5%~10%，微粒之间的距离比直径大许多，可以近似地不考虑微粒群在运动中的互相干扰。这就意味着每个微粒都似乎是在无界流中运动。所以，低 Re 数下圆球在无界流体运动的解，对于铁磁流体相当适合。故本节将对其作较详细的阐释。

## 2.12.2　圆球体在粘性无界流中的低 Re 数平动运动
### 1.运动方程之解

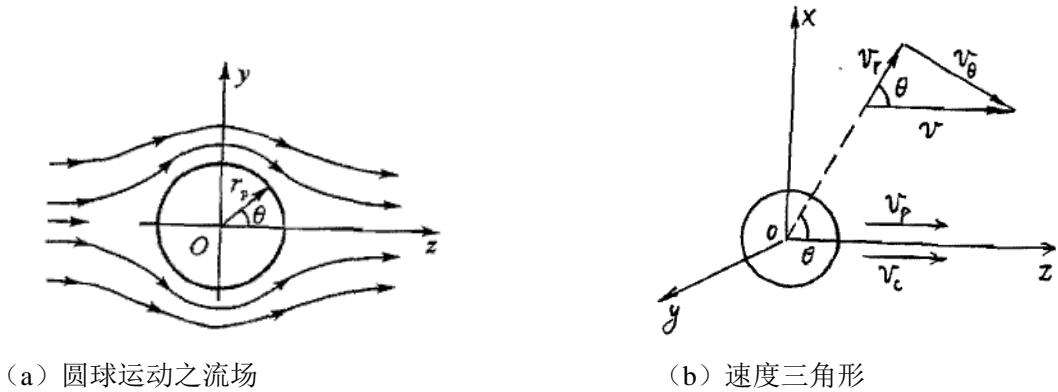

（a）圆球运动之流场　　　　　　　（b）速度三角形

图 2-8　　低 Re 数下圆球体之平动运动

如图 2-8 所示，圆球在无界流中沿 $z$ 轴方向运动。设流动完全对于 $z$ 轴，则在平行于 $xOy$ 坐标面的平行圆之圆周上，流动状况与角度 $\varphi$ 无关，并且没有 $\varphi$ 方向的流速，即 $v_\varphi = 0$。于是方程（2.75c）成为不必要，在方程（2.74）、（2.75a）和（2.75b）中，所有 $\partial(\ )/\partial\varphi = 0$。同时，因 Re 很低，方程（2.75a）和（2.75b）左方的惯性项可全部舍弃，于是有：

连续方程

$$\frac{\partial v_r}{\partial r} + \frac{\partial v_\theta}{r\partial\theta} + \frac{2v_r}{r} + \frac{v_\theta}{r}\cot\theta = 0$$

$r$ 方向的运动方程

$$\frac{\partial p}{\partial r} = \eta_c\left(\frac{\partial^2 v_r}{\partial r^2} + \frac{\partial^2 v_r}{r^2\partial\theta^2} + \frac{2}{r}\frac{\partial v_r}{\partial r} + \frac{\cot\theta}{r}\frac{\partial v_r}{r\partial\theta} - \frac{2}{r}\frac{\partial v_\theta}{r\partial\theta} - \frac{2v_r}{r^2} - \frac{2v_\theta}{r^2}\cot\theta\right)$$

$\theta$ 方向的运动方程

$$\frac{\partial p}{r\partial\theta} = \eta_c\left(\frac{\partial^2 v_\theta}{\partial r^2} + \frac{\partial^2 v_\theta}{r^2\partial\theta^2} + \frac{2}{r}\frac{\partial v_\theta}{\partial r} + \frac{\cot\theta}{r}\frac{\partial v_\theta}{r\partial\theta} + \frac{2}{r}\frac{\partial v_r}{r\partial\theta} - \frac{v_\theta}{r^2\sin^2\theta}\right)$$

式中 $\eta_c$ 是基载液的粘性系数。参照图 2-8（b），边界条件是

$$\left.\begin{array}{llll} r\to\infty & v_r = v_c\cos\theta & v_\theta = -v_c\sin\theta & p = p_0 \\ r = r_p & v_r = v_p\cos\theta & v_\theta = -v_p\sin\theta & p = p_0 + p' \end{array}\right\} \tag{A}$$

式中 $v_c$ 是基载液的速度，$v_p$ 是微粒的运动速度。$r\to\infty$ 的边界条件反映未受扰的状态，$r = r_p$ 的边界条件就是粘附条件。考虑到三个微分方程的线性性质，以及边界条件中有三角正弦和余弦，故取分离变

量形式的试探解：

$$\left.\begin{array}{l} v_r = f_r(r)\cos\theta \\ v_\theta = -f_\theta(r)\sin\theta \\ p = p_0 + \eta_c f_p(r)\cos\theta \end{array}\right\} \tag{2.84}$$

将试探解依次代入到连续方程、$r$ 方向和 $\theta$ 方向的运动方程中，便得

$$f_r' = -\frac{2}{r}(f_r - f_\theta) \tag{B}$$

$$f_p' = f_r'' + \frac{2}{r}f_r' - \frac{4}{r^2}(f_r - f_\theta) \tag{C}$$

$$f_p = rf_\theta'' + 2f_\theta' + \frac{2}{r}(f_r - f_\theta) \tag{D}$$

联立以上三个方程，消去 $f_\theta$ 和 $f_p$，得到 $f_r$ 的四阶常微分方程

$$r^3 f_r'''' + 8r^2 f_r''' + 8rf_r'' - 8f_r' = 0$$

只要令 $y = f_r'$，就可以将此方程化为 Euler 方程，不难求解。但更直接的办法是取试探解为

$$f_r(r) = r^m$$

代入方程中，得

$$m(m-1)(m-2)(m-3) + 8m(m-1)(m-2) + 8m(m-1) - 8m = 0$$

此方程析因式之后，得到

$$m(m-2)(m+1)(m+3) = 0$$

于是得到 $f_r$ 的函数形式是

$$f_r(r) = a + br^2 + \frac{c}{r} + \frac{d}{r^3} \tag{E}$$

将式（E）代入式（B），得到 $f_\theta(r)$ 是

$$f_\theta(r) = a + 2br^2 + \frac{c}{2r} - \frac{d}{2r^3} \tag{F}$$

将式（E）和式（F）代入式（D）之右方，结果为

$$f_p(r) = 10br + \frac{c}{r^2} \tag{G}$$

由边界条件式（A）给出 $f(r)$ 函数的边界条件，以确定积分常数 $a$、$b$、$c$、$d$。由 $r \to \infty$，$f_r(r) = v_c$，

$f_\theta(r) = v_c$，得到积分常数 $a$ 和 $b$ 是

$$a = v_c, \qquad b = 0$$

又由 $r = r_p$，$f_r(r) = v_p$，$f_\theta(r) = v_p$，则式（E）和式（F）得

$$v_p = v_c + \frac{c}{r_p} + \frac{d}{r_p^3}, \qquad v_p = v_c + \frac{c}{2r_p} - \frac{d}{2r_p^3}$$

上两式解出积分常数 $c$ 和 $d$ 为

$$c = \frac{3}{2}r_p(v_p - v_c), \qquad d = -\frac{1}{2}r_p^3(v_p - v_c)$$

于是由式（E）、（F）、（G）得出各 $f(r)$ 的函数为

$$\left.\begin{aligned}
f_r(r) &= v_c + \frac{1}{2}\left(3\frac{r_p}{r} - \frac{r_p^3}{r^3}\right)(v_p - v_c) \\
f_\theta(r) &= v_c + \frac{1}{4}\left(3\frac{r_p}{r} + \frac{r_p^3}{r^3}\right)(v_p - v_c) \\
f_p(r) &= \frac{3}{2}\frac{r_p}{r}(v_p - v_c)
\end{aligned}\right\} \tag{2.85}$$

用式（2.85）代入式（2.84），则有速度和压力的函数形式

$$v_r = v_c\cos\theta + \frac{1}{2}\left(3\frac{r_p}{r} - \frac{r_p^3}{r^3}\right)(v_p - v_c)\cos\theta \tag{2.86a}$$

$$v_\theta = -v_c\sin\theta - \frac{1}{4}\left(3\frac{r_p}{r} + \frac{r_p^3}{r^3}\right)(v_p - v_c)\sin\theta \tag{2.86b}$$

$$p = p_0 + \frac{3}{2}\eta_c\frac{r_p}{r^2}(v_p - v_c)\cos\theta \tag{2.86c}$$

上面三个式子右方的第一项是离圆球无限远处的速度和压力之值，它们都是未受干扰的状态。三个式子的第二项就是扰动值，显然扰动的大小与速度滞后（即 $v_p - v_c$）成正比。

2.圆球平动运动的粘性阻力

在基载液体中取一个球形控制体，它的半径为 $r$ 并且和圆球同心。

（1）正应力

由式（2.77），正应力是

$$\tau_{rr} = -p + 2\eta_c\frac{\partial v_r}{\partial r}$$

用式（2.86a）与式（2.86c）代入，即得

$$\tau_{rr} = -p_0 + \tau'_{rr}\cos\theta \tag{2.87a}$$

式中

$$\tau'_{rr} = \frac{3}{2}\eta_c\frac{r_p}{r^2}\left(1 - 2\frac{r_p^2}{r^2}\right)(v_p - v_c) \tag{2.87b}$$

（2）切应力

由式（2.77），切应力是

$$\tau_{r\theta} = \eta_c \left( \frac{\partial v_r}{r\partial \theta} + \frac{\partial v_\theta}{\partial r} - \frac{v_\theta}{r} \right)$$

用式（2.86a）与式（2.86b）代入，得

$$\tau_{r\theta} = \tau'_{r\theta} \sin \theta \qquad (2.88a)$$

式中

$$\tau'_{r\theta} = \frac{3}{2} \eta_c \frac{r_p^3}{r^4} (v_p - v_c) \qquad (2.88b)$$

（3）球形控制体表面上的阻力

表面应力 $\tau$ 为

$$\tau = \overrightarrow{r^0} \tau_{rr} \overrightarrow{r^0} + \overrightarrow{r^0} \tau_{r\theta} \overrightarrow{\theta^0}$$

在控制体球形表面 $S$ 上积分。如图 2-9 所示，将微元环带面积 $d\overrightarrow{S}$ 上的力在 $\overrightarrow{v^0}$ 和 $\overrightarrow{u^0}$ 方向投影而得两个

方向的分力 $F_u$ 和 $F_v$。

$$F_u = \int_s (d\overrightarrow{S} \cdot \tau) \cdot \overrightarrow{u^0} = \int_s \overrightarrow{r^0} dS \cdot (\overrightarrow{r^0} \tau_{rr} \overrightarrow{r^0} + \overrightarrow{r^0} \tau_{r\theta} \overrightarrow{\theta^0}) \cdot \overrightarrow{u^0} = \int_s (\tau_{rr} \overrightarrow{r^0} + \tau_{r\theta} \overrightarrow{\theta^0}) \cdot \overrightarrow{u^0} dS$$

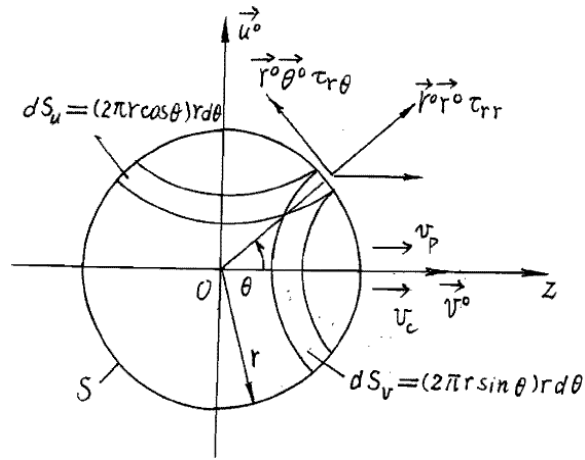

图 2-9  球形控制体的表面应力

由图 2-9 可见：

$$dS_u = (2\pi r \cos \theta) r d\theta, \qquad dS_v = (2\pi r \sin \theta) r d\theta$$

以及 $\overrightarrow{r^0} \cdot \overrightarrow{u^0} = \sin \theta$，$\overrightarrow{\theta^0} \cdot \overrightarrow{u^0} = \cos \theta$，于是粘性阻力在 $\overrightarrow{u^0}$ 方向的分力是

$$F_u = \int_s (\tau_{rr} \sin \theta + \tau_{r\theta} \cos \theta) dS_u$$

将式（2.87a）与式（2.88a）代入积分中，就有

$$F_u = \int_{-\pi/2}^{\pi/2} (-p_0 \sin \theta + \tau'_{rr} \cos \theta \sin \theta + \tau'_{r\theta} \sin \theta \cos \theta)(2\pi r \cos \theta) r d\theta$$

注意到 $p_0$ 是常数，$\tau'_{rr}$ 和 $\tau'_{\theta r}$ 仅是 $r$ 的函数，故上式对 $\theta$ 的及积分结果是零，即

$$F_u = 0$$

这个结果是意料之中的事，因为一开始在运动方程中已假定在 $\varphi$ 方向的对称性。

由图 2-9 可见，$\overline{r^0} \cdot \overline{v^0} = \cos\theta$，$\overline{\theta^0} \cdot \overline{v^0} = -\sin\theta$，粘性力在 $\overline{v^0}$ 方向分量积分式是

$$F_v = \int_s (d\vec{S} \cdot \tau) \cdot \overline{v^0} = \int_s (\tau_{rr}\cos\theta - \tau_{r\theta}\sin\theta)dS_v$$

用式（2.87a）与式（2.88a）代入积分中，就有

$$F_V = \int_0^\pi (-p_0\cos\theta + \tau'_{rr}\cos\theta\cos\theta - \tau'_{r\theta}\sin\theta\sin\theta)(2\pi r\sin\theta)rd\theta$$

显然有关于 $p_0$ 的第一项积分结果是零，只有扰动量 $\tau'_{rr}$ 和 $\tau'_{r\theta}$ 才引起粘性阻力。于是积分的结果为

$$F_v = 2\pi r^2\left(\frac{2}{3}\tau'_{rr} - \frac{4}{3}\tau'_{r\theta}\right)$$

用式（2.87b）与式（2.88b）代入，得

$$F_v = 2\pi r_p\eta_c\left[\left(1 - 2\frac{r_p^2}{r^2}\right) - 2\frac{r_p^2}{r^2}\right](v_p - v_c) \tag{2.89a}$$

在运动的圆球面上，取 $r = r_p$，并注意到 $F_v$ 与 $v_p$、$v_c$ 是同方向的，写成矢量形式：

$$\overrightarrow{F_v} = 2\pi\eta_c r_p(-1-2)(\overrightarrow{v_p} - \overrightarrow{v_c}) = -6\pi\eta_c(\overrightarrow{v_p} - \overrightarrow{v_c}) \tag{2.89b}$$

由上式可见，正应力扰动对粘性阻力的贡献占 1/3，切应力扰动的贡献占 2/3。

定义粘性阻力系数为 $C_d$，且

$$C_d = 6\pi\eta_c r_p, \qquad \overrightarrow{F_v} = -C_d(\overrightarrow{v_p} - \overrightarrow{v_c}) \tag{2.89c}$$

式（2.89c）表明小 Re 数下运动的粘性阻力与速度滞后或相对速度成正比。

### 2.12.3 圆球在粘性无界流中的低 Re 数转动运动
1.运动方程之解

设在粘性流体中，未受扰动处的旋转速度是 $\overrightarrow{\omega_c}$（$2\overrightarrow{\omega_c} = \overrightarrow{\Omega_c} = \nabla \times \overrightarrow{v_c}$），浸入流体中的圆球的转动速度为 $\overrightarrow{\omega_p}$，矢量 $\overrightarrow{\omega_c}$ 与 $\overrightarrow{\omega_p}$ 共线。这种低 Re 数的圆球绕 $z$ 轴转动，可以认为不引起 $r$ 方向和 $\theta$ 方向的扰动。

于是只考虑 $\varphi$ 方向的速度 $v_\varphi$，由于在 $\varphi$ 方向的对称性，故运动方程中对 $\varphi$ 的各阶导数项均为零。于是方程（2.75c）在略去左方的惯性项之后，成为下面比较简单的形式：

$$\frac{\partial^2 v_\varphi}{\partial r^2} + \frac{\partial^2 v_\varphi}{r^2\partial\theta^2} + \frac{2}{r}\frac{\partial v_\varphi}{\partial r} + \frac{\cot\theta}{r}\frac{\partial v_\varphi}{r\partial\theta} - \frac{v_\varphi}{r^2\sin^2\theta} = 0$$

边界条件：　　　　　　　①无界流 $r \to \infty$ ，　　　　 $v_\varphi = v_c = (r\sin\theta)\omega_c$ 　　　　　　　　(H)

　　　　　　　　　②球面粘附条件 $r = r_p$ ，　　　　 $v_\varphi = (r_p\sin\theta)\omega_p$ 　　　　　　　　(I)

设方程之解具有分离变量的形式：

$$v_\varphi = f(r)\sin\theta \tag{J}$$

代入到运动方程中，约去 $\theta$ 的三角函数，得到 $f(r)$ 的常微分方程为

$$r^2 f'' + 2rf' - 2f = 0 \tag{K}$$

这个方程显然可以预料其解为幂函数

$$f(r) = r^m \tag{L}$$

代入常微分方程（K）中，得到 $m$ 的代数方程为

$$m(m-1) + 2m - 2 = 0$$

析因式即得

$$(m-1)(m+2) = 0$$

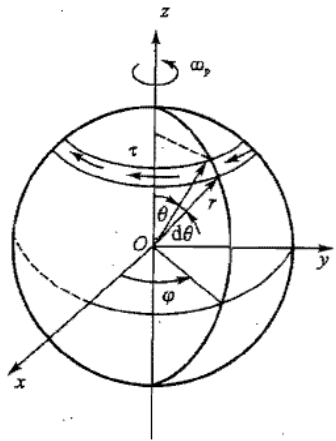

图 2-10　圆球形微粒旋转运动的表面切应力 $\tau$

于是式（L）的函数形式就是

$$f(r) = ar + \frac{b}{r^2} \tag{M}$$

边界条件式（H）给出 $f(r)\big|_{r\to\infty}\sin\theta = (r\sin\theta)\omega_c$ ，得出

$$f(r)\big|_{r\to\infty} = r\omega_c$$

由式（M），当 $r \to \infty$ 时有

$$f(r)\big|_{r\to\infty} = ar$$

将以上两式对照，即得

$$a = \omega_c$$

由式（J）和边界条件式（I）给出圆球表面上

$$v_\varphi\big|_{r=r_p} = f(r_p)\sin\theta = (r_p\sin\theta)\omega_p$$

因而有

$$f(r_p) = r_p\omega_p$$

式（M）给出，当 $r = r_p$ 时，有

$$f(r_p) = r_p\omega_c + \frac{b}{r_p^2}$$

以上两式联立，即得

$$b = r_p^3(\omega_p - \omega_c)$$

将得到的 $a$、$b$ 代入式（M）中，而后由式（J），就有

$$f(r) = r\omega_c + \frac{r_p^3}{r^2}(\omega_p - \omega_c) \tag{2.90a}$$

$$v_\varphi = \left[r\omega_c + \frac{r_p^3}{r^2}(\omega_p - \omega_c)\right]\sin\theta \tag{2.90b}$$

在旋转的圆球表面上，正应力 $\tau_{rr}$ 是通过球中心的，它不会引起力矩。能产生力矩的只是平行圆周上的切应力 $\tau_{r\varphi}$。在式（2.77）中，因 $\varphi$ 方向对称而 $\partial v_r/\partial\varphi = 0$，故有

$$\tau_{r\varphi} = \eta_c\left(\frac{\partial v_\varphi}{\partial r} - \frac{v_\varphi}{r}\right)$$

将式（2.90b）代入，得到

$$\tau_{r\varphi} = -\tau'_{r\varphi}\sin\theta \tag{2.91a}$$

其中右方的 $\tau'_{r\varphi}$ 是

$$\tau'_{r\varphi} = 3\eta_c\frac{r_p^3}{r^3}(\omega_p - \omega_c) \tag{2.91b}$$

由图 2-10 可见，微元面 $dS$ 上的粘性力矩 $d\overrightarrow{L_\tau} = \overrightarrow{r_c} \times d\overrightarrow{F_\varphi}$，$\overrightarrow{r_c}$ 是垂直于 $\vec{k}$（即 $z$ 坐标轴）的平行圆的半径，$\overrightarrow{r_c} = \overrightarrow{r_c^0}(r\sin\theta)$。于是粘性力矩为

$$\overrightarrow{L_\tau} = \int_s \overrightarrow{r_c} \times d\overrightarrow{F_\varphi} = \int_s \overrightarrow{r_c^0}(r\sin\theta) \times [\overrightarrow{r^0}dS \cdot (\overrightarrow{r^0}\tau_{r\varphi}\overrightarrow{\varphi^0})]$$

参照图 2-10，$dS = (2\pi r\sin\theta)rd\theta$，代入上式并且使用式（2.91a），就得

$$\overrightarrow{L_\tau} = \vec{k}\int_0^\pi (r\sin\theta)(-\tau'_{r\varphi}\sin\theta)(2\pi r\sin\theta)rd\theta$$

积分结果是

$$\overline{L_\tau} = -\vec{k} 2\pi r^3 \tau'_{r\varphi} \left(\frac{4}{3}\right)$$

用式（2.91b）代入，就有

$$\overline{L_\tau} = -\vec{k} 8\pi \eta_c r_p^3 (\omega_p - \omega_c) = -C_\tau (\overrightarrow{\omega_p} - \overrightarrow{\omega_c}) \tag{2.92a}$$

式中 $C_\tau$ 是粘性阻力矩系数

$$C_\tau = 8\pi \eta_c r_p^3 = 6\eta_c V_{p1} \tag{2.92b}$$

将式（2.92b）代入式（1.16）关于 $t_B$ 的定义式中，就得

$$t_B = \frac{3\eta_c V_{p1}}{k_0 T} \tag{2.93}$$

## 2.13　正交曲线坐标系中的流体力学方程
### 2.13.1　概述

就参照系而言，直角坐标系是最简单明了的，它的主要简便之处，就是三个作坐标轴的单位矢量 $\vec{i}$，

$\vec{j}$，$\vec{k}$ 都是常矢。任何矢量在直角坐标系的三个坐标轴上的分量，只有模的变化而无方向变化。但是，在求解某些实际问题中，例如分析圆球在流体中的低 Re 数运动，使用圆球坐标系就远比直角坐标系方便。

流体动力学方程的原始形式，是在直角坐标系中，用微元控制体分析方法建立的，而后将这种原始形式改换成描述场（流场是场的一种）变化的微分函数，如梯度、散度、旋度以及它们的组合，这样组成的流体动力学方程组，形式简单而且普适于改换到其他的坐标系。本节主要的目的是阐明如何将这样的方程转到一种一般的正交坐标系上，以经过不复杂的手续就得到所需要的坐标系。

这个一般的坐标系就是正交曲线坐标系。直角坐标系、圆柱坐标系、圆球坐标系等都是它的特殊情况。曲线坐标系与直角坐标系最大的不同点在于曲线坐标的单位矢量不是常矢，虽然它的模是常数 1，但方向是变化的。曲线坐标系的坐标变量与坐标系的曲线弧长之间的系数称为尺度系数或 Lame 系数。在求出描述流场的微分函数梯度、散度、旋度等在曲线坐标系中的表达形式之后，就可以通过已知的 Lame 系数，迅速地换算到所需要的特殊坐标系中，于是得到在该坐标系中的流体力学方程组。

### 2.13.2　正交曲线坐标系

如图 2-11 所示，在直角坐标系 $Oxyz$ 中有三族正交的曲面，其方程是

$$\left.\begin{array}{l} x_1(x, y, z) = C_1 \\ x_2(x, y, z) = C_2 \\ x_3(x, y, z) = C_3 \end{array}\right\} \tag{2.94}$$

当 $C_1$，$C_2$，$C_3$ 确定以后，这三张曲面也就被确定。以此三张曲面为坐标面，则构成一个曲面坐标体系。

三张曲面的交线就是曲线坐标轴。图 2-11 表示了这个曲线坐标系。

显然，直线坐标系 $Oxyz$ 与曲线坐标系 $O'x_1x_2x_3$ 之间有函数关系，由式（2.94）有

$$\left.\begin{array}{l} x_1 = x_1(x, y, z) \\ x_2 = x_2(x, y, z) \\ x_3 = x_3(x, y, z) \end{array}\right\} \tag{2.95a}$$

其反函数是

$$\left.\begin{array}{l} x = x(x_1, x_2, x_3) \\ y = y(x_1, x_2, x_3) \\ z = z(x_1, x_2, x_3) \end{array}\right\} \tag{2.95b}$$

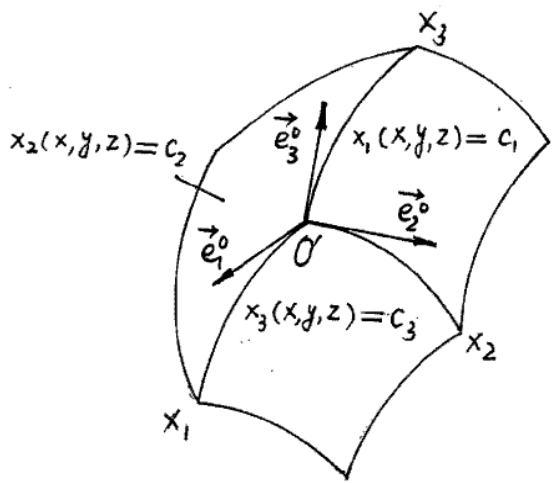

图 2-11　曲线坐标系

设空间一点从 $P$ 移动到 $P+dP$，其经过的弧长为 $dL$。若点 $P$ 的矢径是 $\vec{r}$，点 $P+dP$ 的矢径为 $\vec{r}+d\vec{r}$。

将 $d\vec{r}$ 分解到 $x_1$，$x_2$，$x_3$ 的曲线坐标轴上，则弧长 $dL$ 在三个曲线坐标轴上的投影弧长分别为 $dL_1$，$dL_2$，$dL_3$，于是在坐标系 $O'x_1x_2x_3$ 中有

$$d\vec{r} = \vec{e_1^0}dL_1 + \vec{e_2^0}dL_2 + \vec{e_3^0}dL_3 \tag{2.96a}$$

或缩写成

$$d\vec{r} = \vec{e_i^0}dL_i = \vec{e_i^0}\frac{dL_i}{dx_i}dx_i, \qquad i = 1, 2, 3 \tag{2.96b}$$

再将 $d\vec{r}$ 写成微分形式

$$d\vec{r} = \frac{\partial \vec{r}}{\partial L_1}dL_1 + \frac{\partial \vec{r}}{\partial L_2}dL_2 + \frac{\partial \vec{r}}{\partial L_3}dL_3 \tag{2.96c}$$

或缩写成

$$d\vec{r} = \frac{\partial \vec{r}}{\partial L_i} dL_i \tag{2.96d}$$

比较式（2.96a）和（2.96c），或比较式（2.96b）和（2.96d），即有

$$\vec{e_i^0} = \frac{\partial \vec{r}}{\partial L_i} \tag{2.97}$$

若将 $d\vec{r}$ 在坐标系 $Oxyz$ 内分解，则有

$$d\vec{r} = \vec{i}\,dx + \vec{j}\,dy + \vec{k}\,dz \tag{2.98a}$$

将式（2.95b）两边取微分，得

$$\left. \begin{aligned} dx &= \frac{\partial x}{\partial x_1} dx_1 + \frac{\partial x}{\partial x_2} dx_2 + \frac{\partial x}{\partial x_3} dx_3 \\ dy &= \frac{\partial y}{\partial x_1} dx_1 + \frac{\partial y}{\partial x_2} dx_2 + \frac{\partial y}{\partial x_3} dx_3 \\ dz &= \frac{\partial z}{\partial x_1} dx_1 + \frac{\partial z}{\partial x_2} dx_2 + \frac{\partial z}{\partial x_3} dx_3 \end{aligned} \right\} \tag{2.98b}$$

将（2.98b）代入式（2.98a）中，而后按 $dx_1$、$dx_2$、$dx_3$ 因子整合，就得

$$d\vec{r} = \left( \vec{i}\frac{\partial x}{\partial x_1} + \vec{j}\frac{\partial y}{\partial x_1} + \vec{k}\frac{\partial z}{\partial x_1} \right) dx_1 + \left( \vec{i}\frac{\partial x}{\partial x_2} + \vec{j}\frac{\partial y}{\partial x_2} + \vec{k}\frac{\partial z}{\partial x_2} \right) dx_2 + \left( \vec{i}\frac{\partial x}{\partial x_3} + \vec{j}\frac{\partial y}{\partial x_3} + \vec{k}\frac{\partial z}{\partial x_3} \right) dx_3 \tag{2.98c}$$

引用记号

$$\overrightarrow{H_i} = \vec{i}\frac{\partial x}{\partial x_i} + \vec{j}\frac{\partial y}{\partial x_i} + \vec{k}\frac{\partial z}{\partial x_i} \tag{2.99}$$

于是式（2.98c）缩写成

$$d\vec{r} = \overrightarrow{H_i} dx_i \tag{2.100}$$

将式（2.96b）与式（2.100）相比较，可见

$$\overrightarrow{H_i} = \vec{e_i^0} \frac{dL_i}{dx_i} \tag{2.101a}$$

由式（2.101a）知道，矢量 $\overrightarrow{H_i}$ 的方向是 $\vec{e_i^0}$，而其模由式（2.98c）或式（2.99）得

$$H_i = \frac{dL_i}{dx_i} = \sqrt{\left(\frac{\partial x}{\partial x_i}\right)^2 + \left(\frac{\partial y}{\partial x_i}\right)^2 + \left(\frac{\partial z}{\partial x_i}\right)^2} \tag{2.101b}$$

于是，正交曲线坐标中的坐标轴曲线弧长 $dL_i$ 与直角坐标系中坐标轴的直线长度 $dx_i$ 的比值就是 $H_i$，即 Lame 系数或尺度系数。

在曲线坐标系中的坐标弧长元素直接由式（2.101b）得

$$dL_i = H_i dx_i = \left[ \sqrt{\left(\frac{\partial x}{\partial x_i}\right)^2 + \left(\frac{\partial y}{\partial x_i}\right)^2 + \left(\frac{\partial z}{\partial x_i}\right)^2} \right] dx_i \tag{2.102a}$$

曲线坐标系中，$x_i = \text{const}$ 的曲面面积元素 $d\vec{S_i}$ 是

$$d\vec{S_i} = d\overline{L_j} \times d\overline{L_k} = (\overrightarrow{e_j^0} H_j dx_j) \times (\overrightarrow{e_k^0} H_k dx_k)$$

此即

$$d\vec{S_i} = \overrightarrow{e_i^0} H_j H_k dx_j dx_k \tag{2.102b}$$

曲面的面积元素是矢量，它的方向就是其法线方向。曲线坐标系中的体积元素是数量，即

$$dV = dL_1 dL_2 dL_3 = H_1 H_2 H_3 dx_1 dx_2 dx_3 \tag{2.102c}$$

### 2.13.3　曲线坐标系中，单位矢量对位置坐标的导数

曲线坐标系的单位矢量，由于方向变化而非常矢，所以对位置坐标的导数不等于零。利用正交的特性，有

$$\overrightarrow{e_i^0} \cdot \overrightarrow{e_i^0} = 1, \qquad \overrightarrow{e_i^0} \cdot \overrightarrow{e_j^0} = 0, \quad i \neq j \tag{2.103a}$$

$$\overrightarrow{e_i^0} \times \overrightarrow{e_i^0} = 0, \qquad \overrightarrow{e_i^0} \times \overrightarrow{e_j^0} = \overrightarrow{e_k^0}, \quad i \neq j \neq k \tag{2.103b}$$

$$1. \cdot \frac{\partial \overrightarrow{e_i^0}}{\partial x_j}$$

将式（2.103a）的第二式乘以 $H_i H_j$，写成

$$(H_i \overrightarrow{e_i^0}) \cdot (H_j \overrightarrow{e_j^0}) = 0$$

上式对 $x_k$ 取导数，并且限定 $i \neq j \neq k$，则有

$$\frac{\partial}{\partial x_k}[(H_i \overrightarrow{e_i^0}) \cdot (H_j \overrightarrow{e_j^0})] = (H_i \overrightarrow{e_i^0}) \cdot \frac{\partial}{\partial x_k}(H_j \overrightarrow{e_j^0}) + (H_j \overrightarrow{e_j^0}) \cdot \frac{\partial}{\partial x_k}(H_i \overrightarrow{e_i^0}) = 0 \tag{A}$$

将式（A）轮换下标，有

$$(H_j \overrightarrow{e_j^0}) \cdot \frac{\partial}{\partial x_i}(H_k \overrightarrow{e_k^0}) + (H_k \overrightarrow{e_k^0}) \cdot \frac{\partial}{\partial x_i}(H_j \overrightarrow{e_j^0}) = 0 \tag{B}$$

$$(H_k \overrightarrow{e_k^0}) \cdot \frac{\partial}{\partial x_j}(H_i \overrightarrow{e_i^0}) + (H_i \overrightarrow{e_i^0}) \cdot \frac{\partial}{\partial x_j}(H_k \overrightarrow{e_k^0}) = 0 \tag{C}$$

将以上三式相加，则有

$$(H_i \overrightarrow{e_i^0}) \cdot \left[ \frac{\partial}{\partial x_j}(H_k \overrightarrow{e_k^0}) + \frac{\partial}{\partial x_k}(H_j \overrightarrow{e_j^0}) \right] + (H_j \overrightarrow{e_j^0}) \cdot \left[ \frac{\partial}{\partial x_k}(H_i \overrightarrow{e_i^0}) + \frac{\partial}{\partial x_i}(H_k \overrightarrow{e_k^0}) \right] +$$
$$(H_k \overrightarrow{e_k^0}) \cdot \left[ \frac{\partial}{\partial x_i}(H_j \overrightarrow{e_j^0}) + \frac{\partial}{\partial x_j}(H_i \overrightarrow{e_i^0}) \right] = 0 \tag{D}$$

将式（2.102a）代入式（2.97），就有

$$H_i \overrightarrow{e_i^0} = \frac{\partial \vec{r}}{\partial x_i}$$

用于式（D）的第一个中括号之内，就有

$$\frac{\partial}{\partial x_j}(H_k \overrightarrow{e_k^0}) = \frac{\partial}{\partial x_j}\left(\frac{\partial \vec{r}}{\partial x_k}\right), \qquad \frac{\partial}{\partial x_k}(H_j \overrightarrow{e_j^0}) = \frac{\partial}{\partial x_k}\left(\frac{\partial \vec{r}}{\partial x_j}\right)$$

因为微分与次序无关，上两式之右方是相等的，故左方亦相等，即

$$\frac{\partial}{\partial x_j}(H_k \overrightarrow{e_k^0}) = \frac{\partial}{\partial x_k}(H_j \overrightarrow{e_j^0}) \tag{2.104}$$

由式（2.104）可知式（D）中，各中括号之内两项是相等的，于是有

$$(H_i \overrightarrow{e_i^0}) \cdot \frac{\partial}{\partial x_j}(H_k \overrightarrow{e_k^0}) + (H_j \overrightarrow{e_j^0}) \cdot \frac{\partial}{\partial x_i}(H_k \overrightarrow{e_k^0}) + (H_k \overrightarrow{e_k^0}) \cdot \frac{\partial}{\partial x_i}(H_j \overrightarrow{e_j^0}) = 0 \tag{E}$$

将式（E）与式（B）相减，就得

$$(H_i \overrightarrow{e_i^0}) \cdot \frac{\partial}{\partial x_j}(H_k \overrightarrow{e_k^0}) = 0$$

展开左方，有

$$(H_i \overrightarrow{e_i^0}) \cdot \left(\overrightarrow{e_k^0} \frac{\partial H_k}{\partial x_j} + H_k \frac{\partial \overrightarrow{e_k^0}}{\partial x_j}\right) = 0$$

因 $\overrightarrow{e_i^0} \cdot \overrightarrow{e_k^0} = 0$，故只有

$$\overrightarrow{e_i^0} \cdot \frac{\partial \overrightarrow{e_k^0}}{\partial x_j} = 0 \tag{2.105a}$$

式（2.105a）表明 $\frac{\partial \overrightarrow{e_k^0}}{\partial x_j}$ 与 $\overrightarrow{e_i^0}$ 垂直。又由

$$\frac{\partial}{\partial x_j}(\overrightarrow{e_k^0} \cdot \overrightarrow{e_k^0}) = \frac{\partial}{\partial x_j}(1) = 0$$

或将左方展开，即

$$\frac{\partial}{\partial x_j}(\overrightarrow{e_k^0} \cdot \overrightarrow{e_k^0}) = \overrightarrow{e_k^0} \cdot \frac{\partial \overrightarrow{e_k^0}}{\partial x_j} + \overrightarrow{e_k^0} \cdot \frac{\partial \overrightarrow{e_k^0}}{\partial x_j} = 2\overrightarrow{e_k^0} \cdot \frac{\partial \overrightarrow{e_k^0}}{\partial x_j}$$

从而有

$$\overrightarrow{e_k^0} \cdot \frac{\partial \overrightarrow{e_k^0}}{\partial x_j} = 0 \tag{2.105b}$$

此式表明 $\dfrac{\partial \overrightarrow{e_k^0}}{\partial x_j}$ 与 $\overrightarrow{e_k^0}$ 垂直。

既然 $\dfrac{\partial \overrightarrow{e_k^0}}{\partial x_j}$ 同时与 $\overrightarrow{e_k^0}$ 和 $\overrightarrow{e_i^0}$ 垂直，所以它的方向必定是 $\overrightarrow{e_j^0}$。将式（2.104）两边同时展开，得

$$H_k \frac{\partial \overrightarrow{e_k^0}}{\partial x_j} + \overrightarrow{e_k^0}\frac{\partial H_k}{\partial x_j} = \overrightarrow{e_j^0}\frac{\partial H_j}{\partial x_k} + H_j\frac{\partial \overrightarrow{e_j^0}}{\partial x_k}$$

上式左边第一项和右边第一项都是 $\overrightarrow{e_j^0}$ 方向，而左边第二项和右边第二项都是 $\overrightarrow{e_k^0}$ 方向。按方向拆开就有

$$\frac{\partial \overrightarrow{e_k^0}}{\partial x_j} = \overrightarrow{e_j^0} \cdot \frac{\partial H_j}{H_k \partial x_k}, \qquad \frac{\partial \overrightarrow{e_j^0}}{\partial x_k} = \overrightarrow{e_k^0}\frac{\partial H_k}{H_j \partial x_j} \tag{2.106}$$

2. $\dfrac{\partial \overrightarrow{e_i^0}}{\partial x_i}$

由式（2.103b）

$$\overrightarrow{e_i^0} = \overrightarrow{e_j^0} \times \overrightarrow{e_k^0}$$

两边对 $x_i$ 取导数

$$\frac{\partial \overrightarrow{e_i^0}}{\partial x_i} = \frac{\partial}{\partial x_i}(\overrightarrow{e_j^0} \times \overrightarrow{e_k^0}) = \overrightarrow{e_j^0} \times \frac{\partial \overrightarrow{e_k^0}}{\partial x_i} + \frac{\partial \overrightarrow{e_j^0}}{\partial x_i} \times \overrightarrow{e_k^0}$$

用式（2.106）代入

$$\frac{\partial \overrightarrow{e_i^0}}{\partial x_i} = \overrightarrow{e_j^0} \times \overrightarrow{e_i^0}\frac{\partial H_i}{H_k \partial x_k} + \overrightarrow{e_i^0}\frac{\partial H_i}{H_j \partial x_j} \times \overrightarrow{e_k^0}$$

上式给出 $\overrightarrow{e_i^0}$ 对 $x_i$ 的导数是

$$\frac{\partial \overrightarrow{e_i^0}}{\partial x_i} = -\overrightarrow{e_k^0}\frac{\partial H_i}{H_k \partial x_k} - \overrightarrow{e_j^0}\frac{\partial H_i}{H_j \partial x_j} \tag{2.107}$$

式（2.106）与式（2.107）给出正交曲线坐标系的单位矢量对坐标的导数。其中下标 $i$，$j$，$k$ 可以按 1，2，3 循环轮换取值，并且限定 $i \neq j \neq k$。

2.13.4   描述场的微分函数在正交曲线坐标系中的形式

1.梯度

$$\nabla \varphi = \overrightarrow{e_i^0}\frac{\partial \varphi}{\partial L_i} + \overrightarrow{e_j^0}\frac{\partial \varphi}{\partial L_j} + \overrightarrow{e_k^0}\frac{\partial \varphi}{\partial L_k} = \frac{\overrightarrow{e_i^0}}{H_i}\frac{\partial \varphi}{\partial x_i} + \frac{\overrightarrow{e_j^0}}{H_j}\frac{\partial \varphi}{\partial x_j} + \frac{\overrightarrow{e_k^0}}{H_k}\frac{\partial \varphi}{\partial x_k} \tag{2.108a}$$

式中 $\varphi$ 是数性函数，它沿曲线坐标系的坐标轴弧长的变化就是梯度。式（2.108a）可以缩写成简单形式

为

$$\nabla\varphi = \frac{\overrightarrow{e_i^0}}{H_i}\frac{\partial\varphi}{\partial x_i} \tag{2.108b}$$

曲线坐标系中 Hamilton 算子 $\nabla$ 的缩写形式是

$$\nabla = \frac{\overrightarrow{e_i^0}}{H_i}\frac{\partial}{\partial x_i} \tag{2.109}$$

若 $\varphi$ 是坐标 $x_i$，则

$$\nabla x_i = \frac{\overrightarrow{e_i^0}}{H_i} \tag{2.110}$$

由矢量恒等式 $\nabla\times(\nabla\varphi) = 0$，则有

$$\nabla\times(\nabla x_i) = \nabla\times\left(\frac{\overrightarrow{e_i^0}}{H_i}\right) = 0 \tag{2.111}$$

2.散度

由式（2.103b），$\overrightarrow{e_i^0} = \overrightarrow{e_j^0}\times\overrightarrow{e_k^0}$，两边同时除以 $H_jH_k$，则有

$$\frac{\overrightarrow{e_i^0}}{H_jH_k} = \frac{\overrightarrow{e_j^0}}{H_j}\times\frac{\overrightarrow{e_k^0}}{H_k}$$

两边取散度

$$\nabla\cdot\left(\frac{\overrightarrow{e_i^0}}{H_jH_k}\right) = \nabla\cdot\left(\frac{\overrightarrow{e_j^0}}{H_j}\times\frac{\overrightarrow{e_k^0}}{H_k}\right)$$

右方按矢量恒等式展开

$$\nabla\cdot\left(\frac{\overrightarrow{e_i^0}}{H_jH_k}\right) = \left(\frac{\overrightarrow{e_k^0}}{H_k}\right)\cdot\nabla\times\left(\frac{\overrightarrow{e_j^0}}{H_j}\right) - \left(\frac{\overrightarrow{e_j^0}}{H_j}\right)\cdot\nabla\times\left(\frac{\overrightarrow{e_k^0}}{H_k}\right)$$

由式（2.111）立即可见上式之右方是零，故

$$\nabla\cdot\left(\frac{\overrightarrow{e_i^0}}{H_jH_k}\right) = 0 \tag{2.112}$$

设有矢量 $\overrightarrow{V}$，其散度 $\nabla\cdot\overrightarrow{V}$ 在正交曲线坐标系中的形式，按式（2.109）写成

$$\nabla\cdot\overrightarrow{V} = \left(\overrightarrow{e_i^0}\frac{\partial}{H_i\partial x_i} + \overrightarrow{e_j^0}\frac{\partial}{H_j\partial x_j} + \overrightarrow{e_k^0}\frac{\partial}{H_k\partial x_k}\right)\cdot(\overrightarrow{e_i^0}V_i + \overrightarrow{e_j^0}V_j + \overrightarrow{e_k^0}V_k) =$$

$$\left(\frac{\overrightarrow{e_i^0}}{H_i}\frac{\partial}{\partial x_i} + \frac{\overrightarrow{e_j^0}}{H_j}\frac{\partial}{\partial x_j} + \frac{\overrightarrow{e_k^0}}{H_k}\frac{\partial}{\partial x_k}\right)\cdot\left(\frac{\overrightarrow{e_i^0}}{H_jH_k}H_jH_kV_i + \frac{\overrightarrow{e_j^0}}{H_kH_i}H_kH_iV_j + \frac{\overrightarrow{e_k^0}}{H_iH_j}H_iH_jV_k\right)$$

注意到式（2.112），就有

$$\nabla \cdot \vec{V} = \frac{1}{H_i H_j H_k}\left[\frac{\partial}{\partial x_i}(H_j H_k V_i) + \frac{\partial}{\partial x_j}(H_k H_i V_j) + \frac{\partial}{\partial x_k}(H_i H_j V_k)\right] \tag{2.113a}$$

式中 $i$, $j$, $k$ 按 1, 2, 3 循环取值，并且限定 $i \neq j \neq k$。于是式（2.113a）可缩写为

$$\nabla \cdot \vec{V} = \frac{1}{H_i H_j H_k}\left[\frac{\partial}{\partial x_i}(H_j H_k V_i)\right] \tag{2.113b}$$

3. 旋度

将矢量 $\vec{V}$ 写成下面的形式

$$\vec{V} = \frac{\overrightarrow{e_i^0}}{H_i}(H_i V_i) + \frac{\overrightarrow{e_j^0}}{H_j}(H_j V_j) + \frac{\overrightarrow{e_k^0}}{H_k}(H_k V_k)$$

取旋度得

$$\nabla \times \vec{V} = \nabla \times \left[\frac{\overrightarrow{e_i^0}}{H_i}(H_i V_i) + \frac{\overrightarrow{e_j^0}}{H_j}(H_j V_j) + \frac{\overrightarrow{e_k^0}}{H_k}(H_k V_k)\right]$$

按矢量恒等式展开上式之右方

$$\nabla \times \vec{V} = (H_i V_i)\nabla \times \left(\frac{\overrightarrow{e_i^0}}{H_i}\right) + \nabla(H_i V_i) \times \left(\frac{\overrightarrow{e_i^0}}{H_i}\right) + (H_j V_j)\nabla \times \left(\frac{\overrightarrow{e_j^0}}{H_j}\right) + \nabla(H_j V_j) \times \left(\frac{\overrightarrow{e_j^0}}{H_j}\right) + $$

$$(H_k V_k)\nabla \times \left(\frac{\overrightarrow{e_k^0}}{H_k}\right) + (H_k V_k) \times \left(\frac{\overrightarrow{e_k^0}}{H_k}\right)$$

注意到式（2.111），上式成为

$$\nabla \times \vec{V} = \nabla(H_i V_i) \times \left(\frac{\overrightarrow{e_i^0}}{H_i}\right) + \nabla(H_j V_j) \times \left(\frac{\overrightarrow{e_j^0}}{H_j}\right) + \nabla(H_k V_k) \times \left(\frac{\overrightarrow{e_k^0}}{H_k}\right) \tag{F}$$

先求出右方的第一项，然后用下标轮换求出第二项和第三项。利用式（2.109），有

$$\nabla(H_i V_i) \times \left(\frac{\overrightarrow{e_i^0}}{H_i}\right) = \left(\frac{\overrightarrow{e_i^0}}{H_i}\frac{\partial}{\partial x_i} + \frac{\overrightarrow{e_j^0}}{H_j}\frac{\partial}{\partial x_j} + \frac{\overrightarrow{e_k^0}}{H_k}\frac{\partial}{\partial x_k}\right)(H_i V_i) \times \left(\frac{\overrightarrow{e_i^0}}{H_i}\right) = $$

$$\left(-\frac{\overrightarrow{e_k^0}}{H_j H_i}\frac{\partial}{\partial x_j} + \frac{\overrightarrow{e_j^0}}{H_k H_i}\frac{\partial}{\partial x_k}\right)(H_i V_i) = \frac{1}{H_i H_j H_k}\left(H_j \overrightarrow{e_j^0}\frac{\partial}{\partial x_k} - H_k \overrightarrow{e_k^0}\frac{\partial}{\partial x_j}\right)(H_i V_i) \tag{G}$$

将式（G）代入式（F）的右方第一项，并依下标轮换写出第二项和第三项，就得

$$\nabla \times \vec{V} = \frac{1}{H_i H_j H_k}\left[\left(H_j \overrightarrow{e_j^0}\frac{\partial}{\partial x_k} - H_k \overrightarrow{e_k^0}\frac{\partial}{\partial x_j}\right)(H_i V_i) + \left(H_k \overrightarrow{e_k^0}\frac{\partial}{\partial x_i} - H_i \overrightarrow{e_i^0}\frac{\partial}{\partial x_k}\right)(H_j V_j) + \right.$$

$$\left. \left(H_i \overrightarrow{e_i^0}\frac{\partial}{\partial x_j} - H_j \overrightarrow{e_j^0}\frac{\partial}{\partial x_i}\right)(H_k V_k)\right] \tag{2.114a}$$

上式按方向归项成为

$$\nabla \times \vec{V} = \frac{1}{H_i H_j H_k} \left\{ \left[ \frac{\partial (H_k V_k)}{\partial x_j} - \frac{\partial (H_j V_j)}{\partial x_k} \right] H_i \overrightarrow{e_i^0} + \left[ \frac{\partial (H_i V_i)}{\partial x_k} - \frac{\partial (H_k V_k)}{\partial x_i} \right] H_j \overrightarrow{e_j^0} + \right.$$

$$\left. \left[ \frac{\partial (H_j V_j)}{\partial x_i} - \frac{\partial (H_i V_i)}{\partial x_j} \right] H_k \overrightarrow{e_k^0} \right\} \tag{2.114b}$$

上式下标 $i$，$j$，$k$ 按 1，2，3 循环取值，并限定 $i \neq j \neq k$，则可缩写成

$$\nabla \times \vec{V} = \frac{1}{H_i H_j H_k} \left[ \frac{\partial (H_k V_k)}{\partial x_j} - \frac{\partial (H_j V_j)}{\partial x_k} \right] H_i \overrightarrow{e_i^0} \tag{2.114c}$$

4. Laplace 算子 $\nabla^2$

在式（1.113b）中，用 $\nabla$ 取代 $\vec{V}$，并且注意式（2.109），就有

$$\nabla^2 = \nabla \cdot \nabla = \frac{1}{H_i H_j H_k} \left[ \frac{\partial}{\partial x_i} \left( \frac{H_j H_k}{H_i} \frac{\partial}{\partial x_i} \right) \right] \tag{2.115}$$

式中，$i$，$j$，$k$ 按 1，2，3 循环取值，并且 $i \neq j \neq k$。

5. $\nabla^2 \varphi$

直接使用式（2.115）就得

$$\nabla^2 \varphi = \frac{1}{H_i H_j H_k} \left[ \frac{\partial}{\partial x_i} \left( H_j H_k \frac{\partial \varphi}{H_i \partial x_i} \right) \right] \tag{2.116}$$

式（2.116）右方的求和约定与式（2.115）相同。

6. $\nabla^2 \vec{V}$

由矢量恒等式

$$\nabla^2 \vec{V} = \nabla(\nabla \cdot \vec{V}) - \nabla \times (\nabla \times \vec{V})$$

对右方第一项，先使用式（2.113b），而后使用式（2.108a）；对右方第二项两次使用式（2.114c），最后按方向整理归项，就得

$$\nabla^2 \vec{V} = \left\{ \frac{1}{H_i} \frac{\partial}{\partial x_i} \left[ \frac{1}{H_i H_j H_k} \left( \frac{\partial H_j H_k V_i}{\partial x_i} + \frac{\partial H_k H_i V_j}{\partial x_j} + \frac{\partial H_i H_j V_k}{\partial x_k} \right) \right] - \right.$$

$$\frac{1}{H_j H_k} \frac{\partial}{\partial x_j} \left[ \frac{H_k}{H_i H_j} \left( \frac{\partial H_j V_j}{\partial x_i} - \frac{\partial H_i V_i}{\partial x_j} \right) \right] - \frac{1}{H_k H_j} \frac{\partial}{\partial x_k} \left[ \frac{H_j}{H_k H_i} \left( \frac{\partial H_i V_i}{\partial x_k} - \frac{\partial H_k V_k}{\partial x_i} \right) \right] \right\} \overrightarrow{e_i^0} +$$

$$\left\{ \frac{1}{H_j} \frac{\partial}{\partial x_j} \left[ \frac{1}{H_i H_j H_k} \left( \frac{\partial H_k V_k V_i}{\partial x_i} + \frac{\partial H_k H_i V_j}{\partial x_j} + \frac{\partial H_i H_j V_k}{\partial x_k} \right) \right] - \right.$$

$$\frac{1}{H_k H_i} \frac{\partial}{\partial x_k} \left[ \frac{H_i}{H_j H_k} \left( \frac{\partial H_k V_k}{\partial x_j} - \frac{\partial H_j V_j}{\partial x_k} \right) \right] - \frac{1}{H_k H_i} \frac{\partial}{\partial x_i} \left[ \frac{H_k}{H_i H_j} \left( \frac{\partial H_j V_j}{\partial x_i} - \frac{\partial H_i V_i}{\partial x_j} \right) \right] \right\} \overrightarrow{e_j^0} +$$

$$\left\{\frac{1}{H_k}\frac{\partial}{\partial x_k}\left[\frac{1}{H_iH_jH_k}\left(\frac{\partial H_jH_kV_i}{\partial x_i}+\frac{\partial H_kH_iV_j}{\partial x_j}+\frac{\partial H_iH_jH_k}{\partial x_k}\right)\right]-\right.$$
$$\left.\frac{1}{H_iH_j}\frac{\partial}{\partial x_i}\left[\frac{H_j}{H_kH_i}\left(\frac{\partial H_iV_i}{\partial x_k}-\frac{\partial H_kV_k}{\partial x_i}\right)\right]-\frac{1}{H_iH_j}\frac{\partial}{\partial x_j}\left[\frac{H_i}{H_jH_k}\left(\frac{\partial H_kV_k}{\partial x_j}-\frac{\partial H_jV_j}{\partial x_k}\right)\right]\right\}\overline{e_k^0} \qquad (2.117)$$

上式中取 $i=1$，$j=2$，$k=3$。若缩减写成右方第一个大括号，则按求和约定，$i$，$j$，$k$ 依次轮换取值，并且 $i\neq j\neq k$。

7. $\overrightarrow{U}\cdot\nabla$

使用式（2.109）得

$$\overrightarrow{U}\cdot\nabla=\overline{e_i^0}u_i\cdot\left(\overline{e_i^0}\frac{\partial}{H_i\partial x_i}\right)$$

从而有

$$\overrightarrow{U}\cdot\nabla=\frac{u_i}{H_i}\frac{\partial}{\partial x_i} \qquad (2.118)$$

式（2.118）按求和约定，取 $i=1$，2，3。

8. $\overrightarrow{U}\cdot\nabla\overrightarrow{V}$

使用式（2.118），就有

$$\overrightarrow{U}\cdot\nabla\overrightarrow{V}=\frac{u_i}{H_i}\frac{\partial}{\partial x_i}(\overline{e_j^0}V_j)$$

展开上式之右方

$$\overrightarrow{U}\cdot\nabla\overrightarrow{V}=\overline{e_j^0}\frac{u_i}{H_i}\frac{\partial V_j}{\partial x_i}+\frac{u_iV_j}{H_i}\frac{\partial \overline{e_j^0}}{\partial x_i}=\overline{e_j^0}\overrightarrow{U}\cdot\nabla V_j+\frac{u_jV_j}{H_i}\frac{\partial \overline{e_j^0}}{\partial x_i}$$

当 $j=i$ 时，上式右方第二项按式（2.107）成为

$$\frac{u_iV_i}{H_i}\left(-\overline{e_k^0}\frac{\partial H_i}{H_u\partial x_k}-\overline{e_j^0}\frac{\partial H_i}{H_j\partial x_j}\right)$$

当 $j\neq i$ 时，上式右方第二项按式（2.106）成为

$$\overline{e_i^0}\frac{u_iV_j}{H_i}\frac{\partial H_i}{H_j\partial x_j}+\overline{e_i^0}\frac{u_iV_k}{H_i}\frac{\partial H_i}{H_k\partial x_k}$$

于是得到

$$\overrightarrow{U}\cdot\nabla\overrightarrow{V}=\overline{e_j^0}\overrightarrow{U}\cdot\nabla V_j+\overline{e_i^0}\frac{u_iV_j}{H_i}\frac{\partial H_i}{H_j\partial x_j}+\overline{e_i^0}\frac{u_iV_k}{H_i}\frac{\partial H_i}{H_k\partial x_k}-\overline{e_k^0}\frac{u_iV_i}{H_i}\frac{\partial H_i}{H_k\partial x_k}-\overline{e_j^0}\frac{u_iV_i}{H_i}\frac{\partial H_i}{H_j\partial x_j} \qquad (2.119a)$$

式（2.119a）按照求和约定，$i$，$j$，$k$ 依次轮换取值，并且 $i\neq j\neq k$，就得出

$$\vec{U} \cdot \nabla \vec{V} = \left( \vec{e_1^0} \vec{U} \cdot \nabla V_2 + \vec{e_1^0} \frac{u_1 V_2}{H_1} \frac{\partial H_1}{H_2 \partial x_2} + \vec{e_1^0} \frac{u_1 V_3}{H_1} \frac{\partial H_1}{H_3 \partial x_3} - \vec{e_3^0} \frac{u_1 V_1}{H_1} \frac{\partial H_1}{H_3 \partial x_3} - \vec{e_2^0} \frac{u_1 V_1}{H_1} \frac{\partial H_1}{H_2 \partial x_2} \right) +$$

$$\left( \vec{e_2^0} \vec{U} \cdot \nabla V_3 + \vec{e_2^0} \frac{u_2 V_3}{H_2} \frac{\partial H_2}{H_3 \partial x_3} + \vec{e_2^0} \frac{u_2 V_1}{H_2} \frac{\partial H_2}{H_1 \partial x_1} - \vec{e_1^0} \frac{u_2 V_2}{H_2} \frac{\partial H_2}{H_1 \partial x_1} - \vec{e_3^0} \frac{u_2 V_2}{H_2} \frac{\partial H_2}{H_3 \partial x_3} \right) +$$

$$\left( \vec{e_3^0} \vec{U} \cdot \nabla V_1 + \vec{e_3^0} \frac{u_3 V_1}{H_3} \frac{\partial H_3}{H_1 \partial x_1} + \vec{e_3^0} \frac{u_3 V_2}{H_3} \frac{\partial H_3}{H_2 \partial x_2} - \vec{e_2^0} \frac{u_3 V_3}{H_3} \frac{\partial H_3}{H_2 \partial x_2} - \vec{e_1^0} \frac{u_3 V_3}{H_3} \frac{\partial H_3}{H_1 \partial x_1} \right)$$

上式右方三个括号是由式（2.119a）依次取 $i=1$，$j=2$，$k=3$；$j=1$，$k=2$，$i=3$；$k=1$，$i=2$，$j=3$ 得出。按方向将其归项，就有

$$\vec{U} \cdot \nabla \vec{V} = \vec{e_1^0} \left[ \vec{U} \cdot \nabla V_1 + V_2 \left( \frac{u_1}{H_1} \frac{\partial H_1}{H_2 \partial x_2} - \frac{u_2}{H_2} \frac{\partial H_2}{H_1 \partial x_1} \right) + V_3 \left( \frac{u_1}{H_1} \frac{\partial H_1}{H_3 \partial x_3} - \frac{u_3}{H_3} \frac{\partial H_3}{H_1 \partial x_1} \right) \right] +$$

$$\vec{e_2^0} \left[ \vec{U} \cdot \nabla V_2 + V_3 \left( \frac{u_2}{H_2} \frac{\partial H_2}{H_3 \partial x_3} - \frac{u_3}{H_3} \frac{\partial H_3}{H_2 \partial x_2} \right) + V_1 \left( \frac{u_2}{H_2} \frac{\partial H_2}{H_1 \partial x_1} - \frac{u_1}{H_1} \frac{\partial H_1}{H_2 \partial x_2} \right) \right] + \quad (2.119b)$$

$$\vec{e_3^0} \left[ \vec{U} \cdot \nabla V_3 + V_1 \left( \frac{u_3}{H_3} \frac{\partial H_3}{H_1 \partial x_1} - \frac{u_1}{H_1} \frac{\partial H_1}{H_3 \partial x_3} \right) + V_2 \left( \frac{u_3}{H_3} \frac{\partial H_3}{H_2 \partial x_2} - \frac{u_2}{H_2} \frac{\partial H_2}{H_3 \partial x_3} \right) \right]$$

9.惯性力

单位体积流体的惯性力可以分成当地惯性力与迁移惯性力之和，即

$$\rho \frac{d\vec{U}}{dt} = \rho \frac{\partial \vec{U}}{\partial t} + \rho \vec{U} \cdot \nabla \vec{U}$$

将速矢写成 $\vec{U} = \vec{e_i^0} u_i$ 的缩简形式，而后在式（2.119b）中取 $V_i = u_i$，就得惯性力的缩写形式为

$$\rho \frac{d\vec{U}}{dt} = \vec{e_i^0} \left[ \rho \frac{\partial u_i}{\partial t} + \rho \vec{U} \cdot \nabla u_i + \rho u_j \left( \frac{u_i}{H_i} \frac{\partial H_i}{H_j \partial x_j} - \frac{u_j}{H_j} \frac{\partial H_j}{H_i \partial x_i} \right) + \right.$$

$$\left. \rho u_k \left( \frac{u_i}{H_i} \frac{\partial H_i}{H_k \partial x_k} - \frac{u_k}{H_k} \frac{\partial H_k}{H_i \partial x_i} \right) \right] \quad (2.120a)$$

或将右边中括号内第一项和第二项合并，得

$$\rho \frac{d\vec{U}}{dt} = \vec{e_i^0} \left[ \rho \frac{du_i}{dt} + \rho u_j \left( \frac{u_i}{H_i} \frac{\partial H_i}{H_j \partial x_j} - \frac{u_j}{H_j} \frac{\partial H_j}{H_i \partial x_i} \right) + \rho u_k \left( \frac{u_i}{H_i} \frac{\partial H_i}{H_k \partial x_k} - \frac{u_k}{H_k} \frac{\partial H_k}{H_i \partial x_i} \right) \right] \quad (2.120b)$$

按求和约定，$i$，$j$，$k$ 依次轮换取值，并且 $i \neq j \neq k$。

## 2.13.5 流体微团的变形和应力在曲线坐标中的表达形式

1.正变形（即伸缩变形）$\varepsilon_{ii}$

伸缩变形率，由式（2.7）写成

$$\varepsilon_{ii} = \frac{\partial u_i}{\partial x_i} = \vec{e_i^0} \cdot (\vec{e_i^0} \cdot \nabla \vec{U})$$

Hamilton 算子 $\nabla$ 用式（2.109）代入，并且将 $\vec{U}$ 展开，就有

$$\varepsilon_{ii} = \vec{e_i^0} \cdot \left[ \vec{e_i^0} \cdot \left( \frac{\vec{e_i^0}}{H_i} \frac{\partial}{\partial x_i} \right) (\vec{e_i^0} u_i + \vec{e_j^0} u_j + \vec{e_k^0} u_k) \right] = \vec{e_i^0} \cdot \left[ \frac{\partial}{H_i \partial x_i} (\vec{e_i^0} u_i + \vec{e_j^0} u_j + \vec{e_k^0} u_k) \right] =$$

$$\vec{e_i^0} \cdot \left( \vec{e_i^0} \frac{\partial u_i}{\partial x_i} + u_i \frac{\partial \vec{e_i^0}}{\partial x_i} + \vec{e_j^0} \frac{\partial u_j}{\partial x_i} + u_j \frac{\partial \vec{e_j^0}}{\partial x_i} + \vec{e_k^0} \frac{\partial u_k}{\partial x_i} + u_k \frac{\partial \vec{e_k^0}}{\partial x_i} \right) \frac{1}{H_i}$$

使用曲线坐标系单位矢量微分的关系式（2.106）与（2.107），则上式成为

$$\varepsilon_{ii} = \frac{\vec{e_i^0}}{H_i} \cdot \left[ \vec{e_i^0} \frac{\partial u_i}{\partial x_i} + u_i \left( -\vec{e_k^0} \frac{\partial H_i}{H_k \partial x_k} - \vec{e_j^0} \frac{\partial H_i}{H_j \partial x_j} \right) + \vec{e_j^0} \frac{\partial u_j}{\partial x_i} + u_j \left( \vec{e_i^0} \frac{\partial H_i}{H_j \partial x_j} \right) + \vec{e_k^0} \frac{\partial u_k}{\partial x_i} + u_k \left( \vec{e_i^0} \frac{\partial H_i}{H_k \partial x_k} \right) \right]$$

于是得到

$$\varepsilon_{ii} = \frac{1}{H_i} \frac{\partial u_i}{\partial x_i} + \frac{u_j}{H_i H_j} \frac{\partial H_i}{\partial x_j} + \frac{u_k}{H_i H_k} \frac{\partial H_i}{\partial x_k} \tag{2.121}$$

式中 $i$, $j$, $k$ 依次轮换取值，并且 $i \neq j \neq k$。

2.剪切变形 $\varepsilon_{ij}$

由式（2.12），剪切变形率可以写成

$$2\varepsilon_{ij} = \frac{\partial u_i}{\partial x_j} + \frac{\partial u_j}{\partial x_i} = \vec{e_i^0} \cdot (\vec{e_j^0} \cdot \nabla \vec{U}) + \vec{e_j^0} \cdot (\vec{e_i^0} \cdot \nabla \vec{U})$$

同样，使用式（2.109）并展开 $\vec{U}$，就得

$$2\varepsilon_{ij} = \vec{e_i^0} \cdot \left[ \vec{e_j^0} \cdot \left( \frac{\vec{e_j^0}}{H_j} \frac{\partial}{\partial x_j} \right) (\vec{e_i^0} u_i + \vec{e_j^0} u_j + \vec{e_k^0} u_k) \right] + \vec{e_j^0} \cdot \left[ \vec{e_i^0} \cdot \left( \frac{\vec{e_i^0}}{H_i} \frac{\partial}{\partial x_i} \right) (\vec{e_i^0} u_i + \vec{e_j^0} u_j + \vec{e_k^0} u_k) \right] =$$

$$\vec{e_i^0} \cdot \left( \vec{e_i^0} \frac{\partial u_i}{\partial x_j} + u_i \frac{\partial \vec{e_i^0}}{\partial x_j} + \vec{e_j^0} \frac{\partial u_j}{\partial x_j} + u_j \frac{\partial \vec{e_j^0}}{\partial x_j} + \vec{e_k^0} \frac{\partial u_k}{\partial x_j} + u_k \frac{\partial \vec{e_k^0}}{\partial x_j} \right) \frac{1}{H_j} +$$

$$\vec{e_j^0} \cdot \left( \vec{e_i^0} \frac{\partial u_i}{\partial x_i} + u_i \frac{\partial \vec{e_i^0}}{\partial x_i} + \vec{e_j^0} \frac{\partial u_j}{\partial x_i} + u_j \frac{\partial \vec{e_j^0}}{\partial x_i} + \vec{e_k^0} \frac{\partial u_k}{\partial x_i} + u_k \frac{\partial \vec{e_k^0}}{\partial x_i} \right) \frac{1}{H_i}$$

注意正交性质，当 $i \neq j$ 时，$\vec{e_i^0} \cdot \vec{e_j^0} = 0$，则上式减项成

$$2\varepsilon_{ij} = \frac{\partial u_i}{H_j \partial x_j} + \frac{\vec{e_i^0}}{H_j} \cdot \left( u_i \frac{\partial \vec{e_i^0}}{\partial x_j} + u_j \frac{\partial \vec{e_j^0}}{\partial x_j} + u_k \frac{\partial \vec{e_k^0}}{\partial x_j} \right) + \frac{\partial u_j}{H_i \partial x_i} + \frac{\vec{e_j^0}}{H_i} \cdot \left( u_i \frac{\partial \vec{e_i^0}}{\partial x_i} + u_j \frac{\partial \vec{e_j^0}}{\partial x_i} + u_k \frac{\partial \vec{e_k^0}}{\partial x_i} \right)$$

用式（2.106）与式（2.107）代入，得

$$2\varepsilon_{ij} = \frac{\partial u_i}{H_j \partial x_j} + \frac{\vec{e_i^0}}{H_j} \cdot \left[ u_i \left( \vec{e_j^0} \frac{\partial H_j}{H_i \partial x_i} \right) + u_j \left( -\vec{e_i^0} \frac{\partial H_j}{H_i \partial x_i} - \vec{e_k^0} \frac{\partial H_j}{H_k \partial x_k} \right) + u_k \left( \vec{e_j^0} \cdot \frac{\partial H_j}{H_k \partial x_k} \right) \right] +$$

$$\frac{\partial u_j}{H_i \partial x_i} + \frac{\vec{e_j^0}}{H_i} \cdot \left[ u_i \left( -\vec{e_j^0} \frac{\partial H_i}{H_k \partial x_k} - \vec{e_j^0} \frac{\partial H_i}{H_j \partial x_j} \right) + u_j \left( \vec{e_i^0} \frac{\partial H_i}{H_j \partial x_j} \right) + u_k \left( \vec{e_i^0} \cdot \frac{\partial H_i}{H_k \partial x_k} \right) \right]$$

于是最终有

$$2\varepsilon_{ij} = \frac{\partial u_i}{H_j \partial x_j} + \frac{\partial u_j}{H_i \partial x_i} - \frac{u_j}{H_j} \frac{\partial H_j}{H_i \partial x_i} - \frac{u_i}{H_i} \frac{\partial H_i}{H_j \partial x_j} \tag{2.122}$$

3.流体中的正应力和切应力

由式（2.47），对于不可压缩流体的应力通式是

$$\tau_{ij} = -p\delta_{ij} + 2\eta\varepsilon_{ij} - \frac{2}{3}\eta\nabla\cdot\vec{U}\delta_{ij}$$

当 $i = j$ 时是正应力，而 $\delta_{ii} = 1$，则

$$\tau_{ii} = -p + 2\eta\varepsilon_{ii} - \frac{2}{3}\eta\nabla\cdot\vec{U}$$

当 $i \neq j$ 时是切应力，此时 $\delta_{ij} = 0$，即

$$\tau_{ij} = 2\eta\varepsilon_{ij}$$

使用式（2.121）得正应力是

$$\tau_{ii} = -p + 2\eta\left( \frac{1}{H_i}\frac{\partial u_i}{\partial x_i} + \frac{u_j}{H_iH_j}\frac{\partial H_i}{\partial x_j} + \frac{u_k}{H_iH_k}\frac{\partial H_i}{\partial x_k} \right) - \frac{2}{3}\eta\nabla\cdot\vec{U} \tag{2.123}$$

由式（2.122）得切应力是

$$\tau_{ij} = \eta\left( \frac{\partial u_i}{H_j \partial x_j} + \frac{\partial u_j}{H_i \partial x_i} - \frac{u_j}{H_j}\frac{\partial H_j}{H_i \partial x_i} - \frac{u_i}{H_i}\frac{\partial H_i}{H_j \partial x_j} \right) \tag{2.124}$$

10.表面力 $\vec{f}_s$

式（2.44）表明，力 $\vec{f}_s$ 是表面应力 $\tau$ 的散度，于是有

$$\vec{f}_s = \nabla\cdot\tau = \left( \vec{e_i^0}\frac{\partial}{H_i \partial x_i} + \vec{e_j^0}\frac{\partial}{H_j \partial x_j} + \vec{e_k^0}\frac{\partial}{H_k \partial x_k} \right)\cdot(\vec{e_i^0}\vec{e_i^0}\tau_{ii} + \vec{e_j^0}\vec{e_i^0}\tau_{ji} + \vec{e_k^0}\vec{e_i^0}\tau_{ki}) =$$

$$\left( \vec{e_i^0}\frac{\partial}{H_i \partial x_i} + \vec{e_j^0}\frac{\partial}{H_j \partial x_j} + \vec{e_k^0}\frac{\partial}{H_k \partial x_k} \right)\cdot\left[ \frac{\vec{e_i^0}}{H_jH_k}(\vec{e_i^0}H_jH_k\tau_{ii}) + \frac{\vec{e_j^0}}{H_kH_i}(\vec{e_i^0}H_kH_i\tau_{ji}) + \frac{\vec{e_k^0}}{H_iH_j}(\vec{e_i^0}H_iH_j\tau_{ki}) \right] =$$

$$\frac{1}{H_iH_jH_k}\left[ \frac{\partial}{\partial x_i}(\vec{e_i^0}H_jH_k\tau_{ii}) + \frac{\partial}{\partial x_j}(\vec{e_i^0}H_kH_i\tau_{ji}) + \frac{\partial}{\partial x_k}(\vec{e_i^0}H_iH_j\tau_{ki}) \right] =$$

$$\frac{1}{H_iH_jH_k}\left[ \vec{e_i^0}\frac{\partial}{\partial x_i}(H_jH_k\tau_{ii}) + \vec{e_i^0}\frac{\partial}{\partial x_j}(H_kH_i\tau_{ji}) + \vec{e_i^0}\frac{\partial}{\partial x_k}(H_iH_j\tau_{ki}) + \right.$$

$$\left. H_jH_k\tau_{ii}\frac{\partial\vec{e_i^0}}{\partial x_i} + H_kH_i\tau_{ji}\frac{\partial\vec{e_i^0}}{\partial x_j} + H_iH_j\tau_{ki}\frac{\partial\vec{e_i^0}}{\partial x_k} \right] =$$

$$\frac{1}{H_iH_jH_k}\left\{ \vec{e_i^0}\left[ \frac{\partial}{\partial x_i}(H_jH_k\tau_{ii}) + \frac{\partial}{\partial x_j}(H_kH_i\tau_{ji}) + \frac{\partial}{\partial x_k}(H_iH_j\tau_{ki}) \right] - \right.$$

$$\left. \overline{e_k^0}H_jH_k\tau_{ii}\frac{\partial H_i}{H_k\partial x_k} - \overline{e_j^0}H_jH_k\tau_{ii}\frac{\partial H_i}{H_j\partial x_j} + \overline{e_j^0}H_kH_i\tau_{ji}\frac{\partial H_j}{H_i\partial x_i} + \overline{e_k^0}H_iH_j\tau_{ki}\frac{\partial H_k}{H_i\partial x_i} \right\}$$

于是得到 $\vec{f_s}$ 的简缩形式为

$$\vec{f_s} = \frac{1}{H_iH_jH_k}\left\{\overline{e_i^0}\left[\frac{\partial}{\partial x_i}(H_jH_k\tau_{ii}) + \frac{\partial}{\partial x_j}(H_kH_i\tau_{ji}) + \frac{\partial}{\partial x_k}(H_iH_j\tau_{ki})\right] + \right.$$
$$\left. \overline{e_j^0}H_k\left(\tau_{ji}\frac{\partial H_j}{\partial x_i} - \tau_{ii}\frac{\partial H_i}{\partial x_j}\right) + \overline{e_k^0}H_j\left(\tau_{ki}\frac{\partial H_k}{\partial x_i} - \tau_{ii}\frac{\partial H_i}{\partial x_k}\right)\right\} \tag{2.125}$$

按照求和约定，上式 $i$，$j$，$k$ 依次轮换取值 1，2，3，则得

$$\vec{f_s} = \frac{1}{H_1H_2H_3}\left\{\overline{e_1^0}\left[\frac{\partial}{\partial x_1}(H_2H_3\tau_{11}) + \frac{\partial}{\partial x_2}(H_3H_1\tau_{21}) + \frac{\partial}{\partial x_3}(H_1H_2\tau_{31})\right] + \right.$$
$$\left. \overline{e_2^0}H_3\left(\tau_{21}\frac{\partial H_2}{\partial x_1} - \tau_{11}\frac{\partial H_1}{\partial x_2}\right) + \overline{e_3^0}H_2\left(\tau_{31}\frac{\partial H_3}{\partial x_1} - \tau_{11}\frac{\partial H_1}{\partial x_3}\right)\right\} +$$

$$\frac{1}{H_2H_3H_1}\left\{\overline{e_2^0}\left[\frac{\partial}{\partial x_2}(H_3H_1\tau_{22}) + \frac{\partial}{\partial x_3}(H_1H_2\tau_{32}) + \frac{\partial}{\partial x_1}(H_2H_3\tau_{12})\right] + \right.$$
$$\left. \overline{e_3^0}H_1\left(\tau_{32}\frac{\partial H_3}{\partial x_2} - \tau_{22}\frac{\partial H_2}{\partial x_3}\right) + \overline{e_1^0}H_3\left(\tau_{12}\frac{\partial H_1}{\partial x_2} - \tau_{22}\frac{\partial H_2}{\partial x_1}\right)\right\} +$$

$$\frac{1}{H_3H_1H_2}\left\{\overline{e_3^0}\left[\frac{\partial}{\partial x_3}(H_1H_2\tau_{33}) + \frac{\partial}{\partial x_1}(H_2H_3\tau_{13}) + \frac{\partial}{\partial x_2}(H_3H_1\tau_{23})\right] + \right.$$
$$\left. \overline{e_1^0}H_2\left(\tau_{13}\frac{\partial H_1}{\partial x_3} - \tau_{33}\frac{\partial H_3}{\partial x_1}\right) + \overline{e_2^0}H_1\left(\tau_{23}\frac{\partial H_2}{\partial x_3} - \tau_{33}\frac{\partial H_3}{\partial x_2}\right)\right\}$$

上式右方第一个大括号项是取 $i=1$，$j=2$，$k=3$；第二个大括号项是取 $i=2$，$j=3$，$k=1$；第三个大括号项是取 $i=3$，$j=1$，$k=2$。再将上式按方向整理归项，就得以三个坐标分量表示的 $\vec{f_s}$，即

$$\vec{f_s} = \frac{\overline{e_1^0}}{H_1H_2H_3}\left[\frac{\partial}{\partial x_1}(H_2H_3\tau_{11}) + \frac{\partial}{\partial x_2}(H_3H_1\tau_{21}) + \frac{\partial}{\partial x_3}(H_1H_2\tau_{31}) + \right.$$
$$\left. H_3\left(\tau_{12}\frac{\partial H_1}{\partial x_2} - \tau_{22}\frac{\partial H_2}{\partial x_1}\right) + H_2\left(\tau_{13}\frac{\partial H_1}{\partial x_3} - \tau_{33}\frac{\partial H_3}{\partial x_1}\right)\right] +$$

$$\frac{\overline{e_2^0}}{H_1H_2H_3}\left[\frac{\partial}{\partial x_1}(H_2H_3\tau_{12}) + \frac{\partial}{\partial x_2}(H_3H_1\tau_{22}) + \frac{\partial}{\partial x_3}(H_1H_2\tau_{32}) + \right.$$
$$\left. H_1\left(\tau_{23}\frac{\partial H_2}{\partial x_3} - \tau_{33}\frac{\partial H_3}{\partial x_2}\right) + H_3\left(\tau_{21}\frac{\partial H_2}{\partial x_1} - \tau_{11}\frac{\partial H_1}{\partial x_2}\right)\right] + \tag{2.126a}$$

$$\frac{\overline{e_3^0}}{H_1H_2H_3}\left[\frac{\partial}{\partial x_1}(H_2H_3\tau_{13}) + \frac{\partial}{\partial x_2}(H_3H_1\tau_{23}) + \frac{\partial}{\partial x_3}(H_1H_2\tau_{33}) + \right.$$
$$\left. H_2\left(\tau_{31}\frac{\partial H_3}{\partial x_1} - \tau_{11}\frac{\partial H_1}{\partial x_3}\right) + H_1\left(\tau_{32}\frac{\partial H_3}{\partial x_2} - \tau_{22}\frac{\partial H_2}{\partial x_3}\right)\right]$$

将上式写成简缩的形式，则有

$$\vec{f}_s = \frac{\vec{e_i^0}}{H_i H_j H_k}\left[ \frac{\partial}{\partial x_i}(H_j H_k \tau_{ii}) + \frac{\partial}{\partial x_j}(H_k H_i \tau_{ji}) + \frac{\partial}{\partial x_k}(H_i H_j \tau_{ki}) + \right.$$
$$\left. H_k\left( \tau_{ij}\frac{\partial H_i}{\partial x_j} - \tau_{jj}\frac{\partial H_j}{\partial x_i} \right) + H_j\left( \tau_{ik}\frac{\partial H_i}{\partial x_k} - \tau_{kk}\frac{\partial H_k}{\partial x_i} \right) \right]$$

(2.126b)

式（2.126b）按求和约定，$i$，$j$，$k$ 依次轮流取值 1，2，3，并且 $i \neq j \neq k$。

## 2.13.6 正交曲线坐标系中的流体力学方程

前面已经得出流体力学微分方程的组成元素，包括某些物理参数的导数和一些描述流场的微分算子在正交曲线坐标系中的表达形式。将它们分别对应地代入连续性方程（2.42a）、运动方程（2.43）和能量方程（2.63）中，就能得到曲线坐标系中的流体力学方程。

1.连续性方程——质量守恒方程

由式（2.42a）

$$\frac{\partial \rho}{\partial t} + \nabla \cdot (\rho \vec{U}) = 0$$

在式（2.113b）中，取矢量 $\vec{V} = \rho \vec{U}$，就有

$$\frac{\partial \rho}{\partial t} + \frac{1}{H_i H_j H_k}\left[ \frac{\partial}{\partial x_i}(H_j H_k u_i) \right] = 0$$

(2.127a)

式（2.127a）按求和约定，$i$，$j$，$k$ 依次轮换取值 1，2，3 并且 $i \neq j \neq k$。

对于不可压缩流，$\rho = \mathrm{const}$，则有

$$\frac{\partial}{\partial x_i}(H_j H_k u_i) = 0$$

(2.127b)

2.运动方程——动量守恒方程

由式（2.43）与式（2.44），即

$$\rho \frac{d\vec{U}}{dt} = \vec{f_b} + \vec{f_s}$$

上式之左方是流体的惯性力，右方是作用于流体上的彻体力和表面力。上式亦可改写成

$$\rho \frac{\partial \vec{U}}{\partial t} + \rho \vec{U} \cdot \nabla \vec{U} = \vec{f_b} + \nabla \cdot \tau$$

将式（2.120a）代入上式值左方，式（2.126b）代入上式之右方，并将 $\vec{f_b}$ 写成 $\vec{e_i^0} f_{bi}$ 的形式，于是就有

$$\rho\left[ \frac{\partial u_i}{\partial t} + \vec{U}\cdot\nabla u_i + u_j\left( \frac{u_i}{H_i}\frac{\partial H_i}{H_j \partial x_j} - \frac{u_j}{H_j}\frac{\partial H_j}{H_i \partial x_i} \right) + u_k\left( \frac{u_i}{H_i}\frac{\partial H_i}{H_k \partial x_k} - \frac{u_k}{H_k}\frac{\partial H_k}{H_i \partial x_i} \right) \right] =$$
$$f_{bi} + \frac{1}{H_i H_j H_k}\left[ \frac{\partial}{\partial x_i}(H_j H_k \tau_{ii}) + \frac{\partial}{\partial x_j}(H_k H_i \tau_{ji}) + \frac{\partial}{\partial x_k}(H_i H_j \tau_{ki}) + \right.$$
$$\left. H_k\left( \tau_{ij}\frac{\partial H_i}{\partial x_j} - \tau_{jj}\frac{\partial H_j}{\partial x_i} \right) + H_j\left( \tau_{ik}\frac{\partial H_i}{\partial x_k} - \tau_{kk}\frac{\partial H_k}{\partial x_i} \right) \right]$$

(2.128)

式（2.128）中，表面正应力用式（2.123）代入，表面力用式（2.124）代入。按照求和约定，式中 $i$，$j$，$k$ 依次轮流取值 1，2，3，并且 $i \neq j \neq k$。

3.能量守恒方程

由式（2.63），能量方程是

$$\rho \frac{\partial (c_V' T)}{\partial t} + \rho \vec{U} \cdot \nabla (c_V' T) = -p \nabla \cdot \vec{U} + \nabla \cdot (K_H \nabla T) + \Phi$$

用式（2.118）代入左方第二项，用式（2.113b）代入右方第一项，再用式（2.113b）与式（2.108b）代入右方第二项，就得

$$\rho \frac{\partial (c_V' T)}{\partial t} + \rho \frac{u_i}{H_i} \frac{\partial}{\partial x_i} (c_V' T) = -\frac{p}{H_i H_j H_k} \frac{\partial}{\partial x_i} (H_j H_k u_i) + \frac{1}{H_i H_j H_k} \frac{\partial}{\partial x_i} \left( H_j H_k K_H \frac{\partial T}{H_i \partial x_i} \right) + \Phi \quad (2.129a)$$

对于稳定状态的不可压缩流体，用式（2.127b）代入，就有

$$\frac{u_i}{H_i} \frac{\partial}{\partial x_i} (c_V T) = \frac{1}{H_i H_j H_k} \frac{\partial}{\partial x_i} \left( H_j H_k K_H \frac{\partial T}{H_i \partial x_i} \right) + \Phi \quad (2.129b)$$

式中 $c_V$ 是体积比热容，它与质量比热容 $c_V'$ 的关系是 $c_V = \rho c_V'$。方程（2.129a）、（2.129b）都服从求和约定。

### 2.13.7 曲线坐标与常用坐标系的转换

曲线坐标系与常用坐标系之间的转换，主要取决于 Lame 系数。式（2.102a）给出坐标弧长和坐标值的关系是 $dL_i = H_i dx_i$。在常用坐标系中，坐标弧长和坐标值之间的关系常常是知道的，因而不难确定 Lame 系数。

1.直角坐标系

其坐标是

$$x_1 = x, \qquad x_2 = y, \qquad x_3 = z$$

直角坐标系的坐标轴都是直线，所以坐标弧长等于坐标值，即

$$dL_1 = dx_1 = dx, \qquad dL_2 = dx_2 = dy, \qquad dL_3 = dx_3 = dz$$

由此可知

$$H_1 = 1, \qquad H_2 = 1, \qquad H_3 = 1$$

2.圆柱坐标系

圆柱坐标系由圆柱半径、轴线和圆柱的断面圆周组成。所以它只有一根曲线轴。其坐标几何如图 2-6 所示。

$$x_1 = r, \qquad x_2 = \theta, \qquad x_3 = z$$

坐标弧长是

$$dL_1 = dx_1 = dr, \qquad dL_2 = rdx_2 = rd\theta, \qquad dL_3 = dx_3 = dz$$

故其 Lame 系数为

$$H_1 = 1, \qquad H_2 = r, \qquad H_3 = z$$

3.圆球坐标系

圆球坐标系由圆球半径，圆球表面的经线、纬线所组成，它有两根坐标轴是曲线，即经线、纬线。如图 2-7 所示，坐标值是

$$x_1 = r, \qquad x_2 = \theta, \qquad x_3 = \varphi$$

坐标弧长为

$$dL_1 = dx_1 = dr, \qquad dL_2 = rdx_2 = rd\theta, \qquad dL_3 = r\sin\theta dx_3 = r\sin\theta d\varphi$$

可见圆球坐标系的 Lame 系数是

$$H_1 = 1, \qquad H_2 = r, \qquad H_3 = r\sin\theta$$

# 第三章　Dirac 函数和铁磁流体连续性问题

## 3.1 概述

　　铁磁流体基本上是一种流态物质，所以自然使用流体力学的方程来分析它的运动规律。但是，在使用流体力学方程时，首先遇到的一个问题，就是铁磁流体不是连续的介质。因为在铁磁流体中任何一点处，只可能存在一个相，不可能有两相并存于一点的现象。所以无论是两相合在一起，或是将两相拆开，都是不连续的介质。将不连续的介质转化成"连续的"，就是用平均的办法处理。将两相混合在一起的作法是混合物的物理性质取两相的平均值，将两相分开的做法是每一相的物理性质都在两相的空间中取平均。在两相流的流场中，取一点 $P$ 的邻域，这个邻域的尺寸远大于固相粒的直径和微粒之间的距离，这样，邻域中包含了足够的固相微粒，从而使其物理性质的平均值具有比较光滑的连续性。但邻域与流场的特征尺寸相比又很微小，以致其中的平均值又能代表一点处（即点 $P$）的状态。于是两相流，无论是混合流还是分相流，以连续性为基础的流体力学方程都适用。取平均值的数学工具就是 Dirac 函数，即 $\delta$ 函数。$\delta$ 函数法是一种数学处理方法，但其过程与物理概念互相结合，它的基本思维和通常的微元控制体方法有类似之处。

## 3.2　$\delta$ 函数的定义和性质[①]

### 3.2.1　$\delta$ 函数的定义

　　设 $f(x)$ 是在点 $x$ 的邻域 $[a,b]$ 上连续的任一函数，则 $\delta$ 函数的定义式

$$\int_a^b f(\xi)\delta(\xi-x)d\xi = \begin{cases} 0, & x<a,\ x>b \\ \dfrac{1}{2}f(x+0), & x=a \\ \dfrac{1}{2}f(x-0), & x=b \\ \dfrac{1}{2}[f(x-0)+f(x+0)], & a<x<b \end{cases} \tag{3.1}$$

若 $f(x)=\text{const}$，则式（3.1）左方之 $f(\xi)$ 可提到积分号之外，则有

$$f(x)\int_a^b \delta(\xi-x)d\xi = \frac{1}{2}[f(x-0)+f(x+0)]$$

从而

$$\int_a^b \delta(\xi-x)d\xi = 1 \tag{3.2}$$

　　在三维问题中，空间邻域 $V$ 包围点 $P(x,y,z)$，对于域 $V$ 内任一连续函数 $f(x,y,z;t)$，$\delta$ 函数应当满足

$$\int_V f(\xi,\eta,\varsigma;t)\delta(\xi-x,\eta-y,\varsigma-z)\,d\xi\,d\eta\,d\varsigma = $$
$$\frac{1}{2}[f(x-0,y-0,z-0;t)+f(x+0,y+0,z+0;t)] \tag{3.3}$$

式中 $\xi$，$\eta$，$\varsigma$ 是在邻域 $V$ 中另一点 $P'$ 的坐标值，即 $P'(\xi,\eta,\varsigma)$。图 3-1 表示出参照坐标系的原点 $O$ 指向点 $P(x,y,z)$ 的矢径是 $\vec{r}$，而由点 $P(x,y,z)$ 指向点 $P'(\xi,\eta,\varsigma)$ 的矢径是 $\vec{R'}$，于是由参照系原点 $O$ 指向点

---

① 本节内容主要取自参考文献[1]

$P'(\xi,\eta,\varsigma)$ 的矢径 $\vec{R}$ 为

$$\vec{R} = \vec{r} + \vec{R}' \tag{3.4}$$

式（3.4）中，$\vec{R}$、$\vec{r}$、$\vec{R}'$ 都是三维空间的矢量，若三维矢性函数 $\vec{f}(\vec{r})$ 在空间域 $V$ 内是连续的，则式（3.3）可写成

$$\int_V \vec{f}(\vec{R},t)\delta(\vec{R}-\vec{r})dV = \vec{f}(\vec{r},t) \tag{3.5a}$$

若函数 $\vec{f}(\vec{r},t)$ 对坐标是常矢，则式（3.5a）给出

$$\int_V \delta(\vec{R}-\vec{r})dV = 1 \tag{3.5b}$$

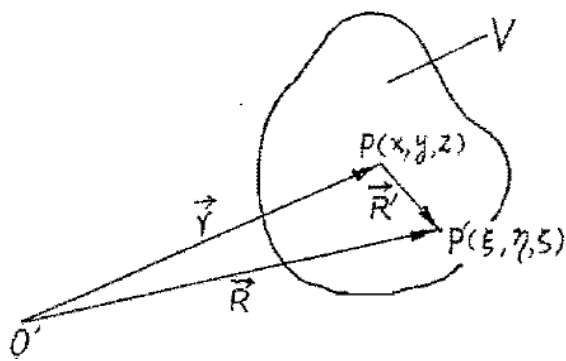

图 3-1　点 $P$ 的空间邻域 $V$

3.2.2　$\delta$ 函数的某些性质

1.$\delta(x)$ 是脉冲函数

$$\left.\begin{array}{ll} \delta(x) = 0 & (x \neq 0) \\ \delta(x) \to \infty & (x = 0) \end{array}\right\} \tag{3.6a}$$

2.$\delta(x)$ 是偶函数

$$\delta(-x) = \delta(x) \tag{3.6b}$$

3.其它

$$\delta(ax) = \frac{1}{a}\delta(x) \tag{3.6c}$$

$$f(x)\delta(x-a) = \frac{1}{2}\big[f(a-0) + f(a+0)\big]\delta(x-a) \tag{3.6d}$$

$$x\delta(x) = x\delta(x-0) = 0 \tag{3.6e}$$

3.3 铁磁流体所含成分的体积分数 $\phi$ 和质量分数 $\varphi$

3.3.1　体积分数 $\phi$

设在铁磁流体中，任取一体积 $V$，在 $V$ 内含有以微粒形式存在的固相体积 $V_p$ 和液相基础载体的体

积 $V_c$。两者混合在一起呈胶体形态，它们彼此独立，没有化学反应，也不互相溶解。因此有

$$V = V_p + V_c \tag{3.7a}$$

两边微分之

$$dV = dV_p + dV_c \tag{3.7b}$$

定义体积分数为

$$\phi_p(\vec{r},t) = \frac{dV_p}{dV}, \qquad \phi_c(\vec{r},t) = \frac{dV_c}{dV} \tag{3.8}$$

式中，下标 $p$ 和 $c$ 分别表示固相和液相。只有在稳定并且均匀状态下，$\phi_p$ 和 $\phi_c$ 才是常数。

将式（3.7b）代入式（3.5b）中，而后使用式（3.8），就有

$$\int_V \delta(\vec{R} - \vec{r}) dV = \int_{V_p} \delta(\vec{R} - \vec{r}) dV_p + \int_{V_c} \delta(\vec{R} - \vec{r}) dV_c =$$
$$\int_V \delta(\vec{R} - \vec{r}) \phi_p(\vec{R},t) dV + \int_V \delta(\vec{R} - \vec{r}) \phi_c(\vec{R},t) dV = 1 \tag{3.9}$$

于是在体积 $V$ 内 $\phi_p$ 和 $\phi_c$ 的平均值为：

$$\left. \begin{array}{l} \phi_p(\vec{r},t) = \int_{V_p} \delta(\vec{R} - \vec{r}) \, dV_p = \int_V \delta(\vec{R} - \vec{r}) \phi_p(\vec{R},t) dV \\ \phi_c(\vec{r},t) = \int_{V_c} \delta(\vec{R} - \vec{r}) \, dV_c = \int_V \delta(\vec{R} - \vec{r}) \phi_c(\vec{R},t) dV \end{array} \right\} \tag{3.10}$$

由式（3.9）与式（3.5b）显然可见，$\phi_p$ 和 $\phi_c$ 的平均值之和等于 1，即

$$\phi_p(\vec{r},t) + \phi_c(\vec{r},t) = 1 \tag{3.11}$$

设以 $\vec{f}(\vec{r},t)$ 表示铁磁流体的某种物理量，并且 $\vec{f}(\vec{r},t)$ 是以单位体积计量的，则必定存在关系

$$\vec{f}(\vec{r},t) dV = \overline{f_{NP}}(\vec{r},t) dV_p + \overline{f_{NC}}(\vec{r},t) dV_c \tag{3.12}$$

式中 $\overline{f_{NP}}(\vec{r},t)$ 是固相物质的体积物理量，$\overline{f_{NC}}(\vec{r},t)$ 是液相物质的体积物理量。将式（3.12）两边同乘以 $\delta(\vec{R} - \vec{r})$ 而后在体积 $V$ 内积分

$$\int_V \vec{f}(\vec{R},t)\delta(\vec{R} - \vec{r}) dV = \int_{V_p} \overline{f_{NP}}(\vec{R},t)\delta(\vec{R} - \vec{r}) dV_p + \int_{V_c} \overline{f_{NC}}(\vec{R},t)\delta(\vec{R} - \vec{r}) dV_c \tag{3.13}$$

将式（3.8）用于式（3.13）的右方，就有

$$\int_V \vec{f}(\vec{R},t)\delta(\vec{R} - \vec{r}) dV = \int_V \phi_p(\vec{R},t)\overline{f_{NP}}(\vec{R},t)\delta(\vec{R} - \vec{r}) dV + \int_V \phi_c(\vec{R},t)\overline{f_{NC}}(\vec{R},t)\delta(\vec{R} - \vec{r}) dV \tag{3.14}$$

由式（3.13）和式（3.14）相比较，可见

$$\left.\begin{array}{l}\int_{V_p}\overline{f_{NP}}(\vec{R},t)\delta(\vec{R}-\vec{r})dV_p=\int_V\phi_p(\vec{R},t)\overline{f_{NP}}(\vec{R},t)\delta(\vec{R}-\vec{r})dV\\[2mm]\int_{V_c}\overline{f_{NC}}(\vec{R},t)\delta(\vec{R}-\vec{r})dV_c=\int_V\phi_c(\vec{R},t)\overline{f_{NC}}(\vec{R},t)\delta(\vec{R}-\vec{r})dV\end{array}\right\}\qquad(3.15)$$

式（3.15）两式之左方表示物理量 $\overline{f_{NP}}$ 和 $\overline{f_{NC}}$ 在体积 $V$ 内的总量，右方则分别表示将 $\overline{f_{NP}}$ 和 $\overline{f_{NC}}$ "稀化" 或 "均摊" 于整个体积 $V$ 之中，这就是所谓的平均值的意思。按平均值的定义式（3.5a），则式（3.14）给出

$$\vec{f}(\vec{r},t)=\phi_p(\vec{r},t)\overline{f_{NP}}(\vec{r},t)+\phi_c(\vec{r},t)\overline{f_{NC}}(\vec{r},t)\qquad(3.16)$$

定义固相 $p$ 和液相 $c$ 的体积分物理量 $\overline{f_p}(\vec{r},t)$ 和 $\overline{f_c}(\vec{r},t)$ 为

$$\left.\begin{array}{l}\overline{f_p}(\vec{r},t)=\phi_p(\vec{r},t)\overline{f_{NP}}(\vec{r},t)=\int_V\phi_p(\vec{R},t)\overline{f_{NP}}(\vec{R},t)\delta(\vec{R}-\vec{r})dV\\[2mm]\overline{f_c}(\vec{r},t)=\phi_c(\vec{r},t)\overline{f_{NC}}(\vec{r},t)=\int_V\phi_c(\vec{R},t)\overline{f_{NC}}(\vec{R},t)\delta(\vec{R}-\vec{r})dV\end{array}\right\}\qquad(3.17)$$

代入式（3.16），就有

$$\vec{f}(\vec{r},t)=\overline{f_p}(\vec{r},t)+\overline{f_c}(\vec{r},t)\qquad(3.18)$$

式（3.18）两边各项都是独自充满相同体积 $V$ 的物理量。

### 3.3.2  质量分数 $\varphi$

有些物理量，如动能、动量等，直接和质量相关联，尤其对于可压缩流体，其体积是可以变化的，但其质量不因体积变化而改变。在这类场合下使用质量分数往往更为方便。设在铁磁流体混合物的质量 $dm$ 内包含成分 $p$ 和成分 $c$ 的质量为 $dm_p$ 和 $dm_c$，即

$$dm=dm_p+dm_c\qquad(3.19)$$

定义质量分数记号为

$$\varphi_p(\vec{r},t)=\frac{dm_p}{dm},\qquad\varphi_c(\vec{r},t)=\frac{dm_c}{dm}\qquad(3.20)$$

设 $\overline{f'}(\vec{r},t)$ 是以单位质量计量的物理量，则在质量 $dm$ 内的所含的物理量是 $\overline{f'}(\vec{r},t)dm$，对于混合物而言，必定存在

$$\overline{f'}(\vec{r},t)dm=\overline{f'_{NP}}(\vec{r},t)dm_p+\overline{f'_{NC}}(\vec{r},t)dm_c\qquad(3.21)$$

式中 $\overline{f'_{NP}}(\vec{r},t)$ 和 $\overline{f'_{NC}}(\vec{r},t)$ 分别属于组成成分 $p$ 和 $c$，并且 $\overline{f'_{NP}}(\vec{r},t)$ 和 $\overline{f'_{NC}}(\vec{r},t)$ 也都是以单位质量计量的。

质量和体积的关系是 $dm=\rho dV$，$\rho$ 是密度。于是式（3.21）可以写成

$$\overline{f'}(\vec{r},t)\rho(\vec{r},t)dV=\overline{f'_{NP}}(\vec{r},t)\rho_{NP}(\vec{r},t)dV_p+\overline{f'_{NC}}(\vec{r},t)\rho_{NC}(\vec{r},t)dV_c\qquad(3.22)$$

上式中，$\rho_{NP}(\vec{r},t)$ 和 $\rho_{NC}(\vec{r},t)$ 是组成成分 $p$ 和 $c$ 的真密度，即材料密度。若将式（3.22）在任一体积 $V$

内取平均，则两边通乘以 $\delta(\vec{R}-\vec{r})$，然后在体积 $V$ 内积分，即

$$\int_V \rho(\vec{R},t)\overrightarrow{f'}(\vec{R},t)\delta(\vec{R}-\vec{r})dV = \int_{V_p} \rho_{NP}(\vec{R},t)\overrightarrow{f'_{NP}}(\vec{R},t)\delta(\vec{R}-\vec{r})dV_p +$$
$$\int_{V_c} \rho_{NC}(\vec{R},t)\overrightarrow{f'_{NC}}(\vec{R},t)\delta(\vec{R}-\vec{r})dV_c \qquad (3.23a)$$

或写成

$$\int_V \rho(\vec{R},t)\overrightarrow{f'}(\vec{R},t)\delta(\vec{R}-\vec{r})dV = \int_V \rho_{NP}(\vec{R},t)\phi_p(\vec{R},t)\overrightarrow{f'_{NP}}(\vec{R},t)\delta(\vec{R}-\vec{r})dV +$$
$$\int_V \rho_{NC}(\vec{R},t)\phi_c(\vec{R},t)\overrightarrow{f'_{NC}}(\vec{R},t)\delta(\vec{R}-\vec{r})dV \qquad (3.23b)$$

定义在矢径端点 $(x,y,z)$ 的邻域 $V$ 内的以单位质量计量的物理量之平均值：

$$\left.\begin{array}{l} \phi_p(\vec{r},t)\rho_{NP}(\vec{r},t)\overrightarrow{f'_{NP}}(\vec{r},t) = \int_{V_p} \rho_{NP}(\vec{R},t)\overrightarrow{f'_{NP}}(\vec{R},t)\delta(\vec{R}-\vec{r})dV_p = \\ \qquad\qquad \int_V \rho_{NP}(\vec{R},t)\phi_p(\vec{R},t)\overrightarrow{f'_{NP}}(\vec{R},t)\delta(\vec{R}-\vec{r})dV \\ \phi_c(\vec{r},t)\rho_{NC}(\vec{r},t)\overrightarrow{f'_{NC}}(\vec{r},t) = \int_{V_c} \rho_{NC}(\vec{R},t)\overrightarrow{f'_{NC}}(\vec{R},t)\delta(\vec{R}-\vec{r})dV_c = \\ \qquad\qquad \int_V \rho_{NC}(\vec{R},t)\phi_c(\vec{R},t)\overrightarrow{f'_{NC}}(\vec{R},t)\delta(\vec{R}-\vec{r})dV \end{array}\right\} \qquad (3.24)$$

若式（3.17）和式（3.24）所表示的物理量是同一的，则以单位体积计量和以单位质量计量的平均值之间的关系为

$$\overrightarrow{f_{NP}}(\vec{r},t) = \rho_{NP}(\vec{r},t)\overrightarrow{f'_{NP}}(\vec{r},t), \qquad \overrightarrow{f_{NC}}(\vec{r},t) = \rho_{NC}(\vec{r},t)\overrightarrow{f'_{NC}}(\vec{r},t) \qquad (3.25a)$$

当然也有

$$\vec{f}(\vec{r},t) = \rho(\vec{r},t)\overrightarrow{f'}(\vec{r},t) \qquad (3.25b)$$

此即两种平均值之间的关联系数是密度。

## 3.4 铁磁流体的密度
### 3.4.1 铁磁流体的密度和分密度

密度关联体积和质量，它本身是以单位体积计量的物理量。在式（3.16）中，取矢径 $\vec{r}$ 端点 $(x,y,z)$ 的邻域 $V$ 内的平均值 $f(\vec{r},t) = \rho_f(\vec{r},t)$，$f_{NP}(\vec{r},t) = \rho_{NP}(\vec{r},t)$，$f_{NC}(\vec{r},t) = \rho_{NC}(\vec{r},t)$，则有

$$\rho_f(\vec{r},t) = \phi_p(\vec{r},t)\rho_{NP}(\vec{r},t) + \phi_c(\vec{r},t)\rho_{NC}(\vec{r},t) \qquad (3.26)$$

式中，$\rho_f(\vec{r},t)$ 是铁磁流体的密度。对于现有的绝大多数铁磁流体而言，其成分的 $\rho_{NP}$ 和 $\rho_{NC}$ 均是常数。但这不说明铁磁流体的密度 $\rho_f$ 也是常数，它还取决于两相的体积分数。只有在稳定的、均匀的状况下，$\rho_f$ 才是常数。式（3.17）所定义的体积分物理量 $\overrightarrow{f_p}(\vec{r},t)$ 和 $\overrightarrow{f_c}(\vec{r},t)$ 在此就是 $\rho_p(\vec{r},t)$ 和 $\rho_c(\vec{r},t)$，即

$$\rho_p(\vec{r},t) = \phi_p(\vec{r},t)\rho_{NP}, \qquad \rho_c(\vec{r},t) = \phi_c(\vec{r},t)\rho_{NC} \qquad (3.27a)$$

$\rho_p(\vec{r},t)$ 和 $\rho_c(\vec{r},t)$ 分别称为固相的分密度和液相的分密度。式（3.26）可以写成铁磁流体在邻域 $V$ 内的密度平均值等于其成分的分密度之和，即

$$\rho_f(\vec{r},t) = \rho_p(\vec{r},t) + \rho_c(\vec{r},t) \tag{3.27b}$$

### 3.4.2 体积分数 $\phi$ 和质量分数 $\varphi$ 的关系

分密度乃是铁磁流体的某一种成分在邻域 $V$ 内的质量"充满"体积 $V$ 时的平均密度，即

$$\left.\begin{aligned}
\rho_p(\vec{r},t) &= \frac{m_p}{V} = \frac{m_f}{V}\frac{m_p}{m_f} = \rho_f(\vec{r},t)\varphi_p(\vec{r},t) \\
\rho_c(\vec{r},t) &= \frac{m_c}{V} = \frac{m_f}{V}\frac{m_c}{m_f} = \rho_f(\vec{r},t)\varphi_c(\vec{r},t)
\end{aligned}\right\} \tag{3.28}$$

由式（3.27a）和式（3.28）得

$$\varphi_p(\vec{r},t) = \frac{\rho_{NP}}{\rho_f(\vec{r},t)}\phi_p(\vec{r},t), \qquad \varphi_c(\vec{r},t) = \frac{\rho_{NC}}{\rho_f(\vec{r},t)}\phi_c(\vec{r},t) \tag{3.29}$$

相加

$$\varphi_p(\vec{r},t) + \varphi_c(\vec{r},t) = \frac{1}{\rho_f(\vec{r},t)}\Big[\phi_p(\vec{r},t)\rho_{NP} + \phi_c(\vec{r},t)\rho_{NC}\Big]$$

使用式（3.26）立即就得

$$\varphi_p(\vec{r},t) + \varphi_c(\vec{r},t) = 1 \tag{3.30}$$

联立（3.29）的两式，解出

$$\varphi_p(\vec{r},t) = \frac{\phi_p(\vec{r},t)}{\phi_p(\vec{r},t) + [1-\phi_p(\vec{r},t)]\dfrac{\rho_{NC}}{\rho_{NP}}} \tag{3.31a}$$

$$\phi_p(\vec{r},t) = \frac{\varphi_p(\vec{r},t)}{\varphi_p(\vec{r},t) + [1-\varphi_p(\vec{r},t)]\dfrac{\rho_{NP}}{\rho_{NC}}} \tag{3.31b}$$

### 3.4.3 用 $\phi_p(\vec{r},t)$ 和 $\varphi_p(\vec{r},t)$ 计算铁磁流体的密度 $\rho_f(\vec{r},t)$

以 $\phi_c(\vec{r},t) = 1-\phi_p(\vec{r},t)$ 代入式（3.26）即有

$$\rho_f(\vec{r},t) = \phi_p(\vec{r},t)\rho_{NP} + [1-\phi_p(\vec{r},t)]\rho_{NC} \tag{3.32a}$$

用式（3.31b）代入（3.29）的第一式之右方，就得

$$\rho_f(\vec{r},t) = \frac{\rho_{NP}}{\varphi_p(\vec{r},t) + [1 - \varphi_p(\vec{r},t)]\dfrac{\rho_{NP}}{\rho_{NC}}} \qquad (3.32b)$$

### 3.5  平均值的导数和导数的平均值

### 3.5.1  平均值对时间的导数

设一代表物理量的矢性函数 $\vec{f}(\vec{r},t) = \vec{f}(x,y,z;t)$，其中 $(x,y,z)$ 是矢径 $\vec{r}$ 的端点坐标。邻域 $V$ 的体积

元 $dV = dxdydz$，则在点 $(x,y,z)$ 的邻域 $V$ 内 $\vec{f}(\vec{r},t)$ 的平均值是

$$\vec{f}(\vec{r},t) = \int_V \vec{f}(\vec{R},t)\delta(\vec{R}-\vec{r})dV = \int_V \vec{f}(\xi,\eta,\varsigma;t)\delta(\xi-x,\eta-y,\varsigma-z)d\xi d\eta d\varsigma$$

将上式两边对时间取导数

$$\frac{d}{dt}\vec{f}(\vec{r},t) = \frac{d}{dt}\int_V \vec{f}(\xi,\eta,\varsigma;t)\delta(\xi-x,\eta-y,\varsigma-z)\,d\xi d\eta d\varsigma =$$

$$\int_V d\xi d\eta d\varsigma \frac{d}{dt}\Big[\vec{f}(\xi,\eta,\varsigma;t)\delta(\xi-x,\eta-y,\varsigma-z)\Big] + $$

$$\int_V \vec{f}(\xi,\eta,\varsigma;t)\delta(\xi-x,\eta-y,\varsigma-z)\frac{d(d\xi d\eta d\varsigma)}{(d\xi d\eta d\varsigma)t}\,d\xi d\eta d\varsigma \qquad (A)$$

式（A）右方第一个积分内

$$\frac{d}{dt}\Big[\vec{f}(\xi,\eta,\varsigma;t)\delta(\xi-x,\eta-y,\varsigma-z)\Big] =$$

$$\frac{\partial}{\partial t}\Big[\vec{f}(\xi,\eta,\varsigma;t)\delta(\xi-x,\eta-y,\varsigma-z)\Big] + \frac{\partial}{\partial \xi}\Big[\vec{f}(\xi,\eta,\varsigma;t)\delta(\xi-x,\eta-y,\varsigma-z)\Big]\frac{d\xi}{dt} +$$

$$\frac{\partial}{\partial \eta}\Big[\vec{f}(\xi,\eta,\varsigma;t)\delta(\xi-x,\eta-y,\varsigma-z)\Big]\frac{d\eta}{dt} + \frac{\partial}{\partial \varsigma}\Big[\vec{f}(\xi,\eta,\varsigma;t)\delta(\xi-x,\eta-y,\varsigma-z)\Big]\frac{d\varsigma}{dt} =$$

$$\frac{\partial}{\partial t}\Big[\vec{f}(\xi,\eta,\varsigma;t)\delta(\xi-x,\eta-y,\varsigma-z)\Big] + \Big[\frac{d\xi}{dt}\frac{\partial}{\partial \xi} + \frac{d\eta}{dt}\frac{\partial}{\partial \eta} + \frac{d\varsigma}{dt}\frac{\partial}{\partial \varsigma}\Big]\vec{f}(\xi,\eta,\varsigma;t)\delta(\xi-x,\eta-y,\varsigma-z) =$$

$$\frac{\partial}{\partial t}\Big[\vec{f}(\vec{R},t)\delta(\vec{R}-\vec{r})\Big] + \Big[\vec{U}(\vec{R},t)\cdot\nabla_R\Big]\vec{f}(\vec{R},t)\delta(\vec{R}-\vec{r}) \qquad (B)$$

上式中，

$$\frac{d\xi}{dt} = u(\xi,\eta,\varsigma;t) = u(\vec{R},t), \qquad \frac{d\eta}{dt} = v(\xi,\eta,\varsigma;t) = v(\vec{R},t), \qquad \frac{d\varsigma}{dt} = w(\xi,\eta,\varsigma;t) = w(\vec{R},t)$$

$$\vec{U}(\vec{R},t) = \vec{i}\,u(\vec{R},t) + \vec{j}\,v(\vec{R},t) + \vec{k}\,w(\vec{R},t), \qquad \nabla_R = \vec{i}\frac{\partial}{\partial \xi} + \vec{j}\frac{\partial}{\partial \eta} + \vec{r}\frac{\partial}{\partial \varsigma}$$

式（A）右方第二个积分内

$$\frac{d(d\xi d\eta d\varsigma)}{(d\xi d\eta d\varsigma)t} = \frac{d}{d\xi}\frac{d\xi}{dt} + \frac{d}{d\eta}\frac{d\eta}{dt} + \frac{d}{d\varsigma}\frac{d\varsigma}{dt} = \frac{d}{d\xi}u(\xi,\eta,\varsigma;t) + \frac{d}{d\eta}v(\xi,\eta,\varsigma;t) + \frac{d}{d\varsigma}w(\xi,\eta,\varsigma;t) =$$

$$\frac{\partial}{\partial \xi}u(\vec{R},t) + \frac{\partial}{\partial \eta}v(\vec{R},t) + \frac{\partial}{\partial \varsigma}w(\vec{R},t) = \nabla_R\cdot\vec{U}(\vec{R},t) \qquad (C)$$

将式（B）和式（C）分别代入式（A）右方的两个积分中，就得

$$\frac{d}{dt}\vec{f}(\vec{r},t) = \int_V \left\{ \frac{\partial}{\partial t}\left[\vec{f}(\vec{R},t)\delta(\vec{R}-\vec{r}) + \vec{U}(\vec{R},t)\cdot\nabla_R\vec{f}(\vec{R},t)\delta(\vec{R}-\vec{r})\right]\right\}d\xi d\eta d\varsigma +$$
$$\int_V \vec{f}(\vec{R},t)\delta(\vec{R}-\vec{r})\left[\nabla_R\cdot\vec{U}(\vec{R},t)\right]d\xi d\eta d\varsigma$$

相关项合并以后得

$$\frac{d}{dt}\vec{f}(\vec{r},t) = \int_V \frac{\partial}{\partial t}\left[\vec{f}(\vec{R},t)\delta(\vec{R}-\vec{r})\right]dV + \int_V \nabla_R\cdot\left[\vec{U}(\vec{R},t)\vec{f}(\vec{R},t)\delta(\vec{R}-\vec{r})\right]dV \quad (3.33a)$$

使用散度定理于上式右方第二个积分，就有

$$\frac{d}{dt}\vec{f}(\vec{r},t) = \int_V \frac{\partial}{\partial t}\left[\vec{f}(\vec{R},t)\delta(\vec{R}-\vec{r})\right]dV + \int_S d\vec{S}\cdot\left[\vec{U}(\vec{R},t)\vec{f}(\vec{R},t)\delta(\vec{R}-\vec{r})\right] \quad (3.33b)$$

式中 $S$ 是包围邻域 $V$ 的面积。

在铁磁流体中，包含固相和液相两种成分，式（3.13）给出：

$$\int_V \vec{f}(\vec{R},t)\delta(\vec{R}-\vec{r})dV = \int_{V_p}\overline{f_{NP}}(\vec{R},t)\delta(\vec{R}-\vec{r})dV_p + \int_{V_c}\overline{f_{NC}}(\vec{R},t)\delta(\vec{R}-\vec{r})dV_c$$

两边对时间 $t$ 取导数，参照式（3.33a），得到

$$\frac{d}{dt}\vec{f}(\vec{r},t) = \int_V \frac{\partial}{\partial t}\left[\vec{f}(\vec{R},t)\delta(\vec{R}-\vec{r})\right]dV + \nabla_R\cdot\left[\vec{U}(\vec{R},t)\vec{f}(\vec{R},t)\delta(\vec{R}-\vec{r})\right]dV =$$
$$\int_{V_p}\frac{\partial}{\partial t}\left[\overline{f_{NP}}(\vec{R},t)\delta(\vec{R}-\vec{r})\right]dV_p + \int_{V_p}\nabla_R\cdot\left[\overline{U_p}(\vec{R},t)\overline{f_{NP}}(\vec{R},t)\delta(\vec{R}-\vec{r})\right]dV_p + \quad (3.34a)$$
$$\int_{V_c}\frac{\partial}{\partial t}\left[\overline{f_{NC}}(\vec{R},t)\delta(\vec{R}-\vec{r})\right]dV_c + \int_{V_c}\nabla_R\cdot\left[\overline{U_c}(\vec{R},t)\overline{f_{NC}}(\vec{R},t)\delta(\vec{R}-\vec{r})\right]dV_c$$

或参照式（3.33b）得使用散度定理后的形式

$$\int_V \frac{\partial}{\partial t}\left[\vec{f}(\vec{R},t)\delta(\vec{R}-\vec{r})\right]dV + \int_S d\vec{S}\cdot\left[\vec{U}(\vec{R},t)\vec{f}(\vec{R},t)\delta(\vec{R}-\vec{r})\right] =$$
$$\int_{V_p}\frac{\partial}{\partial t}\left[\overline{f_{NP}}(\vec{R},t)\delta(\vec{R}-\vec{r})\right]dV_p + \int_{S_p}d\vec{S_p}\cdot\left[\overline{U_p}(\vec{R},t)\overline{f_{NP}}(\vec{R},t)\delta(\vec{R}-\vec{r})\right] + \quad (3.34b)$$
$$\int_{V_c}\frac{\partial}{\partial t}\left[\overline{f_{NC}}(\vec{R},t)\delta(\vec{R}-\vec{r})\right]dV_c + \int_{S_c}d\vec{S_c}\cdot\left[\overline{U_c}(\vec{R},t)\overline{f_{NC}}(\vec{R},t)\delta(\vec{R}-\vec{r})\right]$$

在铁磁流体中，固相以微粒形式弥散分布于基载液体中，所以在体积 $V$ 内两相的表面积是

$$S_p = S_{p0} + \sum_N S_{p1}, \qquad S_c = S_{c0} + \sum_N S_{c1} \quad (3.35a)$$

式中 $S_{p0}$ 和 $S_{c0}$ 是体积 $V$ 外表面上的固相和液相的面积，$S_{p1}$ 是单个微粒的表面积，$S_{c1}$ 是基载液体与单个固相微粒的交接面积，显然在数值上 $S_{p1} = S_{c1}$，$N$ 是体积 $V$ 中包含的固相微粒数目，由式（3.35a）取微元面矢：

$$d\vec{S_p} = d\vec{S_{p0}} + \sum_N d\vec{S_{p1}}, \qquad d\vec{S_c} = d\vec{S_{c0}} + \sum_N d\vec{S_{c1}} \quad (3.35b)$$

若跨过固相微粒与基载液的接界面取一微元体，底面积为 $dS_{p1}$，顶面积为 $dS_{c1}$，高度为 $h$，且 $h \to 0$，如图 3-2 所示。微元面矢 $d\vec{S_{p1}} = \overline{n_{p1}^0}dS_{p1}$，$d\vec{S_{c1}} = \overline{n_{c1}^0}dS_{c1}$，由图 3-2 可见，单位法向矢 $\overline{n_{p1}^0}$ 对界面微元体

是其底面的内向法线方向，而 $\overrightarrow{n_{c1}^0}$ 是其顶面的外向法线方向，所以 $\overrightarrow{n_{p1}^0} = -\overrightarrow{n_{c1}^0}$ ，在数值上 $dS_{p1} = dS_{c1}$ ，从而

$$-d\overrightarrow{S_{p1}} = d\overrightarrow{S_{c1}} \tag{3.35c}$$

体积 $V$ 的全面积

$$S = S_p + S_c = S_{p0} + \sum_N S_{p1} + S_{c0} + \sum_N S_{c1} \tag{3.35d}$$

矢性面元

$$d\overrightarrow{S} = d\overrightarrow{S_{p0}} + \sum_N d\overrightarrow{S_{p1}} + d\overrightarrow{S_{c0}} + \sum_N d\overrightarrow{S_{c1}} = d\overrightarrow{S_{p0}} + d\overrightarrow{S_{c0}} \tag{3.35e}$$

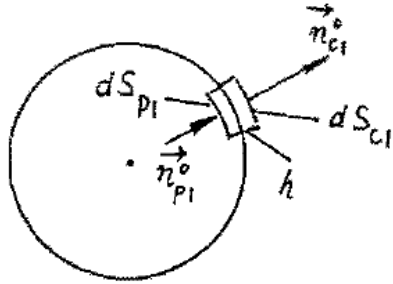

图 3-2　固相微粒表面上的跨界面微元体

用式（3.35b）代入式（3.34b）中，就有

$$
\int_V \frac{\partial}{\partial t}\left[\vec{f}(\vec{R},t)\delta(\vec{R}-\vec{r})\right]dV + \int_S d\vec{S} \cdot \left[\vec{U}(\vec{R},t)\vec{f}(\vec{R},t)\delta(\vec{R}-\vec{r})\right] =
$$
$$
\int_{V_p} \frac{\partial}{\partial t}\left[\overline{f_{NP}}(\vec{R},t)\delta(\vec{R}-\vec{r})\right]dV_p + \int_{S_{p0}} d\overrightarrow{S_{p0}} \cdot \left[\overrightarrow{U_{p0}}(\vec{R},t)\overline{f_{NP}}(\vec{R},t)\delta(\vec{R}-\vec{r})\right] +
$$
$$
\int_{V_c} \frac{\partial}{\partial t}\left[\overline{f_{NC}}(\vec{R},t)\delta(\vec{R}-\vec{r})\right]dV_c + \int_{S_{c0}} d\overrightarrow{S_{c0}} \cdot \left[\overrightarrow{U_{c0}}(\vec{R},t)\overline{f_{NC}}(\vec{R},t)\delta(\vec{R}-\vec{r})\right] +
$$
$$
\sum_N \int_{S_{p1}} d\overrightarrow{S_{p1}} \cdot \left[\overrightarrow{U_{p1}}(\vec{R},t)\overline{f_{NP}}(\vec{R},t) - \overrightarrow{U_{c1}}(\vec{R},t)\overline{f_{NC}}(\vec{R},t)\right]\delta(\vec{R}-\vec{r}) \tag{3.36a}
$$

在固相微粒与基载液接触的界面上恒有粘附条件，即

$$\overrightarrow{U_{p1}}(\vec{R},t) = \overrightarrow{U_{c1}}(\vec{R},t)$$

若在界面上同时存在

$$\overline{f_{NP}}(\vec{R},t)\Big|_{S_{p1}} = \overline{f_{NC}}(\vec{R},t)\Big|_{S_{c1}}$$

则式（3.36a）右方之 $\sum_N$ 项等于零，即

$$
\int_V \frac{\partial}{\partial t}\left[\vec{f}(\vec{R},t)\delta(\vec{R}-\vec{r})\right]dV + \int_S d\vec{S} \cdot \left[\vec{U}(\vec{R},t)\vec{f}(\vec{R},t)\delta(\vec{R}-\vec{r})\right] =
$$
$$
\int_{V_p} \frac{\partial}{\partial t}\left[\overline{f_{NP}}(\vec{R},t)\delta(\vec{R}-\vec{r})\right]dV_p + \int_{S_{p0}} d\overrightarrow{S_{p0}} \cdot \left[\overrightarrow{U_{p0}}(\vec{R},t)\overline{f_{NP}}(\vec{R},t)\delta(\vec{R}-\vec{r})\right] +
$$
$$
\int_{V_c} \frac{\partial}{\partial t}\left[\overline{f_{NC}}(\vec{R},t)\delta(\vec{R}-\vec{r})\right]dV_c + \int_{S_{c0}} d\overrightarrow{S_{c0}} \cdot \left[\overrightarrow{U_{c0}}(\vec{R},t)\overline{f_{NC}}(\vec{R},t)\delta(\vec{R}-\vec{r})\right] \tag{3.36b}
$$

3.5.2 平均值的散度和散度的平均值

平均值的散度是

$$\nabla_r \cdot \overline{f_r}(\vec{r},t) = \nabla_r \cdot \int_V \overline{f}(\vec{R},t)\delta(\vec{R}-\vec{r})dV$$

注意到上式的积分变量是 $dV = d\xi d\eta d\varsigma$，而算子 $\nabla_r$ 是对 $(x,y,z)$ 运算。所以 $\nabla_r$ 可以写入积分号之内，即

$$\nabla_r \cdot \int_V \overline{f}(\vec{R},t)\delta(\vec{R}-\vec{r})dV = \int_V \nabla_r \cdot \left[\overline{f}(\vec{R},t)\delta(\vec{R}-\vec{r})\right]dV = \int_V \overline{f}(\vec{R},t)\cdot\nabla_r\delta(\vec{R}-\vec{r})dV \quad (3.37)$$

由于[①]

$$\nabla_r\delta(\vec{R}-\vec{r}) = -\nabla_R\delta(\vec{R}-\vec{r})$$

于是式（3.37）可改写成

$$\nabla_r \cdot \int_V \overline{f}(\vec{R},t)\delta(\vec{R}-\vec{r})dV = -\int_V \overline{f}(\vec{R},t)\cdot\nabla_R\delta(\vec{R}-\vec{r})dV$$

由矢量恒等式给出

$$\nabla_R \cdot \left[\overline{f}(\vec{R},t)\delta(\vec{R}-\vec{r})\right] = \overline{f}(\vec{R},t)\cdot\nabla_R\delta(\vec{R}-\vec{r}) + \delta(\vec{R}-\vec{r})\nabla_R \cdot \overline{f}(\vec{R},t)$$

代入上式之右方就得

$$\nabla_r \cdot \int_V \overline{f}(\vec{R},t)\delta(\vec{R}-\vec{r})dV = \int_V \delta(\vec{R}-\vec{r})\nabla_R \cdot \overline{f}(\vec{R},t)dV - \int_V \nabla_R \cdot \left[\overline{f}(\vec{R},t)\delta(\vec{R}-\vec{r})\right]dV \quad (3.38a)$$

使用散度定于上式右方第二个积分，则有

$$\nabla_r \cdot \int_V \overline{f}(\vec{R},t)\delta(\vec{R}-\vec{r})dV = \int_V \delta(\vec{R}-\vec{r})\nabla_R \cdot \overline{f}(\vec{R},t)dV - \int_S d\vec{S} \cdot \overline{f}(\vec{R},t)\delta(\vec{R}-\vec{r}) \quad (3.38b)$$

上式左边是平均值 $\overline{f}(\vec{r},t)$ 的散度，右边第一个积分是其散度的平均值。两者之差别就在右边第二个积分上，即体积 $V$ 表面上的法向通量。

铁磁流体是两相混合物，由式（3.13）和式（3.14）两边取散度，得

$$\nabla_r \cdot \int_V \overline{f}(\vec{R},t)\delta(\vec{R}-\vec{r})dV = \nabla_r \cdot \int_{V_p} \overline{f_{NP}}(\vec{R},t)\delta(\vec{R}-\vec{r})dV_p + \nabla_r \cdot \int_{V_c} \overline{f_{NC}}(\vec{R},t)\delta(\vec{R}-\vec{r})dV_c \quad (3.39a)$$

和

---

[①] $\nabla_r = \vec{i}\dfrac{\partial}{\partial x} + \vec{j}\dfrac{\partial}{\partial y} + \vec{k}\dfrac{\partial}{\partial z}$, $\nabla_R = \vec{i}\dfrac{\partial}{\partial \xi} + \vec{j}\dfrac{\partial}{\partial \eta} + \vec{k}\dfrac{\partial}{\partial \varsigma}$, $\delta(\vec{R}-\vec{r}) = \delta(\xi-x,\eta-y,\varsigma-z)$, $\vec{R}-\vec{r}=\vec{R'}$,

$\nabla_r\delta = \vec{i}\dfrac{\partial\delta}{\partial x} + \vec{j}\dfrac{\partial\delta}{\partial y} + \vec{k}\dfrac{\partial\delta}{\partial z} = \vec{i}\dfrac{\partial\delta}{\partial(\xi-x)}\dfrac{\partial(\xi-x)}{\partial x} + \vec{j}\dfrac{\partial\delta}{\partial(\eta-y)}\dfrac{\partial(\eta-y)}{\partial y} + \vec{k}\dfrac{\partial\delta}{\partial(\varsigma-z)}\dfrac{\partial(\varsigma-z)}{\partial z} =$

$\qquad -\vec{i}\dfrac{\partial\delta}{\partial(\xi-x)} - \vec{j}\dfrac{\partial\delta}{\partial(\eta-y)} - \vec{k}\dfrac{\partial\delta}{\partial(\varsigma-z)} = -\nabla_R\delta(\vec{R}-\vec{r})$,

$\nabla_R\delta = \vec{i}\dfrac{\partial\delta}{\partial\xi} + \vec{j}\dfrac{\partial\delta}{\partial\eta} + \vec{k}\dfrac{\partial\delta}{\partial\varsigma} = \vec{i}\dfrac{\partial\delta}{\partial(\xi-x)}\dfrac{\partial(\xi-x)}{\partial\xi} + \vec{j}\dfrac{\partial\delta}{\partial(\eta-y)}\dfrac{\partial(\eta-y)}{\partial\eta} + \vec{k}\dfrac{\partial\delta}{\partial(\varsigma-z)}\dfrac{\partial(\varsigma-z)}{\partial\varsigma} =$

$\qquad \vec{i}\dfrac{\partial\delta}{\partial(\xi-x)} + \vec{j}\dfrac{\partial\delta}{\partial(\eta-y)} + \vec{k}\dfrac{\partial\delta}{\partial(\varsigma-z)} = \nabla_R\delta(\vec{R}-\vec{r})$

故 $\nabla_r\delta = -\nabla_R\delta$。

$$\nabla_r \cdot \int_V \overline{f}(\vec{R},t)\delta(\vec{R}-\vec{r})dV = \nabla_r \cdot \int_V \phi_p(\vec{R},t)\overline{f_{NP}}(\vec{R},t)\delta(\vec{R}-\vec{r})dV +$$
$$\nabla_r \cdot \int_V \phi_c(\vec{R},t)\overline{f_{NC}}(\vec{R},t)\delta(\vec{R}-\vec{r})dV \tag{3.39b}$$

由式（3.39b）可以从形式上写出

$$\nabla_r \cdot \overline{f}(\vec{r},t) = \nabla_r \cdot \left[\phi_p(\vec{r},t)\overline{f_{NP}}(\vec{r},t)\right] + \nabla_r \cdot \left[\phi_c(\vec{r},t)\overline{f_{NC}}(\vec{r},t)\right] \tag{3.39c}$$

参照式（3.38b），则式（3.39a）可以写成为

$$\int_V \delta(\vec{R}-\vec{r})\nabla_R \cdot \overline{f}(\vec{R},t)dV - \int_S d\vec{S} \cdot \overline{f}(\vec{R},t)\delta(\vec{R}-\vec{r}) =$$
$$\int_{V_p} \delta(\vec{R}-\vec{r})\nabla_R \cdot \overline{f_{NP}}(\vec{R},t)dV_p - \int_{S_p} d\vec{S_p} \cdot \overline{f_{NP}}(\vec{R},t)\delta(\vec{R}-\vec{r}) +$$
$$\int_{V_c} \delta(\vec{R}-\vec{r})\nabla_R \cdot \overline{f_{NC}}(\vec{R},t)dV_c - \int_{S_c} d\vec{S_c} \cdot \overline{f_{NC}}(\vec{R},t)\delta(\vec{R}-\vec{r})$$

考虑到式（3.35b）和式（3.35c），则上式可写成

$$\nabla_r \cdot \int_V \overline{f}(\vec{R},t)\delta(\vec{R}-\vec{r})dV = \int_V \delta(\vec{R}-\vec{r})\nabla_R \cdot \overline{f}(\vec{R},t)dV - \int_S d\vec{S} \cdot \overline{f}(\vec{R},t)\delta(\vec{R}-\vec{r}) =$$
$$\int_{V_p} \delta(\vec{R}-\vec{r})\nabla_R \cdot \overline{f_{NP}}(\vec{R},t)dV_p - \int_{S_{p0}} d\vec{S_{p0}} \cdot \overline{f_{NP}}(\vec{R},t)\delta(\vec{R}-\vec{r}) +$$
$$\int_{V_c} \delta(\vec{R}-\vec{r})\nabla_R \cdot \overline{f_{NC}}(\vec{R},t)dV_c - \int_{S_{c0}} d\vec{S_{c0}} \cdot \overline{f_{NC}}(\vec{R},t)\delta(\vec{R}-\vec{r}) -$$
$$\sum_N \int_{S_{p1}} d\vec{S_{p1}} \cdot \left[\overline{f_{NP}}(\vec{R},t) - \overline{f_{NC}}(\vec{R},t)\right]\delta(\vec{R}-\vec{r}) \tag{3.40a}$$

若在两相的界面上存在

$$\overline{f_{NP}}(\vec{R},t)\Big|_{S_{p1}} = \overline{f_{NC}}(\vec{R},t)\Big|_{S_{c1}}$$

则式（3.40a）右方之 $\Sigma_N$ 项等于零，从而有

$$\nabla_r \cdot \int_V \overline{f}(\vec{R},t)\delta(\vec{R}-\vec{r})dV = \int_V \delta(\vec{R}-\vec{r})\nabla_R \cdot \overline{f}(\vec{R},t)dV - \int_S d\vec{S} \cdot \overline{f}(\vec{R},t)\delta(\vec{R}-\vec{r}) =$$
$$\int_{V_p} \delta(\vec{R}-\vec{r})\nabla_R \cdot \overline{f_{NP}}(\vec{R},t)dV_p - \int_{S_{p0}} d\vec{S_{p0}} \cdot \overline{f_{NP}}(\vec{R},t)\delta(\vec{R}-\vec{r}) +$$
$$\int_{V_c} \delta(\vec{R}-\vec{r})\nabla_R \cdot \overline{f_{NC}}(\vec{R},t)dV_c - \int_{S_{c0}} d\vec{S_{c0}} \cdot \overline{f_{NC}}(\vec{R},t)\delta(\vec{R}-\vec{r}) \tag{3.40b}$$

### 3.5.3 平均值的旋度与旋度的平均值

上节平均值的散度，使用的是矢量恒等式 $\nabla \cdot (\varphi\vec{A}) = \varphi\nabla \cdot \vec{A} + \vec{A} \cdot \nabla\varphi$ 和散度定理，即 $\int_V \nabla \cdot \vec{A}dV = \int_S d\vec{S} \cdot \vec{A}$，而本节平均值的旋度，将使用恒等式 $\nabla \times (\varphi\vec{A}) = \varphi\nabla \times \vec{A} + (\nabla\varphi) \times \vec{A}$ 和旋度定理，即 $\int_V \nabla \times \vec{A}dV = \int_S d\vec{S} \times \vec{A}$。所以上节 3.5.2 中的所有公式只需将点乘改为叉乘，即可用于有关旋度的问题。参照式（3.38a）与式（3.38b），有

$$\nabla_r \times \int_V \overline{f}(\vec{R},t)\delta(\vec{R}-\vec{r})dV = \int_V \delta(\vec{R}-\vec{r})\nabla_R \times \overline{f}(\vec{R},t)dV - \int_V \nabla_R \times \left[\overline{f}(\vec{R},t)\delta(\vec{R}-\vec{r})\right]dV \tag{3.41a}$$

和

$$\nabla_r \times \int_V \overrightarrow{f}(\overrightarrow{R},t)\delta(\overrightarrow{R}-\overrightarrow{r})dV = \int_V \delta(\overrightarrow{R}-\overrightarrow{r})\nabla_R \times \overrightarrow{f}(\overrightarrow{R},t)dV - \int_S d\overrightarrow{S} \times \overrightarrow{f}(\overrightarrow{R},t)\delta(\overrightarrow{R}-\overrightarrow{r}) \tag{3.41b}$$

以及形式上的关系类似于式（3.39a）～式（3.39c）：

$$\nabla_r \times \int_V \overrightarrow{f}(\overrightarrow{R},t)\delta(\overrightarrow{R}-\overrightarrow{r})dV = \nabla_r \times \int_{V_p} \overline{f_{NP}}(\overrightarrow{R},t)\delta(\overrightarrow{R}-\overrightarrow{r})dV_p + \nabla_r \times \int_{V_c} \overline{f_{NC}}(\overrightarrow{R},t)\delta(\overrightarrow{R}-\overrightarrow{r})dV_c \tag{3.42a}$$

$$\begin{aligned}\nabla_r \times \int_V \overrightarrow{f}(\overrightarrow{R},t)\delta(\overrightarrow{R}-\overrightarrow{r})dV = &\nabla_r \times \int_V \phi_p(\overrightarrow{R},t)\overline{f_{NP}}(\overrightarrow{R},t)\delta(\overrightarrow{R}-\overrightarrow{r})dV + \\ &\nabla_r \times \int_V \phi_c(\overrightarrow{R},t)\overline{f_{NC}}(\overrightarrow{R},t)\delta(\overrightarrow{R}-\overrightarrow{r})dV\end{aligned} \tag{3.42b}$$

和

$$\nabla_r \times \overrightarrow{f}(\overrightarrow{r},t) = \nabla_r \times \left[\phi_p(\overrightarrow{r},t)\overline{f_{NP}}(\overrightarrow{r},t)\right] + \nabla_r \times \left[\phi_c(\overrightarrow{r},t)\overline{f_{NC}}(\overrightarrow{r},t)\right] \tag{3.42c}$$

参照式（3.40a）就有

$$\begin{aligned}\nabla_r \times \int_V \overrightarrow{f}(\overrightarrow{R},t)\delta(\overrightarrow{R}-\overrightarrow{r})dV = &\int_V \delta(\overrightarrow{R}-\overrightarrow{r})\nabla_R \times \overrightarrow{f}(\overrightarrow{R},t)dV - \int_S d\overrightarrow{S} \times \overrightarrow{f}(\overrightarrow{R},t)\delta(\overrightarrow{R}-\overrightarrow{r}) = \\ &\int_{V_p} \delta(\overrightarrow{R}-\overrightarrow{r})\nabla_R \times \overline{f_{NP}}(\overrightarrow{R},t)dV_p - \int_{S_{p0}} d\overrightarrow{S_{p0}} \times \overline{f_{NP}}(\overrightarrow{R},t)\delta(\overrightarrow{R}-\overrightarrow{r}) + \\ &\int_{V_c} \delta(\overrightarrow{R}-\overrightarrow{r})\nabla_R \times \overline{f_{NC}}(\overrightarrow{R},t)dV_c - \int_{S_{c0}} d\overrightarrow{S_{c0}} \times \overline{f_{NC}}(\overrightarrow{R},t)\delta(\overrightarrow{R}-\overrightarrow{r}) - \\ &\sum_N \int_S d\overrightarrow{S_{p1}} \times \left[\overline{f_{NP}}(\overrightarrow{R},t) - \overline{f_{NC}}(\overrightarrow{R},t)\right]\delta(\overrightarrow{R}-\overrightarrow{r})\end{aligned} \tag{3.42d}$$

若在两相界面上存在

$$\overline{f_{NP}}(\overrightarrow{R},t)\Big|_{S_{p1}} = \overline{f_{NC}}(\overrightarrow{R},t)\Big|_{S_{c1}}$$

则有

$$\begin{aligned}\nabla_r \times \int_V \overrightarrow{f}(\overrightarrow{R},t)\delta(\overrightarrow{R}-\overrightarrow{r})dV = &\int_V \delta(\overrightarrow{R}-\overrightarrow{r})\nabla_R \times \overrightarrow{f}(\overrightarrow{R},t)dV - \int_S d\overrightarrow{S} \times \overrightarrow{f}(\overrightarrow{R},t)\delta(\overrightarrow{R}-\overrightarrow{r}) = \\ &\int_{V_p} \delta(\overrightarrow{R}-\overrightarrow{r})\nabla_R \times \overline{f_{NP}}(\overrightarrow{R},t)dV_p - \int_{S_{p0}} d\overrightarrow{S_{p0}} \times \overline{f_{NP}}(\overrightarrow{R},t)\delta(\overrightarrow{R}-\overrightarrow{r}) + \\ &\int_{V_c} \delta(\overrightarrow{R}-\overrightarrow{r})\nabla_R \times \overline{f_{NC}}(\overrightarrow{R},t)dV_c - \int_{S_{c0}} d\overrightarrow{S_{c0}} \times \overline{f_{NC}}(\overrightarrow{R},t)\delta(\overrightarrow{R}-\overrightarrow{r})\end{aligned} \tag{3.42e}$$

### 3.5.4 平均值的梯度和梯度的平均值

设 $f(\overrightarrow{r},t)$ 是数性函数，其平均值之梯度为

$$\nabla_r f(\overrightarrow{r},t) = \nabla_r \int_V f(\overrightarrow{R},t)\delta(\overrightarrow{R}-\overrightarrow{r})dV$$

$\nabla_r$ 只对 $r$ 运作，和 $dV$ 没有关系，故 $\nabla_r$ 可写入积分号之内，即

$$\begin{aligned}\nabla_r \int_V f(\overrightarrow{R},t)\delta(\overrightarrow{R}-\overrightarrow{r})dV = &\int_V \nabla_r f(\overrightarrow{R},t)\delta(\overrightarrow{R}-\overrightarrow{r})dV = \\ &\int_V f(\overrightarrow{R},t)\nabla_r \delta(\overrightarrow{R}-\overrightarrow{r})dV = -\int_V f(\overrightarrow{R},t)\nabla_R \delta(\overrightarrow{R}-\overrightarrow{r})dV = \\ &-\int_V \nabla_R \left[f(\overrightarrow{R},t)\delta(\overrightarrow{R}-\overrightarrow{r})\right]dV + \int_V \delta(\overrightarrow{R}-\overrightarrow{r})\nabla_R f(\overrightarrow{R},t)dV\end{aligned}$$

由梯度定理给出

$$\int_V \nabla_R \left[ f(\vec{R},t)\delta(\vec{R}-\vec{r}) \right] dV = \int_S f(\vec{R},t)\delta(\vec{R}-\vec{r}) d\vec{S}$$

于是有

$$\int_V \delta(\vec{R}-\vec{r})\nabla_R f(\vec{R},t) dV = \nabla_r \int_V f(\vec{R},t)\delta(\vec{R}-\vec{r}) dV + \int_S f(\vec{R},t)\delta(\vec{R}-\vec{r}) d\vec{S} \qquad (3.43a)$$

对于两相流动，则有

$$\nabla_R \int_V \delta(\vec{R}-\vec{r}) f(\vec{R},t) dV = \int_V \delta(\vec{R}-\vec{r})\nabla_R f(\vec{R},t) dV - \int_S f(\vec{R},t)\delta(\vec{R}-\vec{r}) d\vec{S} =$$
$$\int_{V_p} \delta(\vec{R}-\vec{r})\nabla_R f_{NP}(\vec{R},t) dV_p - \int_{S_p} f_{NP}(\vec{R},t)\delta(\vec{R}-\vec{r}) d\vec{S_p} +$$
$$\int_{V_c} \delta(\vec{R}-\vec{r})\nabla_R f_{NC}(\vec{R},t) dV_c - \int_{S_c} f_{NC}(\vec{R},t)\delta(\vec{R}-\vec{r}) d\vec{S_c}$$

或写成

$$\nabla_r \int_V \delta(\vec{R}-\vec{r}) f(\vec{R},t) dV = \int_{V_p} \delta(\vec{R}-\vec{r})\nabla_R f_{NP}(\vec{R},t) dV_p - \int_{S_{p0}} f_{NP}(\vec{R},t)\delta(\vec{R}-\vec{r}) d\vec{S_{p0}} +$$
$$\int_{V_c} \delta(\vec{R}-\vec{r})\nabla_R f_{NC}(\vec{R},t) dV_c - \int_{S_{c0}} f_{NC}(\vec{R},t)\delta(\vec{R}-\vec{r}) d\vec{S_{c0}} - \qquad (3.43b)$$
$$\sum_N \int_{S_{p1}} d\vec{S_{p1}} \left[ f_{NP}(\vec{R},t) - f_{NC}(\vec{R},t) \right] \delta(\vec{R}-\vec{r})$$

## 3.6　铁磁流体输运通量的平均值

### 3.6.1　铁磁流体两相流对流输运通量的平均值

对流输运的特点是必须存在载体的宏观运动，而物理量之间的交换必须由其载体在流场中的对流交换来实现。铁磁流体中的载体当然就是固体微粒群和基础液体。

设 $\vec{U}(\vec{r},t)$ 是铁磁流体的两相混合物的速度，而其所"携带"的物理量以 $f(\vec{r},t)$ 表示。这里 $f(\vec{r},t)$ 是以单位体积计量的数性函数。穿过微元面积的输运通量是

$$f(\vec{r},t)\vec{U}(\vec{r},t)\cdot\vec{n^0} dS = f_{NP}(\vec{r},t)\vec{U_p}(\vec{r},t)\cdot\vec{n_p^0} dS_p + f_{NC}(\vec{r},t)\vec{U_c}(\vec{r},t)\cdot\vec{n_c^0} dS_c \qquad (3.44)$$

式中 $\vec{n^0}$，$\vec{n_p^0}$，$\vec{n_c^0}$ 分别是微元面积 $dS$，$dS_p$，$dS_c$ 的法向单位矢量，而 $S$，$S_p$，$S_c$ 分别是包围邻域 $V$，$V_p$，$V_c$ 的面积，并且 $V_p + V_c = V$。将式（3.44）两边在 $S$ 上取平均值，则有

$$\int_S \delta(\vec{R}-\vec{r}) f(\vec{R},t)\vec{U}(\vec{R},t)\cdot\vec{n^0} dS = \int_{S_p} \delta(\vec{R}-\vec{r}) f_{NP}(\vec{R},t)\vec{U_p}(\vec{R},t)\cdot\vec{n_p^0} dS_p +$$
$$\int_{S_c} \delta(\vec{R}-\vec{r}) f_{NC}(\vec{R},t)\vec{U_c}(\vec{R},t)\cdot\vec{n_c^0} dS_c \qquad (3.45)$$

上式两边使用散度定理，得

$$\int_V \nabla_R \cdot \left[ \delta(\vec{R}-\vec{r}) f(\vec{R},t)\vec{U}(\vec{R},t) \right] dV = \int_{V_p} \nabla_R \cdot \left[ \delta(\vec{R}-\vec{r}) f_{NP}(\vec{R},t)\vec{U_p}(\vec{R},t) \right] dV_p +$$
$$\int_{V_c} \nabla_R \cdot \left[ \delta(\vec{R}-\vec{r}) f_{NC}(\vec{R},t)\vec{U_c}(\vec{R},t) \right] dV_c \qquad (3.46)$$

先改变式（3.46）右方的第一个积分，使用定义式（3.8）和矢量恒等式 $\nabla \cdot (\varphi\vec{A}) = \varphi\nabla \cdot \vec{A} + \vec{A} \cdot \nabla\varphi$，就有

$$\int_{V_p} \nabla_R \cdot \left[ \delta(\vec{R}-r) f_{NP}(\vec{R},t) \overline{U_p}(\vec{R},t) \right] dV_p = \int_{V_p} \phi_p \nabla_R \cdot \left[ \delta(\vec{R}-r) f_{NP}(\vec{R},t) \overline{U_p}(\vec{R},t) \right] dV =$$

$$\int_{V} \nabla_R \cdot \left[ \delta(\vec{R}-r) \phi_p(\vec{R},t) f_{NP}(\vec{R},t) \overline{U_p}(\vec{R},t) \right] dV - \int_{V} \delta(\vec{R}-r) f_{NP}(\vec{R},t) \overline{U_p}(\vec{R},t) \cdot \nabla_R \phi_p(\vec{R},t) dV$$

式（3.46）右方的第二个积分亦仿照上式改变，并且注意到 $\nabla \phi_c(\vec{r},t) = -\nabla \phi_p(\vec{r},t)$，于是式（3.46）成为下面的形式：

$$\int_{V} \nabla_R \cdot \left[ \delta(\vec{R}-r) f(\vec{R},t) \overline{U}(\vec{R},t) \right] dV = \int_{V} \nabla_R \cdot \left[ \delta(\vec{R}-r) \phi_p(\vec{R},t) f_{NP}(\vec{R},t) \overline{U_p}(\vec{R},t) \right] dV +$$

$$\int_{V} \nabla_R \cdot \left[ \delta(\vec{R}-r) \phi_c(\vec{R},t) f_{NC}(\vec{R},t) \overline{U_c}(\vec{R},t) \right] dV - \qquad (3.47a)$$

$$\int_{V} \delta(\vec{R}-r) \left[ f_{NP}(\vec{R},t) \overline{U_p}(\vec{R},t) - f_{NC}(\vec{R},t) \overline{U_c}(\vec{R},t) \right] \cdot \nabla_R \phi_p(\vec{R},t) dV$$

再使用散度定理，则上式成为

$$\int_{S} \delta(\vec{R}-t) f(\vec{R},t) \overline{U}(\vec{R},t) \cdot \vec{n}^0 dS = \int_{S} \delta(\vec{R}-r) \phi_p(\vec{R},t) f_{NP}(\vec{R},t) \overline{U_p}(\vec{R},t) \cdot \vec{n}^0 dS +$$

$$\int_{S} \delta(\vec{R}-t) \phi_{NC}(\vec{R},t) f_{NC}(\vec{R},t) \overline{U_c}(\vec{R},t) \cdot \vec{n}^0 dS - \qquad (3.47b)$$

$$\int_{V} \delta(\vec{R}-t) \left[ f_{NP}(\vec{R},t) \overline{U_p}(\vec{R},t) - f_{NC}(\vec{R},t) \overline{U_c}(\vec{R},t) \right] \cdot \nabla_R \phi_p(\vec{R},t) dV$$

将式（3.16）两边的函数均取为数性函数，而后代入式（3.47b）的等号左边，以取代掉 $f(\vec{R},t)$。此时若存在

$$\vec{U}(\vec{r},t) = \overline{U_p}(\vec{r},t) = \overline{U_c}(\vec{r},t) \qquad (3.48a)$$

则立即可见

$$\int_{V} \delta(\vec{R}-r) \left[ f_{NP}(\vec{R},t) - f_{NC}(\vec{R},t) \right] \vec{U}(\vec{R},t) \cdot \nabla_R \phi_p(\vec{R},t) dV = 0$$

上面积分为零有三种可能：一是 $f_{NP}(\vec{r},t) = f_{NC}(\vec{r},t)$，但这是纯一的单相流而不是铁磁流体的两相流。二是 $\vec{U}(\vec{r},t) = 0$ 这是静止不流动的状况，当然谈不上对流输运的问题。最后只能是

$$\nabla \phi_p(\vec{r},t) = 0 \qquad (3.48b)$$

将式（3.48b）代入（3.47b）中，然后将其剩余部分与式（3.45）比较，即可看到

$$d\vec{S_p} = \phi_p(t) d\vec{S}, \qquad d\vec{S_c} = \phi_c(t) d\vec{S} \qquad (3.48c)$$

综观本节所述，可以归纳出几个概念：①式（3.47b）表示两相混合物对流输运通量的平均值，并不等于各自对流输运通量的平均值之和，其中的差别在于因成分分布不均匀而导致的内部对流输运。②式（3.48a）表示铁磁流体在流动中两相之间没有速度滞后，如果没有很大的外磁场梯度存在，铁磁流体两相之间的速度滞后很小可略。在此情况下，式（3.48b）表示无速度滞后，铁磁流体才能保持均匀性。③式（3.48c）表示在均匀铁磁流体的任一剖面上，两相各占的面积分数就等于其体积分数。

### 3.6.2 铁磁流体两相流传导输运通量的平均值

传导输运与对流输运不同，它不需要载体的宏观运动，实际上它是一种微观的分子运动所导致的交换过程。分子运动中的动量交换、动能交换和质量交换，就是传导输运的三种基本形式。说是交换

过程，就因其实际过程是双向的。而传导输运的定理与其数学表达式，只是指净效应而言，所以似乎是单向过程。

1.动量传递

在有速度梯度的层流流动中，低速层的分子与高速层的分子因碰撞而引起的动量交换，阻滞高速层而推动低速层，这表示流层间有力的作用，这种力即粘性应力。牛顿粘性定律给出粘性应力的数学关系式为

$$\tau_{ij} = \eta \left( \frac{\partial u_i}{\partial x_i} + \frac{\partial u_j}{\partial x_i} \right)$$

式中 $\tau_{ij}$ 是表面粘性应力，它是一种二阶张量。$\tau_{ij}$ 与速度梯度成正比，比例系数 $\eta$ 就是粘性系数。

2.能量传递

分子间的动能交换就是热传导的过程。它遵守 Fourier 导热定律，即

$$\overrightarrow{q_H} = -K_H \nabla T$$

式中 $K_H$ 称为导热系数，它是物质的一种属性。

3.质量传递

分子本身的掺和就是质量传递的过程。Fick 定律指出质量传递的流量和密度的梯度成正比，即

$$\overrightarrow{q_m} = -D \nabla \rho$$

比例系数 $D$ 称为扩散系数。

设以矢性函数 $\vec{q}(\vec{r},t)$ 表示传导输运的通量。则铁磁流体混合物通过微元面积 $d\vec{S}$ 的传导输运的通量是它的固相成分和液相成分各自通过 $d\overrightarrow{S_p}$ 和 $d\overrightarrow{S_c}$ 的传导通量之和，即

$$\vec{q}(\vec{r},t) \cdot d\vec{S} = \overrightarrow{q_{NP}}(\vec{r},t) \cdot d\overrightarrow{S_p} + \overrightarrow{q_{NC}}(\vec{r},t) \cdot d\overrightarrow{S_c} \tag{3.49}$$

上式在面积上取平均值，则有

$$\int_S \delta(\vec{R}-\vec{r})\vec{q}(\vec{R},t) \cdot \overrightarrow{n^0} dS = \int_{S_p} \delta(\vec{R}-\vec{r})\overrightarrow{q_{NP}}(\vec{R},t) \cdot \overrightarrow{n_p^0} dS_p + \int_{S_c} \delta(\vec{R}-\vec{r})\overrightarrow{q_{NC}}(\vec{R},t) \cdot \overrightarrow{n_c^0} dS_c \tag{3.50}$$

式（3.50）使用散度定理后，得

$$\int_V \nabla_R \cdot \left[ \delta(\vec{R}-\vec{r})\vec{q}(\vec{R},t) \right] dV = \int_{V_p} \nabla_R \cdot \left[ \delta(\vec{R}-\vec{r})\overrightarrow{q_{NP}}(\vec{R},t) \right] dV_p + \int_{V_c} \nabla_R \cdot \left[ \delta(\vec{R}-\vec{r})\overrightarrow{q_{NC}}(\vec{R},t) \right] dV_c =$$

$$\int_V \phi_p(\vec{R},t)\nabla_R \cdot \left[ \delta(\vec{R}-\vec{r})\overrightarrow{q_{NP}}(\vec{R},t) \right] dV + \int_V \phi_c(\vec{R},t)\nabla_R \cdot \left[ \delta(\vec{R}-\vec{r})\overrightarrow{q_{NC}}(\vec{R},t) \right] dV$$

由矢量恒等式 $\nabla \cdot (\varphi \vec{A}) = \varphi \nabla \cdot \vec{A} + \vec{A} \cdot \nabla \varphi$，并且注意到 $\nabla \phi_c = \nabla(1-\phi_p) = -\nabla \phi_p$，于是有

$$\int_V \nabla_R \cdot \left[ \delta(\vec{R} - \vec{r}) \vec{q}(\vec{R}, t) \right] dV =$$

$$\int_V \nabla_R \cdot \left[ \delta(\vec{R} - \vec{r}) \phi_p(\vec{R}, t) \overrightarrow{q_{NP}}(\vec{R}, t) \right] dV - \int_V \delta(\vec{R} - \vec{r}) \overrightarrow{q_{NP}}(\vec{R}, t) \cdot \nabla_R \phi_p(\vec{R}, t) dV +$$

$$\int_V \nabla_R \cdot \left[ \delta(\vec{R} - \vec{r}) \phi_c(\vec{R}, t) \overrightarrow{q_{NC}}(\vec{R}, t) \right] dV - \int_V \delta(\vec{R} - \vec{r}) \overrightarrow{q_{NC}}(\vec{R}, t) \cdot \nabla_R \phi_c(\vec{R}, t) dV = \qquad (3.51a)$$

$$\int_V \nabla_R \cdot \left[ \delta(\vec{R} - \vec{r}) \phi_p(\vec{R}, t) \overrightarrow{q_{NP}}(\vec{R}, t) \right] dV + \int_V \nabla_R \cdot \left[ \delta(\vec{R} - \vec{r}) \phi_c(\vec{R}, t) q_{NC}(\vec{R}, t) \right] dV -$$

$$\int_V \delta(\vec{R} - \vec{r}) \left[ \overrightarrow{q_{NP}}(\vec{R}, t) - \overrightarrow{q_{NC}}(\vec{R}, t) \right] \cdot \nabla_R \phi_p(\vec{R}, t) dV$$

若对式（3.51a）再使用散度定理，就有

$$\int_S \delta(\vec{R} - \vec{r}) \vec{q}(\vec{R}, t) \cdot \overrightarrow{n^0} dS = \int_S \delta(\vec{R} - \vec{r}) \phi_p(\vec{R}, t) \overrightarrow{q_{NP}}(\vec{R}, t) \cdot \overrightarrow{n^0} dS +$$

$$\int_S \delta(\vec{R} - \vec{r}) \phi_c(\vec{R}, t) \overrightarrow{q_{NC}}(\vec{R}, t) \cdot \overrightarrow{n^0} dS - \qquad (3.51b)$$

$$\int_V \delta(\vec{R} - \vec{r}) \left[ \overrightarrow{q_{NP}}(\vec{R}, t) - \overrightarrow{q_{NC}}(\vec{R}, t) \right] \cdot \nabla_R \phi_p(\vec{R}, t) dV$$

式（3.51a）与式（3.51b）均表示铁磁流体混合物的传导输运通量平均值并不等于其两种成分的传导输运通量的平均值之和，差别就在因为成分不均匀而引起的内部传导的失衡。

若铁磁流体是均匀的，即 $\nabla \phi_p = 0$，内部传导为零。由式（3.15b）余部与式（3.50）比较，同样得出 $d\overrightarrow{S_p} = \phi_p(t) d\vec{S}$，$d\overrightarrow{S_c} = \phi_c(t) d\vec{S}$。

### 3.6.3 铁磁流体输运通量平均值的形式表达

无论是对流输运还是传导输运，它们的通量都是指穿过面积的量，所以，①通量均沿面积的法线方向；②它们的平均值是在曲面上的。式（3.47b）与式（3.51b）可以按照平均值的概念，形式上写成下面的式子：

1.对流输运通量的平均值

由式（3.47b）得

$$f(\vec{r}, t) \vec{U}(\vec{r}, t) \cdot \overrightarrow{n^0} = \phi_p(\vec{r}, t) f_{NP} \overrightarrow{U_p}(\vec{r}, t) \cdot \overrightarrow{n^0} + \phi_c(\vec{r}, t) f_{NC}(\vec{r}, t) \overrightarrow{U_c}(\vec{r}, t) \cdot \overrightarrow{n^0} - \left[ f_{NP}(\vec{r}, t) \overrightarrow{U_p}(\vec{r}, t) - f_{NC}(\vec{r}, t) \overrightarrow{U_c}(\vec{r}, t) \right] \cdot \nabla \phi_p(\vec{r}, t) \qquad (3.52)$$

对于均匀铁磁流体，取 $\vec{U}(\vec{r}, t) = \overrightarrow{U_p}(\vec{r}, t) = \overrightarrow{U_c}(\vec{r}, t)$ 和 $\nabla \phi_p(\vec{r}, t) = 0$，就得到与式（3.16）相同的结果。

2.传导输运通量的平均值

由式（3.51b）得

$$\vec{q}(\vec{r}, t) \cdot \overrightarrow{n^0} = \phi_p(\vec{r}, t) \overrightarrow{q_{NP}}(\vec{r}, t) \cdot \overrightarrow{n^0} + \phi_c(\vec{r}, t) \overrightarrow{q_{NC}}(\vec{r}, t) \cdot \overrightarrow{n^0} - \left[ \overrightarrow{q_{NP}}(\vec{r}, t) - \overrightarrow{q_{NC}}(\vec{r}, t) \right] \cdot \nabla \phi_p(\vec{r}, t) \qquad (3.53a)$$

对于均匀铁磁流体，取 $\nabla \phi_p(\vec{r}, t) = 0$，就有

$$\vec{q}(\vec{r}, t) = \phi_p(t) \overrightarrow{q_{NP}}(\vec{r}, t) + \phi_c(t) \overrightarrow{q_{NC}}(\vec{r}, t) \qquad (3.53b)$$

这就是式（3.16）中的 $\vec{f}$ 换为 $\vec{q}$ 的结果。式（3.53b）亦可改写成分通量形式，即

$$\vec{q}(\vec{r}, t) = \overrightarrow{q_p}(\vec{r}, t) + \overrightarrow{q_c}(\vec{r}, t), \qquad \overrightarrow{q_p}(\vec{r}, t) = \phi_p(\vec{r}, t) \overrightarrow{q_{NP}}(\vec{r}, t), \qquad \overrightarrow{q_c}(\vec{r}, t) = \phi_c(\vec{r}, t) \overrightarrow{q_{NC}}(\vec{r}, t) \qquad (3.53c)$$

### 3.6.4 铁磁流体粘性系数 $\eta$ 的平均值

将式（3.53a）中的函数 $\vec{q}$ 具体化为铁磁流体及其成分的表面应力张量 $\tau$，则有

$$(\overline{e_i^0 \tau_{ij} e_j^0}) \cdot \vec{n}^0 = \phi_p(\vec{r},t)(\overline{e_i^0 \tau_{ij} e_j^0})_p \cdot \vec{n}^0 + \phi_c(\vec{r},t)(\overline{e_i^0 \tau_{ij} e_j^0}) \cdot \vec{n}^0 - \overline{e_i^0}\left[(\tau_{ij})_p - (\tau_{ij})_c\right]\overline{e_j^0} \cdot \nabla\phi_p(\vec{r},t) \tag{3.54a}$$

用 $\tau_{ij} = \eta\left(\dfrac{\partial u_i}{\partial x_j} + \dfrac{\partial u_j}{\partial x_i}\right)$ 代入上式得，

$$\left[\overline{e_i^0}\eta(\vec{r},t)\left(\frac{\partial u_i}{\partial x_j} + \frac{\partial u_j}{\partial x_i}\right)\overline{e_j^0}\right] \cdot \vec{n}^0 = \phi_p(\vec{r},t)\left[\overline{e_i^0}\eta_p(\vec{r},t)\left(\frac{\partial u_i}{\partial x_j} + \frac{\partial u_j}{\partial x_i}\right)\overline{e_j^0}\right] \cdot \vec{n}^0 +$$

$$\phi_c(\vec{r},t)\left[\overline{e_i^0}\eta_c(\vec{r},t)\left(\frac{\partial u_i}{\partial x_j} + \frac{\partial u_j}{\partial x_i}\right)\overline{e_j^0}\right] \cdot \vec{n}^0 - \tag{3.54b}$$

$$\overline{e_i^0}\left[\eta_p(\vec{r},t)\left(\frac{\partial u_i}{\partial x_j} + \frac{\partial u_j}{\partial x_i}\right)_p - \eta_c(\vec{r},t)\left(\frac{\partial u_i}{\partial x_j} + \frac{\partial u_j}{\partial x_i}\right)_c\right]\overline{e_j^0} \cdot \nabla\phi_p(\vec{r},t)$$

若铁磁流体的流动是稳定的，并且保持成分均匀，则必有 $\nabla\phi_p(\vec{r},t) = 0$，以及 $\partial u_i/\partial x_j + \partial u_j/\partial x_i = (\partial u_i/\partial x_j + \partial u_j/\partial x_i)_p = (\partial u_i/\partial x_j + \partial u_j/\partial x_i)_c$，于是式（3.54b）成为

$$\eta = \phi_p \eta_p + \phi_c \eta_c = \phi_p \eta_p + (1 - \phi_p)\eta_c \tag{3.54c}$$

由于铁磁流体的均匀性质，上式中 $\eta_p$、$\eta_c$ 和 $\phi_p$ 均为常数。

在铁磁流体中，以微粒状分散存在的固体物质，虽然其体积积分数一般不及 10%，但对整体粘度影响十分巨大。实测结果表明，尽管铁磁流体的体积 90% 是基载液体，但铁磁流体的粘性系数是其基载液的数倍到十几倍。原因是基载液带动固相微粒运动的粘性力和粘性力矩所折算出来的粘性系数远大于基载液自身的粘性系数。

设流动是单纯的一维剪切流（Couett 流）如图 3-3 所示。

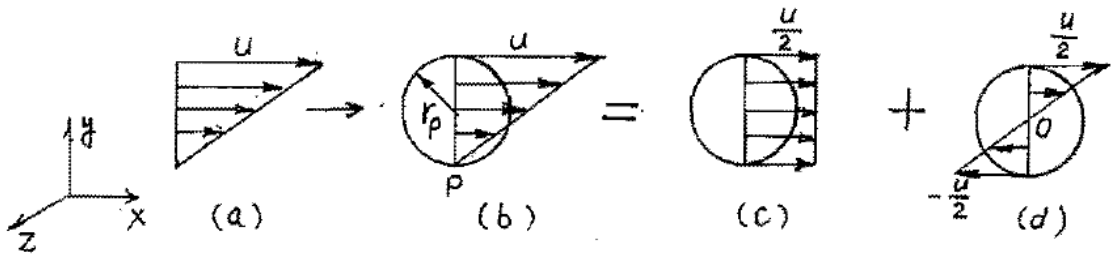

图 3-3　固相微粒在基载液中的运动

图（a）是纯剪切流动中的一个液体微团。如果铁磁流体的两相没有任何运动上的滞后，则固相微粒具有和流体微团相同的速度分布。如图（b）所示，对于一个刚体圆球来说，这样的线性速度分布模式，就是整个圆球绕瞬心点 P 作纯滚动运动。其绕瞬心点 P 的转动角速度是

$$\omega_p = \frac{\partial u}{\partial y} = \frac{u}{d_p} = \frac{u}{2r_p} = \omega_c \tag{A}$$

式中 $\omega_c$ 是基载液的涡旋角速度。于是固相微粒表面上的粘性应力 $\tau_p$ 是

$$\tau_p = -\eta_p \frac{\partial u}{\partial y} = -\eta_p \omega_p \tag{B}$$

对于图（b）的绕点 $P$ 的纯滚动运动可以分解为如图（c）的以速度 $u/2$ 的平移运动和如图（d）的以角速度 $\omega_p$ 绕球心 $O$ 的旋转运动。

以速度 $u/2$ 平移的阻力，由式（2.89c）给出为

$$F_u = -6\pi\eta_c r_p \left(\frac{u}{2}\right) = -6\pi\eta_c r_p^2 \omega_p \tag{C}$$

平移阻力的合力 $F_u$ 之作用点在球心 $O$，故它对瞬心 $P$ 的力矩 $L_u$ 是

$$L_u = r_p F_u = (-6\pi\eta_c r_p^2 \omega_p) r_p \tag{D}$$

绕球心 $O$ 旋转运动的粘性阻力矩 $L_\omega$ 由式（2.92a）给出为

$$L_\omega = -8\pi\eta_c r_p^3 \omega_p = (-8\pi\eta_c r_p^2 \omega_p) r_p \tag{E}$$

因为 $L_\omega$ 实际上是一个力偶，其矢量为自由矢，将它从 $O$ 移到 $P$，并不改变球形微粒的运动状态。于是绕瞬心 $P$ 的总力矩是

$$L = L_u + L_\omega = (-14\pi\eta_c r_p^2 w_p) r_p \tag{F}$$

式（F）右方括号中相当于作用在球心 $O$ 的合力，它是由表面应力合成而来的。将它除以圆球的表面积 $4\pi r_p^2$，就得固相微粒表面上的平均粘性应力 $\tau_p$，即

$$\tau_p = -\frac{14\pi\eta_c r_p^2 \omega_p}{4\pi r_p^2} = -3.5\eta_c \omega_p \tag{3.55a}$$

由上式可见，推动固相微粒以等同于基载液的速度旋转，其所需的表面应力是基载液体的 3.5 倍。将上式和式（B）相比较，立即可知

$$\eta_p = 3.5\eta_c \tag{3.55b}$$

将此式代入式（3.54c），就得铁磁流体混合物的粘性系数 $\eta$ 为

$$\eta = \left(1 + 2.5\phi_p\right)\eta_c \tag{3.55c}$$

式（3.55c）就是 Einstein 公式。在实际应用中，这个公式的误差较大。其中一个重要原因是，所采用的圆球阻力和阻力矩计算式均建立在固相体积分量很小，即稀疏相的假设之上。有鉴于此，Rosensweig

作了修正，即设定铁磁流体的粘性系数 $\eta$ 与基载液的粘性系数 $\eta_c$ 成二次项的倒数关系，即

$$\eta = \frac{\eta_c}{1 + a\phi_p + b\phi_p^2} \tag{G}$$

用两种极端状况决定常数 $a$ 和 $b$：

①稀疏相：

此时 $\phi_p \ll 1$，故 $\phi_p^2$ 是可略的二阶小量，从而有

$$\eta \approx (1 - a\phi_p)\eta_c$$

与式（3.55c）相比较，就有 $a = -2.5$。

②刚性堆砌

随固相微粒体积分数 $\phi_p$ 不断地增加，铁磁流体的流动性越来越差。最后，固相不能流动，而基载液体成为滤流。不能流动而聚在一起的固相称为刚性堆砌。此时堆积密度最大，孔隙率最小。由于铁磁流体不能整体流动，就可认为其粘性系数 $\eta$ 趋于无限大。于是式（G）的分母等于零，即

$$1 - 2.5\phi_p + b\phi_p^2 = 0$$

Graton 和 Fraser 计算出圆球体刚性堆砌的体积分数 $\phi_p$ 为 0.7405。代入上式得出 $b = 1.55$，于是修正后的铁磁流体粘性系数为

$$\eta = \frac{\eta_c}{1 - 2.5\phi_p + 1.55\phi_p^2} \tag{3.56}$$

③分散剂的影响

分散剂是长链分子，其链长 $\delta$ 一般不超过固相微粒的平均半径。分散剂一端吸附在固相微粒的表面，而整个分子链绕其吸附端作空间摆动。这种摆动运动是一种热运动，其热运动能形成一种能垒以防止固相微粒过分接近而聚集成团。由于分散剂的空间摆动，相当于微粒尺寸增大，即

$$\frac{4}{3}\pi\left(r_p + \delta\right)^3 = \frac{4}{3}\pi r_p^3\left(1 + \frac{\delta}{r_p}\right)^3 = V_{p1}\left(1 + \frac{\delta}{r_p}\right)^3$$

若铁磁流体的体积为 $V$，其中有 $N$ 个微粒，则固相的当量体积分数为

$$\frac{NV_{p1}}{V}\left(1 + \frac{\delta}{r_p}\right)^3 = \phi_p\left(1 + \frac{\delta}{r_p}\right)^3$$

代入到式（3.56）之分母中，于是 Rosensweig 修正公式最终是

$$\eta = \frac{1}{1 - 2.5\left(1 + \frac{\delta}{r_p}\right)^3\phi_p + 1.55\left(1 + \frac{\delta}{r_p}\right)^6\phi_p^2}\eta_c \tag{3.57}$$

在一般情况下，式（3.57）算出的结果与实测值相差不很大，至少数量级是吻合的。

### 3.6.5 铁磁流体热传导系数 $K_H$ 的平均值

将热传导的 Fourier 定律式代入式（3.53a）中以取代 $\vec{q}$，就有

$$\vec{e_j^0}K_H(\vec{r},t)\left(\frac{\partial T}{\partial x_j}\right)\cdot\vec{n^0}=\vec{e_j^0}\phi_p(\vec{r},t)K_{HP}(\vec{r},t)\left(\frac{\partial T_p}{\partial x_j}\right)\cdot\vec{n^0}+\vec{e_j^0}\phi_c(\vec{r},t)K_{HC}(\vec{r},t)\left(\frac{\partial T_c}{\partial x_j}\right)\cdot\vec{n^0}-$$
$$\vec{e_j^0}\left[K_{HP}(\vec{r},t)\frac{\partial T_p}{\partial x_j}-K_{HC}(\vec{r},t)\frac{\partial T_c}{\partial x_j}\right]\cdot\nabla\phi_p(\vec{r},t)$$

(3.58a)

对于均匀铁磁流体 $\nabla\phi_p(\vec{r},t)=0$，在稳定状态下就有

$$K_H\frac{\partial T}{\partial x_j}=\phi_p K_{HP}\frac{\partial T_p}{\partial x_j}+\phi_c K_{HC}\frac{\partial T_c}{\partial x_j}$$

(3.58b)

式（3.58b）实际上是一种能量守恒的方程，在所有情况下都成立。当然在 $\partial T/\partial x_j=\partial T_p/\partial x_j=\partial T_c/\partial x_j$ 时也是成立的。于是就有

$$K_H=\phi_p K_{HP}+\phi_c K_{HC}=\phi_p K_{HP}+(1-\phi_p)K_{HC}$$

(3.58c)

将式（3.58c）代入式（3.58b）中，给出铁磁流体的温度梯度为

$$\frac{\partial T}{\partial x_j}=\frac{\phi_p K_{HP}\dfrac{\partial T_p}{\partial x_j}+(1-\phi_p)K_{HC}\dfrac{\partial T_c}{\partial x_j}}{\phi_p K_{HP}+(1-\phi_p)K_{HC}}$$

(3.58d)

### 3.6.6 铁磁流体扩散系数 D 的平均值

铁磁流体两相流的扩散与粘性应力传递和热传导有很大的不同。粘性应力和热的传递是分子或粒子运动动量和能量的交换。既使在均匀铁磁流体中，只要存在分子或粒子动量和能量的差异，就将发生粘性应力和热的传递。但扩散是传质问题，没有质量浓度梯度的均匀铁磁流体内，是不可能发生质量的净传递的。所以，扩散的基本条件是必须存在浓度梯度。

在铁磁流体中，固相以单个微粒形态分散存在。所以它的质量浓度可用微粒数目的多少来计量。单位体积的铁磁流体内含有固相微粒的个数称为数密度 $n$。如果铁磁流体中，各处的数密度 $n$ 不一样，则必然出现在数密度高的地方向数密度低的地方迁移微粒的净效应。这就是一种质量扩散过程。所以对固相微粒而言，扩散的驱动力是数密度的梯度 $\nabla n$。但数密度 $n$ 和固相密度 $\rho_p$ 是直接相关的。设铁磁流体的体积 $V$ 内含有 $N$ 个固相微粒，则数密度 $n$ 是

$$n=\frac{N}{V}$$

(3.59a)

两边同乘以单个微粒的质量 $\rho_{NP}V_{p1}$，就有

$$\rho_{NP}V_{p1}n=\frac{NV_{p1}}{V}\rho_{NP}=\phi_p\rho_{NP}$$

由式（3.27a）知，上式之右方即固相之分密度 $\rho_p$，故得

$$\rho_{NP}V_{p1}n = \rho_p \tag{3.59b}$$

所以式（2.23b）用于铁磁流体的固相成分时，密度 $\rho$ 应当是与 $n$ 相对应的分密度 $\rho_p$，即

$$\overrightarrow{q_{mp}} = \rho_p\overrightarrow{U_{pd}} = -D_p\nabla\rho_p \tag{3.60a}$$

式中 $\overrightarrow{U_{pd}}$ 是固相扩散流速的平均值。若将式（3.59b）代入式（3.60a），并注意在扩散过程中 $\rho_{NP}$ 和 $V_{p1}$ 是不变的常量，于是有

$$n\overrightarrow{U_{pd}} = -D_p\nabla n \tag{3.60b}$$

比较以上的式（3.60a）和式（3.60b）可见分密度确实反映扩散的实际过程。对于铁磁流体的液相扩散，也是分子迁移的结果，与固相微粒的扩散，机制上没有本质的差异。所以用分密度 $\rho_c$ 代入式（2.23b），有

$$\overrightarrow{q_{mc}} = \rho_c\overrightarrow{U_{dc}} = -D_c\nabla\rho_c \tag{3.60c}$$

由式（3.27a）可知分密度 $\rho_p(\vec{r},t)$ 和 $\rho_c(\vec{r},t)$ 是矢径 $\vec{r}$ 端点 $(x,y,z)$ 的邻域内的平均值。所以，对铁磁流体来说，质量扩散的驱动力是分密度这种平均值的梯度。由于按 $\rho_p$ 和 $\rho_c$ 扩散的两种成分具有相同的扩散面积，从而有

$$\overrightarrow{q_m}(\vec{r},t) = \overrightarrow{q_{mp}}(\vec{r},t) + \overrightarrow{q_{mc}}(\vec{r},t) \tag{3.61a}$$

当铁磁流体和另一种液体互相接触而且互相扩散时，此种液体将扩散而进入到铁磁流体的体积 $V_f$ 之内，同时也有部分铁磁流体因扩散而离开体积 $V_f$，这样一来，在 $V_f$ 内的铁磁流体的体积便从 $V_f$ 减小到 $V_f'$，从而它的平均密度将从 $\rho_f$ 减小为 $\rho_f'$，即

$$\rho_f' = \frac{V_f'}{V_f}\rho_f = \phi_f\rho_f \tag{3.61b}$$

式中 $\phi_f = V_f'/V_f$。随 $\rho_f$ 降低为 $\rho_f'$，铁磁流体内固相和液相的分密度也相应地从 $\rho_p$ 和 $\rho_c$ 降低为 $\rho_p'$ 和 $\rho_c'$。即固相体积从 $V_p$ 减小到 $V_p'$，液相体积从 $V_c$ 减小到 $V_c'$，则两者的分密度成为

$$\left.\begin{array}{l} \rho'_p = \dfrac{V'_p}{V_f}\rho_{NP} = \dfrac{V'_f}{V_f}\dfrac{V'_p}{V'_f}\rho_{NP} = \phi_f \phi'_p \rho_{NP} \\[3mm] \rho'_c = \dfrac{V'_c}{V_f}\rho_{NC} = \dfrac{V'_f}{V_f}\dfrac{V'_c}{V'_f}\rho_{NC} = \phi_f \phi'_c \rho_{NC} \end{array}\right\} \tag{3.61c}$$

于是式（3.61a）就成为

$$-D\nabla\rho'_f = -D_p\nabla\rho'_p - D_c\nabla\rho'_c \tag{3.62a}$$

将式（3.61b）与式（3.61c）分别代入上式之两边

$$-D\nabla\left(\phi_f\rho_f\right) = -D_p\rho_{NP}\nabla\left(\phi_f\phi'_p\right) - D_c\rho_{NC}\nabla\left(\phi_f\phi'_c\right) \tag{3.62b}$$

若铁磁流体均匀地扩散，即固相和液相均与铁磁流体按相同比例稀释，则有

$$\frac{V'_p}{V_p} = \frac{V'_c}{V_c} = \frac{V'_f}{V_f} \tag{3.63a}$$

上式可拆开写成

$$\frac{V'_p}{V'_f} = \frac{V_p}{V_f}, \qquad \frac{V'_c}{V'_f} = \frac{V_c}{V_f}$$

于是有

$$\phi'_p = \phi_p, \qquad \phi'_c = \phi_c \tag{3.63b}$$

设在扩散相溶之前铁磁流体是均匀的，即 $\phi_p$ 和 $\phi_c$ 均为常数。将式（3.63b）代入式（3.61c）中，就得扩散过程中的分密度 $\rho'_p$、$\rho'_c$ 与原始的分密度 $\rho_p$ 和 $\rho_c$ 的关系为

$$\left.\begin{array}{l} \rho'_p(\vec{r},t) = \phi_f(\vec{r},t)\phi_p\rho_{NP} = \phi_f(\vec{r},t)\rho_p \\[2mm] \rho'_c(\vec{r},t) = \phi_f(\vec{r},t)\phi_c\rho_{NC} = \phi_f(\vec{r},t)\rho_c \end{array}\right\} \tag{3.63c}$$

将式（3.61b）与式（3.63b）代入式（3.62b），并且记住 $\phi_p$ 和 $\phi_c$ 是常数，因而 $\rho_f$ 也是常数。于是就有

$$-D\rho_f\nabla\phi_f(\vec{r},t) = -D_p\phi_p\rho_{NP}\nabla\phi_f(\vec{r},t) - D_c\phi_c\rho_{NC}\nabla\phi_f(\vec{r},t)$$

约去 $\nabla\phi_f(\vec{r},t)$ 给出铁磁流体的扩散系数为

$$D = \left(D_p\phi_p\rho_{NP} + D_c\phi_c\rho_{NC}\right)\big/\rho_f$$

利用式（3.26），上式写成

$$D = \frac{D_p\phi_p\rho_{NP} + D_c\phi_c\rho_{NC}}{\phi_p\rho_{NP} + \phi_c\rho_{NC}} \tag{3.64a}$$

或使用式（3.29）得

$$D = \varphi_p D_p + \varphi_c D_c \tag{3.64b}$$

铁磁流体的基载液和另一种液体之间的扩散是液相分子相互交换的行为，而固相扩散则是微粒在液体中 Brown 运动的结果。所以维持均匀扩散是不能做到的。这里所论的均匀扩散只是一种理想化的假设。

### 3.6.7 固相微粒在铁磁流体内的扩散系数 $D_p$

铁磁流体中固相微粒的 Brown 运动，原本是随机的杂乱无章的热运动，其平动速度的统计平均值等于零。但是，如果存在固相微粒的浓度梯度，则由浓度高的地方向浓度低的地方运动的微粒数，比同时由浓度低处向浓度高处运动的微粒数多，从而出现由高浓度处向低浓度处的定向的净质量迁移的宏观现象。此时微粒平动速度的统计平均值不再是零，而是被称为扩散速度的 $(u_i)_e$，随着浓度梯度的降低，$(u_i)_e$ 逐渐减小，直到扩散终止，固相微粒均匀分布，$(u_i)_e$ 就成为零。

由动力学第二定律 $F = ma$ 给出的单位体积固相微粒流动的运动方程为

$$\rho_p \frac{d(u_i)_e}{dt} = -nC_d(u_i)_e - \frac{\partial p_p}{\partial x_i} \tag{3.65}$$

式（3.65）的左方是惯性项，右方第一项是阻力项，由式（2.89c）所给出，其中 $n$ 是单位体积铁磁流体内所含有的微粒数目，即固相微粒的数密度，阻力系数 $C_d = 6\pi \eta_c r_p$，$(u_i)_e$ 的下标 $e$ 表示数学期望或统计平均值，下标 $i$ 相应于坐标 $x_i$。压力梯度 $\partial p_p / \partial x_i$ 是一种表面力，可参见式（2.50）。对扩散过程而言，$\partial p_p / \partial x_i$ 是物质迁移的驱动力。由于微粒的热运动属于分子行为，应当遵循气体状态方程，即

$$p = \rho RT = \frac{m}{V} RT \tag{3.66}$$

将右方的质量 $m$ 写成摩尔质量 M 和摩尔数 $v$ 之乘积，即 $m = Mv$，摩尔数 $v = N/N_0$，$N$ 是体积 $V$ 内的气体分子数目，$N_0$ 是一个摩尔内的气体分子数目，它是一个与气体种类无关的普适的物理常数，即 Avogadro 常数。于是

$$\frac{m}{V} = \frac{Mv}{V} = \frac{MN}{N_0 V} = \frac{M}{N_0} n$$

代入状态方程（3.66）中，有

$$p = \frac{M}{N_0} RTn$$

$MR = R_u$ 称为通用气体常数，它也是不取决于气体种类的普适的物理常数。定义

$$k_0 = \frac{MR}{N_0} = \frac{R_u}{N_0} \tag{3.67}$$

$k_0$ 即 Boltzmann 常数，它的物理意义是一个气体分子的通用气体常数。于是气体状态方程就成为

$$p = k_0 T n \qquad (3.68)$$

固相微粒在基载液体中扩散时，必须作功以克服液体的粘性阻力，即式（3.65）右方的第一项。因为微粒的运动是热运动，所以微粒克服液体的粘性摩擦阻力所作之功，必将消耗其热内能。但是摩擦功所转化成的摩擦热，大部分即刻被液体吸收。整个扩散过程中，铁磁流体可视为与外界无能量交换的孤立系统，从而摩擦热只能通过两相间的传热，又回到固相微粒之中。固相微粒与液相间的相对接触面积极为巨大，每毫升体积的固相微粒与液相的交界面积竟能达到 45 平方米之谱。其传热速度之快远远超过扩散速度和 Brown 运动速度，所以固相微粒在基载液中的扩散和 Brown 运动，确实是热平衡状态下的等温过程（不仅是铁磁流体，任何一种固液两相胶体混合物均如此）。将气体状态方程式（3.68）左方的 $p$ 取为固相微粒流的分压 $p_p$，右方的 $n$ 理解为微粒的数密度。然后两边对 $x_i$ 取导数，就有

$$\frac{\partial p_p}{\partial x_i} = k_0 T \frac{\partial n}{\partial x_i} \qquad (3.69)$$

将式（3.69）代入运动方程式（3.65）之右方，即得

$$\rho_p \frac{d(u_i)_e}{dt} = -n C_d (u_i)_e - k_0 T \frac{\partial n}{\partial x_i} \qquad (3.70a)$$

固相微粒在基载液中的运动是低 Re 数的运动，所以式（3.70a）左方的惯性项可以略去，从而有

$$n(u_i)_e = -\frac{k_0 T}{C_d} \frac{\partial n}{\partial x_i} \qquad (3.70b)$$

或两边同乘以 $\rho_{NP} V_{p1}$，并且由式（3.59b）知 $n \rho_{NP} V_{p1} = \rho_p$，于是上式成为

$$\rho_p (u_i)_e = -\frac{k_0 T}{C_d} \frac{\partial \rho_p}{\partial x_i} \qquad (3.70c)$$

将式（3.70b）、式（3.70c）与 Fick 定律式（3.60a）、式（3.60b）比较，立即可知固相微粒的扩散系数 $D_p$ 是

$$D_p = \frac{k_0 T}{C_d} = \frac{k_0 T}{6 \pi \eta_c r_p} \qquad (3.71)$$

## 3.7 铁磁流体热力学参数的平均值

### 3.7.1 压力 $p$ 的平均值

在铁磁流体内，任何一点的邻域之内只能有一个压力。无论是固相还是液相，它们都处于相同的压力之下，即

$$p_f = p_{NP} = p_{NC} \qquad (3.72)$$

将式（3.16）中的函数取为数性函数，且令 $f(\vec{r}, t) = p_f(\vec{r}, t)$、$f_{NP}(\vec{r}, t) = p_{NP}(\vec{r}, t)$、$f_{NC}(\vec{r}, t) = p_{NC}(\vec{r}, t)$，则平均值由式（3.16）得

$$p_f(\vec{r},t) = \phi_p(\vec{r},t)p_{NP}(\vec{r},t) + \phi_c(\vec{r},t)p_{NC}(\vec{r},t)$$

用式（3.72）代入，就有

$$p_f(\vec{r},t) = \phi_p(\vec{r},t)p_f(\vec{r},t) + \phi_c(\vec{r},t)p_f(\vec{r},t)$$

定义固相和液相的分压是

$$p_p(\vec{r},t) = \phi_p(\vec{r},t)p_f(\vec{r},t), \qquad p_c(\vec{r},t) = \phi_c(\vec{r},t)p_f(\vec{r},t) \tag{3.73a}$$

从而有

$$p_f(\vec{r},t) = p_p(\vec{r},t) + p_c(\vec{r},t) \tag{3.73b}$$

### 3.7.2 铁磁流体的定容比热容的平均值

1.体积比热容 $c_V$

体积比热容是以单位体积计量的。在式（3.16）中将 $\vec{f}(\vec{r},t)$ 取为数性函数热量 $c_V T$ ，则有

$$c_V(\vec{r},t)T_f(\vec{r},t) = \phi_p(\vec{r},t)c_s(\vec{r},t)T_p(\vec{r},t) + \phi_c(\vec{r},t)c_c(\vec{r},t)T_c(\vec{r},t) \tag{3.74}$$

式中 $c_s$ 和 $c_c$ 分别是固相物质和液相物质的比热容。由于 $c_s$ 和 $c_c$ 至少是温度的函数，而铁磁流体内的温度不一定均匀，并且有可能随时间改变，故 $c_s$ 和 $c_c$ 写成位置和时间的函数。上面式（3.74）中取 $T_f(\vec{r},t) = T_p(\vec{r},t) = T_c(\vec{r},t)$ ，于是有

$$c_V(\vec{r},t) = \phi_p(\vec{r},t)c_s(\vec{r},t) + \phi_c(\vec{r},t)c_c(\vec{r},t) \tag{3.75a}$$

将式（3.75a）代入式（3.74）之左方，就得

$$T_f(\vec{r},t) = \frac{\phi_p(\vec{r},t)c_s(\vec{r},t)T_p(\vec{r},t) + \phi_c(\vec{r},t)c_c(\vec{r},t)T_c(\vec{r},t)}{\phi_p(\vec{r},t)c_s(\vec{r},t) + \phi_c(\vec{r},t)c_c(\vec{r},t)} \tag{3.75b}$$

2.质量比热容 $c'_V$

$c'_V$ 是以单位质量计量的，在式（3.23b）中取 $\vec{f}(\vec{R},t)$ 为数性函数 $c'_V(\vec{R},t)T(\vec{R},t)$ ，并且使用式（3.24），就得

$$\rho_f(\vec{r},t)c'_V(\vec{r},t)T_f(\vec{r},t) = \phi_p(\vec{r},t)\rho_{NP}c'_s(\vec{r},t)T_p(\vec{r},t) + \phi_c(\vec{r},t)\rho_{NC}c'_c(\vec{r},t)T_c(\vec{r},t) \tag{3.76a}$$

在均匀温度下，即 $T_f(\vec{r},t) = T_p(\vec{r},t) = T_c(\vec{r},t)$ ，则上式给出

$$\rho_f(\vec{r},t)c'_V(\vec{r},t) = \phi_p(\vec{r},t)\rho_{NP}c'_s(\vec{r},t) + \phi_c(\vec{r},t)\rho_{NC}c'_c(\vec{r},t) \tag{3.76b}$$

使用式（3.26）以后，就有

$$c'_V(\vec{r},t) = \frac{\phi_p(\vec{r},t)\rho_{NP}c'_s(\vec{r},t) + \phi_c(\vec{r},t)\rho_{NC}c'_c(\vec{r},t)}{\phi_p(\vec{r},t)\rho_{NP} + \phi_c(\vec{r},t)\rho_{NC}}$$  (3.77a)

或利用式（3.27a）和式（3.27b），则上式可写成分密度的形式

$$c'_v(\vec{r},t) = \frac{\rho_p(\vec{r},t)c'_s(\vec{r},t) + \rho_c(\vec{r},t)c'_c(\vec{r},t)}{\rho_p(\vec{r},t) + \rho_c(\vec{r},t)}$$  (3.77b)

若使用式（3.29）于式（3.77a），则得到

$$c'_V(\vec{r},t) = \varphi_p(\vec{r},t)c'_s(\vec{r},t) + \varphi_c(\vec{r},t)c'_c(\vec{r},t)$$  (3.77c)

将式（3.29）与式（3.77c）用于式（3.76a），就得铁磁流体的混合温度 $T_f(\vec{r},t)$

$$T_f(\vec{r},t) = \frac{\varphi_p(\vec{r},t)c'_s(\vec{r},t)T_p(\vec{r},t) + \varphi_c(\vec{r},t)c'_c(\vec{r},t)T_c(\vec{r},t)}{\varphi_p(\vec{r},t)c'_s(\vec{r},t) + \varphi_c(\vec{r},t)c'_c(\vec{r},t)}$$  (3.77d)

### 3.7.3 铁磁流体的熵的平均值

熵是热力学的状态参数，它的定义是

$$dS = \frac{\delta Q}{T}$$  (3.78a)

式中 $\delta Q$ 是热量，它和热力学过程有关，所以 $\delta Q$ 不是状态参数。单位体积的熵称为体积比熵 $ds$，单位质量的熵称为质量比熵 $ds'$。即

$$ds = \frac{\delta q}{T}, \qquad ds' = \frac{\delta q'}{T}$$  (3.78b)

1.铁磁流体体积比熵的平均值

在式（3.16）中，将 $\vec{f}(\vec{r},t)$ 取为数性函数体积比熵 $ds(\vec{r},t)$，则其平均值是

$$ds_f(\vec{r},t) = \phi_p(\vec{r},t)ds_p(\vec{r},t) + \phi_c(\vec{r},t)ds_c(\vec{r},t)$$  (3.79)

式中 $s_f$ 是铁磁流体混合物的比熵，$s_p$ 和 $s_c$ 分别是固相和液相的比熵。

将式（3.78b）代入式（3.79）中，并且注意 $\delta q_p = c_s dT_p$，$\delta q_c = c_c dT_c$，就得

$$ds_f(\vec{r},t) = \phi_p(\vec{r},t)\frac{c_s(\vec{r},t)dT_p(\vec{r},t)}{T_p(\vec{r},t)} + \phi_c(\vec{r},t)\frac{c_c(\vec{r},t)dT_c(\vec{r},t)}{T_c(\vec{r},t)}$$  (3.80a)

若铁磁流体是均匀的则 $\phi_p$ 和 $\phi_c$ 是常数，并且 $c_s$ 和 $c_c$ 均取平均值，则此时铁磁流体的体积比熵是全微分，即它是与热力学过程无关的状态参数。积分式（3.80a）有

$$s_f(\vec{r},t) = \phi_p c_s \ln T_p(\vec{r},t) + \phi_c c_c \ln T_c(\vec{r},t)$$  (3.80b)

2.铁磁流体质量比熵的平均值

在式（3.23b）中，将 $\vec{f}'(\vec{R},t)$ 取为质量比熵 $ds'(\vec{R},t)$，则给出平均值的关系为

$$\rho_f(\vec{r},t)ds_f'(\vec{r},t) = \phi_p(\vec{r},t)\rho_{NP}ds_p'(\vec{r},t) + \phi_c(\vec{r},t)\rho_{NC}ds_c'(\vec{r},t) \tag{3.81a}$$

将式（3.29）用于上式中，就有

$$ds_f'(\vec{r},t) = \varphi_p(\vec{r},t)ds_p'(\vec{r},t) + \varphi_c(\vec{r},t)ds_c'(\vec{r},t) \tag{3.81b}$$

将 $ds' = (\delta q')/T = c'dT/T$ 代入上式，得

$$ds_f'(\vec{r},t) = \varphi_p(\vec{r},t)c_s'(\vec{r},t)d\ln T_p(\vec{r},t) + \varphi_c(\vec{r},t)c_c'(\vec{r},t)d\ln T_c(\vec{r},t) \tag{3.82a}$$

若 $\varphi_p$、$\varphi_c$、$c_s'$、$c_c'$ 均为常数，则铁磁流体混合物的质量比熵之平均值是热力学的状态参数。积分式（3.82a）得

$$s_f'(\vec{r},t) = \varphi_p c_s' \ln T_p(\vec{r},t) + \varphi_c c_c' \ln T_c(\vec{r},t) \tag{3.82b}$$

### 3.7.4 铁磁流体的焓和总焓的平均值

焓 $i$ 实际上是一种广义的内能，在定压条件下和定容条件下，将可压缩流体加热到相同的温升所需的热量是不一样的。定容下流体的体积不可膨胀，而定压条件下可压缩流体的体积随温度升高而膨胀，因而加入到流体中的热量有一部分变成膨胀功。所以，定压下加热需要的热量更多一些。计及膨胀功在内的热容量系数就是定压比热容 $c_p$，即

$$c_p = c_V + R$$

式中 $c_V$ 是定容比热容，$R$ 是气体常数。上式两边同乘以温度 $T$，得

$$c_p T = c_V T + RT$$

上式右边第一项 $c_V T$ 是流体的热内能，第二项由气体状态方程 $RT = pV$，在定压下，$p$ 不变，$V$ 由于加热而增大，所以 $RT$ 此时表示膨胀功。上式左边 $c_p T$ 是一种热力学状态参数，称为焓 $i$，即

$$i = c_p T \tag{3.83a}$$

常常将 $i$ 称作静焓。流体在流动中具有动能，在第 2.9 节中将动能归纳到内能里，称为总内能。同样将动能归纳到焓中，就称为总焓。即

$$i^* = c_p T^* \tag{3.83b}$$

式中的 $T^*$ 称为总温或滞止温度，它体现动能和温度之间的变换，即

$$T^* = T + \frac{u^2}{2c_V} \tag{3.83c}$$

式中 $u$ 是流体的运动速度。总焓 $i^*$ 和静焓 $i$ 一样都是热力学状态参数。

1. 铁磁流体体积比焓 $i_f(\vec{r},t)$ 的平均值

在式（3.16）中将 $\vec{f}(\vec{r},t)$ 取为数性函数 $i(\vec{r},t)$，就得平均值

$$i_f(\vec{r},t) = \phi_p(\vec{r},t)i_s(\vec{r},t) + \phi_c(\vec{r},t)i_c(\vec{r},t) \tag{3.84a}$$

显然式（3.84a）是一种能量守恒方程，在均匀温度下依然成立，所以取 $T_f = T_p = T_c$ 就得铁磁流体的定压体积比热容 $c_{pf}$ 的平均值是

$$c_{pf}(\vec{r},t) = \phi_p(\vec{r},t)c_{ps}(\vec{r},t) + \phi_c(\vec{r},t)c_{pc}(\vec{r},t) \tag{3.84b}$$

式中 $c_{ps}$ 和 $c_{pc}$ 分别是固相微粒流和基载流体的定压体积比热容。将式（3.84b）代入式（3.84a）即可得出铁磁流体在定压条件下混合温度之平均值是

$$T_f(\vec{r},t) = \frac{\phi_p(\vec{r},t)c_{ps}(\vec{r},t)T_p(\vec{r},t) + \phi_c(\vec{r},t)c_{pc}(\vec{r},t)T_c(\vec{r},t)}{\phi_p(\vec{r},t)c_{ps}(\vec{r},t) + \phi_c(\vec{r},t)c_{pc}(\vec{r},t)} \tag{3.84c}$$

式（3.84c）给出的 $T_f$ 与式（3.75b）所给出的 $T_f$ 不相等，原因很明显，前者是在定压条件下，而后者是在定容条件下，差别就在于是否存在膨胀功。

2. 铁磁流体质量比焓 $i'_f(\vec{r},t)$ 的平均值

在式（3.23b）中取 $\vec{f}(\vec{r},t)$ 为数性函数质量比焓 $i'(\vec{r},t)$，则有

$$\rho_f(\vec{r},t)i'_f(\vec{r},t) = \phi_p(\vec{r},t)\rho_{NP}i'_s(\vec{r},t) + \phi_c(\vec{r},t)\rho_{NC}i'_c(\vec{r},t) \tag{3.85a}$$

上式两边同除以 $\rho_f(\vec{r},t)$，而后用式（3.29）代入，即得

$$i'_f(\vec{r},t) = \varphi_p(\vec{r},t)i'_s(\vec{r},t) + \varphi_c(\vec{r},t)i'_c(\vec{r},t) \tag{3.85b}$$

在 $T_f = T_p = T_c$ 情况下，得铁磁流体的定压质量比热容 $c'_{pf}(\vec{r},t)$ 之平均值为

$$c'_{pf}(\vec{r},t) = \varphi_p(\vec{r},t)c'_{ps}(\vec{r},t) + \varphi_c(\vec{r},t)c'_{pc}(\vec{r},t) \tag{3.85c}$$

对于不可压缩的铁磁流体，则有 $c_p = c_V$，$c'_p = c'_V$；$c_{ps} = c_s$，$c'_{ps} = c'_s$；$c_{pc} = c_c$，$c'_{pc} = c'_c$。

## 3.8 铁磁流体磁参数的平均值
### 3.8.1 概述

就目前已有的铁磁流体来看，其磁化性能完全来自于固相微粒体，而液相基础载体是不能磁化的。但是本节仍然假定基载液体具有磁化性能，目的仅在于保持结果的普适性。对铁磁流体最有用的磁学

参数就是铁磁流体内的磁感应强度 $\overline{B}$、磁场强度 $\overline{H}$ 和铁磁流体自身的磁化强度 $\overline{M}$。

磁感应强度 $\overline{B}$ 遵守 Gauss 定理。Gauss 定理是说无论有无电流穿过磁场，$\overline{B}$ 的散度都等于零，即 $\nabla \cdot \overline{B} = 0$，所以 $\overline{B}$ 场是无源场。磁场强度 $\overline{H}$ 遵守 Ampere 环路定理，即 $\overline{H}$ 的旋度等于穿过磁场的传导电流的面密度，$\nabla \times \overline{H} = \overrightarrow{j_c}$。若无传导电流穿过磁场 $\overrightarrow{j_c} = 0$，则 $\overline{H}$ 场是无旋场。

磁化强度 $\overline{M}$ 是单位体积内的有效磁矩，所以它是以单位体积计量的参数，求其体积平均值当然是顺理成章的。但 $\overline{B}$ 和 $\overline{H}$ 并非是以体积计量的。然而它们都是位置坐标的函数，即 $\overline{B} = \overline{B}(\vec{r},t)$，$\overline{H} = \overline{H}(\vec{r},t)$。既然它们分布于空间之中，当然也就可以从数学上来求它们在空间中的平均值，即体积平均值。这种平均值应当而且必须分别符合 Gauss 定理或者 Ampere 环路定理。

磁场在空间中的分布，并不依赖空间是否有物质而存在。即使在真空中仍然可以有磁场分布。所以讨论磁场的质量平均值是没有意义的。

### 3.8.2 铁磁流体中磁感应强度 $\overline{B}$ 的体积平均值

设在铁磁流体内一点 $P(x,y,z)$ 的邻域 $V$ 内，$\overline{B}(\vec{r},t)$ 的平均值是

$$\int_V \overline{B}(\vec{R},t)\delta(\vec{R}-\vec{r})dV = \int_{V_p} \overline{B_p}(\vec{R},t)\delta(\vec{R}-\vec{r})dV_p + \int_{V_c} \overline{B_c}(\vec{R},t)\delta(\vec{R}-\vec{r})dV_c \qquad (3.86a)$$

上式的物理涵义是，$\overline{B_p}$ 只存在于固相微粒流的体积 $V_p$ 中，而 $\overline{B_c}$ 只存在于基载液的体积 $V_c$ 中。$V_p$ 和 $V_c$ 是并集，组成邻域 $V$。

将式（3.86a）改写成

$$\int_V \overline{B}(\vec{R},t)\delta(\vec{R}-\vec{r})dV = \int_V \overline{B_p}(\vec{R},t)\phi_p(\vec{R},t)\delta(\vec{R}-\vec{r})dV + \int_V \overline{B_c}(\vec{R},t)\phi_c(\vec{R},t)\delta(\vec{R}-\vec{r})dV$$

此式给出形式上的平均值关系为

$$\overline{B}(\vec{r},t) = \phi_p(\vec{r},t)\overline{B_p}(\vec{r},t) + \phi_c(\vec{r},t)\overline{B_c}(\vec{r},t) \qquad (3.86b)$$

式（3.86b）亦可由式（3.16）直接写出。将式（3.86a）两边取散度，显然所有的项都是平均值的散度，即

$$\nabla_r \cdot \int_V \overline{B}(\vec{R},t)\delta(\vec{R}-\vec{r})dV = \nabla_r \cdot \int_{V_p} \overline{B_p}(\vec{R},t)\delta(\vec{R}-\vec{r})dV_p + \nabla_r \cdot \int_{V_c} \overline{B_c}(\vec{R},t)\delta(\vec{R}-\vec{r})dV_c \qquad (3.87)$$

在式（3.40a）中，将 $\vec{f}(\vec{R},t)$ 取为 $\overline{B}(\vec{R},t)$，则上式成为

$$\nabla_r \cdot \vec{B}(\vec{r},t) = \nabla_r \cdot \int_V \vec{B}(\vec{R},t)\delta(\vec{R}-\vec{r})dV =$$
$$\int_{V_p} \delta(\vec{R}-\vec{r})\nabla_R \cdot \vec{B_p}(\vec{R},t)dV_p + \int_{V_c} \delta(\vec{R}-\vec{r})\nabla_R \cdot \vec{B_c}(\vec{R},t)dV_c -$$
$$\int_{c_{p0}} dS_{p0}\overline{n_s^0} \cdot \vec{B_p}(\vec{R},t)\delta(\vec{R}-\vec{r}) - \int_{S_{c0}} dS_{c0}\overline{n_s^0} \cdot \vec{B_c}(\vec{R},t)\delta(\vec{R}-\vec{r}) -$$
$$\sum_{N'} \int_{S_{p1}} dS_{p1}\overline{n_{p1}^0} \cdot \left[ \vec{B_p}(\vec{R},t) - \vec{B_c}(\vec{R},t) \right] \delta(\vec{R}-\vec{r}) \tag{3.88}$$

以下逐项考察式（3.88）之右方：

①在式（3.88）右方第一和第二两项积分内，被积函数包含有磁感应强度 $\vec{B_p}$ 和 $\vec{B_c}$ 的散度。由 Gauss 定理知道

$$\nabla_R \cdot \vec{B_p}(\vec{R},t) = 0, \qquad \nabla_R \cdot \vec{B_c}(\vec{R},t) = 0$$

②式（3.88）右方第三项和第四项的被积函数可按本章 3.2.2 节中的式（3.6d）处理，即

$$f(x)\delta(x-a) = \frac{1}{2}[f(a-0) + f(a+0)]\delta(x-a)$$

因为式（3.88）右方的第三和第四积分项，都是在包围邻域体积 $V$ 的表面 $S$ 上的积分，所以其中的位置矢径 $\vec{r}$ 可取为 $\vec{r}(x_s, y_s, z_s) = \vec{r_s}$，于是由式（3.6d）有

$$\overline{n_s^0} \cdot \vec{B}(\vec{R},t)\delta(\vec{R}-\vec{r_s}) = \frac{1}{2}\left[ \overline{n_{s-0}^0} \cdot \vec{B}(\vec{r_s}-0,t) + \overline{n_{s+0}^0} \cdot \vec{B}(\vec{r_s}+0,t) \right]\delta(\vec{R}-\vec{r_s}) \tag{3.89}$$

如图 3-4 所示，在体积 $V$ 的边界 $S$ 上取一扁柱形微元体，其上顶面紧贴 $S$ 的外侧，下底面紧贴 $S$ 之内侧。图中的 $\vec{B}(\vec{r_s}-0,t)$ 代表 $\vec{B_p}$ 或 $\vec{B_c}$，而 $\vec{B}(\vec{r_s}+0,t)$ 取决于 $S$ 表面之外部的介质状况。

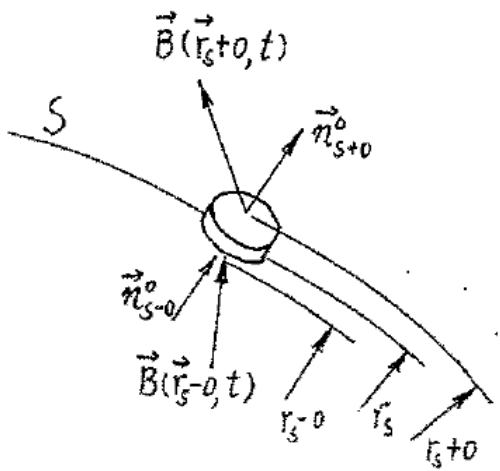

图 3-4  邻域 $V$ 外边界 $S$ 上的磁感应强度

对于此边界微元体来说，$\overline{n_{s+0}^0}$ 是其上顶面的外向法线单位矢量，而 $\overline{n_{s-0}^0}$ 是其下底面的内向法线单位矢量，故有

$$\overline{n_{s-0}^0} = -\overline{n_{s+0}^0} = -\overline{n_s^0}$$

于是式（3.89）成为

$$\overrightarrow{n_s^0} \cdot \overline{B}(\vec{R},t)\delta(\vec{R}-\vec{r_s}) = \frac{1}{2}\overrightarrow{n_s^0} \cdot \left[\overline{B}(\vec{r_s}+0,t) - \overline{B}(\vec{r_s}-0,t)\right]\delta(\vec{R}-\vec{r_s}) =$$
$$\frac{1}{2}\left[B_n(\vec{r_s}+0,t) - B_n(\vec{r_s}-0,t)\right]\delta(\vec{R}-\vec{r_s}) = 0$$

所以在表面 $S$ 上有

$$\overrightarrow{n_s^0} \cdot \overline{B_p}(\vec{R},t)\delta(\vec{R}-\vec{r}) = 0, \qquad \overrightarrow{n_s^0} \cdot \overline{B_c}(\vec{R},t)\delta(\vec{R}-\vec{r}) = 0 \tag{3.90}$$

③在式（3.88）右方末项，求和号 $\Sigma$ 内的积分是单个微粒表面内外的情况。注意到在铁磁流体内部，每个微粒的表面 $S_{p1}$ 之内的磁场是 $\overline{B_p}$，而 $S_{p1}$ 之外的磁场是 $\overline{B_c}$。点乘 $\overrightarrow{n_{p1}^0} \cdot \overline{B_p} = B_{pn}$，$\overrightarrow{n_{p1}^0} \cdot \overline{B_c} = B_{cn}$。由磁边界的 Gauss 定理给出 $B_{pm} = B_{cn}$，此即

$$\overrightarrow{n_{p1}^0} \cdot \left[\overline{B_p}(\vec{R},t) - \overline{B_c}(\vec{R},t)\right] = B_{pm}(\vec{R},t) - B_{cn}(\vec{R},t) = 0$$

故整个求和号 $\Sigma$ 内各项均是零。

将以上①、②、③的结果代入式（3.88）就得铁磁流体混合物的磁感应强度之平均值遵守 Gauss 定理，即

$$\nabla \cdot \vec{B}(\vec{r},t) = \nabla_r \cdot \int_V \overline{B}(\vec{R},t)\delta(\vec{R}-\vec{r})dV = 0 \tag{3.91}$$

### 3.8.3 铁磁流体中磁场强度 $\overline{H}$ 的体积平均值

设在铁磁流体内一点 $P(x,y,z)$ 的空间邻域 $V$ 内，$\overline{H}(\vec{r},t)$ 场的平均值是

$$\int_V \overline{H}(\vec{R},t)\delta(\vec{R}-\vec{r})dV = \int_{V_p} \overline{H_p}(\vec{R},t)\delta(\vec{R}-\vec{r})dV_p + \int_{V_c} \overline{H_c}(\vec{R},t)\delta(\vec{R}-\vec{r})dV_c \tag{3.92a}$$

式（3.92）给出形式上的平均值关系为

$$\overline{H}(\vec{r},t) = \phi_p(\vec{r},t)\overline{H_p}(\vec{r},t) + \phi_c(\vec{r},t)\overline{H_c}(\vec{r},t) \tag{3.92b}$$

将式（3.92a）两边取旋度，而后将式（3.43a）中的 $\vec{f}(\vec{r},t)$ 取为 $\overline{H}(\vec{r},t)$，就有

$$\begin{aligned}
\nabla \times \overline{H}(\vec{r},t) = \nabla_r \times \int_V \overline{H}(\vec{R},t)\delta(\vec{R}-\vec{r})dV = \\
\int_{V_p} \delta(\vec{R}-\vec{r})\nabla_R \times \overline{H_p}(\vec{R},t)dV_p + \int_{V_c} \delta(\vec{R}-\vec{r})\nabla_R \times \overline{H_c}(\vec{R},t)dV_c - \\
\int_{s_{p0}} dS_{p0} \overrightarrow{n_s^0} \times \overline{H_p}(\vec{R},t)\delta(\vec{R}-\vec{r}) - \int_{s_{c0}} dS_{c0} \overrightarrow{n_s^0} \times \overline{H_c}(\vec{R},t)\delta(\vec{R}-\vec{r}) - \\
\sum_N \int_{s_{p1}} dS_{p1} \overrightarrow{n_{p1}^0} \times \left[\overline{H_p}(\vec{R},t) - \overline{H_c}(\vec{R},t)\right]\delta(\vec{R}-\vec{r})
\end{aligned} \tag{3.93}$$

以下逐项考察式（3.93）之右方：

①式（3.93）右方第一项和第二项积分内，被积函数包含 $\overline{H_p}$ 和 $\overline{H_c}$ 的旋度。当没有传导电流穿过磁场时，按 Ampere 环路定理，此时 $\overline{H_p}$ 和 $\overline{H_c}$ 是无旋的，即

$$\nabla_R \times \overrightarrow{H_p}(\vec{R},t) = 0, \qquad \nabla_R \times \overrightarrow{H_c}(\vec{R},t) = 0$$

故第一和第二两体积分均是零。

②式（3.93）右方第三项和第四项是在包围邻域 $V$ 的表面 $S$ 上的积分，由本章 3.2.2 节的式（3.6d），它们的被积函数可以写成

$$\vec{n}_s^0 \times \overrightarrow{H}(\vec{R},t)\delta(\vec{R}-\vec{r_s}) = \frac{1}{2}\left[\overrightarrow{n_{s-0}^0} \times \overrightarrow{H}(\vec{r_s}-0,t) + \overrightarrow{n_{s+0}^0} \times \overrightarrow{H}(\vec{r_s}+0,t)\right]\delta(\vec{R}-\vec{r_s}) \tag{3.94}$$

如图 3-5 所示，在体积 $V$ 的边界面上取一个扁矩形的回路 $abcd$，其中边 $ab$ 和 $cd$ 平行于表面 $S$，当另外两边 $ad$ 和 $bc$ 趋于零时，$ab$ 就紧贴 $S$ 的内表面，而 $cd$ 就紧贴 $S$ 的外表面。式（3.94）中的 $\overrightarrow{H}(\vec{r_s}-0,t)$ 不是 $\overrightarrow{H_p}$ 就是 $\overrightarrow{H_c}$，而 $\overrightarrow{H}(\vec{r_s}+0,t)$ 则取决于 $V$ 外的介质。对于回路 $abcd$，$\overrightarrow{n_{s-0}^0}$ 是 $ab$ 上的内向法线，$\overrightarrow{n_{s+0}^0}$ 是 $cd$ 上的外向法线，故有 $\overrightarrow{n_{s-0}^0} = -\overrightarrow{n_{s+0}^0} = -\vec{n}_s^0$。此外，将式（3.94）右方括号内的矢量 $\overrightarrow{H}$ 分解为 $\overrightarrow{H_n}$ 和 $\overrightarrow{H_\tau}$，如图 3-5（b）所示。于是式（3.94）可写成

$$\vec{n}_s^0 \times \overrightarrow{H}(\vec{R},t)\delta(\vec{R}-\vec{r_s}) =$$
$$\frac{1}{2}\left\{-\vec{n}_s^0 \times \left[\overrightarrow{H_n}(\vec{r_s}-0,t) + \overrightarrow{H_\tau}(\vec{r_s}-0,t)\right] + \vec{n}_s^0 \times \left[\overrightarrow{H_n}(\vec{r_s}+0,t) + \overrightarrow{H_\tau}(\vec{r_s}+0,t)\right]\right\}\delta(\vec{R}-\vec{r_s})$$

由于 $\vec{n}_s^0 \times \overrightarrow{H_n} = 0$，$\vec{n}_s^0 \times \overrightarrow{H_\tau} = \vec{n}_s^0 \times \vec{\tau}^0 H_\tau = -\vec{v}_0^0 H_\tau$，从而上式成为

$$\vec{n}_s^0 \times \overrightarrow{H}(\vec{R},t)\delta(\vec{R}-\vec{r_s}) = \frac{1}{2}\vec{v}_0\left[H_\tau(\vec{r_s}-0,t) - H_\tau(\vec{r_s}+0,t)\right]\delta(\vec{R}-\vec{r_s}) \tag{3.95a}$$

由于 $ad \to 0$，$bc \to 0$，则 Ampere 环路定理的积分是

$$\oiint_{abcd} \overrightarrow{H} \cdot d\vec{l} = \int_{ab}\overrightarrow{H}(\vec{r_s}-0,t)\cdot d\vec{l} + \int_{cd}\overrightarrow{H}(\vec{r_s}+0,t)\cdot d\vec{l} = [H_\tau(\vec{r_s}-0,t) - H_\tau(\vec{r_s}+0,t)]l = I_c \tag{3.95b}$$

式中 $l = ab = cd$，$I_c$ 是穿过环路 $abcd$ 所围成的矩形面积之电流强度。若将 $I_c$ 在 $abcd$ 环路线上取平均，

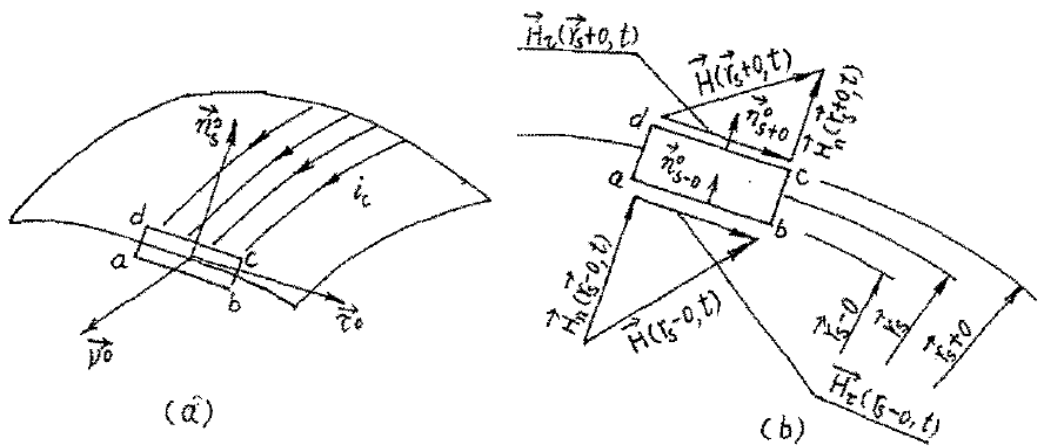

图 3-5　磁场强度 $\overrightarrow{H}$ 的切向分量在边界上的连续性

其平均线密度为 $i_c$，则有

$$I_c = \oiint_{abcd} i_c \, dl = 2l i_c$$

将此式代入式（3.95b），就得

$$H_\tau(\vec{r_s} - 0, t) - H_\tau(\vec{r_s} + 0, t) = 2i_c \qquad (3.95c)$$

再将式（3.95c）代入式（3.95a），即有

$$\vec{n_s^0} \times \vec{H}(\vec{R}, t)\delta(\vec{R} - \vec{r_s}) = \vec{v^0} i_c \delta(\vec{R} - \vec{r_s}) = \vec{i_c} \delta(\vec{R} - \vec{r_s})$$

如图 3-5（a）所示。矢量 $\vec{i_c}$ 的方向就是 $\vec{v^0}$。当没有传导电流穿过面积 $abcd$ 时，$\vec{i_c} = 0$，则

$$\vec{n_s^0} \times \vec{H}(\vec{R}, t)\delta(\vec{R} - \vec{r_s}) = 0$$

于是，式（3.93）右方第三项和第四项积分均为零。

③式（3.93）右方末项求和号内积分是在单个微粒表面 $S_{p1}$ 上进行的，由被积函数中的 $\overline{H_p}(\vec{R}, t)$ 和

$\overline{H_c}(\vec{R}, t)$ 在 $\overline{n_{p1}^0}$ 方向与 $\overline{\tau_{p1}^0}$ 方向分解就得

$$\overline{n_{p1}^0} \times \left[ \overline{H_p}(\vec{R}, t) - \overline{H_c}(\vec{R}, t) \right] = \overline{n_{p1}^0} \times \left[ \overline{n_{p1}^0} H_{pn}(\vec{R}, t) + \overline{\tau_{p1}^0} H_{p\tau}(\vec{R}, t) - \overline{n_{p1}^0} H_{cn}(\vec{R}, t) - \overline{\tau_{p1}^0} H_{c\tau}(\vec{R}, t) \right]$$

由 $\overline{n_{p1}^0} \times \overline{n_{p1}^0} = 0$ 和 $H_{p\tau}(\vec{R}, t) = H_{c\tau}(\vec{R}, t)$，故有[①]

$$\overline{n_{p1}^0} \times \left[ \overline{H_p}(\vec{R}, t) - \overline{H_c}(\vec{R}, t) \right] = 0$$

将已得的①、②、③的结果代入式（3.93）的右方，就给出铁磁流体混合物之磁场强度平均值遵守 Ampere 环路定理，即

$$\nabla \times \overline{H}(\vec{r}, t) = \nabla_r \times \int_V \overline{H}(\vec{R}, t)\delta(\vec{R} - \vec{r})dV = 0 \qquad (3.96)$$

### 3.8.4 铁磁流体磁化强度 $\overline{M}$ 的体积平均值

由于尺寸的微小，铁磁流体内的固相微粒都是磁单畴的，它们的磁矩 $\overline{m_{p1}}$ 在外磁场作用下，受到转向外磁场方向的力矩。但是因为杂乱无序的热运动（Brown 运动）之干扰，每个微粒的磁矩转向外磁场的程度不尽一致。大量微粒的磁矩与外磁场 $\overline{H}$ 的夹角余弦之统计平均值，就是式（1.22）给出的 Langevin 函数 $L(\alpha)$。所以对大量微粒而言，其磁矩在外磁场方向的实际有效分量 $(\overline{m_{p1}})_e$ 是

$$(\overline{m_{p1}})_e = \overline{m_{p1}} L(\alpha)$$

上式两边同除以单个微粒的体积 $V_{p1}$ 就得磁化强度的关系

---

[①] $\overline{n_{p1}^0}$ 是在固相微粒与基载液界面的内侧。若在此界面上也划出一个像 $abcd$ 那样的矩形回路，则 $\overline{n_{p1}^0}$ 就是图 3-5 中的 $\overline{n_{s-0}^0}$，而不是 $\overline{n_{s+0}^0}$ 或 $\overline{n_s^0}$。

$$\overline{M_{pe}} = \overline{M_p} L(\alpha)$$

因为磁化强度是单位体积内的磁矩，所以它是以体积计量的物理量，从而有铁磁流体混合物的磁化强度是

$$\overline{M}(\vec{r},t)dV = \overline{M_{pe}}(\vec{r},t)dV_p + \overline{M_c}(\vec{r},t)dV_c$$

两边取平均

$$\int_V \overline{M}(\vec{R},t)\delta(\vec{R}-\vec{r})dV = \int_{V_p} \overline{M_{pe}}(\vec{R},t)\delta(\vec{R}-\vec{r})dV_p + \int_{V_c} \overline{M_c}(\vec{R},t)\delta(\vec{R}-\vec{r})dV_c$$

从而得到形式上的平均值的关系

$$\overline{M}(\vec{r},t) = \phi_p(\vec{r},t)\overline{M_{pe}}(\vec{r},t) + \phi_c(\vec{r},t)\overline{M_c}(\vec{r},t) \qquad (3.97)$$

式中 $\overline{M_c}(\vec{r},t)$ 是基载液体的磁化强度。在现有的体磁流体中，基载液体是不可磁化的，于是铁磁流体的磁化强度 $\overline{M}(\vec{r},t)$ 等于

$$\overline{M}(\vec{r},t) = \phi_p(\vec{r},t)\overline{M_{pe}}(\vec{r},t)$$

式中，$\overline{M_{pe}}(\vec{r},t) = \overline{M_p}L(\alpha) = \overline{M_p}(\coth\alpha - 1/\alpha)$，$\alpha = \mu_0 m_{p1}H(\vec{r},t)/[k_0 T(\vec{r},t)]$，当外磁场强度 $H(\vec{r},t)$ 很高时，$L(\alpha) \to 1$，此时铁磁流体达到饱和磁化，即全部固相微粒的磁矩均与外磁场方向一致，从而有

$$\overline{M}(\vec{r},t) = \overline{M_s}(\vec{r},t) = \phi_p(\vec{r},t)M_p\overline{H^0} \qquad (3.98)$$

式中 $\overline{H^0}$ 是外磁场的单位矢量。当然，不独式（3.98），在式（3.97）中各项的方向都与外磁场一致，即沿 $\overline{H^0}$ 的方向。

# 第四章　在旋转外磁场中铁磁流体的粘度—牛顿流理论

## 4.1 概述

当静止的铁磁流体处于静止的外磁场之中，由于外磁场对固相微粒的磁矩 $\overrightarrow{m_{p1}}$ 作用有磁力矩 $\overrightarrow{m_{p1}} \times \overrightarrow{B_0}$，并且这样的磁力矩都力图使固相微粒的磁矩矢量转向外磁场方向。因为纷乱无序的热运动之干扰，外磁场作用于固相微粒群体上的磁力矩的方向是随机的，且各方向均有相同的几率，所以其总体合成是零。此时，外磁场对铁磁流体唯一的作用就是磁化。磁化的结果是出现磁化强度 $\overrightarrow{M}$。铁磁流体磁化强度 $\overrightarrow{M}$ 的本质是热运动中的大量固相微粒之磁矩，在外磁场方向投影的统计平均值。所以，$\overrightarrow{M}$ 总是和外磁场 $\overrightarrow{B_0}$ 相平行，从而也没有宏观的磁力矩。这和微观上固相微粒群的磁力矩总体合成是零完全一致。正因为如此，磁化过程不引起铁磁流体宏观的旋转。

使铁磁流体发生旋转的因素有两个：一个是外磁场旋转，另一个是铁磁流体在流动中产生的涡旋。当外磁场旋转速度 $\overrightarrow{\omega_H}$ 与基载液体的涡旋速度 $\overrightarrow{\omega_C}$ 不相等时，外磁场强度 $\overrightarrow{B_0}$ 与铁磁流体磁化强度 $\overrightarrow{M}$ 不再平行，而两个矢量之间出现夹角，即相位角 $\alpha_m$，于是磁力矩 $\overrightarrow{L_m} = \overrightarrow{M} \times \overrightarrow{B_0} = \overrightarrow{L_0^m} M B_0 \sin \alpha_m$ 不为零。与此同时也存在微粒本体与基载液体之间的相对转动，从而有粘性力矩 $\overrightarrow{L_\tau} = C_\tau (\overrightarrow{\omega_H} - \overrightarrow{\omega_C})$。在 $\overrightarrow{L_m}$ 和 $\overrightarrow{L_\tau}$ 中，其中必有一个是驱动力矩，另一个则是阻力矩。当这两个力矩平衡时，固相微粒的磁矩和本体都保持等速旋转。

大量固相微粒磁矩转速的统计平均值就是铁磁流体磁化强度 $\overrightarrow{M}$ 的旋转速度 $\overrightarrow{\omega_M}$。$\overrightarrow{\omega_M}$ 必定和外磁场转速 $\overrightarrow{\omega_H}$ 方向相同，频率相等。否则相位角 $\alpha_m$ 将从 0 到 $2\pi$ 之间周而复始地循环变化，而外磁场作用于微粒磁矩上的磁力矩，在 $[0, \pi]$ 和 $[\pi, 2\pi]$ 两个区间内，大小相等而方向相反。所以每循环一周，外磁场对固相微粒所作之功为零。这就等于说，外磁场不会影响铁磁流体的粘度。显然，这是与事实相悖的。唯一的可能就是 $\overrightarrow{\omega_M} = \overrightarrow{\omega_H}$。

为什么在热运动的纷扰下，固相微粒的磁矩能和外磁场同频旋转？原因就在磁松弛速度上。所谓松弛过程，就是从一个平衡状态转到另一个平衡状态的过程。当磁力矩与涡粘力矩相等时，固相微粒的磁矩和本体均作等速旋转，就是一种平衡状态。若 $\alpha_m$ 因外磁场 $\overrightarrow{B_0}$ 的旋转而改变时，磁力矩就要随之改变，从而打破了平衡状态。松弛过程于是发生。固相微粒磁矩转动的松弛，是固相材料电子自旋轴的偏转或是电子轨道运动平面的偏转。它就是 Neel 扩散过程，这个过程进行非常之快，对于单畴或亚单畴的固相微粒，其扩散的特征时间 $t_N$ 为 $10^{-9}$ 秒的量级。Neel 扩散速度比外磁场的转速要快几个量级。

所以，当外磁场转动一个角度时，固相微粒的磁矩总能及时地松弛到保持 $\alpha_m$ 不变的位置。这在宏观的

表现上就是 $\overrightarrow{\omega_M}$ 与 $\overrightarrow{\omega_H}$ 相等。

$\overrightarrow{\omega_P}$ 是大量固相微粒本体旋转速度的统计平均值。在铁磁流体中，$\overrightarrow{\omega_P}$ 的作用兼有两个方面：①由于微粒是其磁矩的载体，故微粒转动必牵连磁矩旋转；②$\overrightarrow{\omega_P}$ 与液相涡旋速度 $\overrightarrow{\omega_C}$ 之间的滞后正比于粘性力矩。固相微粒在铁磁流体中的松弛过程依赖于 Brown 扩散，它是微粒之间的相互碰撞传递动量和动量矩的过程，由于碰撞频率极高，过程进行很快，其特征时间是 Brown 扩散时间 $t_B$。对于常用的微粒尺寸，$t_B$ 也是 $10^{-9} \sim 10^{-7}$ 秒的量级。由于磁力矩和粘性力矩同时作用于固相微粒上，其中一个是驱动力矩，另一个是阻力矩，所以 $\overrightarrow{\omega_P}$ 只能介于 $\overrightarrow{\omega_H}$ 和 $\overrightarrow{\omega_C}$ 之间。

因为纷乱无序的热运动的干扰，铁磁流体中固相微粒的行为特点是：①并非全部微粒的磁矩都遵循外磁场的制约，受到制约的也存在程度上的差别；②并非所有的固相微粒本体均顺从液相涡旋而运动，即使受到涡旋作用的微粒本体，也同样存在程度上的不同。

取 $\alpha_m$ 和 $\overrightarrow{\omega_P}$ 的统计平均值，就是将上述①和②两方面的效应，同时而分别地摊派到每一个微粒的磁矩和微粒的本体上。于是，就出现在同一个固相微粒上，磁矩的转速是 $\overrightarrow{\omega_H}$，而微粒本体的转速是 $\overrightarrow{\omega_P}$ 的现象。

如图 4-1（a）所示，取空间坐标系 $Ox'y'z$ 的 $Oz$ 轴与外磁场旋转速度矢量 $\overrightarrow{\omega_H}$ 共线，$Ox'y'z$ 是在空间中固定不动的坐标系。在初始的平衡状况下，外磁场 $\overrightarrow{H_0}$ 没有旋转，并且位于 $x'Oz$ 坐标平面上，铁磁流体也处于静止之中，这时 $\overrightarrow{\omega_H}$、$\overrightarrow{\omega_C}$ 和 $\overrightarrow{\omega_P}$ 均为零，磁力矩 $\overrightarrow{L_m}$ 和粘性力矩 $\overrightarrow{L_\tau}$ 都不存在。而铁磁流体的磁化强度为 $\overrightarrow{M_0}$ 且与外磁场 $\overrightarrow{H_0}$ 共线，它们只有分量 $H_{0x}$，$H_{0z}$，$M_{0x}$ 和 $M_{0z}$，外磁场 $\overrightarrow{H_0}$ 与 $Oz$ 轴的夹角是 $\theta_0$，则有

$$H_{0x} = H_0 \sin\theta_0, \qquad H_{0z} = H_0 \cos\theta_0 \qquad (4.1)$$
$$M_{0x} = M_0 \sin\theta_0, \qquad M_{0z} = M_0 \cos\theta_0$$

若外磁场不改变姿态地绕 $Oz$ 轴旋转，其轨迹就是一个正圆锥，如图 4-1（a）所画的那样，锥顶半角 $\theta_0 = \text{const}$。若同时也存在铁磁流体绕 $Oz$ 轴的涡旋。显然，磁力矩 $\overrightarrow{L_m}$ 和粘性力矩 $\overrightarrow{L_\tau}$ 一定会出现，而铁磁流体的磁化强度矢量 $\overrightarrow{M}$ 不再和外磁场矢量 $\overrightarrow{H}$ 共线，它们之间形成相位角 $\alpha_m$，于是有

旋转速度

$$\overrightarrow{\omega_H} = \overrightarrow{\omega_{Hz}}, \qquad \overrightarrow{\omega_M} = \overrightarrow{\omega_{Mz}}, \qquad \overrightarrow{\omega_C} = \overrightarrow{\omega_{Cz}}, \qquad \overrightarrow{\omega_P} = \overrightarrow{\omega_{Pz}} \qquad (4.2)$$

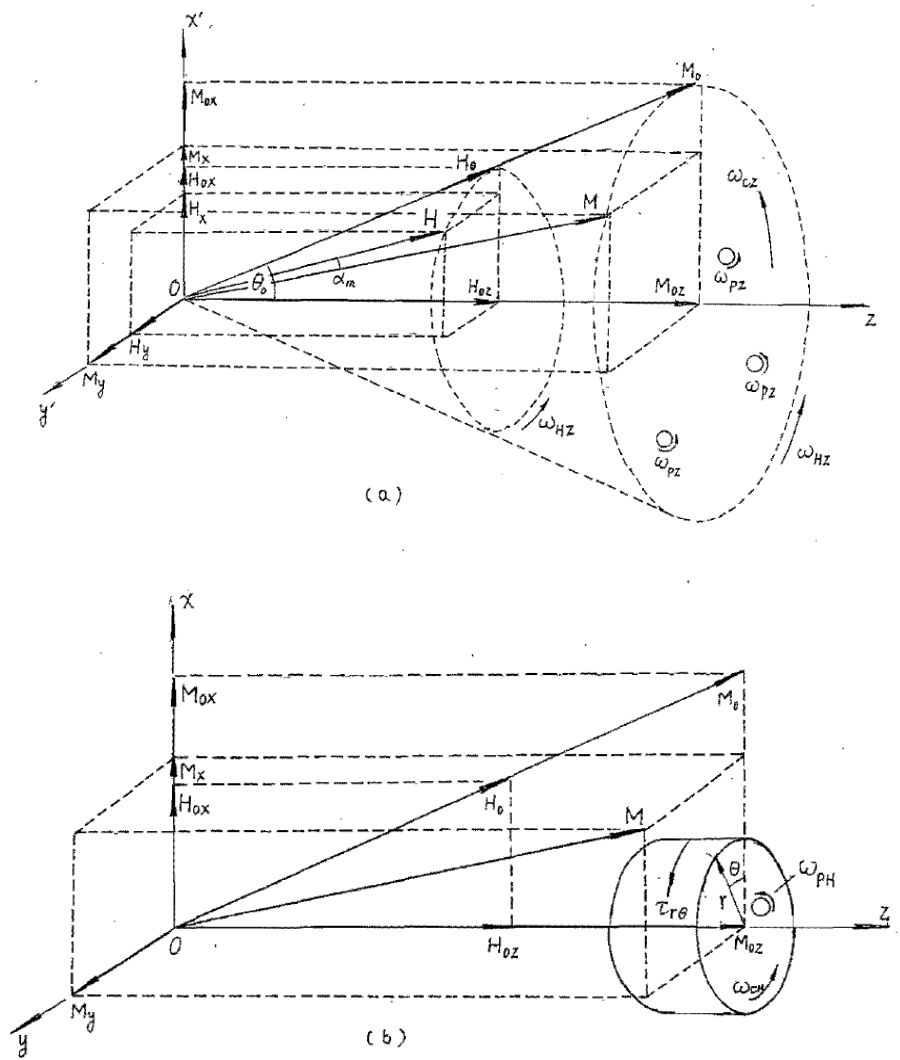

图4-1 静止坐标系和旋转坐标系中的磁场、磁化强度和涡旋转速

外磁场

$$\overrightarrow{H} = \overrightarrow{H_x} + \overrightarrow{H_y} + \overrightarrow{H_z}$$ (4.3a)

$$\left.\begin{array}{l} H_x = H_{0x}\cos(\omega_{Hz}t) = H_0\sin\theta_0\cos(\omega_{Hz}t) \\ H_y = H_{0x}\sin(\omega_{Hz}t) = H_0\sin\theta_0\sin(\omega_{Hz}t) \\ H_z = H_{0z} = H_0\cos\theta_0 \end{array}\right\}$$ (4.3b)

铁磁流体的磁化强度

$$\overrightarrow{M} = \overrightarrow{M_x} + \overrightarrow{M_y} + \overrightarrow{M_z}$$ (4.4a)

$$\left.\begin{array}{l} M_x = M_{0x}\cos(\omega_{Hz}t + \alpha_m) = M_0\sin\theta_0\cos(\omega_{Hz}t + \alpha_m) \\ M_y = M_{0x}\sin(\omega_{Hz}t + \alpha_m) = M_0\sin\theta_0\sin(\omega_{Hz}t + \alpha_m) \\ M_z = M_{0z} = M_0\cos\theta_0 \end{array}\right\}$$ (4.4b)

作用于固相微粒上的磁力矩和粘性力矩

$$\left.\begin{array}{l} \overline{L_m} = \overline{L_{mx}} + \overline{L_{my}} + \overline{L_{mz}} \\ \overline{L_\tau} = \overline{L_{\tau z}} \end{array}\right\} \tag{4.5}$$

在 $\overline{L_m}$ 的分量中，只有 $\overline{L_{mz}}$ 是旋转矢量，而 $\overline{L_{mx}}$ 和 $\overline{L_{my}}$ 是不旋转的。在 $\overline{L_\tau}$ 的分量中，$\overline{L_{\tau x}} = \overline{L_{\tau y}} = 0$。规定 $\alpha_m$ 的方向是 $\overline{M}$ 顺着 $\overline{\omega_{Hz}}$ 转向 $\overline{H}$ 的位置。

磁力矩和粘性力矩的相对关系有两种情况：

① $\overline{\omega_{Hz}} < \overline{\omega_{Cz}}$

此时粘性力矩 $\overline{L_{\tau z}}$ 是驱动力矩，磁力矩 $\overline{L_{mz}}$ 是阻力矩。$\alpha_m$ 是超前相位角，其值为 $\pi \le \alpha_m \le 2\pi$，并且有

$$\overline{\omega_{Hz}} < \overline{\omega_{Pz}} < \overline{\omega_{Cz}}, \qquad \overline{\omega_{Mz}} = \overline{\omega_{Hz}} \tag{4.6a}$$

上面旋转速度 $\overline{\omega}$ 的第一下标 "$H$" 表示外磁场、"$P$" 表示固相微粒、"$C$" 表示基载液体、"$M$" 表示铁磁流体的磁化程度。第二下标表示方向，"$z$" 即 $Oz$ 轴方向。

② $\overline{\omega_{Cz}} < \overline{\omega_{Hz}}$

此时磁力矩 $\overline{L_{mz}}$ 是驱动力矩，粘性力矩 $\overline{L_{\tau z}}$ 是阻力矩。$\alpha_m$ 是滞后相位角，即 $0 \le \alpha_m \le \pi$，并且有

$$\overline{\omega_{Cz}} < \overline{\omega_{Pz}} < \overline{\omega_{Hz}}, \qquad \overline{\omega_{Mz}} = \overline{\omega_{Hz}} \tag{4.6b}$$

若将式（4.6a）与式（4.6b）通减以 $\overline{\omega_{Hz}}$，则有

① $0 < (\overline{\omega_{CH}})_z$

$$0 < (\overline{\omega_{PH}})_z < (\overline{\omega_{CH}})_z, \qquad (\overline{\omega_{MH}})_z = 0 \tag{4.7a}$$

此时粘性力矩 $\overline{L_{\tau z}}$ 没有改变，而且仍然是驱动力矩。同样，由于 $\alpha_m$ 不受坐标系旋转的影响，磁力矩 $\overline{L_{mz}}$ 还是阻力矩。但它在旋转坐标系 $Oxyz$ 中是静止的力矩。

② $(\overline{\omega_{CH}})_z < 0$

$$(\overline{\omega_{CH}})_z < (\overline{\omega_{PH}})_z < 0, \qquad (\overline{\omega_{MH}})_z = 0 \tag{4.7b}$$

此时粘性力矩 $\overline{L_{\tau z}}$ 与外磁场旋转方向相反，所以它是负方向力矩，因而也就是阻力矩。于是静止的磁力矩 $\overline{L_{mz}}$ 被认为是驱动力矩。

在式（4.7a）与式（4.7b）中，相对转速的定义是

$$(\overrightarrow{\omega_{CH}})_z = \omega_{Cz} - \omega_{Hz}, \qquad (\overrightarrow{\omega_{PH}})_z = \omega_{Pz} - \omega_{Hz}, \qquad (\overrightarrow{\omega_{MH}})_z = \omega_{Mz} - \omega_{Hz} \qquad (4.7c)$$

式（4.7a）～式（4.7c）相当于赋予静止坐标系 $Ox'y'z$ 以旋转速度 $\overrightarrow{\omega_{Hz}}$，这个和外磁场 $\overrightarrow{H}$ 同方向同频率绕 $Oz$ 轴旋转的坐标系，用 $Oxyz$ 表示。在坐标系 $Oxyz$ 中，外磁场成为静止不转的。此时在式（4.3b）中取 $\omega_{Hz} = 0$，则有

$$\overrightarrow{H_x} = \overrightarrow{H_{0x}}, \qquad \overrightarrow{H_y} = 0, \qquad \overrightarrow{H_z} = \overrightarrow{H_{0z}} \qquad (4.8a)$$

在式（4.4b）中取 $\omega_{Hz} = 0$，就得

$$\left. \begin{array}{l} \overrightarrow{M_x} = \vec{i} M_{0x} \cos \alpha_m = \vec{i} M_0 \sin \theta_0 \cos \alpha_m \\ \overrightarrow{M_y} = \vec{j} M_{0x} \sin \alpha_m = \vec{j} M_0 \sin \theta_0 \sin \alpha_m \\ \overrightarrow{M_z} = \vec{k} M_{0z} \end{array} \right\} \qquad (4.8b)$$

由式（4.8a）可见，在旋转坐标 $Oxyz$ 中，外磁场是位于 $Oxz$ 平面上的平面矢量 $\overrightarrow{H_0}$，而由式（4.8b）可见，铁磁流体的磁化强度 $\overrightarrow{M}$ 是一个静止的空间矢量。如图 4-1（b）所示。

## 4.2 铁磁流体中作用于固相微粒上的粘性力矩

### 4.2.1 概述

通常用于铁磁流体的固相微粒尺寸平均级数为 $10^{-8}\,\mathrm{m}$，对于普通的固液两相流，如此细小的微粒，无论是平动运动，还是旋转运动，它们与液相之间的滞后完全可以忽略。但在铁磁流体中，固相微粒都是铁磁性的物质，它们的运动不可避免地受到外磁场的制约。外磁场对固相微粒群体的作用力和力矩，使固相与液相之间运动不同步，从而发生作用于两相的粘性力和粘性力矩，其效应有两个：一个是将磁作用传递到液相上，改变整个铁磁流体的运动状态，另一个就是改变铁磁流体的粘度。后者就是本章的讨论内容。

### 4.2.2 固相体积分量 $\phi_p$ 的影响

在第二章式（2.92）中，已经导出圆球在流体中旋转时所受到的粘性力矩。导出式（2.92）中有四个假设前提：①所有固相微粒都是大小相等的圆球，并且均匀地分布于流体之中，都以相等的速度 $\overrightarrow{\omega_p}$ 旋转；②微粒尺寸极其小，旋转是很低的 Re 数的运动；③在圆球坐标系中，只有 $\varphi$ 方向的旋转，而 $r$ 和 $\theta$ 方向的速度是零。所以，只有在垂直于旋转轴的平行圆周上（即纬线圆上）形成切应力环流。此即平面问题；④固相微粒分散在铁磁流体混合物中是稀疏相，互相没有干扰，从而可以按圆球在无界流中的转动处理。

审视上述四个假设，前三个无疑是适宜的。但是第四个假设，对于通常铁磁流体的固相体积分量 $\phi_p$，不能很好地满足无界流的要求，需要修正。

设在铁磁流体的流场中，任取一个体积为 $V_f$ 的控制体，其中含有 $N$ 个体积为 $V_{p1}$ 的固相微粒，于是

固相微粒的体积分量 $\phi_p$ 是

$$\phi_p = \frac{NV_{p1}}{V_f} = \frac{V_{p1}}{\left(\dfrac{V_f}{N}\right)} = \frac{V_{p1}}{V_{f1}} \tag{4.9}$$

式中 $V_{f1}$ 是每个固相微粒平均占有的铁磁流体的体积。将控制体 $V_f$ 平均分割成 $N$ 个小体积 $V_{f1}$，将 $V_{f1}$ 当量成相同体积的正方六面体，其棱长均为 $S$，则有

$$V_{f1} = S^3 \tag{4.10}$$

如图 4-2 所示，任意两个相邻的 $V_{f1}$ 的当量六面体中，固相微粒之间的平均中心距离就是 $S$，将固相微粒体积 $V_{p1} = \pi d_p^3/6$ 和式（4.10），代入式（4.9）中，就得①

$$\frac{s}{d_p} = \left(\frac{\pi}{6\phi_p}\right)^{1/3} \tag{4.11}$$

式（4.11）表明，铁磁流体内固相微粒之间的相对距离 $s/d_p$ 仅取决于固相的体积分量 $\phi_p$。

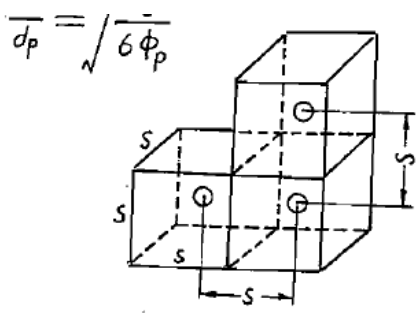

图4-2　$V_{f1}$ 的当量六面体和固相微粒间的平均距离 $s$

固相体积分量受分散剂的影响很大，分散剂的链分子在固相微粒表面附着，其摆动热运动形成防止微粒聚积的能垒，但同时增大了微粒的实际尺寸。设分散剂链分子长度为 $\delta$，则微粒的半径为

$$r_\delta = r_p + \delta = \left(1 + \frac{\delta}{r_p}\right)r_p \tag{4.12a}$$

微粒体积

$$(V_{p1})_\delta = \left(1 + \frac{\delta}{r_p}\right)^3 V_{p1} \tag{4.12b}$$

---

① 式（4.11）在文献[2]可查到。

体积分量

$$\phi_\delta = \left(1 + \frac{\delta}{r_p}\right)^3 \phi_p \tag{4.12c}$$

微粒间的相对距离

$$\frac{s_\delta}{d_\delta} = \left(\frac{\pi}{6\phi_\delta}\right)^{1/3} = \frac{1}{1 + \delta/r_p}\left(\frac{\pi}{6\phi_p}\right)^{1/3} = \frac{1}{1 + \delta/r_p}\frac{s}{d} = \frac{s}{d_\delta} \tag{4.13a}$$

由此可见

$$s_\delta = s \tag{4.13b}$$

即分散剂在固相微粒表面的附着并不改变微粒之间的平均距离，只使相对间距减小。

固相微粒的数密度 $n$ 是单位体积铁磁流体内所含有的固相微粒的数目。设体积为 $V_f$ 的铁磁流体内含有 $N$ 个固相微粒，则

$$n = N/V_f$$

右方分子分母通乘以 $V_{p1}$，得

$$n = \frac{NV_{p1}}{V_f V_{p1}} = \frac{\phi_p}{V_{p1}} = \frac{6\phi_p}{\pi d_p^3} = \frac{1}{s^3} \tag{4.14}$$

因为 $s$ 不受分散剂的影响，故数密度 $n$ 也和分散剂的附着无关。

为了提供数量概念，按式（4.11）、（4.14）和（4.12c）计算一种典型情况，结果列于表 4-1 中。

| $\phi_p$ | 0.01 | 0.05 | 0.10 | 0.15 | 0.1908 | 0.25 | 0.2699 | 0.30 | 0.40 | 0.5235 |
|---|---|---|---|---|---|---|---|---|---|---|
| $s/d_p$ | 3.74 | 2.19 | 1.74 | 1.52 | 1.40 | 1.28 | 1.25 | 1.20 | 1.09 | 1.00 |
| $N/cm^3$ | $3.73 \times 10^{16}$ | $1.87 \times 10^{17}$ | $3.73 \times 10^{17}$ | $5.60 \times 10^{17}$ | $7.12 \times 10^{17}$ | $9.33 \times 10^{17}$ | $1.00 \times 10^{18}$ | $1.12 \times 10^{18}$ | $1.49 \times 10^{18}$ | $1.95 \times 10^{18}$ |
| $\phi_\delta$ | 0.027 | 0.137 | 0.274 | 0.415 | 0.5235 | 0.69 | 0.7406 | —— | —— | —— |

表 4-1  $s/d$、$n$、$\phi_p$ 的典型数值，$d_p = 8 \times 10^{-7}$ cm，$\delta/d_p = 0.2$

由表 4-1 所列的数值可见

①固相微粒之间的平均距离与其直径的数量级一样，并且大得不很多。所以在铁磁流体中固相微粒的运动不能近似为无界流内的运动；

②铁磁流体中固相微粒的数密度很大，每毫升铁磁流体内含有固相微粒个数的数量级在 $10^{16} \sim 10^{18}$ 之间；

③分散剂的附着使固相微粒的有效体积分量增大很多，在表 4-1 的条件下，$\phi_\delta$ 是 $\phi_p$ 的 2.7 倍；

④固相微粒的体积分量为 0.5235 和 0.7406 是两个临界状况。 0.5235 对应于圆球形微粒的立方堆

砌，由于互相接触而不能旋转，0.7406 对应于圆球形微粒的菱形堆砌，此时铁磁流体不能流动。

### 4.2.3 固相微粒之间的相互干涉

**1.概述**

表 4-1 所列的典型数值表明，微粒之间的平均相对间距并不很大，远不是无界流的状况，故铁磁流体流动中微粒间的互相干涉，将是不可避免的。

对于液相自身流动的速度和产生的切应力，均平等地作用于每个微粒之上，无关于微粒之间的相互影响。而微粒相对于液相的运动，引起额外的速度和切应力，就是扰动。这两种扰动都是通过液相的粘性向周围传递的。所谓微粒间相互干涉，就仅限于速度扰动和切应力扰动施加于临近微粒的影响。讨论微粒间的相互干涉，需要事先知道它们引起的速度扰动和切应力扰动的传播规律，于是不得不借助无界流中圆球转动运动的解。虽然这样的解在数密度很大的环境中未必合适。但是借用无界流的结果，不会带来实质性的误差。因为在球面附近，远方边界条件的影响很微弱。

**2.圆球形微粒旋转的速度扰动**

为了简单，设液体是静止的，圆球形微粒在无界流中低 Re 数旋转，其所引起的速度扰动可以近似地由式（2.90b）中取 $\omega_C = 0$ 得到

$$v_\varphi = \frac{r_p^3}{r^2} \omega_p \sin\theta \tag{4.15}$$

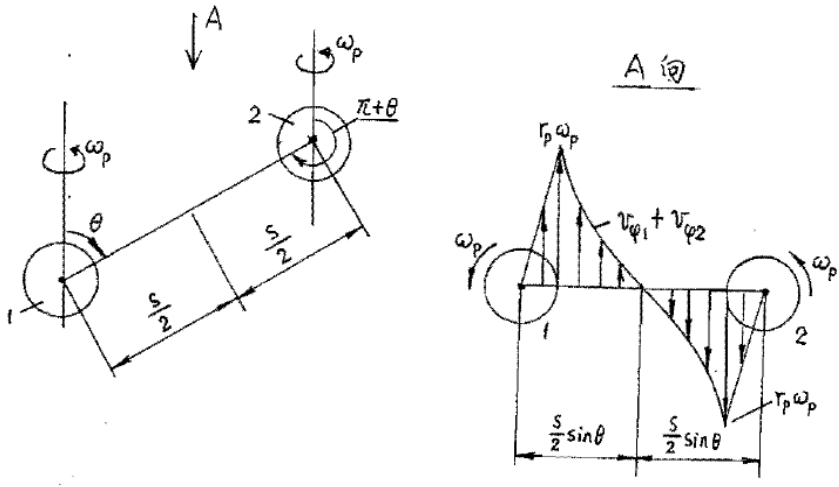

**图4-3 圆相微粒间的速度扰动**

在图 4-3 中，按照所有微粒都以统计平均速度 $\overline{\omega_p}$ 旋转的假设，以微粒 1 中心为原点 $O$，以微粒 2 中心为 $S$，则在两微粒之间的速度扰动量的叠加为

$$v_{\varphi 1} + v_{\varphi 2} = \frac{r_p^3}{r^2} \omega_p \sin\theta + \frac{r_p^3}{(s-r)^2} \omega_p \sin(\pi + \theta)$$

或写成

$$v_{\varphi 1} + v_{\varphi 2} = \left[ \frac{1}{r^2} - \frac{1}{(s-r)^2} \right] r_p^3 \omega_p \sin\theta \tag{4.16}$$

式中 $r_p \le r \le s - r_p$，显然，当 $r = s/2$ 时，有

$$[v_{\varphi 1} + v_{\varphi 2}]_{s/2} = 0 \tag{4.17}$$

式（4.17）表示，在相邻两微粒距离的中点处，$v_{\varphi 1}$ 与 $v_{\varphi 2}$ 大小相等而方向相反，故此点处没有速度的扰动。

3. 切应力扰动

切应力的扰动，表现为液相速度梯度的变化。将式（4.15）代入圆球坐标系内沿纬线圆周的切向应力式

$$\tau_{r\varphi} = \eta_c \left( \frac{\partial v_\varphi}{\partial r} - \frac{v_\varphi}{r} \right)$$

就得出

$$\tau_{r\varphi} = -\eta_c \frac{3r_p^3}{r^3} \omega_p \sin\theta \tag{4.18}$$

式（4.18）亦可直接从式（2.91a）与式（2.91b）取 $\omega_C = 0$ 得到。

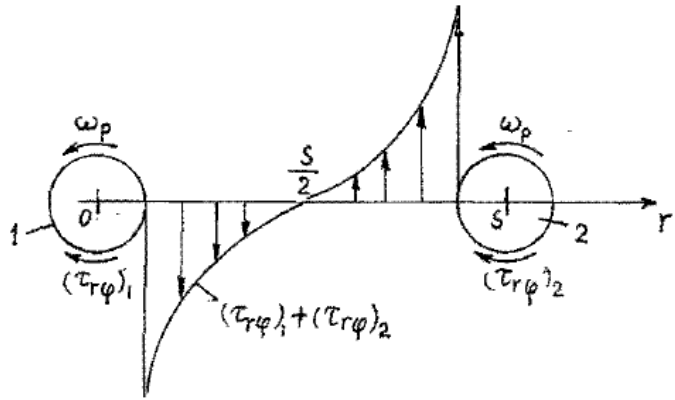

图4-4 微粒间距中的应力扰动

在图 4-4 中，原点 $O$ 是微粒 1 的中心，$S$ 是微粒 2 的中心，则有

$$(\tau_{r\varphi})_1 = -\eta_c \frac{3r_p^3}{r^3} \omega_p \sin\theta, \qquad (\tau_{r\varphi})_2 = -\eta_c \frac{3r_p^3}{(s-r)^3} \omega_p \sin(\pi + \theta)$$

于是，在两微粒间距内切应力的叠加为

$$(\tau_{r\varphi})_1 + (\tau_{r\varphi})_2 = -\left[ \frac{1}{r^3} - \frac{1}{(s-r)^3} \right] 3\eta_c r_p^3 \omega_p \sin\theta \tag{4.19}$$

式中 $r_p \le r \le s - r_p$，当 $r = s/2$，就有

$$[(\tau_{r\varphi})_1 + (\tau_{r\varphi})_2]_{s/2} = 0 \tag{4.20}$$

图 4-4 表示了两相邻微粒之间切应力扰动叠加的示意曲线。从式（4.17）与式（4.20），或从图 4-3 与图 4-4 均可得知，两相邻微粒之距离中点，即 $r = s/2$ 处是零扰动的位置。

4.铁磁流体中包围微粒的零扰动圆球壳形相空间曲面

注意到式（4.17）与式（4.20）关于零扰动除指定 $r = s/2$ 外，对于任何 $\theta$ 值均成立，同时在整个问题求解时已经明确与角度 $\varphi$ 无关。这就预示必有一个零扰动球形空间相曲面的存在。

由表 4-1 所列的典型数值表示出铁磁流体中固相微粒的数密度极其巨大，每毫升中微粒的数目是 $10^{16} \sim 10^{18}$ 的量级。并且由于固相微粒的纳米级尺寸，所以在铁磁流体中固相微粒有十分活跃的热运动。设在铁磁流体中，任意指定一个微粒 $P$，则其它微粒在热运动中向 $P$ 接近的平均距离就是 $S$，因为微粒数目巨大，以及热运动的随机性质，则从所有的方向趋近于 $P$ 的机率是相同的，而且任何一个方向接近 $P$ 的频率均以每秒数十万次到数百万次计。当然，每一个方向，每个瞬时都有大量微粒接近 $P$，也都有大量微粒离开 $P$，这是一种动态过程。在稳定的平衡状态下，从宏观上看来，似乎形成空间上和时间上均是稳定"连续"的圆球形空间曲面，它的中心是 $P$，半径是 $S$。在这个球壳空间曲面上"连续地布满"着固相微粒，这些微粒全都与 $P$ 以同样的速度 $\overrightarrow{\omega_p}$ 旋转。在 $r = s/2$ 处存在一个同心球面，它就是零扰动面。如果液相是静止的，这个零扰动球面提供微粒 $P$ 旋转运动定解的一个边界条件，就是

$$r = s/2, \qquad v_\varphi = 0 \tag{4.21}$$

4.2.4 在 Couett 流中圆球形微粒的有界旋转运动

牛顿粘性定律是在一维线性的 Couett 流基础上的得到的，它表明流体应力与速度梯度成正比，其比例系数就是粘性系数，即

$$\tau = \eta \frac{\partial u}{\partial y}$$

如果流体不是单相的液体，其中含有大量的球形固相微粒，微粒的旋转运动必将引起粘性的改变。图 4-5 表示在 Couett 流中微粒的旋转。

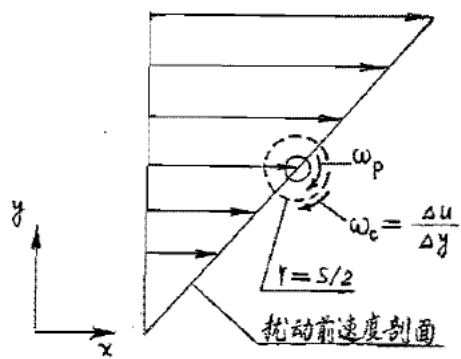

图4-5  旋转的微粒处于Couett流中

Couett 流虽然是简单的速度线性分布的流动，但其流场中到处都有涡旋速度 $\omega_c$，即图 4-5 中所示的 $\Delta u/\Delta y$，在 $r = s/2$ 处，微粒旋转运动的扰动互相抵消，如式（4.21）所示。但现在液相不是静止的，而是处处都存在涡旋的 Couett 流动，所以在半径为 $S/2$ 的球面仍然有 $\overrightarrow{\omega_c}$，于是有

$$r = s/2, \qquad v_\varphi = \left(\frac{s}{2}\sin\theta\right)\omega_C \tag{4.22}$$

在前面 4.2.2 节中，已经提到过，关于式（2.92）的四项假设前提，除无界流之外，其余三项均适用于铁磁流体，所以微粒旋转运动的泛定方程仍然是

$$\frac{\partial^2 v_\varphi}{\partial r^2} + \frac{\partial^2 v_\varphi}{r^2 \partial\theta^2} + \frac{2}{r}\frac{\partial v_\varphi}{\partial r} + \frac{\cot\theta}{r}\frac{\partial v_\varphi}{r\partial\theta} - \frac{v_\varphi}{r^2\sin^2\theta} = 0 \tag{4.23}$$

定解条件。在微粒的表面，无论有无微粒间的相互干扰和液相涡旋 $\omega_C$ 的存在，都不能改变表面粘附状态。所以在微粒表面上的边界条件仍然是

$$r = r_p, \qquad v_\varphi = (r_p \sin\theta)\omega_p \tag{4.24}$$

而在微粒之外的边界条件，就是式（4.22）。

由泛定方程（4.23）和定解条件式（4.22）、式（4.24）的线性性质，则可以使用分离变量法求解。两个边界条件中都含有因子 $\sin\theta$，所以取试探解的形式为

$$v_\varphi = f(r)\sin\theta \tag{4.25}$$

将试探解式（4.25）代入泛定方程（4.23），得出 $f(r)$ 的函数形式，已在第二章 2.12.3 节中给出过，它是

$$f(r) = ar + \frac{b}{r^2} \tag{4.26a}$$

代入边界条件（4.22）中有

$$a\frac{s}{2} + \frac{b}{\left(\frac{s}{2}\right)^2} = \frac{s}{2}\omega_C \tag{4.26b}$$

代入边界条件式（4.24）中有

$$ar_p + \frac{b}{r_p^2} = r_p\omega_P \tag{4.26c}$$

由式（4.26b）与式（4.26c）解出

$$b = \frac{r_p^3}{1 - (d_p/s)^3}(\omega_P - \omega_C) \tag{4.26d}$$

$$a = \omega_P - \frac{\omega_P - \omega_C}{1 - (d_p/s)^3} \tag{4.26e}$$

于是由式（4.26d）、式（4.26e）、式（4.26a）及式（4.25）得速度 $v_\varphi$ 为

$$v_\varphi = r\left[\omega_P - \frac{1}{1 - (d_p/s)^3}\left(1 - \frac{r_p^3}{r^3}\right)(\omega_P - \omega_C)\right]\sin\theta \tag{4.27}$$

在只有 $v_\varphi$ 的情况下，切应力是

$$\tau_{r\varphi} = \eta_c \left( \frac{\partial v_\varphi}{\partial r} - \frac{v_\varphi}{r} \right)$$

式中 $\eta_c$ 是基载液体的粘性系数，用式（4.27）代入，就有

$$\tau_{r\varphi} = -\frac{1}{1-(d_p/s)^3} \eta_c \frac{3r_p^3}{r^3} (\omega_p - \omega_C) \sin\theta \tag{4.28}$$

由式（4.11）知 $(d_p/s)^3 = 6\phi_p/\pi$，于是可见式（4.28）右方的因子 $1/[1-(d_p/s)^3]$ 仅取决于固相的体积分量 $\phi_p$，引用记号

$$C_\phi = \frac{1}{1-(6\phi_p/\pi)} \tag{4.29}$$

以及

$$\eta_\phi = C_\phi \eta_c \tag{4.30}$$

则式（4.28）成为

$$\tau_{r\varphi} = -\eta_\phi \frac{3r_p^3}{r^3} (\omega_p - \omega_C) \sin\theta \tag{4.31}$$

考虑到分散剂链分子在固相微粒表面附着的影响，使用式（4.12a）~式（4.12c）给出

$$C_\delta = \frac{1}{1-(6\phi_\delta/\pi)}, \qquad \eta_\delta = C_\delta \eta_c \tag{4.32}$$

此时切应力为

$$\tau_{r\varphi} = -\eta_\delta \frac{3r_\delta^3}{r^3} (\omega_p - \omega_C) \sin\theta \tag{4.33}$$

在表 4-2 中列出了 $\eta_\delta/\eta_c$ 的常用值。当 $\delta = 0$ 时，$\phi_\delta = \phi_p$，$\eta_\delta = \eta_\phi$。

4.2.5 单位体积铁磁流体中，固相微粒所受到的粘性力矩

参照式（2.92a）与式（2.92b），单个固相微粒受到的粘性力矩是

$$\overrightarrow{L_{\tau1}} = 6\eta_\delta (V_{p1})_\delta (\overrightarrow{\omega_C} - \overrightarrow{\omega_p})$$

若在铁磁流体的微元体积 $dV_f$ 中，含有 $dN$ 个固相微粒，则粘性力矩的总和是

$$d\overrightarrow{L_\tau} = \overrightarrow{L_{\tau1}} dN = 6\eta_\delta (V_{p1})_\delta (\overrightarrow{\omega_C} - \overrightarrow{\omega_p}) dN$$

上式右方 $(V_{p1})_\delta dN = dV_{p\delta}$，由于 $\phi_\delta = dV_{p\delta}/dV_f$，故上式两边同除以 $dV_f$ 就得单位体积铁磁流体中的粘性力矩

$$\overrightarrow{L_\tau} = \frac{d\overrightarrow{L_\tau}}{dV_f} = 6\phi_\delta \eta_\delta (\overrightarrow{\omega_C} - \overrightarrow{\omega_p}) \tag{4.34}$$

| $\delta/r_p$ | 0.2 | | | | | 0.4 | | | | |
|---|---|---|---|---|---|---|---|---|---|---|
| $\phi_p$ | 0.02 | 0.04 | 0.06 | 0.08 | 0.10 | 0.02 | 0.04 | 0.06 | 0.08 | 0.10 |
| $\phi_\delta$ | 0.035 | 0.069 | 0.104 | 0.140 | 0.173 | 0.055 | 0.110 | 0.165 | 0.220 | 0.274 |
| $\eta_\delta/\eta_c$ | 1.071 | 1.152 | 1.247 | 1.359 | 1.493 | 1.117 | 1.265 | 1.459 | 1.722 | 2.101 |
| $\delta/r_p$ | 0.6 | | | | | 0.8 | | | | |
| $\phi_p$ | 0.02 | 0.04 | 0.06 | 0.08 | 0.10 | 0.02 | 0.04 | 0.06 | 0.08 | 0.10 |
| $\phi_\delta$ | 0.082 | 0.164 | 0.246 | 0.328 | 0.410 | 0.117 | 0.233 | 0.350 | 0.467 | —— |
| $\eta_\delta/\eta_c$ | 1.185 | 1.455 | 1.885 | 2.673 | 4.593 | 1.287 | 1.804 | 3.014 | 9.251 | —— |

表 4-2　　$\eta_\delta/\eta_c$ 与 $\phi_p$ 和 $\delta/r_p$ 的关系

## 4.3 铁磁流体内的粘性力矩与涡旋速度的关系

如图 4-1（b）所示，在铁磁流体中取一微元圆柱控制体，其对称轴是 $Oz$，在其表面上作用有粘性切应力环流 $\tau_{r\theta}$，此切应力环流在微元圆柱体 $dz$ 段上产生的粘性力矩是 $d\overrightarrow{L_{\tau z}}$

$$d\overrightarrow{L_{\tau z}} = \vec{r} \times (d\vec{s} \cdot \tau_{r\theta}) = r\overrightarrow{r^0} \times [(2\pi r dz)\overrightarrow{r^0} \cdot (\overrightarrow{r^0}\tau_{r\theta}\overrightarrow{\theta^0})] = \vec{k}\tau_{r\theta}(2\pi r^2 dz) = \vec{k}\tau_{r\theta}(2dV_f)$$

于是单位体积上的粘性力矩是

$$\overrightarrow{L_{\tau z}} = \frac{d\overrightarrow{L_{\tau z}}}{dV_f} = \vec{k}\,2\tau_{r\theta} \tag{4.35}$$

设铁磁流体两相混合物的涡粘系数是 $\eta_v$，则由式（2.73）知其 $\tau_{r\theta}$ 是

$$\tau_{r\theta} = \eta_v\left[ r\frac{\partial}{\partial r}\left(\frac{v_\theta}{r}\right) + \frac{\partial v_r}{r\partial\theta} \right]$$

现在所讨论的问题是在圆柱坐标系中对称于 $z$ 轴的应力环流，所以它与坐标 $\theta$ 无关，即 $\partial v_r/(r\partial\theta) = 0$，于是

$$\tau_{r\theta} = \eta_v\, r\frac{\partial}{\partial r}\left(\frac{v_\theta}{r}\right)$$

在圆柱坐标系中，铁磁流体的涡旋速度 $\omega_{f_z}$ 是

$$\omega_{fz} = \frac{1}{2}\Omega_{fz} = \frac{1}{2}\left[ r\frac{\partial}{\partial r}\left(\frac{v_\theta}{r}\right) - \frac{\partial v_r}{r\partial\theta} \right]$$

同样因为对称于 $z$ 轴，故

$$\omega_{fz} = \frac{1}{2}r\frac{\partial}{\partial r}\left(\frac{v_\theta}{r}\right)$$

于是立即可见

$$\tau_{r\theta} = 2\eta_v\omega_{fz} \tag{4.36}$$

将式（4.36）代入式（4.35），就有

$$\overrightarrow{L_{\tau z}} = \vec{k}4\eta_v\omega_{fz} = 4\eta_v\overrightarrow{\omega_{fz}} \tag{4.37}$$

式中 $\eta_v$ 是铁磁流体混合物的涡粘系数。参照图 3-3 与式（3.55b）的导出过程，并且纳入球形微粒绕中

轴旋转运动时相互干涉的效应，取液相的粘性系数为 $\eta_\delta$，即 $C_\delta\eta_c$，于是绕轴旋转的阻力矩是

$$L_\omega = -(8\pi\eta_\delta r_p^2\omega_P)r_p$$

而对于平动运动阻力对瞬心点 $P$ 造成的力矩，其液相的粘性系数仍然是 $\eta_c$，故

$$L_u = -(6\pi\eta_c r_p^2\omega_P)r_p$$

合成力矩是

$$L = L_u + L_\omega = -(1.5\eta_c + 2\eta_\delta)4\pi r_p^2\omega_P r_p$$

则微粒表面上的平均粘性应力是

$$\tau_p = (1.5 + 2C_\delta)\eta_c\omega_P \tag{4.38a}$$

式中 $C_\delta$ 的由式（4.32）确定。将式（4.38a）代入式（3.54c）之中，就有

$$\eta_v = (1 - \phi_\delta)\eta_c + \phi_\delta(1.5 + 2C_\delta)\eta_c$$

或写成

$$\eta_v = \xi_\delta\eta_c, \qquad \xi_\delta = 1 + 0.5\phi_\delta + 2C_\delta\phi_\delta \tag{4.38b}$$

往后，对于铁磁流体混合物的涡粘系数，均将使用 $\eta_v$ 来取代 Rosensweig 的修正公式（3.57）所给出的 $\eta_{fo}$。

在表 4-3 中列出 $\eta_v/\eta_c$ 之计算值，为了比较同时也列出按式（3.57）计算的 $\eta_{f0}/\eta_c$。

| $\phi_\delta$ | 0.02 | 0.04 | 0.06 | 0.08 | 0.10 | 0.20 | 0.30 | 0.40 | 0.45 |
|---|---|---|---|---|---|---|---|---|---|
| $C_\delta$ | 1.040 | 1.083 | 1.129 | 1.180 | 1.236 | 1.618 | 2.342 | 4.236 | 7.114 |
| $\eta_v/\eta_c$ | 1.052 | 1.107 | 1.165 | 1.229 | 1.297 | 1.747 | 2.555 | 4.589 | 7.628 |
| $\eta_{f0}/\eta_c$ | 1.052 | 1.108 | 1.169 | 1.235 | 1.306 | 1.779 | 2.567 | 4.032 | 5.295 |

表 4-3 铁磁流体混合物的涡粘系数

4.4 在外磁场中铁磁流体粘性系数 $\eta_H$ 的定义

当铁磁流体处于外磁场中，其涡旋运动牵连固相微粒旋转，以致固相微粒的磁矩偏离外磁场方向，从而产生磁力矩 $\overrightarrow{L_{mz}}$，磁力矩作用于固相微粒上，力图使固相微粒回到外磁场方向。于是磁力矩成为涡旋运动的阻力矩。此时铁磁流体所受到的合力矩是

$$\overrightarrow{L_z} = \overrightarrow{L_{\tau z}} - \overrightarrow{L_{mz}} = 4\eta_v \overrightarrow{\omega_{fz}} - \overrightarrow{L_{mz}} \tag{4.39}$$

仿照式（4.37），定义 $\overrightarrow{L_z}$ 与 $\overrightarrow{\omega_{fz}}$ 的关联系数为 $\eta_H$，即

$$\overrightarrow{L_z} = 4\eta_H \overrightarrow{\omega_{fz}} \tag{4.40}$$

式（4.40）是铁磁流体在外磁场中的粘性系数 $\eta_H$ 之定义式。将式（4.40）代入式（4.39）中，就得到

$$\eta_H = \eta_v - \eta_m \tag{4.41}$$

式中 $\eta_m$ 是铁磁流体的磁粘系数

$$\eta_m = \frac{\overrightarrow{L_{mz}}}{4\overrightarrow{\omega_{fz}}} \tag{4.42}$$

4.5 磁力矩 $\overrightarrow{L_m}$ 在 $z$ 方向的分量 $\overrightarrow{L_{mz}}$

4.5.1 作用于单位体积铁磁流体上的磁力矩

设 $\overrightarrow{L_{m1}}$ 是作用于一个微粒上的磁力矩，它是微粒的磁矩 $\overrightarrow{m_{p1}}$ 与外磁场 $\overrightarrow{B_0}$ 的矢性积，即

$$\overrightarrow{L_{m1}} = \overrightarrow{m_{p1}} \times \overrightarrow{B_0} = \mu_0 \overrightarrow{m_{p1}} \times \overrightarrow{H} = \mu_0 V_{p1} \overrightarrow{M_p} \times \overrightarrow{H}$$

式中 $\overrightarrow{M_p}$ 是微粒材料的磁化强度。若体积为 $dV_f$ 的铁磁流体内含有 $dN$ 个固相微粒，则作用于 $dV_f$ 上的磁力矩 $d\overrightarrow{L_m}$ 是

$$d\overrightarrow{L_m} = \overrightarrow{L_{m1}} dN = \mu_0 V_{p1}(dN)\overrightarrow{M_p} L(\alpha) \times \overrightarrow{H}$$

式中 $L(\alpha)$ 是 Langevin 函数，它反映的事实是，在热运动中，各个微粒的方向不相同，而它们与外磁场产生的磁力矩也不一致，$L(\alpha)$ 是一个统计平均的因数。将上式两边同除以 $dV_f$，并且注意到右方的 $V_{p1} dN = dV_p$，于是就得单位体积铁磁流体中的磁力矩是

$$\overrightarrow{L_m} = \frac{d\overrightarrow{L_m}}{dV_p} = \mu_0 \phi_p \overrightarrow{M_p} L(\alpha) \times \overrightarrow{H} = \mu_0 \overrightarrow{M} \times \overrightarrow{H} \tag{4.43}$$

式中 $\overrightarrow{M} = \phi_p \overrightarrow{M_p} L(\alpha)$，是铁磁流体的磁化强度。

4.5.2 磁力矩 $\overrightarrow{L_m}$ 的分解

由图 4-1（b）看到，外磁场 $\overrightarrow{H_0}$ 位于坐标平面 $zOx$ 内，它仅有两个分量，即

$$\overrightarrow{H_0} = \vec{i}H_{0x} + \vec{k}H_{0z} = \vec{i}H_x + \vec{k}H_z = \overrightarrow{H} \tag{4.44a}$$

式（4.44a）表明 $\overrightarrow{H_0}$ 是一静止的矢量，所以恒有

$$\overrightarrow{H_x} = \overrightarrow{H_{0x}}, \qquad \overrightarrow{H_z} = \overrightarrow{H_{0z}} \tag{4.44b}$$

未受旋转运动干扰的铁磁流体磁化强度 $\overrightarrow{M_0}$ 和外磁场 $\overrightarrow{H_0}$ 共线，所以也有

$$\overrightarrow{M_0} = \vec{i}M_{0x} + \vec{k}M_{0z} \tag{4.45}$$

由于铁磁流体涡旋牵连固相微粒旋转造成的干扰，使得磁化强度 $\overrightarrow{M}$ 不再与 $\overrightarrow{H_0}$ 共线，此时 $\overrightarrow{M}$ 旋转而不位于 $zOx$ 坐标平面之内，故矢量 $\overrightarrow{M}$ 的方向和模均与 $\overrightarrow{M_0}$ 不同，但是因为 $\overrightarrow{M}$ 绕 $Oz$ 轴旋转且保持 $\theta_0$ 不变，从而分量 $\overrightarrow{M_z} = \overrightarrow{M_{0z}}$ 不变。所以

$$\overrightarrow{M} = \vec{i}M_x + \vec{j}M_y + \vec{k}M_z \tag{4.46}$$

将式（4.44a）与式（4.46）代入式（4.43）之右方，即得

$$\overrightarrow{L_m} = \mu_0(\vec{i}M_x + \vec{j}M_y + \vec{k}M_z) \times (\vec{i}H_x + \vec{k}H_z) = \mu_0[\vec{i}M_yH_z + \vec{j}(M_zH_x - M_xH_z) - \vec{k}M_yH_x]$$

矢量 $\overrightarrow{L_m}$ 的分量是

$$\overrightarrow{L_m} = \vec{i}L_{mx} + \vec{j}L_{my} + \vec{k}L_{mz}$$

两式对照，按方向拆开就有

$$\overrightarrow{L_{mx}} = \vec{i}\mu_0 M_y H_z \tag{4.47a}$$

$$\overrightarrow{L_{my}} = \vec{j}\mu_0 (M_z H_x - M_x H_z) \tag{4.47b}$$

$$\overrightarrow{L_{mz}} = -\vec{k}\mu_0 M_y H_x \tag{4.47c}$$

将式（4.47c）代入式（4.42）右方分子，就得磁粘系数 $\eta_m$ 为

$$\eta_m = \frac{\mu_0 M_y H_x}{4\omega_{fz}} \tag{4.48}$$

4.6 当外磁场与固相微粒之间有相对转速 $\overrightarrow{\omega_{PH}}$ 时，铁磁流体磁化强度 $\overrightarrow{M}$ 的 $y$ 向分量 $\overrightarrow{M_y}$

### 4.6.1 概述

由式（4.48）可见磁粘系数 $\eta_m$ 取决于磁化强度的 $y$ 向分量模 $M_y$，因为 $H_x = H_{0x}$ 是已知的。所以导出 $M_y$ 是得到 $\eta_m$ 的关键。

当铁磁流体的固相微粒与外磁场有相对转速 $\overrightarrow{\omega_{PH}} = \overrightarrow{\omega_P} - \overrightarrow{\omega_H}$ 时，铁磁流体的磁化强度之改变取决于两个性质不同的相反过程。一个是宏观上固相微粒的磁矩受微粒本体的牵连以 $\overrightarrow{\omega_{PH}}$ 旋转，而越来越偏离外磁场的方向，即原来磁化强度 $\overrightarrow{M}$ 与外磁场 $\overrightarrow{H}$ 共线的平衡状态被破坏，出现越来越大的磁"紧张"的局面。

另外一方面微观上，固相微粒内部分子的热振动和微粒本体在基载液中的 Brown 运动，它们都促使磁化强度恢复与外磁场之间的平衡，而最终达成一个新的平衡状态，这也就是磁松弛过程。

当上述两方面的作用建立一个新的、稳定的平衡状态时，就相应地有一个稳定的磁化强度，其 $y$ 向分量 $\overrightarrow{M_y}$ 即为所求。

在上述两个过程中，分散剂的影响需要考虑。分散剂是不导磁的物质，所以分散剂链分子的附着只是单纯的使微粒的尺寸增加，而影响微粒本体在铁磁流体中的运动，对于微粒的磁化状态不起作用。

①铁磁流体的磁化强度 $\overrightarrow{M}$

固相磁性微粒的磁矩 $\overrightarrow{m_{p1}}$ 是

$$\overrightarrow{m_{p1}} = \overrightarrow{M_p} V_{p1}$$

式中 $\overrightarrow{M_p}$ 是固相微粒的铁磁性材料的磁化强度，相应地 $V_{p1}$ 也是微粒的铁磁性材料的体积。故铁磁流体的磁化强度 $\overrightarrow{M}$ 仍然是

$$\overline{M} = nV_{p1}\overline{M_p}L(\alpha)$$

已经知道，分散剂链分子附着并不改变数密度 $n$，而 $\alpha$ 是

$$\alpha = \frac{\mu_0 m_{p1}H}{k_0 T} = \frac{\mu_0 V_{p1}M_p H}{k_0 T}$$

显然 $\alpha$ 与分散剂无关。于是最后有 $\overline{M}$ 是

$$\overline{M} = \phi_p \overline{M_p}L(\alpha)$$

②Neel 扩散时间 $t_N$

因为 Neel 扩散完全是微粒的铁磁材料内磁晶性质，所以 Neel 扩散时间

$$t_N = \frac{1}{f_0}\exp\left(\frac{K_1 V_{p1}}{k_0 T}\right)$$

与分散剂无关

③Brown 扩散时间 $t_B$

由式（1.16）

$$t_B = \frac{C_\tau}{2k_0 T}$$

由式（2.92b）在没有分散剂链分子附着和无界流的条件下，$C_\tau$ 为

$$C_\tau = 6\eta_c V_{p1}$$

现在的问题是，既有分散剂链分子的附着，又非无界流，微粒存在相互干扰的局面。所以 $C_\tau$ 为

$$C_{\tau\delta} = 6\eta_\delta (V_{p1})_\delta \tag{4.49}$$

于是 Brown 扩散时间此时是 $t_{B\delta}$

$$t_{B\delta} = \frac{3\eta_\delta (V_{p1})_\delta}{k_0 T} = \frac{\left(1+\dfrac{\delta}{r_p}\right)^3}{1-\dfrac{6\phi_\delta}{\pi}}t_B \tag{4.50}$$

式中 $\phi_\delta$ 由式（4.12c）得出。

④综合扩散时间 $t_{r\delta}$

磁松弛过程是以磁矩旋转来实现的。固相微粒内部的 Neel 扩散是磁矩对固相微粒本体的相对运动，

设其速度为 $\omega_N$，而微粒本体在基载液体中的 Brown 扩散对磁矩而言是牵连运动，设其速度为 $\omega_B$。所以磁矩在铁磁流体中的绝对速度 $\omega_r$ 应当是 $\omega_r = \omega_N + \omega_B$，若磁矩转过的角度为 $\theta$，则有 $\theta/t_r = (\theta/t_N) + (\theta/t_B)$，此即[1]

$$t_r = \frac{t_N t_B}{t_N + t_B} \tag{4.51a}$$

在有分散剂链分子附着的场合，使用 $t_{B\delta}$ 取代上式中的 $t_B$，就得

$$t_{r\delta} = \frac{t_N t_{B\delta}}{t_N + t_{B\delta}} \tag{4.51b}$$

4.6.2 铁磁流体的磁化方程——与时间相关的过程

铁磁流体磁化强度 $\overline{M}$ 随时间的变化率由三个部分组成：宏观涡旋运动所致的牵连变率，微观的 Brown 扩散所致的牵连变率，以及微观的 Neel 扩散所致的相对变率。故其总的绝对变率就是

$$\frac{d\overline{M}(t)}{dt} = \left[\frac{d\overline{M}(t)}{dt}\right]_\omega + \left[\frac{d\overline{M}(t)}{dt}\right]_B + \left[\frac{d\overline{M}(t)}{dt}\right]_N \tag{4.52}$$

上式右方各项的下标：$\omega$ 表示涡旋的作用，$B$ 和 $N$ 分别表示 Brown 和 Neel 松弛的作用。

1. $\left[\dfrac{d\overline{M}(t)}{dt}\right]_\omega$

在铁磁流体中，粘性基载液体的涡旋运动，以粘性力矩驱动固相微粒本体以速度 $\overrightarrow{\omega_P}$ 旋转，$\overrightarrow{\omega_P}$ 是大量固相微粒的统计平均转速。处于微粒体中的磁矩，$\overrightarrow{\omega_P}$ 是其牵连转速，为了不失一般性，取 $\overrightarrow{\omega_P}$ 在坐标系内是空间矢量。

固相微粒与外磁场之间存在相对转速 $\overrightarrow{\omega_{PH}}$ 时，微粒本体牵连其磁矩偏离外磁场方向。由于微粒本体转动而造成的磁化强度分量示于图 4-6 中。在图 4-6 中，取角速度逆时针为正方向。

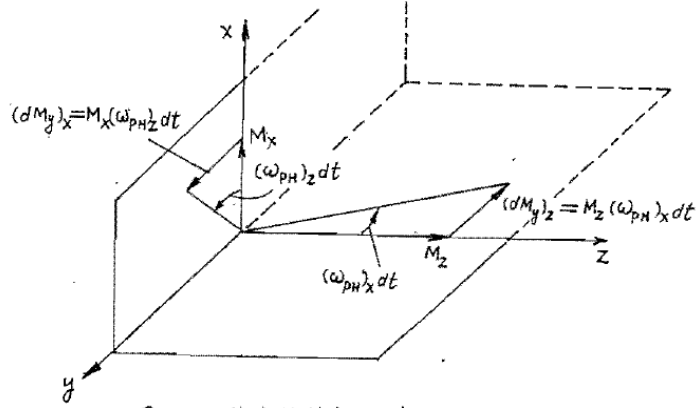

图4-6 微粒旋转产生的磁化强度分量

在图 4-6 中，$(dM_y)_x$ 是磁化强度在 $x$ 方向的分量 $\overrightarrow{M_x}$，绕 $z$ 轴以角速度 $(\omega_{PH})_z$ 在时间 $dt$ 内形成的 $y$ 方向的磁化强度之增量。$(dM_y)_z$ 是 $\overrightarrow{M_z}$ 绕 $x$ 轴以 $(\omega_{PH})_x$ 在 $dt$ 内形成的 $y$ 向增量。已经设定旋转逆时针方向为正，故图 4-6 中 $(dM_y)_x$ 是正，而 $(dM_y)_z$ 是负的，它们各沿 $y$ 轴的正、负方向。故

$$d\overrightarrow{M_y} = (d\overrightarrow{M_y})_x + (d\overrightarrow{M_y})_z = \vec{j}[(\omega_{PH})_z M_x - (\omega_{PH})_x M_z]\, dt \tag{4.53}$$

从而有磁化强度 $\overrightarrow{M}$ 在 $y$ 方向的变化率为

$$\frac{d\overrightarrow{M_y}}{dt} = \vec{j}[(\omega_{PH})_z M_x - (\omega_{PH})_x M_z] \tag{4.54a}$$

同样有

$$\frac{d\overrightarrow{M_z}}{dt} = \vec{k}[(\omega_{PH})_x M_y - (\omega_{PH})_y M_x] \tag{4.54b}$$

$$\frac{d\overrightarrow{M_x}}{dt} = \vec{i}[(\omega_{PH})_y M_z - (\omega_{PH})_z M_y] \tag{4.54c}$$

式（4.54a）～式（4.54c）可以合并写成

$$\left[\frac{d\overrightarrow{M}(t)}{dt}\right]_\omega = \overrightarrow{\omega_{PH}} \times \overrightarrow{M}(t) \tag{4.55}$$

2. $\left[\dfrac{d\overrightarrow{M}(t)}{dt}\right]_B$

非内禀性磁松弛过程，即 Brown 松弛的近似时间过程是[1]

$$\left.\begin{aligned} \overrightarrow{M}(t) &= \phi_p \overrightarrow{M_p} L(\alpha) \frac{t}{t_{B\delta}}, && 0 \le t \le t_{B\delta} \\ \overrightarrow{M}(t) &= M_0 = \phi_P \overrightarrow{M_p} L(\alpha), && t_{B\delta} \le t \end{aligned}\right\} \tag{4.56}$$

上式中 $t_{B\delta}$ 按式（4.50）计算。当外磁场 $H$ 非常大时，即 $\alpha = (\mu_0 M_p V_{p1} H)/(k_0 T) \to \infty$，则 Langevin 函数 $L(\alpha) \approx 1$，就有

$$\overrightarrow{M}(t) = \overrightarrow{M_s} = \phi_p \overrightarrow{M_p} \tag{4.57}$$

上式表示一种极限饱和磁化强度，它的物理意义是 100% 的固相微粒的磁矩都与外磁场共线。当外磁场极其微弱时，此时 $\alpha \approx 0, L(\alpha) \approx 0$，即

$$\overrightarrow{M}(t) = 0$$

此时铁流磁磁体没有被磁化。

式（4.56）表示当 $t \le t_{B\delta}$ 时，$\overrightarrow{M}(t)$ 随时间呈线线增大，当 $t \ge t_{B\delta}$ 之后，$\overrightarrow{M}(t)$ 达到与外磁场强度相适应的饱和磁化强度 $\overrightarrow{M_0}$，它是外磁场对磁矩的约束作用和微粒本体热运动的松弛作用，两者相互平衡的状态。只要外磁场和温度不改变，$\overrightarrow{M}(t)$ 将一直保持为常值 $\overrightarrow{M_0}$。在图 4-7 中表示了这种磁化的折线近似过程。

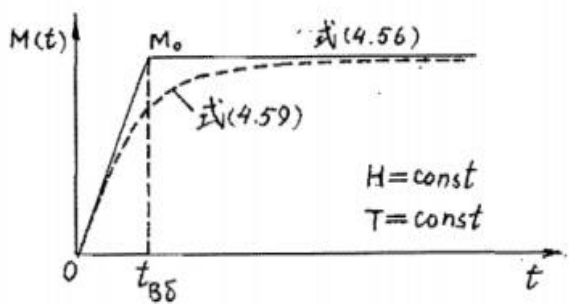

图 4-7　铁磁流体磁化的时间过程

为了分析上的方便，设法使用一连续函数来近似地替代阶代阶段函数式（4.56）。由指数函数的 Taylor 展开

$$\exp\left(-\frac{t}{t_{B\delta}}\right) = 1 - \frac{t}{t_{B\delta}} + \frac{1}{2!}\left(\frac{t}{t_{B\delta}}\right)^2 - \frac{1}{3!}\left(\frac{t}{t_{B\delta}}\right)^3 + \cdots$$

上式的右方近似取头两项代替整个级数，便有

$$\frac{t}{t_{B\delta}} \approx 1 - \exp\left(-\frac{t}{t_{B\delta}}\right) \tag{4.58}$$

将式（4.58）代入式（4.56）中，就得

$$\overrightarrow{M}(t) = \phi_p \overrightarrow{M_p} L(\alpha)\left[1 - \exp\left(-\frac{t}{t_{B\delta}}\right)\right] = \overrightarrow{M_0}\left[1 - \exp\left(-\frac{t}{t_{B\delta}}\right)\right] \tag{4.59}$$

将式（4.59）对时间 $t$ 取导数，给出

$$\frac{d\overrightarrow{M}(t)}{dt} = \frac{1}{t_{B\delta}} \overrightarrow{M_0} \exp\left(-\frac{t}{t_{B\delta}}\right) \tag{4.60 a}$$

将式（4.59）改写成 $\exp(-t/t_{B\delta}) = 1 - [\overrightarrow{M}(t)/\overrightarrow{M_0}]$，而后代入式（4.60 a），得

$$\left[\frac{d\overrightarrow{M}(t)}{dt}\right]_B = \frac{\overrightarrow{M_0} - \overrightarrow{M}(t)}{t_{B\delta}} \tag{4.60b}$$

3. $\left[\dfrac{d\overrightarrow{M}(t)}{dt}\right]_N$

在施加外磁场时，固相微粒的磁矩由纷乱无序转向外磁场的时间，就是 Neel 扩散时间 $t_N$，与铁磁流体流动的特征时间相比，$t_N$ 是极为短暂的，所以在时区 $[0, t_N]$ 中，铁磁流体的磁化强度的变化可以用线性过程来近似。在 $t_N$ 以后，只要外磁场和温度保持不变，磁化强度也保持为平衡值 $\overrightarrow{M_0}$ 不变，形成一条平行于时间轴的水平直线。所以内禀性磁化过程也与图 4-7 中的折线相似。于是可以描绘成下面的数学形式：

$$\left.\begin{array}{l} \overrightarrow{M}(t) = \vec{a}t + \vec{b}, \quad 0 \le t \le t_N \\ \overrightarrow{M}(t) = \overrightarrow{M_0}, \qquad t_N \le t \end{array}\right\} \tag{4.61 a}$$

取定解条件 $t = 0$，$\overrightarrow{M}(t) = 0$；$t = t_N$，$\overrightarrow{M}(t) = \overrightarrow{M_0}$。

定解条件用于式（4.61 a），就得

$$\left.\begin{array}{l} \overrightarrow{M}(t) = \overrightarrow{M_0} \dfrac{t}{t_N}, \quad 0 \le t \le t_N \\ \overrightarrow{M}(t) = \overrightarrow{M_0}, \qquad t_N \le t \end{array}\right\} \tag{4.61 b}$$

借鉴式(5.58)，可以写出

$$\frac{t}{t_N} \approx 1 - \exp\left(-\frac{t}{t_N}\right)$$

于是式（4.61 b）成为

$$\overrightarrow{M}(t) = \overrightarrow{M_0}\left[1 - \exp\left(-\frac{t}{t_N}\right)\right]$$

对时间 $t$ 取导数，就得

$$\left[\frac{d\overrightarrow{M}(t)}{dt}\right]_N = \frac{\overrightarrow{M_0}}{t_N} \exp\left(-\frac{t}{t_N}\right) = \frac{\overrightarrow{M_0} - \overrightarrow{M}(t)}{t_N} \tag{4.62}$$

4.6.3    铁磁流体的磁化方程及其分解

将式（4.55）、式（4.60 b）、式（4.62）代入磁化方程（4.52）之右方，得

$$\frac{d\overrightarrow{M}(t)}{dt} = \overrightarrow{\omega_{PH}} \times \overrightarrow{M}(t) + \frac{\overrightarrow{M_0} - M(t)}{t_{B\delta}} + \frac{\overrightarrow{M_0} - \overrightarrow{M}(t)}{t_N} \tag{4.63}$$

使用式（4.51 b）所给出的综合松弛时间 $t_{r\delta}$ 于上式右方之后两项，即得

$$\frac{d\overrightarrow{M}(t)}{dt} = \overrightarrow{\omega_{PH}} \times \overrightarrow{M}(t) + \frac{\overrightarrow{M_0} - \overrightarrow{M}(t)}{t_{r\delta}} \tag{4.64}$$

式（4.64）右方第二项表明在磁松弛过程中，磁矩相对微粒本体的旋转和磁矩被微粒本体所牵连的旋转，两者同时进行。$t_{r\delta}$ 即是铁磁流体在坐标系中绝对松弛速度的特征时间。

在稳定状态下，$d\overrightarrow{M}(t)/dt = 0$，就有

$$\overrightarrow{\omega_{PH}} \times \overrightarrow{M}(t) = \frac{\overrightarrow{M}(t) - \overrightarrow{M_0}}{t_{r\delta}} \tag{4.65}$$

式(4.65)的物理意义是：由微粒本体旋转引起的磁化强度变化率被 Neel 松弛和 Brown 松弛引起的磁化强度变化率，两相抵消，于是出现一种平衡状态，包括一个不变的相位角 $\alpha_m$ 和统计平均的转速同频于外磁场。同时磁化强度也与时间无关，即应取 $\overrightarrow{M}(t) = \overrightarrow{M}$。

将式（4.65）的右方按坐标方向分解

$$\frac{\overrightarrow{M}(t) - \overrightarrow{M_0}}{t_{r\delta}} = \frac{1}{t_{r\delta}}[\vec{i}(M_x - M_{0x}) + \vec{j}(M_y - M_{0y}) + \vec{k}(M_z - M_{0z})] \tag{4.66}$$

而式（4.65）的左方还原为式（4.54a）~式（4.54c）之右方的形式。于是磁化平衡方程（4.65）成为

$$\begin{aligned}
&\vec{i}[(\omega_{PH})_y M_z - (\omega_{PH})_z M_y] + \vec{j}[(\omega_{PH})_z M_x - (\omega_{PH})_x M_z] + \vec{k}[(\omega_{PH})_x M_y - (\omega_{PH})_y M_x] \\
&= \frac{1}{t_{r\delta}}[\vec{i}(M_x - M_{0x}) + \vec{j}(M_y - M_{0y}) + \vec{k}(M_z - M_{0z})]
\end{aligned} \tag{4.67}$$

由图 4-1（b）的物理模型和坐标系统可见，$(\omega_{PH})_x = (\omega_{PH})_y = 0$，$M_{0y} = 0$ 再将方程式（4.67）按方向拆开，就得

$$\vec{i}: \qquad -(\omega_{PH})_z t_{r\delta} M_y = M_x - M_{0x} \tag{4.68 a}$$

$$\vec{j}: \qquad (\omega_{PH})_z t_{r\delta} M_x = M_y \tag{4.68 b}$$

$$\vec{k}: \qquad 0 = M_z - M_{0z} \tag{4.68 c}$$

## 4.7  固相微粒的旋转运动
### 4.7.1 概述

上述方程式（4.68a）~式（4.68c）只是提供了磁矩旋转与磁松弛之间平衡的数学关系。这三个方程中包含有四个未知数，即 $M_x$、$M_y$、$M_z$ 和 $(\omega_{PH})_z$，所以在数学上它是不完备的。从物理上看，这个问题必定与固相微粒的力学行为相关联，故而补充的方程就是固相微粒的旋转运动方程。

在运动方程的惯性项中，包含有微粒材料的质量密度。本来微粒材料的质量密度 $\rho_{NP}$ 是已知的常数，在分散剂链分子附着的情况下，微粒体积增加，而且链分子形成的微粒外层密度 $\rho_d$ 与中心密度 $\rho_{NP}$ 相差甚远。设此时的平均密度为 $\rho_\delta$，则

$$(V_{p1})_\delta \rho_\delta = V_{P1}\rho_{NP} + [(V_{p1})_\delta - V_{p1}]\rho_d$$

由 $(V_{p1})_\delta = V_{p1}(1+\delta/r_p)^3$，则上式给出

$$\rho_\delta = (\rho_{NP} - \rho_d)\left(1 + \frac{\delta}{r_p}\right)^{-3} + \rho_d \tag{4.69}$$

在固相微粒表面附着的分散剂链分子层的密度 $\rho_d$ 并不等于分散剂的密度，因为附着的链分子之间存在空间，其中充满的是基载液体。所以 $\rho_d$ 的真实值是不知道的，近似的方法是简单地使用基载液的质量密度 $\rho_{NC}$ 取代 $\rho_d$。

### 4.7.2 在外磁场中铁磁流体固相微粒的旋转运动方程

已有的假设是固相微粒为大小一致的圆球形并且都以统计平均速度旋转，单位体积铁磁流内 $n$ 个固相微粒的旋转运动方程为

$$nJ_1\frac{d^2\varphi}{dt^2} = \overrightarrow{L_\tau} + \overrightarrow{L_m} \tag{4.70 a}$$

式中，$n$ 是固相微粒在铁磁流体中的数密度，$J_1$ 是一个圆球形固相微粒的转动惯性矩

$$J_1 = \frac{8}{15}\pi r_\delta^5 \rho_\delta = 0.1\rho_\delta (V_{p1})_\delta d_\delta^2$$

式中，$d_\delta = 2r_\delta$。

在等速旋转下，则

$$\frac{d^2\varphi}{dt^2} = 0$$

此时作用于固相微粒上的粘性力矩 $\overrightarrow{L_\tau}$ 与磁力矩 $\overrightarrow{L_m}$，大小相等而方向相反。即

$$\overrightarrow{L_\tau} + \overrightarrow{L_m} = 0$$

将式（4.34）右方括号内同时加减 $\overrightarrow{\omega_H}$ 而得 $\overrightarrow{\omega_{CH}} - \overrightarrow{\omega_{PH}}$，而后与后式（4.43）一起代入上面的力矩平衡方

程，得到

$$6\phi_\delta \eta_\delta (\overrightarrow{\omega_{CH}} - \overrightarrow{\omega_{PH}}) + \mu_0 \vec{M} \times \vec{H} = 0 \tag{4.70 b}$$

## 4.8 铁磁流体的磁粘度系数 $\eta_m$

### 4.8.1 磁化方程（4.65）与运动方程（4.70 b）联立解 $M_y$

由式（4.48）可见，外磁场对铁磁流体粘性的影响，就是由于 $M_y$ 的存在。$M_y$ 表征磁化强度与外磁场之间的不平衡，造成与涡旋力矩相反的磁力矩 $\overrightarrow{L_{mz}}$，在式（4.39）中，右方的负号就说明磁力矩是阻力矩。

将方程（4.70 b）与 $\vec{M}$ 作矢性积，就有

$$6\phi_\delta \eta_\delta (\overrightarrow{\omega_{CH}} - \overrightarrow{\omega_{PH}}) \times \vec{M} + \mu_0 (\vec{M} \times \vec{H}) \times \vec{M} = 0$$

用式（4.65）代入，就得

$$6\phi_\delta \eta_\delta \overrightarrow{\omega_{CH}} \times \vec{M} - \frac{6\phi_\delta \eta_\delta}{t_{r\delta}} (\vec{M} - \vec{M_0}) + \mu_0 (\vec{M} \times \vec{H}) \times \vec{M} = 0 \tag{4.71}$$

引用记号

$$\beta = \frac{\mu_0 t_{r\delta}}{6\phi_\delta \eta_\delta} \tag{4.72}$$

于是方程（4.71）成

$$t_{r\delta} (\overrightarrow{\omega_{CH}} \times \vec{M}) - (\vec{M} - \vec{M_0}) + \beta (\vec{M} \times \vec{H}) \times \vec{M} = 0 \tag{4.73}$$

将方程（4.73）逐项分解。

1. $\vec{\omega}_{CH} \times \vec{M}$

$$\overrightarrow{\omega_{CH}} \times \vec{M} = \vec{i}[(\omega_{CH})_y M_z - (\omega_{CH})_z M_y] + \vec{j}[(\omega_{CH})_z M_x - (\omega_{CH})_x M_z] + \vec{k}[(\omega_{CH})_x M_y - (\omega_{CH})_y M_x] \tag{4.74 a}$$

2. $\vec{M} - \vec{M_0}$

$$\vec{M} - \vec{M_0} = \vec{i}(M_x - M_{0x}) + \vec{j}(M_y - M_{0y}) + \vec{k}(M_z - M_{0z}) \tag{4.74 b}$$

3. $(\vec{M} \times \vec{H}) \times \vec{M}$

$$\begin{aligned}
(\vec{M} \times \vec{H}) \times \vec{M} &= [(\vec{i}M_x + \vec{j}M_y + \vec{k}M_z) \times (\vec{k}H_z + \vec{i}H_x)] \times (\vec{i}M_x + \vec{j}M_y + \vec{k}M_z) = \\
&\quad [\vec{i}M_y H_z + \vec{j}(M_z H_x - M_x H_z) - \vec{k}M_y H_x] \times (\vec{i}M_x + \vec{j}M_y + \vec{k}M_z) = \\
&\quad \vec{i}(M_z^2 H_x - M_x M_z H_z + M_y^2 H_x) + \vec{j}(-M_y M_z H_z - M_y M_x H_x) + \\
&\quad \vec{k}(M_y^2 H_z - M_z M_x H_x + M_z^2 H_z)
\end{aligned} \tag{4.74c}$$

将式（4.74 a）～式（4.74 c）代入方程式（4.73）中，然后按方向拆开，并且注意到图 4-1（b）有 $M_{0y}=0$，$(\omega_{CH})_y=0$，$(\omega_{CH})_x=0$ 以及式（4.68 c）给出的 $M_z-M_{0z}=0$，于是三个方向的方程为

$$\vec{i}: \quad -(\omega_{CH})_z t_{r\delta}M_y-(M_x-M_{0x})+\beta(M_z^2 H_x-M_x M_z H_z+M_y^2 H_x)=0 \tag{4.75 a}$$

$$\vec{j}: \quad (\omega_{CH})_z t_{r\delta}M_x+M_y+\beta(-M_y M_z H_z-M_y M_x H_x)=0 \tag{4.75 b}$$

$$\vec{k}: \quad M_y^2 H_z-M_z M_x H_x+M_z^2 H_z=0 \tag{4.75 c}$$

现在的目标是集中于求解 $M_y$。观察上述三个方程中，式（4.75a）与式（4.75b）只含 $M_x$ 的一次项，由式（4.68 c）$M_z=M_{0z}$，可以认为是已知的。所以利用这两个方程，消去 $M_x$ 就能获得 $M_y$ 的方程。为此，将式（4.75 a），式（4.75 b）写成下面的形式：

$$(1+\beta M_z H_z)M_x=M_{0x}-(\omega_{CH})_z t_{r\delta}M_y+\beta(M_z^2+M_y^2)H_x \tag{4.76 a}$$

$$[(\omega_{CH})_z t_{r\delta}-\beta M_y H_x]M_x=(1+\beta M_z H_z)M_y \tag{4.76 b}$$

使用下述几个关系式或定义式改写 $\beta$：

$$M_0=\phi_p M_p L(\alpha), \qquad \alpha=\frac{\mu_0 M_p V_{p1}H}{R_0 T}, \qquad t_{r\delta}=\frac{3\eta_\delta(V_{p1})_\delta}{R_0 T}$$

$$\phi_\delta=\phi_p\left(1+\frac{\delta}{r_p}\right)^3, \qquad (V_{p1})_\delta=V_{p1}\left(1+\frac{\delta}{r_p}\right)^3$$

于是有

$$\beta=\frac{\mu_0 t_{r\delta}}{6\phi_\delta\eta_\delta}=\frac{\mu_0 M_0 H}{M_0 H}\cdot\frac{t_{B\delta}}{6\phi_\delta\eta_\delta}\cdot\frac{t_{r\delta}}{t_{B\delta}}=\frac{\mu_0\phi_p M_p L(\alpha)H}{M_0 H}\cdot\frac{1}{6\phi_\delta\eta_\delta}\cdot\frac{3\eta_\delta(V_{p1})_\delta}{k_0 T}\cdot\frac{t_{r\delta}}{t_{B\delta}}=$$
$$\frac{1}{2}\frac{\mu_0 M_p V_{p1}H}{k_0 T}L(\alpha)\frac{t_{r\delta}}{t_{B\delta}}\frac{1}{M_0 H}=0.5\alpha L(\alpha)\frac{t_{r\delta}}{t_{B\delta}}\frac{1}{M_0 H}$$

引用记号

$$\beta_e=0.5\alpha L(\alpha)\frac{t_{r\delta}}{t_{B\delta}} \tag{4.77 a}$$

则有

$$\beta=\frac{\mu_0 t_{r\delta}}{6\phi_\delta\eta_\delta}=\beta_e\frac{1}{M_0 H} \tag{4.77 b}$$

将式（4.77 b）用于方程组（4.76 a）和（4.76 b），并且各式两边通除以 $M_0$，就得

$$\left(1+\beta_e\frac{M_z H_z}{M_0 H}\right)\frac{M_x}{M_0}=\frac{M_{0x}}{M_0}-(\omega_{CH})_z t_{r\delta}\frac{M_y}{M_0}+\beta_e\left(\frac{M_z^2}{M_0^2}+\frac{M_y^2}{M_0^2}\right)\frac{H_x}{H}$$

及

$$\left[(\omega_{CH})_z t_{r\delta} - \beta_e \frac{M_y H_x}{M_0 H}\right] \frac{M_x}{M_0} = \left(1 + \beta_e \frac{M_z H_z}{M_0 H}\right) \frac{M_y}{M_0}$$

两式联立消去 $M_x/M_0$，而后按 $M_x/M_0$ 的方次集项，就得

$$-\beta_e^2 \frac{H_x^2}{H^2}\left(\frac{M_y}{M_0}\right)^3 + \beta_e (\omega_{CH})_z t_{r\delta} \frac{H_x}{H}\left(\frac{M_y}{M_0}\right)^2 -$$

$$\left[\left(1 + \beta_e \frac{M_z H_z}{M_0 H}\right) + \beta_e \frac{M_{0x} H_x}{M_0 H} + \beta_e^2 \frac{H_x^2}{H^2}\frac{M_x^2}{M_0^2} + (\omega_{CH})_z^2 t_{r\delta}^2\right]\left(\frac{M_y}{M_0}\right) + \qquad (4.78)$$

$$\left[(\omega_{CH})_z t_{r\delta} \frac{M_{0x}}{M_0} + \beta_e (\omega_{CH})_z t_{r\delta} \frac{H_x}{H}\frac{M_z^2}{M_0^2}\right] = 0$$

由图 4-1（b）以及式（4.8 a）给出的 $H_x = H_{0x}$，$H_z = H_{0z}$，从而有几何关系

$$\frac{H_x}{H} = \frac{M_{0x}}{M_0} = \sin\theta_0, \qquad\qquad \frac{H_z}{H} = \frac{M_z}{M_0} = \cos\theta_0$$

角度 $\theta_0$ 是已知的常数。则方程式（4.78）成为

$$(-\beta_e^2 \sin^2\theta_0)\left(\frac{M_y}{M_0}\right)^3 + [\beta_e (\omega_{CH})_z t_{r\delta} \sin\theta_0]\left(\frac{M_y}{M_0}\right)^2 -$$

$$[(1 + \beta_e \cos^2\theta_0)^2 + \beta_e \sin^2\theta_0 + \beta_e^2 \sin^2\theta_0 \cos^2\theta_0 + (\omega_{CH})_z^2 t_{r\delta}^2]\left(\frac{M_y}{M_0}\right) + \qquad (4.79)$$

$$[(\omega_{CH})_z t_{r\delta} \sin\theta_0 + \beta_e (\omega_{CH})_z t_{r\delta} \sin\theta_0 \cos^2\theta_0] = 0$$

方程式（4.79）是一维三次方程。由式（4.68 b）知道

$$\frac{M_y}{M_0} = (\omega_{PH})_z t_{r\delta} \frac{M_x}{M_0}$$

由于 $M_x < M_{0x}$，从而 $M_x/M_0 < \sin\theta_0$。即 $M_x/M_0$ 是小于 1 的数，故有 $M_y/M_0 < (\omega_{PH})_z t_{r\delta}$。在通常的情况下，$(\omega_{PH})_z t_{r\delta}$ 的量级是 $10^{-3} \sim 10^{-2}$。所以方程式（4.79）左方的首项是小于 $10^{-9} \sim 10^{-6}$ 的小量。于是，略去三次方的首项以后，式（4.79）成为一维二次方程。引用记号

$$a = \beta_e (\omega_{CH})_z t_{r\delta} \sin\theta_0 \qquad\qquad (4.80a)$$

$$b = (1 + \beta_e \cos^2\theta_0)^2 + \beta_e \sin^2\theta_0 + \beta_e^2 \sin^2\theta_0 \cos^2\theta_0 + (\omega_{CH})_z^2 t_{r\delta}^2 =$$
$$(1 + \beta_e)(1 + \beta_e \cos^2\theta_0) + (\omega_{CH})_z^2 t_{r\delta}^2 \qquad\qquad (4.80\ b)$$

$$c = (1 + \beta_e \cos^2\theta_0)(\omega_{CH})_z t_{r\delta} \sin\theta_0 \qquad\qquad (4.80\ c)$$

于是方程式（4.79）变成的简单形式为

$$a\left(\frac{M_y}{M_0}\right)^2 - b\left(\frac{M_y}{M_0}\right) + c = 0 \tag{4.81}$$

上式的解是

$$\frac{M_y}{M_0} = \frac{b \pm \sqrt{b^2 - 4ac}}{2a} \tag{4.82}$$

由式（4.80a）～式（4.80c）可以看到 $b$ 远大于 $a$ 和 $c$，而 $M_y/M_0$ 不能大于 1，故式（4.82）右方根号前取负号。得到 $M_y/M_0$ 就是得到决定铁磁流体在外磁场中，额外产生的磁粘系数 $\eta_m$ 的关键因素。

### 4.8.2 磁粘系数 $\eta_m$ 的计算关系式

将式（4.48）改写

$$\eta_m = \frac{\mu_0 M_y H_x}{4\omega_{fz}} = -\frac{\mu_0 M_0 H}{4\omega_{fz}}\left(\frac{M_y}{M_0}\right)\left(\frac{H_x}{H}\right)\left(\frac{t_{B\delta}}{t_{B\delta}}\right)$$

利用给出 $\beta_e$ 的式（4.77 a）与式(4.77 b)所使用过的有关 $M_0$、$\alpha$、$t_{B\delta}$、$\phi_\delta$、$(V_{p1})_\delta$ 的式子加上 $\eta_\delta = C_\delta \eta_c$，$H_x/H = \sin\theta_0$，以及式（4.82）代入上式，则有

$$\eta_m = -\frac{3}{2}\phi_p \eta_c C_\delta \frac{0.5\alpha L(\alpha)}{\omega_{fz} t_{B\delta}}\left(1 + \frac{\delta}{r_p}\right)^3 \frac{b - \sqrt{b^2 - 4ac}}{2a}\sin\theta_0 \tag{4.83}$$

### 4.8.3 在一些特定情况下的 $\eta_m$

**1.概述**

本节所述的特定情况，实际上是最常见、最有用的情况。除了铁磁流体作为高速滑动轴承的润滑剂以外，$(\omega_{CH})_z t_{B\delta}$ 是很小的小量，它的二次方可以略去而不会造成不可接受的误差。由式（4.80a）～式（4.80c）可见，$b$ 值的级级是 1，而乘积 $ac$ 值的级级是二阶小量 $(\omega_{CH})_z^2 t_{r\delta}^2$，即 $10^{-6} \sim 10^{-4}$。所以，$4ac \ll b^2$，于是式（4.83）右方的二次方程根项

$$\frac{b - \sqrt{b^2 - 4ac}}{2a} \approx \frac{b - b(1 - 2ac/b^2)}{2a} = \frac{c}{b}$$

将式（4.80 b）、（4.80 c）以及式（4.77 a）代入上式并略去二阶小量 $(\omega_{CH})_z^2 t_{r\delta}^2$，就有

$$\frac{M_y}{M_0} = \frac{b - \sqrt{b^2 - 4ac}}{2a} \approx \frac{c}{b} = \frac{(\omega_{CH})_z t_{r\delta}\sin\theta_0}{1 + \left[0.5\alpha L(\alpha)\dfrac{t_{r\delta}}{t_{B\delta}}\right]} \tag{4.84}$$

于是式（4.83）成为

$$\eta_m = -\frac{3}{2}\phi_p\eta_c C_\delta \frac{0.5\alpha L(\alpha)\frac{t_{r\delta}}{t_{B\delta}}}{1+\left[0.5\alpha L(\alpha)\frac{t_{r\delta}}{t_{B\delta}}\right]}\left(1+\frac{\delta}{r_p}\right)^3 \frac{(\omega_{CH})_z}{\omega_{fz}}\sin^2\theta_0 \tag{4.85a}$$

或用 $(\omega_{CH})_z = \omega_{Cz} - \omega_{Hz}$ 代入，就有

$$\eta_m = -\frac{3}{2}\phi_p\eta_c C_\delta \frac{0.5\alpha L(\alpha)\frac{t_{r\delta}}{t_{B\delta}}}{1+\left[0.5\alpha L(\alpha)\frac{t_{r\delta}}{t_{B\delta}}\right]}\left(1+\frac{\delta}{r_p}\right)^3\left(1-\frac{\omega_{Hz}}{\omega_{Cz}}\right)\frac{\omega_{Cz}}{\omega_{fz}}\sin^2\theta_0 \tag{4.85b}$$

以下，就式（4.85 b）讨论对磁粘系数 $\eta_m$ 的影响因素。

2.外磁场对 $\eta_m$ 的影响

外磁场对 $\eta_m$ 的影响包括三个方面，即：外磁场矢量 $\overline{H}$ 与基载液涡旋速度矢量 $\overrightarrow{\omega_{Cz}}$ 之间的夹角 $\theta_0$；外磁场矢量 $\overline{H}$ 的旋转速度 $\overrightarrow{\omega_{Hz}}$ 与基载液涡旋速度 $\overrightarrow{\omega_{Cz}}$ 的相对关系；外磁场强度 $\overline{H}$ 的模 $H$ 之强弱。

①外磁场矢量 $\overline{H}$ 与基载液涡旋速度矢量 $\overrightarrow{\omega_{Cz}}$ 之间的夹角 $\theta_0$

a. 若 $\overline{H}$ 与 $\overrightarrow{\omega_{Cz}}$ 平行，即 $\theta_0 = 0$；或两者反平行，$\theta_0 = \pi$，则 $\sin\theta_0 = 0$，即

$$\eta_m = 0$$

简单地说，就是平行外磁场对铁磁流体的粘度不产生影响。

b. 若 $\overline{H}$ 与 $\overrightarrow{\omega_{Cz}}$ 垂直，即 $\theta_0 = \pi/2$，则 $\sin\theta_0 = 1$，此时外磁场具最大影响，即

$$\eta_m = -\frac{3}{2}\phi_p\eta_c C_\delta \frac{0.5\alpha L(\alpha)\frac{t_{r\delta}}{t_{B\delta}}}{1+\left[0.5\alpha L(\alpha)\frac{t_{r\delta}}{t_{B\delta}}\right]}\left(1+\frac{\delta}{r_p}\right)^3\left(1-\frac{\omega_{Hz}}{\omega_{Cz}}\right)\frac{\omega_{Cz}}{\omega_{fz}} \tag{4.86 a}$$

简单地说，垂直外磁场对铁磁流体的粘度影响最大。

②磁场矢量的旋转速度 $\overrightarrow{\omega_{Hz}}$

a. 静止外磁场 $\overrightarrow{\omega_{Hz}} = 0$，

$$\eta_m = -\frac{3}{2}\phi_p\eta_c C_\delta \frac{0.5\alpha L(\alpha)\frac{t_{r\delta}}{t_{B\delta}}}{1+\left[0.5\alpha L(\alpha)\frac{t_{r\delta}}{t_{B\delta}}\right]}\left(1+\frac{\delta}{r_p}\right)^3 \frac{\omega_{Cz}}{\omega_{fz}}\sin^2\theta_0 \tag{4.86 b}$$

b. 外磁场与基载液同频向同旋转，即 $\overrightarrow{\omega_{Hz}} = \overrightarrow{\omega_{Cz}}$，则

$$\eta_m = 0$$

由此可见，必须存在外磁场与基载液之间的相对转速，才有 $\eta_m$ 的发生。

　　c. 外磁场的转速快于基载液的涡旋转速，即 $\omega_{Hz} > \omega_{Cz}$，则

$$\eta_m = \frac{3}{2}\phi_p\eta_c C_\delta \frac{0.5\alpha L(\alpha)\frac{t_{r\delta}}{t_{B\delta}}}{1+\left[0.5\alpha L(\alpha)\frac{t_{r\delta}}{t_{B\delta}}\right]}\left(1+\frac{\delta}{r_p}\right)^3\left(\frac{\omega_{Hz}}{\omega_{Cz}}-1\right)\frac{\omega_{Cz}}{\omega_{Hz}}\sin^2\theta_0 \qquad (4.86\ c)$$

由上式可见，这种情况下外磁场使铁磁流体的粘度降低。因为此时磁场造成 $\overline{M}$ 与 $\overline{H}$ 间的滞后相位角，磁力矩成为微粒旋转的驱动力矩。

　　d. 外磁场旋转与基载液涡旋方向相反，则

$$\eta_m = -\frac{3}{2}\phi_p\eta_c C_\delta \frac{0.5\alpha L(\alpha)\frac{t_{r\delta}}{t_{B\delta}}}{1+\left[0.5\alpha L(\alpha)\frac{t_{r\delta}}{t_{B\delta}}\right]}\left(1+\frac{\delta}{r_p}\right)^3\left(1+\frac{|\omega_{Hz}|}{|\omega_{Cz}|}\right)\frac{\omega_{Cz}}{\omega_{fz}}\sin^2\theta_0 \qquad (4.86\ d)$$

在这种情况下，外磁场恒使铁磁流体的粘度增大。

　　③外磁场 $\overline{H}$ 强弱的影响

　　外磁场强度的数值包含在参数 $\alpha$ 之中，即

$$\alpha = \frac{\mu_0 M_p V_{p1} H}{k_0 T}$$

在式（4.85 b）中，含有分式因子 $\beta_e/(1+\beta_e)$，$\beta_e$ 由式（4.77 a）所定义，即

$$\beta_e = 0.5\alpha L(\alpha)\frac{t_{r\delta}}{t_{B\delta}}$$

将 $\beta_e/(1+\beta_e)$ 对 $\beta_e$ 求导，有

$$\frac{d}{d\beta_e}\left(\frac{\beta_e}{1+\beta_e}\right) = \frac{1}{(1+\beta_e)^2} > 0$$

故分式 $\beta_e/(1+\beta_e)$ 随 $\beta_e$ 的增大而增大。在 $\beta_e$ 中，$t_{r\delta}$ 和 $t_{B\delta}$ 均与磁场无关。所以 $\beta_e$ 随 $H$ 的改变取决于 $\alpha L(\alpha)$，于是取导数为

$$\frac{d\beta_e}{d\alpha} = 0.5\frac{t_{r\delta}}{t_{B\delta}}\frac{d}{d\alpha}[\alpha L(\alpha)] = 0.5\frac{t_{r\delta}}{t_{B\delta}}\frac{d}{d\alpha}(\alpha\coth\alpha - 1) = 0.5\frac{t_{r\delta}}{t_{B\delta}}\frac{\cosh\alpha\sinh\alpha - \alpha}{\sinh^2\alpha} \qquad (A)$$

上式的右方分母 $\sinh^2\alpha$ 恒为正值，而分子对 $\alpha$ 求导有

$$\frac{d}{d\alpha}(\cosh\alpha\sinh\alpha - \alpha) = 2\sinh^2\alpha > 0$$

所以分子是随 $\alpha$ 增大的函数。但式（A）右方当 $\alpha = 0$ 时，成为 0/0 的不定式，但 $\alpha \to 0$ 的极限用 L'Hospital 法则得

$$\lim_{\alpha \to 0} \frac{\cosh\alpha \sinh\alpha - \alpha}{\sinh^2\alpha} = \lim_{\alpha \to 0} \frac{2\sinh^2\alpha}{2\cosh\alpha \sinh\alpha} = \lim_{\alpha \to 0} \tanh\alpha = 0 \tag{B}$$

由式（A）与式（B）可知，只要 $\alpha > 0$，则 $d\beta_r / d\alpha$ 恒大于零，此即磁粘系数 $\eta_m$ 随外磁场强度 $H$ 增大而增大。

3.铁磁流体的固相微粒的影响

①固相微粒的尺寸

a.小尺寸微粒，完全内禀性磁松弛

此时 Neel 扩散时间 $t_N$ 远远小于 Brown 扩散时间 $t_B$ 即

$$\frac{t_r}{t_B} \approx \frac{t_N}{t_B} \to 0$$

由式（4.85 b）可见，$\eta_m \to 0$，这个结果说明内禀性铁磁流体的粘度几乎不受外磁场的影响。虽然固相微粒本体受基载液体涡旋的驱动而旋转，但统计平均的磁矩（表现为铁磁流体的磁化强度 $\overline{M}$），却始终保持与外磁场共线而不发生磁力矩。此时铁磁流体只表现出涡粘系数 $\eta_v$。

b.大尺寸微粒完全非内禀性磁松弛

此时 $t_r \approx t_B$，则有

$$\frac{t_r}{t_B} \approx 1$$

于是式（4.85 b）成为

$$\eta_m = -\frac{3}{2}\phi_p \eta_c C_\delta \frac{0.5\alpha L(\alpha)}{1 + 0.5\alpha L(\alpha)}\left(1 + \frac{\delta}{r_p}\right)^3 \left(1 - \frac{\omega_{Hc}}{\omega_{Cc}}\right)\frac{\omega_{Cc}}{\omega_{fc}}\sin^2\theta_0 \tag{4.86 e}$$

微粒的固体体积 $V_{p1}$ 不仅影响磁松弛时间，而且使 $\alpha$ 成比例地改变。微粒表面的分散剂链分子层，不仅使 $\phi_p$ 增大到 $\phi_\delta$，而且也改变 Brown 松弛时间 $t_B$。这些都对 $\eta_m$ 有影响。

②微粒的固相材料的影响

微粒固相材质的作用表现在 $\alpha$ 分子中的 $M_p$。它是固相材料所固有的磁化能力。它对 $\eta_m$ 的影响完全可以套用 $\overline{H}$ 影响的分析。显然，$M_p$ 越高，在外磁场中产生的磁力矩越大，增强了固相微粒抵制液相涡旋的能力，这将表现为铁磁流体的粘度增大。

4.9 在旋转外磁场中，铁磁流体的粘度系数 $\eta_H$

### 4.9.1 概述

式（4.41）表示铁磁流体在外磁场作用下其粘度系数是两部分之叠加，即无外磁场或与外磁场无关的涡旋粘度系数 $\eta_v$（或用 $\eta_{f0}$），与在外磁场作用下产生的磁粘度系数 $\eta_m$（或用 $\Delta\eta$），两者简单地代数相加。用式（4.38 b）与式（4.85 b）代入式（4.41），即得

$$\eta_H = \eta_c\left[(1+0.5\phi_\delta+2C_\delta\phi_\delta)+\frac{3}{2}\phi_p C_\delta\frac{\beta_e}{1+\beta_e}\left(1+\frac{\delta}{r_p}\right)^3\left(1-\frac{\omega_{Hz}}{\omega_{Cz}}\right)\frac{\omega_{Cz}}{\omega_{fz}}\sin^2\theta_0\right] \tag{4.87}$$

式中 $\beta_e$ 见式（4.77 a），如果使用 $\eta_{f0}$，则由式（3.57）代入得

$$\eta_H = \eta_c\left[(1-2.5\phi_\delta+1.55\phi_\delta^2)^{-1}+\frac{3}{2}\phi_p C_\delta\frac{\beta_e}{1+\beta_e}\left(1+\frac{\delta}{r_p}\right)^3\left(1-\frac{\omega_{Hz}}{\omega_{Cz}}\right)\frac{\omega_{Cz}}{\omega_{fz}}\sin^2\theta_0\right] \tag{4.88}$$

式中 $\phi_\delta = \phi_p(1+\delta/r_p)^3$。虽然式（4.87）与式（4.88）比 Shiliomis 公式[2]多包含一些影响因素，但是考虑磁作用的基本思路和处理方法仍然是 Shiliomis 提出的。不过，以增量形式表示磁作用而与涡旋作用相叠加的形式值得推敲，尤其是式（4.88）中的涡粘系数与磁作用增量两部分具有不一致的前提条件。涡粘系数的基础是 Einstein 公式，它以固相微粒等同液相微团涡旋的状态，绕其瞬心点滚动得出的，而外磁场对粘度的贡献必须存在固液两相间旋转速度的滞后。也就是说，存在外磁场时，Einstein 公式成立的条件已不存在。所以式（4.88）的合理性是有疑问的。虽然式（4.87）中，涡粘系数 $\eta_v$ 的导出取消了固液同频的前提，但两部分简单叠加的形式没有改变。

有鉴于此，本节提出一个将涡粘作用和磁力矩作用融合在一起的 $\eta_H$ 之关系式。为此，先讨论两个转速比，即 $(\omega_{PH})_z/(\omega_{CH})_z$ 和 $(\omega_{PH})_z/\omega_{fz}$。

### 4.9.2 固液两相的相对转速比 $(\omega_{PH})_z/(\omega_{CH})_z$

由力矩平衡方程式（4.70 b）

$$6\phi_\delta\eta_\delta(\overrightarrow{\omega_{CH}}-\overrightarrow{\omega_{PH}})+\mu_0(\overrightarrow{M}\times\overrightarrow{H}) = 0$$

左边展开成分量的形式

$$6\phi_\delta\eta_\delta\left\{\vec{i}[(\omega_{CH})_x-(\omega_{PH})_x]+\vec{j}[(\omega_{CH})_y-(\omega_{PH})_y]+\vec{k}[(\omega_{CH})_z-(\omega_{PH})_z]\right\}+$$
$$\mu_0(\vec{i}M_x+\vec{j}M_y+\vec{k}M_z)\times(\vec{i}H_x+\vec{k}H_y) = 0$$

由图 4-1（b），除 $\overline{H}$ 在 $y$ 方向无分量以外，坐标系取向使涡旋在 $x$ 和 $y$ 方向也无分量，于是只取 $z$ 方向的平衡得

$$6\phi_\delta\eta_\delta[(\omega_{CH})_z-(\omega_{PH})_z]-\mu_0 M_y H_x = 0 \tag{C}$$

上式改写成

$$\left[1 - \frac{(\omega_{PH})_z}{(\omega_{CH})_z}\right](\omega_{CH})_z - \frac{\mu_0 M_0 H}{6\phi_\delta \eta_\delta} \frac{M_y H_x}{M_0 H} = 0 \tag{D}$$

由式（4.84）并且注意到式（4.77a），则有

$$\frac{M_y}{M_0} = \frac{(\omega_{CH})_z t_{r\delta}}{1 + \left[0.5\alpha L(\alpha)\dfrac{t_{r\delta}}{t_{B\delta}}\right]}\sin\theta_0 = \frac{(\omega_{CH})_z t_{r\delta}}{1 + \beta_e}\sin\theta_0$$

将上式以及 $H_x/H = \sin\theta_0$ 代入方程式（D），约去公因子 $(\omega_{CH})_z$ 之后，得

$$1 - \frac{(\omega_{PH})_z}{(\omega_{CH})_z} - \frac{\mu_0 M_0 H}{6\phi_\delta \eta_\delta}\frac{t_{r\delta}}{1 + \beta_e}\sin^2\theta_0 = 0$$

将式（4.77 b）用于上式，则有

$$1 - \frac{(\omega_{PH})_z}{(\omega_{CH})_z} - \frac{\beta_e}{1 + \beta_e}\sin^2\theta_0 = 0$$

于是最终有

$$\frac{(\omega_{PH})_z}{(\omega_{CH})_z} = 1 - \frac{\beta_e}{1 + \beta_e}\sin^2\theta_0 = \frac{1 + \beta_e\cos^2\theta_0}{1 + \beta_e} \tag{4.89}$$

4.9.3 固相微粒的相对转速与铁磁流体的涡旋转速之比 $(\omega_{PH})_z/\omega_{fz}$

设包含有一个微粒的铁磁流体之平均体积为 $V_{f1}$，其质量为 $m_{f1}$，则由动量守恒知道，$m_{f1}$ 的动量等于 $V_{f1}$ 内所包含的固相微粒的动量和液相的动量之和，即

$$m_{f1}\overline{U_f} = m_{\delta 1}\overline{U_p} + (m_{f1} - m_{\delta 1})\overline{U_c} = m_{\delta 1}(\overline{U_p} - \overline{U_c}) + m_{f1}\overline{U_c}$$

式中 $m_{\delta 1}$ 是一个具有分散剂附着层的微粒质量。上式可改写成

$$\rho_f V_{f1}\overline{U_f} = \rho_\delta (V_{p1})_\delta(\overline{U_p} - \overline{U_c}) + \rho_f V_{f1}\overline{U_c}$$

式中 $\rho_\delta$ 是具有分散剂附着层的微粒的平均密度，见式（4.69）。上式两边同除以 $\rho_f V_{f1}$，得

$$\overline{U_f} = \phi_\delta \frac{\rho_\delta}{\rho_f}(\overline{U_p} - \overline{U_c}) + \overline{U_c}$$

将铁磁流体中数目极为巨大的固相微粒视为一种"连续"的粒子流，则上式两边取旋度，由 $\nabla\times\overline{U} = \overline{\Omega} = 2\overline{\omega}$，并且取 $\phi_\delta$ 之平均值，就有

$$\overline{\omega_f} = \phi_\delta \frac{\rho_\delta}{\rho_f}(\overline{\omega_p} - \overline{\omega_c}) + \overline{\omega_c}$$

将上式写成相对转速，并仅取其 $z$ 方向分量的方程，就有

$$(\omega_{fH})_z = \phi_\delta \frac{\rho_\delta}{\rho_f}[(\omega_{PH})_z - (\omega_{CH})_z] + (\omega_{CH})_z$$

或改写

$$\frac{(\omega_{fH})_z}{(\omega_{PH})_z} = \phi_\delta \frac{\rho_\delta}{\rho_f}\left[1 - \frac{(\omega_{CH})_z}{(\omega_{PH})_z}\right] + \frac{(\omega_{CH})_z}{(\omega_{PH})_z} = \phi_\delta \frac{\rho_\delta}{\rho_f} + \left(1 - \phi_\delta \frac{\rho_\delta}{\rho_f}\right)\frac{(\omega_{CH})_z}{(\omega_{PH})_z}$$

用式（4.89）代入

$$\frac{(\omega_{fH})_z}{(\omega_{PH})_z} = \phi_\delta \frac{\rho_\delta}{\rho_f} + \left(1 - \phi_\delta \frac{\rho_\delta}{\rho_f}\right)\frac{1 + \beta_e}{1 + \beta_e \cos^2 \theta_0} \tag{4.90}$$

由 $\dfrac{(\omega_{PH})_z}{\omega_{fz}} = \dfrac{(\omega_{PH})_z}{(\omega_{fH})_z}\dfrac{(\omega_{fH})_z}{\omega_{fz}} = \dfrac{(\omega_{PH})_z}{(\omega_{fH})_z}\left(1 - \dfrac{\omega_{Hz}}{\omega_{fz}}\right)$，用式（4.90）代入右方，即得

$$\frac{(\omega_{PH})_z}{\omega_{fz}} = \left[\phi_\delta \frac{\rho_\delta}{\rho_f} + \left(1 - \phi_\delta \frac{\rho_\delta}{\rho_f}\right)\frac{1 + \beta_e}{1 + \beta_e \cos^2 \theta_0}\right]^{-1}\left(1 - \frac{\omega_{Hz}}{\omega_{fz}}\right) \tag{4.91}$$

#### 4.9.4 在旋转外磁场中，铁磁流体粘度的另一种关系式

流体内摩擦力的传递是一种分子动量的输运过程。所以作为一种混合物的铁磁流体，其动量输运通量应当等于它所包含的固液两种成分的动量输运通量之和，于是由式（3.54b），对于均匀铁磁流体有

$$\eta_H \overrightarrow{\omega_f} = (1 - \phi_\delta)\eta_c \overrightarrow{\omega_C} + \phi_\delta \eta_p \overrightarrow{\omega_P}$$

上式可改写

$$\eta_H \overrightarrow{\omega_f} = (1 - \phi_\delta)\eta_c (\overrightarrow{\omega_C} - \overrightarrow{\omega_H} + \overrightarrow{\omega_H}) + \phi_\delta \eta_p (\overrightarrow{\omega_P} - \overrightarrow{\omega_H} + \overrightarrow{\omega_H})$$

于是有

$$\eta_H \overrightarrow{\omega_f} = (1 - \phi_\delta)\eta_c (\overrightarrow{\omega_{CH}} + \overrightarrow{\omega_H}) + \phi_\delta \eta_p (\overrightarrow{\omega_{PH}} + \overrightarrow{\omega_H})$$

上式可分解到三个坐标轴方向，按照图 4-1（a）只需取 z 轴方向，即

$$\eta_H \omega_{fz} = (1 - \phi_\delta)\eta_c (\omega_{CH})_z + \phi_\delta \eta_p (\omega_{PH})_z + [(1 - \phi_\delta)\eta_c + \phi_\delta \eta_p]\omega_{HZ}$$

两边通除 $\omega_{fz}$，就有

$$\eta_H = \left[(1 - \phi_\delta)\eta_c \frac{(\omega_{CH})_z}{(\omega_{PH})_z} + \phi_\delta \eta_p\right]\frac{(\omega_{PH})_z}{\omega_{fz}} + [(1 - \phi_\delta)\eta_c + \phi_\delta \eta_p]\frac{\omega_{Hz}}{\omega_{fz}} \tag{4.92}$$

式中 $(\omega_{CH})_z/(\omega_{PH})_z$ 和 $(\omega_{PH})_z/\omega_{fz}$ 分以式（4.89）与式（4.91）代入，$\eta_p$ 以式（4.38a）代入，结果得出

$$
\eta_H = \eta_c \left\{ \left[ (1-\phi_\delta)\frac{1+\beta_e}{1+\beta_e \cos^2\theta_0} + \phi_\delta(1.5+2C_\delta) \right] \left[ \left( 1-\phi_\delta\frac{\rho_\delta}{\rho_f} \right)\frac{1+\beta_e}{1+\beta_e \cos^2\theta_0} + \phi_\delta\frac{\rho_\delta}{\rho_f} \right]^{-1} \left( 1-\frac{\omega_{Hz}}{\omega_{fz}} \right) + \right.
$$
$$
\left. (1+0.5\phi_\delta + 2C_\delta\phi_\delta)\frac{\omega_{Hz}}{\omega_{fz}} \right\}
\tag{4.93}
$$

在式 (4.93) 中，磁作用表现在 $\beta_e$ 中，涡旋作用表现为 $\omega_{fz}$，外磁场方向和转速反映为 $\cos\theta_0$ 和 $\omega_{Hz}$，此外

两体积分量与质量比也以 $\phi_\delta$，$C_\delta$ 和 $\rho_\delta/\rho_f$ 体现。对于式（4.93）有两方面需要讨论：

1.在下述任一情况下，$\eta_H$ 均将蜕化为 $\eta_v = \eta_c(1+0.5\phi_\delta + 2C_\delta\phi_\delta)$

①不存在外磁场，$H=0$，则 $\beta_e = 0$，$\omega_{Hz} = 0$；

②外磁场 $\overrightarrow{H}$ 与铁磁流体涡旋矢量 $\overrightarrow{\omega_{fz}}$ 平行，则 $\theta_0 = 0$；

③外磁场与铁磁流体同频旋转，即 $\omega_{Hz} = \omega_{fz}$；

④完全内禀铁磁流体，则 $\beta_e \approx 0$，$\eta_H \approx \eta_v$。

2.式（4.93）中，唯一的未知数是 $\omega_{fz}$，将 $\eta_H$ 的式（4.93）用于铁磁流体混合流的涡量方程中，则 $\eta_H$

式中的 $\omega_{fz}$ 参与到涡量方程的求解函数之中。

### 4.9.5  本章结束语

尽管式（4.87）、式（4.88）及式（4.93）在形式上包括了多种因素，但对真正解决铁磁流体在外磁场中的粘性问题，仍然是不够的。因为本章在认为铁磁流体是一种牛顿流体的基础上得到 $\eta_H$ 的关系式。

但实际上，铁磁流体大多数表现为非牛顿流 Bingham 拟塑性体，它既有屈服应力，又存在剪切稀化[1]。外磁场在一定范围内使铁磁流体的表观粘性增加，而剪切率使表观粘性降低，所以在外磁场中铁磁流体的表观粘度不是线性的。文献[1]利用一些公开的实验数据得出铁磁流体的二次型粘度系数。在文献[2]的表 8.1 中所列出的煤油基铁磁流体粘度的实验数据，明显地反映了铁磁流体的非牛顿流性质。表4-4 摘录文献[2]表 8.1 的相关数据。从其中可以看出在表中的磁场强度范围内，铁磁流体的粘度系数随磁场强度增大而增大，但在高剪切率下，表观粘度增量小于低剪切率下的，剪切稀化是显而易见的。关于铁磁流体的非牛顿流问题可参考文献[1]的专门章节。

| $\beta/T$ | $\Omega/S^{-1}$ | $\Delta\eta/\eta$（实验值） |
|---|---|---|
| 0.089 | 4.6 | 0.30 |
| | 11.5 | 0.34 |
| | 230 | 0.26 |
| 0.714 | 46 | 0.60 |
| | 115 | 0.70 |
| | 230 | 0.55 |
| 2.14 | 46 | 1.15 |
| | 115 | 1.02 |
| | 230 | 0.44 |

表 4-4　铁磁流体粘度增量

# 第五章 铁磁流体混合流

## 5.1 概述

铁磁流体混合流是将液相基础载体和固相磁性微粒合在一起，作为一种单一的流态物质来处理。为了解决连续性的问题，将两相的有关物性参数，诸如密度，粘度，比热容等，使用 $\delta$ 函数将两相参数混合取平均值，从而近似地建立铁磁流体是一种"单相"的连续介质的观念。在这样的基础上，普通流体力学的一切处理方法均适用于铁磁流体。

但是铁磁流体与普通两相流有不同之处，最重要的是铁磁流体对外磁场有响应能力。在它的动力学方程组中，包括反映外磁场施加于铁磁流体的作用，如磁力、磁力矩、磁化功等。

在铁磁流体中只有固相成分能够对外磁场做出响应。这些处于纷乱热运动之中的细小的固相微粒，在外磁场作用下，各个微粒克服纷乱的热运动影响，而在不同程度上其磁矩倾向外磁场的方向。所以整体铁磁流体所呈现的宏观磁化强度 $\overline{M}$，就是大量微粒的磁矩在外磁场方向投影的统计平均值，其平均因子就是 Langevin 函数 $L(\alpha)$。

正是因为铁磁流体的磁化强度 $\overline{M}$ 是大量微粒的磁矩在外磁场方向投影的统计平均值。所以当铁磁流体处于静止时，磁化强度 $\overline{M}$ 总是平行于外磁场 $\overline{B_0}$（即 $\mu_0\overline{H}$）。但是在流动状态下，由于铁磁流体是一种粘性较大的液体，所以必定会发生涡旋，从而对固相微粒产生粘性力矩推动固相微粒旋转。当磁力矩与粘性力矩相平衡之后，磁化强度 $\overline{M}$ 与外磁场 $\overline{B_0}$ 之间存在一个相位角 $\alpha_m$。

磁力矩和粘性力矩同时作用于固相微粒上，但两者在物理机制上很不相同。磁力矩只作用于磁矩上，它关系到 Neel 扩散的磁松弛；粘性力矩是基载液体对微粒本体的作用，它关系到 Brown 扩散的磁松弛过程。已经知道内禀性小尺寸固相微粒的磁松弛取决于 Neel 扩散，非内禀性的大尺寸微粒的磁松弛取决于 Brown 扩散[1,2]。

磁松弛对铁磁流体的动力学方程，有很重要的影响。若磁松弛过程相对于铁磁流体的运动是极其快的，以致于其磁化强度 $\overline{M}$，在流场中处处都与外磁场相适应，这就是所谓的"磁平衡流"。若磁松弛过程相对于铁磁流体的运动非常地慢，以致于铁磁流体的磁化强度 $\overline{M}$ 在流场中来不及随外磁场改变，这就是所谓的"磁冻结流"。实际的情况是介于磁平衡流和磁冻结流之间，即"磁松弛流"。可以说，磁平衡流与磁冻结流两者均是磁松弛流的极端特例。一般说来，完全内禀性的铁磁流体很接近磁平衡流，在其动力学方程中，磁化强度可以按实验的磁化曲线计算，或者用 Langevin 函数估计。对于磁冻结流，铁磁流体的磁化强度 $\overline{M}$ 在流场中是常矢。最接近于磁冻结流的情况是，完全非内禀性铁磁流体的高速势流流动。因为在势流中，推动固相微粒旋转的涡旋不存在或很微弱，而非内禀性性质又使磁矩对外磁场的方向变化反应不够快。在高速流动下来不及松弛就随流动离开外磁场所在的区间。于是可以近似地认为铁磁流体通过外磁场区间时，其磁化强度 $\overline{M}$ 保持不变的 $\overline{M_i}$，不过由于外磁场可能是位置的函数，则仍然会有磁力，即 Kelvin 力作用于铁磁流体上。

在磁平衡流中，无论微粒本体是怎样的旋转，其磁矩总是时时处处与外磁场相适应，也就是"不会"发生磁力矩。所以，微粒本体与其磁矩似乎是"轴承"式的联结，只能传递力而不能传递力矩。在磁冻

结流中，微粒本体和磁矩之间似乎是"锁定"在一起的。无论外磁场如何改变，只要微粒本体不旋转，磁矩就在方向上保持一定。

对于普通的两相流，液相的涡粘力矩与固相颗粒的转动惯性相平衡，所以它是一种内力矩，不出现在动力学方程中，不影响两相流体的运动状态。而在铁磁流体中，固相微粒极其细小，它们的旋转惯性可以略去不计，涡粘力矩只与同时作用于微粒上的磁力矩相平衡。磁力矩是外磁场施加的所以是外力矩。作用于固相微粒上的磁力矩，因固相微粒的分布而在铁磁流体中处处存在，所以它是一种彻体力矩。随着外磁场的不均匀，磁力矩也不均匀。从而导致一种附加的彻体力，不言而喻，这是一种外力。

磁力矩造成铁磁流体内部的角动量，Rosensweig 等人认为现有的理论对这个问题的理解不完善，所以提出了"极性流体"的概念[2]。在极性流的分析中，纳入了磁力矩和它诱导出的附加力。

## 5.2 固相微粒的动量矩方程

圆球形固相微粒的绕轴动量矩是

$$\vec{S} = J_{1\delta} \vec{\omega_P} \tag{5.1 a}$$

式中 $J_{1\delta}$ 是一个微粒的惯性矩

$$J_{1\delta} = \frac{8}{15}\pi r_\delta^5 \rho_\delta = \frac{2}{5}(V_{p1})_\delta r_\delta^2 \rho_\delta \tag{5.1 b}$$

上式右方的 $(V_{p1})_\delta$、$r_\delta$、$\rho_\delta$ 依次按式（4.12 b）、式（4.12 a）与式（4.69）计算。

设铁磁流体内固相微粒的数密度为 $n$，则单位体积铁磁流体中固相的平均惯性矩为

$$nJ_{1\delta} = \frac{2}{5}n(V_{p1})_\delta r_\delta^2 \rho_\delta$$

用 $n = \phi_p/V_{p1} = \phi_\delta/(V_{p1})_\delta$ 代入，则有

$$nJ_{1\delta} = \frac{2}{5}\phi_\delta \rho_\delta r_\delta^2 = \frac{2}{5}(\rho_{p\delta})r_\delta^2 \tag{5.1 c}$$

式中 $\rho_{p\delta} = \phi_\delta \rho_\delta$ 是有分散剂附着层的固相在铁磁流体中的分密度。

铁磁流体内单个微粒的动量矩方程是

$$\frac{d\vec{S_1}}{dt} = \overrightarrow{L_{m1}} + \overrightarrow{L_{z1}} \tag{5.2 a}$$

式中各项的下标"1"表示一个微粒。对于单位体积铁磁流体内，固相微粒的平均动量矩方程是

$$\frac{d(n\vec{S_1})}{dt} = n\overrightarrow{L_{m1}} + n\overrightarrow{L_{z1}} + \left[\frac{d(n\vec{S_1})}{dt}\right]_B \tag{5.2 b}$$

上式右方第三项是因 $\vec{S_1}$ 和 $n$ 不均匀而引起的动量矩扩散。这种扩散的机制是 Brown 扩散，其净效应是定向的，即从高 $\vec{S_1}$ 和 $n$ 的区域向低 $\vec{S_1}$ 和 $n$ 的区域扩散。以下给出方程式（5.2 b）的各项。

1. $\dfrac{d(n\vec{S_1})}{dt}$

用 $n = \phi_\delta/(V_{p1})_\delta$ 和 $\overrightarrow{S_1} = J_{1\delta}\overrightarrow{\omega_p}$ 代入，并且注意 $(V_{p1})_\delta$ 和 $J_{1\delta}$ 均为平均值，所以是常数，于是

$$\frac{d(n\overrightarrow{S_1})}{dt} = \frac{J_{1\delta}}{(V_{p1})_\delta}\frac{d(\phi_\delta\overrightarrow{\omega_p})}{dt} \tag{5.3a}$$

2. $n\overrightarrow{L_{m1}}$

$$n\overrightarrow{L_{m1}} = n\overrightarrow{m_{p1}}L(\alpha)\times\overrightarrow{B_0} = \frac{\phi_p}{V_{p1}}\overrightarrow{m_{p1}}L(\alpha)\times\overrightarrow{B_0} = \overrightarrow{M}\times\overrightarrow{B_0} \tag{5.3 b}$$

式中 $\overrightarrow{m_{p1}}$ 是一个微粒的磁矩。

3. $n\overrightarrow{L_{\tau 1}}$

由式（2.92 a）和式（2.72 b），有

$$n\overrightarrow{L_{\tau 1}} = \frac{\phi_\delta}{(V_{p1})_\delta}6(V_{p1})_\delta\eta_\delta(\overrightarrow{\omega_C} - \overrightarrow{\omega_P}) = 6\phi_\delta\eta_\delta(\overrightarrow{\omega_C} - \overrightarrow{\omega_P}) \tag{5.3 c}$$

4. $\left[\dfrac{d(n\overrightarrow{S_1})}{dt}\right]_B$

在一般情况下，Brown 运动是纷乱无序的，随机的，运动在各个方向上都有相同的几率，因而不引起宏观迁移现象。但是，①若在铁磁流体内，固相微粒的浓度（以数密度 $n$ 计量）分布不均匀，则必然出现固相微粒由高浓度区向低浓度区的扩散，从而也携带广义动量迁移；②若铁磁流体内各处的固相微粒所具有的广义动量不相等，则在热运动时，微粒之间持续地换位和不断地碰撞，同样造成广义动量的迁移。虽然物质浓度的净效应是零，但动量的传递是定向的，其净效应是高动量处向低动量处迁移。

上述两种因素引起的动量迁移，都依赖 Brown 扩散而实现其定向的宏观净效应。这两种动量都遵守 Fick 定律，即

$$\overrightarrow{q_s} = -D_{p\delta}\frac{\partial\overrightarrow{S}}{\partial x_i} = -D_{p\delta}\frac{\partial(n\overrightarrow{S_1})}{\partial x_i} \tag{5.4}$$

式中 $\overrightarrow{q_s}$ 是动量矩的迁移流量，即单位时间内穿过单位面积的固相微粒所携带的动量矩。$D_{p\delta}$ 是式（3.71）所给出的扩散系数，$D_{p\delta}$ 是考虑了分散剂附着层的影响，即

$$D_{p\delta} = \frac{k_0 T}{6\pi\eta_c r_\delta} = \frac{2}{3}\frac{r_\delta^2}{t_{B\delta}}\frac{\eta_\delta}{\eta_c} = D_p\frac{1}{1+\dfrac{\delta}{r_p}} \tag{5.5}$$

在图 5-1 中，以平行于坐标面 $xOy$ 的上下两表面的情况为例，导出 $\overrightarrow{S}$ 各分量通过此二平面的净通量。规定进出六面控制体的流量，进为正，出为负，于是通过上下表面的净通量为：

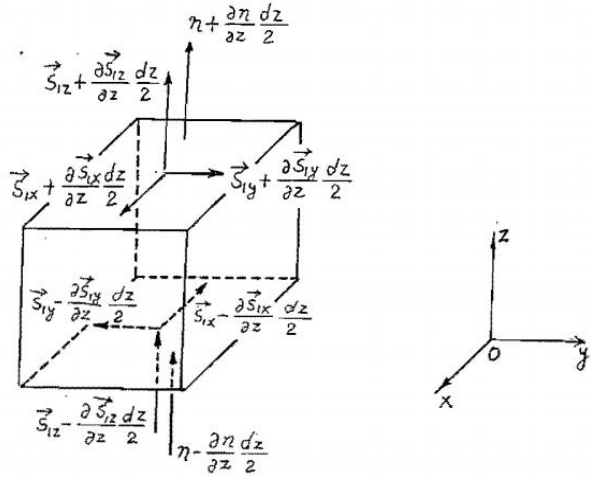

图5-1 通过六面控制体的动量矩

$$-D_{p\delta}\frac{\partial}{\partial z}\left\{\left[-\left(n+\frac{\partial n}{\partial z}\frac{dz}{2}\right)\left(\overrightarrow{S_{1z}}+\frac{\partial\overrightarrow{S_{1z}}}{\partial z}\frac{dz}{2}\right)+\left(n-\frac{\partial n}{\partial z}\frac{dz}{2}\right)\left(\overrightarrow{S_{1z}}-\frac{\partial\overrightarrow{S_{1z}}}{\partial z}\frac{dz}{2}\right)\right]+\right.$$

$$\left[-\left(n+\frac{\partial n}{\partial z}\frac{dz}{2}\right)\left(\overrightarrow{S_{1x}}+\frac{\partial\overrightarrow{S_{1x}}}{\partial z}\frac{dz}{2}\right)+\left(n-\frac{\partial n}{\partial z}\frac{dz}{2}\right)\left(\overrightarrow{S_{1x}}-\frac{\partial\overrightarrow{S_{1x}}}{\partial z}\frac{dz}{2}\right)\right]+$$

$$\left.\left[-\left(n+\frac{\partial n}{\partial z}\frac{dz}{2}\right)\left(\overrightarrow{S_{1y}}+\frac{\partial\overrightarrow{S_{1y}}}{\partial z}\frac{dz}{2}\right)+\left(n-\frac{\partial n}{\partial z}\frac{dz}{2}\right)\left(\overrightarrow{S_{1y}}-\frac{\partial\overrightarrow{S_{1y}}}{\partial z}\frac{dz}{2}\right)\right]\right\}dxdydt$$

略去二阶小量，得

$$-D_{p\delta}\frac{\partial}{\partial z}\left[-\frac{\partial}{\partial z}(n\overrightarrow{S_{1z}})-\frac{\partial}{\partial z}(n\overrightarrow{S_{1x}})-\frac{\partial}{\partial z}(n\overrightarrow{S_{1y}})\right]dzdxdydt=$$

$$D_{p\delta}\frac{\partial^2}{\partial z^2}(n\overrightarrow{S_{1z}}+n\overrightarrow{S_{1x}}+n\overrightarrow{S_{1y}})dzdxdydt=D_{p\delta}\frac{\partial^2\overrightarrow{S}}{\partial z^2}dxdydzdt$$

同样，通过平行 $yOz$ 坐标平面的前后量平面的动量矩净通量是

$$D_{p\delta}\frac{\partial^2\overrightarrow{S}}{\partial x^2}dxdydzdt$$

通过平行 $zOx$ 坐标平面的左、右两侧平面的动量矩净通量是

$$D_{p\delta}\frac{\partial^2\overrightarrow{S}}{\partial y^2}dxdydzdt$$

式中 $\overrightarrow{S}=n\overrightarrow{S_{1x}}+n\overrightarrow{S_{1y}}+n\overrightarrow{S_{1z}}=n\overrightarrow{S_1}$ 是单位体积铁磁流体内的动量矩。通过六个平面的动量矩净通量总和是

$$D_{p\delta}=\left(\frac{\partial^2\overrightarrow{S}}{\partial x^2}+\frac{\partial^2\overrightarrow{S}}{\partial y^2}+\frac{\partial^2\overrightarrow{S}}{\partial z^2}\right)dxdydzdt=D_{p\delta}(\nabla^2\overrightarrow{S})dxdydzdt$$

微元六面控制体界面上动量矩迁移的净通量必引起六面体内部动量矩的变化，在 $dt$ 时间内其变化量为 $(d\overrightarrow{S})_B dxdydz$，由动量矩守恒，得

$$(d\overrightarrow{S})_B dxdydz=D_{p\delta}(\nabla^2\overrightarrow{S})dxdydzdt$$

于是，单位体积铁磁流体内，由于 Brown 扩散引起的动量矩变化率为

$$\left(\frac{d\vec{S}}{dt}\right)_B = D_{p\delta}\nabla^2\vec{S} = \frac{J_{1\delta}}{(V_{p1})_\delta}D_{p\delta}\nabla^2(\phi_\delta\overrightarrow{\omega_P}) \tag{5.6}$$

将式（5.3a）～式（5.3c）及式（5.6）代入方程式（5.2b）中，就有

$$\frac{J_{1\delta}}{(V_{p1})_\delta}\frac{d(\phi_\delta\overrightarrow{\omega_P})}{dt} = \vec{M}\times\vec{B_0} + 6\phi_\delta\eta_\delta(\overrightarrow{\omega_C}-\overrightarrow{\omega_P}) - \frac{J_{1\delta}}{(V_{p1})_\delta}D_{p\delta}\nabla^2(\phi_\delta\overrightarrow{\omega_P}) \tag{5.7}$$

固相微粒在铁磁流体内的旋转是一种 Re 数极低的运动，所以方程式（5.7）的左方惯性项可以略去。其右方第三项系数与第二项系数相比，大约是 $10^{-10}$ 的量级。所以方程式（5.7）提供了磁力矩与粘性力矩的平衡关系，即

$$\vec{M}\times\vec{B_0} = 6\phi_\delta\eta_\delta(\overrightarrow{\omega_P}-\overrightarrow{\omega_C}) \tag{5.8}$$

## 5.3 彻体力矩 $\overrightarrow{L_b}$ 引起的附加力 $\overrightarrow{f_L}$

彻体力矩在铁磁流体内通常造成剪应力环流。若彻体力矩是位置的函数时，则其剪应力环流除合成力矩之外，还可能合成力，即附加力 $\overrightarrow{f_L}$。

如图 5-2，在流场中取出一个微元六面控制体，其内部的彻体力矩引起六个表面上的剪应力环流。由于内部彻体力矩不均匀，表面上的剪应力就有增量。为了和一般剪应力相区别，剪应力环流用 $\tau'$ 表示。

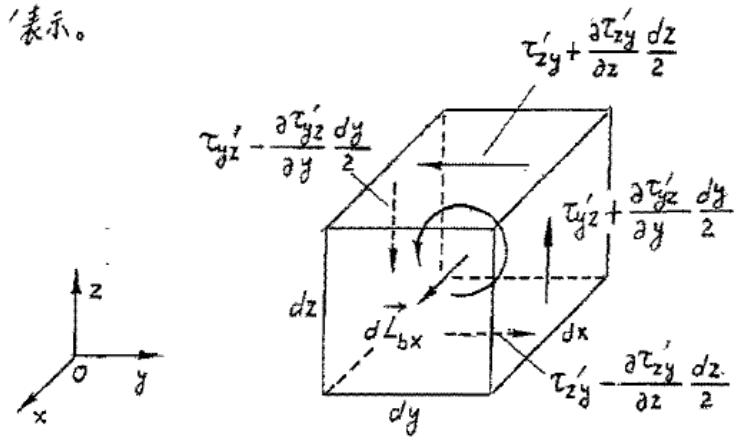

图 5-2　微元六面控制体表面上的剪应力环流

由图 5-2 可见

$$d\overrightarrow{L_{bx}} = \left[\left(\tau'_{yz}+\frac{\partial\tau'_{yz}}{\partial y}\frac{dy}{2}\right)+\left(\tau'_{yz}-\frac{\partial\tau'_{yz}}{\partial y}\frac{dy}{2}\right)\right]dzdx\left(\frac{dy}{2}\right)+$$

$$\left[\left(\tau'_{zy}+\frac{\partial\tau'_{zy}}{\partial z}\frac{dz}{2}\right)+\left(\tau'_{zy}-\frac{\partial\tau'_{yz}}{\partial z}\frac{dz}{2}\right)\right]dxdy\left(\frac{dz}{2}\right)$$

$$= (\tau'_{yz}+\tau'_{zy})dxdydz$$

于是单位体积铁磁流体内的彻体力矩是

$$\overline{L_{bx}} = \frac{d\overline{L_{bx}}}{dxdydz} = \tau'_{yz} + \tau'_{zy} = 2\tau'_{yz} = 2\tau'_{zy} \tag{5.9a}$$

同样可有

$$\overline{L_{by}} = \tau'_{zx} + \tau'_{xz} = 2\tau'_{zx} = 2\tau'_{xz} \tag{5.9b}$$

$$\overline{L_{bz}} = \tau'_{xy} + \tau'_{yx} = 2\tau'_{xy} = 2\tau'_{yx} \tag{5.9c}$$

关于附加力 $\overrightarrow{f_L}$，它是表面剪应力之合成。如图 5-2，在平行于 $xOy$ 坐标面的表面上 $y$ 向剪应力的合成为

$$d\overrightarrow{F}_{zy} = \vec{j}\left[\left(\tau'_{zy} - \frac{\partial\tau'_{zy}}{\partial z}\frac{\partial z}{2}\right) - \left(\tau'_{zy} + \frac{\partial\tau'_{zy}}{\partial z}\frac{dz}{2}\right)\right]dxdy = -\vec{j}\frac{\partial\tau'_{zy}}{\partial z}dxdydz$$

于是对于单位体积的铁磁流体，则有

$$\overrightarrow{f_{zy}} = \frac{d\overrightarrow{F_{zy}}}{dxdydz} = -\vec{j}\frac{\partial\tau'_{zy}}{\partial z}$$

如图 5-2 在平行于 $zOx$ 坐标面的表面上，$z$ 向剪应力的合成结果为

$$\overrightarrow{f_{yz}} = \vec{k}\frac{\partial\tau'_{yz}}{\partial y}$$

仿照上面 $\overrightarrow{f_{zy}}$ 和 $\overrightarrow{f_{yz}}$ 的导出过程，同样可得其它表面上的剪应力的合成结果，归纳写出

$$\left.\begin{array}{ll}\overrightarrow{f_{xy}} = \vec{j}\dfrac{\partial\tau'_{xy}}{\partial x}, & \overrightarrow{f_{yx}} = -\vec{i}\dfrac{\partial\tau'_{yx}}{\partial y} \\[2mm] \overrightarrow{f_{yz}} = \vec{k}\dfrac{\partial\tau'_{yz}}{\partial y}, & \overrightarrow{f_{zy}} = -\vec{j}\dfrac{\partial\tau'_{zy}}{\partial z} \\[2mm] \overrightarrow{f_{zx}} = \vec{i}\dfrac{\partial\tau'_{zx}}{\partial z}, & \overrightarrow{f_{xz}} = -\vec{k}\dfrac{\partial\tau'_{xz}}{\partial x}\end{array}\right\} \tag{5.10}$$

将式（5.10）按方向归项，给出

$$\overrightarrow{f_L} = \vec{i}\left(\frac{\partial\tau'_{zx}}{\partial z} - \frac{\partial\tau'_{yx}}{\partial y}\right) + \vec{j}\left(\frac{\partial\tau'_{xy}}{\partial x} - \frac{\partial\tau'_{zy}}{\partial z}\right) + \vec{k}\left(\frac{\partial\tau'_{yz}}{\partial y} - \frac{\partial\tau'_{xz}}{\partial x}\right) \tag{5.11a}$$

用式（5.9a）～式（5.9c）代入上式，就得附加力 $\overrightarrow{f_L}$ 与彻体力矩 $\overline{L_b}$ 的关系，即

$$\overrightarrow{f_L} = \frac{1}{2}\left[\vec{i}\left(\frac{\partial L_{by}}{\partial z} - \frac{\partial L_{bz}}{\partial y}\right) + \vec{j}\left(\frac{\partial L_{bz}}{\partial x} - \frac{\partial L_{bx}}{\partial z}\right) + \vec{k}\left(\frac{\partial L_{bx}}{\partial y} - \frac{\partial L_{by}}{\partial x}\right)\right] \tag{5.11b}$$

或合并写成

$$\overrightarrow{f_L} = \frac{1}{2}\nabla\times\overline{L_b} \tag{5.11c}$$

在外磁场中，由外部施加到铁磁流体内的彻体力矩就是磁力矩 $\overrightarrow{L_m}$，即

$$\vec{f}_L = \frac{1}{2} \nabla \times \overrightarrow{L_m} = \frac{\mu_0}{2} \nabla \times (\overrightarrow{M} \times \overrightarrow{H})$$ (5.12)

5.4 铁磁流体中，固相微粒的平动运动方程

单个固相微粒在铁磁流体中运动时，所受到的力有基载液的粘性阻力 $F_{v1}$，压差力 $F_{p1}$，外磁场的磁力 $F_{m1}$，磁力矩引起的附加力 $F_{L1}$、所有力的下标"1"表示一个微粒受到的力。

单个微粒的运动方程和微粒流的运动方程不完全相同，但其所涉及的一些关系式，在微粒流中是有用的。

对于单个微粒的牛顿第二定律方程为

$$\rho_\delta (V_p)_\delta \frac{d\overrightarrow{U_p}}{dt} = \overrightarrow{F_{m1}} + \overrightarrow{F_{v1}} + \overrightarrow{F_{L1}} + \overrightarrow{F_{p1}}$$ (5.13)

方程式（5.13）已略去重力。下面对方程式（5.13）逐项给出具体关系式。

1. $\overrightarrow{F_{m1}}$

设在铁磁流体中，到处都没有传导电流穿过，也没有剩余束缚电流存在，则固相微粒受到的磁作用力就是 Kelvin 力，即

$$\overrightarrow{F_{m1}} = \overrightarrow{F_{k1}} = \overrightarrow{m_{p1}} \cdot \nabla \overrightarrow{B_0}$$

2. $\overrightarrow{F_{v1}}$

它是液相作用于固相的粘性力。对于一个固相微粒则式（2.89c）给出

$$\overrightarrow{F_{v1}} = -C_d (\overrightarrow{U_p} - \overrightarrow{U_c})$$

式中 $\overrightarrow{U_p}$ 和 $\overrightarrow{U_c}$ 分别是固相微粒和基载液体的速度矢量，$C_d$ 是阻力系数，即

$$C_d = 6\pi \eta_c r_p$$

若计及分散剂附着层的厚度 $\delta$，则有

$$C_{d\delta} = 6\pi \eta_c r_\delta$$

3. $\overrightarrow{F_{L1}}$

在铁磁流体中，唯一的作为彻体力矩存在的外力矩就是磁力矩，将式（5.12）改写成只对一个固相微粒，则有

$$\overrightarrow{F_{L1}} = \frac{1}{2} \nabla \times (\overrightarrow{m_{p1}} \times \overrightarrow{B_0})$$

对于一个微粒，式（5.8）可以写成

$$\overrightarrow{m_{p1}} \times \overrightarrow{B_0} = 8\pi\eta_\delta r_\delta^3(\overrightarrow{\omega_p} - \overrightarrow{\omega_C})$$

于是作用于一个微粒上的附加力是

$$\overrightarrow{F_{L1}} = \frac{1}{2}\nabla \times (\overrightarrow{m_{p1}} \times \overrightarrow{B_0}) = 4\pi\eta_\delta r_\delta^3 \nabla \times (\overrightarrow{\omega_p} - \overrightarrow{\omega_C}) \tag{5.14}$$

4. $\overrightarrow{F_{p1}}$

注意到 $F_{v1}$ 是固相微粒在铁磁流体中运动所引起的正应力和切应力扰动的合成结果。未受扰动的压力被假定为常数而不包括在力 $F_{v1}$ 中。但是，若未受扰动的压力是不均匀的，则将压力在固相微粒中心 $O$ 点处展成 Taylor 级数，并且只取头两项，就有

$$p = p_0 + \left.\frac{\partial p}{\partial x_i}\right|_0 x_i$$

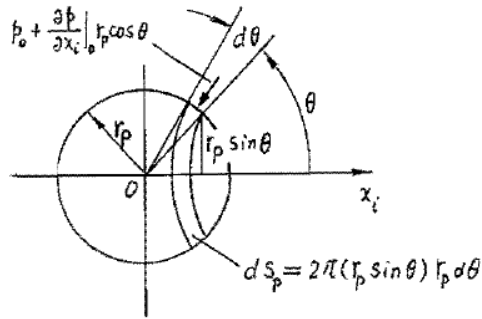

图 5-3　作用于微粒表面上的压力

在固相微粒表面上坐标 $x_i = r_p \cos\theta$，作用于微元球台表面上的力 $dF_{p1}$ 是压力在 $x_i$ 方向之投影乘 $ds_p$，即

$$dF_{p1} = -\left(p_0 + \left.\frac{\partial p}{\partial x_i}\right|_0 r_p \cos\theta\right)\cos\theta ds_p = -\left(p_0 + \left.\frac{\partial p}{\partial x_i}\right|_0 r_p \cos\theta\right)\cos\theta \cdot 2\pi r_p \sin\theta \cdot r_p d\theta$$

于是整个微粒表上的压差力是积分

$$F_{p1} = \int_0^\pi dF_{p1} = -\int_0^\pi \left(p_0 + \left.\frac{\partial p}{\partial x_i}\right|_0 r_p \cos\theta\right) 2\pi r_p^2 \cos\theta \sin\theta d\theta$$

$$= 2\pi r_p^2 \left(\frac{1}{2}p_0 \cos^2\theta + \frac{1}{3}\left.\frac{\partial p}{\partial x_i}\right|_0 r_p \cos^3\theta\right)_0^\pi = -\frac{\partial p}{\partial x_i}\left(\frac{4}{3}\pi r_p^3\right) = -\frac{\partial p}{\partial x_i}V_{p1}$$

上式最右省了压力梯度的下标"0"，若取 $x_i$ 的下标 $i = 1$，2，3，则 $\partial p/\partial x_i$ 就是 $\nabla p$，于是

$$F_{p1} = -\frac{4}{3}\pi r_p^3 \nabla p = -V_{p1}\nabla p \tag{5.15a}$$

考虑分散剂附着层，则上式成为：

$$F_{p1} = -\frac{4}{3}\pi r_\delta^3 \nabla p = -(V_{p1})_\delta \nabla p \tag{5.15b}$$

将以上各项结果代入方程式（5.13）中，就得单个微粒的运动方程为

$$\rho_\delta (V_{p1})_\delta \frac{d\overrightarrow{U_p}}{dt} = \overrightarrow{m_{p1}} \cdot \nabla \overrightarrow{B_0} - 6\pi\eta_c r_\delta (\overrightarrow{U_p} - \overrightarrow{U_c}) + 4\pi\eta_\delta r_\delta^3 \nabla \times (\overrightarrow{\omega_p} - \overrightarrow{\omega_c}) - \frac{4}{3}\pi r_\delta^3 \nabla p \tag{5.16}$$

若考虑到固相微粒在铁磁流体中的运动是很低 Re 数的运动，则上式左方的惯性项可以略去。上式右方的第三项和第四项的系数中含有 $r_\delta^3$，而第二项的系数中只有 $r_\delta$，所以第三项和第四项相对于第二项是很小的，于是就有近似平衡关系，即磁场力与粘性力的平衡

$$\overrightarrow{m_{p1}} \cdot \nabla \overrightarrow{B_0} = 6\pi\eta_c r_\delta (\overrightarrow{U_p} - \overrightarrow{U_c}) \tag{5.17}$$

## 5.5 外磁场转速 $\overrightarrow{\omega_H}$ 和涡旋转速 $\overrightarrow{\omega_C}$ 不相同时，铁磁流体内的磁力矩和附加力

### 5.5.1 概述

无论是磁平衡流或是磁冻结流都是极端的简化情况。实际中的一般情况是，固相微粒在铁磁流体流动过程中，受到磁力矩和涡粘力矩的作用，持续地处于旋转状态，而其磁矩的 Neel 松弛和微粒本体的 Brown 松弛也持续地伴随进行。这样的一般情况可以称之为磁松弛流动。

在铁磁流体流动中的所谓磁松弛过程，仅指铁磁流体磁化强度的变化。所以第四章中的磁化方程式（4.86a)~式（4.68c）是磁松弛分析的基本方程。

磁松弛过程中铁磁流体磁化强度的变化，改变铁磁流体所受到的磁力矩和磁场力。同时，磁松弛还使铁磁流体的粘度改变。最后的结果就将影响铁磁流体的运动状态。

### 5.5.2 磁松弛流的磁力矩和附加力的一般关系式

1. 磁力矩 $\overrightarrow{L_m}$

由图 4-1 的模型，式（4.85）表明，当外磁场转速 $\omega_{H_z}$ 与液相的涡旋速度 $\omega_{C_z}$ 不相等时，必定存在磁力矩 $\overrightarrow{L_{mz}}$，它与作用于固相微粒本体上的涡粘力矩相平衡，从而产生粘性系数的增量，即磁粘系数 $\eta_m$。

将式（4.77a）代入式（4.84）之右方分母中，就有

$$\frac{M_y}{M_0} = \frac{(\omega_{CH})_z}{1+\beta_e} t_{r\delta} \sin\theta_0 \tag{5.18a}$$

再以式（4.89）置换 $(\omega_{CH})_z$，则得

$$\frac{M_y}{M_0} = \frac{(\omega_{PH})_z}{1+\beta_e \cos^2\theta_0} t_{r\delta} \sin\theta_0 \tag{5.18b}$$

将式（4.91）改写成

$$(\omega_{PH})_z = \frac{1+\beta_e \cos^2\theta_0}{1+\beta_e - \phi_\delta \frac{\rho_\delta}{\rho_f}\beta_e \sin^2\theta_0}(\omega_{fc} - \omega_{Hz}) \tag{5.18c}$$

将式（5.18c）代入式（5.18b）右方分子中，给出

$$\frac{M_y}{M_0} = \frac{t_{r\delta}}{1 + \beta_e - \phi_\delta \frac{\rho_\delta}{\rho_f} \beta_e \sin^2 \theta_0} (\omega_{fz} - \omega_{Hz}) \sin \theta_0 \qquad (5.18d)$$

引用记号

$$R_m = \frac{t_{r\delta}}{1 + \left(1 - \phi_\delta \frac{\rho_\delta}{\rho_f} \sin^2 \theta_0\right) \beta_e} \qquad (5.19)$$

$R_m$ 是表示磁松弛的一个综合参数，将式（5.19）代入式（5.18d），则有

$$\frac{M_y}{M_0} = R_m(\omega_{fz} - \omega_{Hz}) \sin \theta_0 = R_m(\omega_{fH})_z \sin \theta_0 \qquad (5.20a)$$

由式（4.68a）与式（4.68b）

$$-(\omega_{PH})_z t_{r\delta} M_y = M_x - M_{0x}, \qquad (\omega_{PH})_z t_{r\delta} M_x = M_y$$

两式相除，消去 $(\omega_{PH})_z t_{r\delta}$，得

$$M_x^2 - M_{0x} M_x = -M_y^2$$

对 $M_x$ 求解，有

$$M_x = \frac{1}{2}(M_{0x} \pm \sqrt{M_{0x}^2 - 4M_y^2}) \qquad (A)$$

又由式（4.68b）代入式（4.68a）消去 $M_y$，得

$$M_{0x} = M_x[1 + (\omega_{PH})_z^2 t_{r\delta}^2] \qquad (B)$$

式中，$(\omega_{PH})_z^2 t_{r\delta}^2$ 与 1 相比是二阶小量，故 $M_x$ 与 $M_{0x}$ 非常接近。于是式（A）右方根号前取正号。此外，由式（4.68b）知道 $M_y$ 与 $M_x$ 相比是一阶小量。故式（A）可近似处理为

$$M_x = \frac{1}{2}\left[M_{0x} + M_{0x}\sqrt{1 - 4\left(\frac{M_y}{M_{0x}}\right)^2}\right] \approx \frac{1}{2}\left[M_{0x} + M_{0x}\left(1 - 2\frac{M_y^2}{M_{0x}^2}\right)\right]$$

于是就有

$$M_x = M_{0x} - \frac{M_y^2}{M_{0x}}$$

或写成

$$\frac{M_x}{M_0} = \frac{M_{0x}}{M_0} - \left(\frac{M_y}{M_0}\right)^2 \frac{M_0}{M_{0x}} \tag{C}$$

由图 4-1（b）可见，$\dfrac{M_{0x}}{M_0} = \dfrac{B_x}{B_0} = \sin\theta_0$，$\dfrac{M_z}{M_0} = \dfrac{B_z}{B_0} = \cos\theta_0$，以及利用式（5.20a），则式（C）可以写成

$$\frac{M_x}{M_0} = \sin\theta - R_m^2(\omega_{fH})_z^2 \sin\theta_0 \tag{5.20b}$$

将磁力矩的关系式（4.47a）～式（4.47c）改写成

$$\overrightarrow{L_{mx}} = \vec{i}\left(\frac{M_y}{M_0}\frac{B_z}{B_0}\right)M_0 B_0 \tag{5.21a}$$

$$\overrightarrow{L_{my}} = \vec{j}\left(\frac{M_z}{M_0}\frac{B_x}{B_0} - \frac{M_x}{M_0}\frac{B_z}{B_0}\right)M_0 B_0 \tag{5.21b}$$

$$\overrightarrow{L_{mz}} = -\vec{k}\left(\frac{M_y}{M_0}\frac{B_x}{B_0}\right)M_0 B_0 \tag{5.21c}$$

将已经用过的图 4-1（b）中的正弦和余弦关系，以及式（5.20a）、式（5.20b）代入上面的式（5.21a）～式（5.21c），就得磁松弛情况下的磁力矩是

$$\overrightarrow{L_{mx}} = \vec{i}R_m(\omega_{fH})_z M_0 B_0 \sin\theta_0 \cos\theta_0 \tag{5.22a}$$

$$\overrightarrow{L_{my}} = \vec{j}R_m^2(\omega_{fH})_z^2 M_0 B_0 \sin\theta_0 \cos\theta_0 \tag{5.22b}$$

$$\overrightarrow{L_{mz}} = -\vec{k}R_m(\omega_{fH})_z M_0 B_0 \sin^2\theta_0 \tag{5.22c}$$

合成的磁力矩是

$$\overrightarrow{L_m} = R_m(\omega_{fH})_z M_0 B_0[\vec{i}\cos\theta_0 + \vec{j}R_m(\omega_{fH})_z \cos\theta_0 - \vec{k}\sin\theta_0]\sin\theta_0 \tag{5.22d}$$

2.附加力 $\overrightarrow{f_L}$

由外部施加到铁磁流体中的彻体力矩就是磁力矩 $\overrightarrow{L_m}$，式（5.12）给出

$$\overrightarrow{f_L} = \frac{1}{2}\nabla\times\overrightarrow{L_m} = \frac{1}{2}\left(\vec{i}\frac{\partial}{\partial x} + \vec{j}\frac{\partial}{\partial y} + \vec{k}\frac{\partial}{\partial z}\right)\times[\vec{i}R_m(\omega_{fH})_z M_0 B_0 \sin\theta_0 \cos\theta_0 +$$
$$\vec{j}R_m^2(\omega_{fH})_z^2 M_0 B_0 \sin\theta_0 \cos\theta_0 - \vec{k}R_m(\omega_{fH})_z M_0 B_0 \sin^2\theta_0]$$

乘开之后，得出

$$\vec{f}_L = \frac{1}{2}\vec{i}\left\{-\frac{\partial}{\partial y}[R_m(\omega_{fH})_z M_0 B_0]\sin\theta_0 - \frac{\partial}{\partial z}[R_m^2(\omega_{fH})_z^2 M_0 B_0]\cos\theta_0\right\}\sin\theta_0 +$$

$$\frac{1}{2}\vec{j}\left\{\frac{\partial}{\partial z}[R_m(\omega_{fH})_z M_0 B_0]\cos\theta_0 + \frac{\partial}{\partial x}[R_m(\omega_{fH})_z M_0 B_0]\sin\theta_0\right\}\sin\theta_0 + \quad (5.23a)$$

$$\frac{1}{2}\vec{k}\left\{\frac{\partial}{\partial x}[R_m^2(\omega_{fH})_z^2 M_0 B_0] - \frac{\partial}{\partial y}[R_m(\omega_{fH})_z M_0 B_0]\right\}\sin\theta_0\cos\theta_0$$

将式（5.23a）按方向拆开，则三个方向的附加力分量是：

$$\overrightarrow{f_{Lx}} = -\frac{1}{2}\vec{i}\left\{\frac{\partial}{\partial y}[R_m(\omega_{fH})_z M_0 B_0]\sin\theta_0 + \frac{\partial}{\partial z}[R_m^2(\omega_{fH})_z^2 M_0 B_0]\cos\theta_0\right\}\sin\theta_0 \quad (5.23b)$$

$$\overrightarrow{f_{Ly}} = \frac{1}{2}\vec{j}\left\{\frac{\partial}{\partial z}[R_m(\omega_{fH})_z M_0 B_0]\cos\theta_0 + \frac{\partial}{\partial x}[R_m(\omega_{fH})_z M_0 B_0]\sin\theta_0\right\}\sin\theta_0 \quad (5.23c)$$

$$\overrightarrow{f_{Lz}} = \frac{1}{2}\vec{k}\left\{\frac{\partial}{\partial x}[R_m^2(\omega_{fH})_z^2 M_0 B_0] - \frac{\partial}{\partial y}[R_m(\omega_{fH})_z M_0 B_0]\right\}\sin\theta_0\cos\theta_0 \quad (5.23d)$$

5.5.3 一些特殊情况下的磁力矩和附加力

1.内禀性铁磁流体或外磁场与铁磁流体相对静止

对于内禀性铁磁流体可以认为是最接近磁平衡的流动。由于 $t_{rS} \approx t_N \approx 0$，则式（5.19）给出 $R_m \approx 0$。

对于外磁场与铁磁流体相对静止的情况，则有 $(\omega_{fH})_z = 0$。以上两种情况，式（5.22d）给出 $\overrightarrow{L_m} = 0$，式（5.23a）给出 $\vec{f}_L = 0$。

2.垂直磁场

所谓垂直磁场，就是 $\overrightarrow{B_0} \perp \overrightarrow{(\omega_{fH})_z}$，此时 $\theta_0 = \pi/2$，$\cos\theta_0 = 0$ 则由式（5.22a）～式（5.22c）得出

$$\overrightarrow{L_{mx}} = 0, \qquad \overrightarrow{L_{my}} = 0, \qquad \overrightarrow{L_{mz}} = -\vec{k}R_m(\omega_{fH})_z M_0 B_0 \quad (5.24a)$$

式（5.23b）～式（5.23d）给出

$$\left.\begin{array}{l}\overrightarrow{f_{Lx}} = -\dfrac{1}{2}\vec{i}\dfrac{\partial}{\partial y}[R_m(\omega_{fH})_z M_0 B_0]\\[2mm]\overrightarrow{f_{Ly}} = \dfrac{1}{2}\vec{j}\dfrac{\partial}{\partial x}[R_m(\omega_{fH})_z M_0 B_0]\\[2mm]\overrightarrow{f_{Lz}} = 0\end{array}\right\} \quad (5.24b)$$

式（5.24a）与式（5.24b）表明垂直磁场仅引起 $z$ 方向的磁力矩和 $x$，$y$ 方向的附加力。

3.平行磁场

平行磁场就是 $\overrightarrow{B_0} \parallel \overrightarrow{(\omega_{fH})_z}$，此时 $\theta_0 = 0$，$\sin\theta_0 = 0$，则式（5.22d）与式（5.23a）给出 $\overrightarrow{L_m} = 0$，$\vec{f}_L = 0$ 即平行磁场不产生磁力矩和附加力。当然不会引起铁磁流体的磁粘性增量。

## 5.6 在旋转磁场中的 Kelvin 力

### 5.6.1 概述

在磁场中没有传导电流穿过，也没有剩余束缚电流的情况下，外磁场作用于铁磁流体上的力，就是 Kelvin 力。磁场力出现的前提条件，是外磁场必须不均匀。在均匀磁场内只能出现磁力矩而不出现磁场力。与磁力矩相关联的功是磁化功，即铁磁流体的磁化强度升高而成为其内能的增量。与磁场力相关联的是移动功。铁磁流体由外磁场较高处移往较低处，其势能将升高，因而必须有外力对铁磁流体作功。显然 Kelvin 力的方向是指向外磁场强度升高的方向，如式（1.32）所给出的那样，即

$$\vec{f_k} = \vec{M} \cdot \nabla \vec{B_0}$$

### 5.6.2 磁松弛流的 Kelvin 力的一般关系式

由图 4-1（b）的物理模型，Kelvin 力可以写成

$$\vec{f_k} = (\vec{i}M_x + \vec{j}M_y + \vec{k}M_z) \cdot \left( \vec{i}\frac{\partial}{\partial x} + \vec{j}\frac{\partial}{\partial y} + \vec{k}\frac{\partial}{\partial z} \right)(\vec{i}B_x + \vec{k}B_z)$$

展开以后就有

$$\vec{f_k} = M_0 \left( \frac{M_x}{M_0}\frac{\partial}{\partial x} + \frac{M_y}{M_0}\frac{\partial}{\partial y} + \frac{M_z}{M_0}\frac{\partial}{\partial z} \right)(\vec{i}B_0 + \vec{k}B_z) \tag{5.25a}$$

上式右方 $M_x/M_0$，用式（5.20b）代入，$M_y/M_0$ 用式（5.20a）代入。并且 $M_z/M_0 = \cos\theta_0$，$B_x = B_0\sin\theta_0$，

$B_z = B_0\cos\theta_0$，于是式（5.25a）成为

$$\vec{f_k} = M_0 \left\{ [1 - R_m^2(\omega_{\beta H})_z^2 \sin\theta_0]\sin\theta_0 \frac{\partial}{\partial x} + R_m(\omega_{\beta H})_z \sin\theta_0 \frac{\partial}{\partial y} + \cos\theta_0 \frac{\partial}{\partial z} \right\}(\vec{i}B_0\sin\theta_0 + \vec{k}B_0\cos\theta_0) \tag{5.25b}$$

按方向拆开，得

$$\vec{i}: \quad f_{kx} = M_0 \left\{ [1 - R_m^2(\omega_{\beta H})_z^2 \sin\theta_0]\sin\theta_0 \frac{\partial B_0}{\partial x} + R_m(\omega_{\beta H})_z \sin\theta_0 \frac{\partial B_0}{\partial y} + \cos\theta_0 \frac{\partial B_0}{\partial z} \right\}\sin\theta_0 \tag{5.25c}$$

$$\vec{k}: \quad f_{kz} = M_0 \left\{ [1 - R_m^2(\omega_{\beta H})_z^2 \sin\theta_0]\sin\theta_0 \frac{\partial B_0}{\partial x} + R_m(\omega_{\beta H})_z \sin\theta_0 \frac{\partial B_0}{\partial y} + \cos\theta_0 \frac{\partial B_0}{\partial z} \right\}\cos\theta_0 \tag{5.25d}$$

### 5.6.3 磁松弛流的 Kelvin 力的一些特殊情况

1. 内禀性铁磁流体或外磁场与铁磁流体相对静止

此时有 $R_m \approx 0$ 或 $(\omega_{\beta H})_z = 0$，于是式（5.25c）、式（5.25d）成为

$$\vec{i}: \qquad f_{kx} = M_0 \left( \sin\theta_0 \frac{\partial B_0}{\partial x} + \cos\theta_0 \frac{\partial B_0}{\partial z} \right)\sin\theta_0 \tag{5.26a}$$

$$\vec{k}: \qquad f_{kz} = M_0 \left( \sin\theta_0 \frac{\partial B_0}{\partial x} + \cos\theta_0 \frac{\partial B_0}{\partial z} \right)\cos\theta_0 \tag{5.26b}$$

注意到 $f_k = f_{kx}\sin\theta_0 + f_{kz}\cos\theta_0$，而

$$f_{kx} = f_k \sin\theta_0 = (f_{kx}\sin\theta_0 + f_{kz}\cos\theta_0)\sin\theta_0$$
$$f_{kz} = f_k \cos\theta_0 = (f_{kx}\sin\theta_0 + f_{kz}\cos\theta_0)\cos\theta_0$$

将此二式与式（5.26a）、式（5.26b）相比照，立即可知

$$\overrightarrow{f_{kx}} = \vec{i}M_0\frac{\partial B_0}{\partial x}, \qquad \overrightarrow{f_{kz}} = \vec{k}M_0\frac{\partial B_0}{\partial z} \tag{5.26c}$$

两式合并，就得

$$\overrightarrow{f_k} = M_0\nabla B_0 \tag{5.26d}$$

这就是已经得出过的式（1.31），即 $\overrightarrow{M_0}$ 与 $\overrightarrow{B_0}$ 相平行时的 Kelvin 力。式中 $M_0$ 是铁磁流体的平衡磁化强度，它可以按实验的磁化曲线取值，亦可以按 Langevin 公式计算。

2.垂直磁场

此时 $\overrightarrow{B_0} \perp \overrightarrow{\omega_{fz}}$，$\theta_0 = \pi/2$，$\cos\theta_0 = 0$。由式（5.25c）、式（5.25d）得

$$\overrightarrow{f_{kx}} = \vec{i}M_0\left\{[1 - R_m^2(\omega_{\beta H})_z^2]\frac{\partial B_0}{\partial x} + R_m(\omega_{\beta H})_z\frac{\partial B_0}{\partial y}\right\}, \qquad \overrightarrow{f_{kz}} = 0 \tag{5.27a}$$

3.平行磁场

此时 $\overrightarrow{B_0} \parallel \overrightarrow{\omega_{fz}}$，$\theta_0 = 0$，$\sin\theta_0 = 0$。式（5.25c）、式（5.25d）给出

$$\overrightarrow{f_{kx}} = 0, \qquad \overrightarrow{f_{kz}} = \vec{k}M_0\frac{\partial B_0}{\partial z} \tag{5.27b}$$

## 5.7 铁磁流体的磁松弛混合流方程组
### 5.7.1 概述

铁磁流体的动力学方程与普通两相流一样，有两种处理方法。一种是混合流，一种是分相流。混合流的思路是将固、液两相的物性参数，如粘性系数、比热容等，按体积分量或质量分量加以平均。采用平均混合参数，从而铁磁流体就按单相流处理。

既然是单相流，则固相微粒和液相在运动上的差异，就不能反映。但在一般情况下固相微粒和外磁场之间存在相对转动，于是固相微粒的磁矩时时刻刻都在偏离外磁场方向，从而磁松弛过程不断地伴随发生。磁松弛仅仅使铁磁流体的磁化强度改变，所以它和平动运动没有直接关联。直到磁力力矩与粘性力矩相平衡，磁松弛过程才停止。

磁松弛过程影响铁磁流体的磁化强度，必然使磁场力、磁力矩和附加力发生变化。磁场力和磁力矩是控制铁磁流体运动的两个决定性因素，所以磁松弛的作用将最终反映到铁磁流体的流动状况上。由于混合流不讨论两相间的运动滞后问题。因此，只作用在固相微粒上的磁场力和磁力矩是如何传递到整个铁磁流体之上的，在混合流中，不可能得到回答，这是混合流的一个缺陷。

在外磁场作用下，铁磁流体不是各向同性的物质，它的物性参数和方向相关联。

### 5.7.2 铁磁流体磁松弛混合流的动量守恒方程——运动方程的矢量形式

由牛顿第二定律，对单位体积铁磁流体有

$$\rho_f \frac{d\overline{U_f}}{dt} = \overrightarrow{f_g} + \overrightarrow{f_p} + \overrightarrow{f_\eta} + \overrightarrow{f_k} + \overrightarrow{f_L} \tag{5.28a}$$

式中，$\rho_f$、$\overline{U_f}$ 是铁磁流体的密度和速度，$\overrightarrow{f_g}$ 是重力，$\overrightarrow{f_p}$ 是压力梯度形成的力，$\overrightarrow{f_\eta}$ 是粘性力，$\overrightarrow{f_k}$ 是 Kelvin 力，$\overrightarrow{f_L}$ 是附加力。这些力也可以归为表面力 $\overrightarrow{f_s}$ 和彻体力 $\overrightarrow{f_b}$，即

$$\rho_f \frac{d\overline{U_f}}{dt} = \overrightarrow{f_s} + \overrightarrow{f_b} \tag{5.28b}$$

表面力 $\overrightarrow{f_s}$ 可直接使用式（2.49）得到

$$\overrightarrow{f_s} = \overrightarrow{f_p} + \overrightarrow{f_\eta} = \nabla \cdot \tau = -\nabla p + \frac{\partial}{\partial x_j}\left[\eta_{Hj}\left(\frac{\partial u_i}{\partial x_j} + \frac{\partial u_j}{\partial x_i}\right)\right] - \frac{2}{3}\nabla(\eta_{Hj}\nabla \cdot \overrightarrow{U_f})$$

彻体力 $f_b$ 由重力 $f_g$、磁场力 $f_k$ 和附加力 $f_L$ 组成，即

$$\overrightarrow{f_b} = \overrightarrow{f_g} + \overrightarrow{f_k} + \overrightarrow{f_L} = \rho_f \vec{g} + \overrightarrow{M} \cdot \nabla \overrightarrow{B_0} + \frac{1}{2}\nabla \times (\overrightarrow{M} \times \overrightarrow{B_0})$$

于是，铁磁流体混合流的运动方程就是

$$\rho_f \frac{d\overline{U_f}}{dt} = \rho_f \vec{g} - \nabla p + \frac{\partial}{\partial x_j}\left[\eta_{Hj}\left(\frac{\partial u_i}{\partial x_j} + \frac{\partial u_j}{\partial x_i}\right)\right] - \frac{2}{3}\nabla(\eta_{Hj}\nabla \cdot \overrightarrow{U_f}) + \overrightarrow{M} \cdot \nabla \overrightarrow{B_0} + \frac{1}{2}\nabla \times (\overrightarrow{M} \times \overrightarrow{B_0}) \tag{5.29}$$

式中，$\eta_{Hj}$ 是铁磁流体在 $j$ 方向的粘性系数，下标 $j = x$，$y$，$z$。

在图 4-1（a）和（b）中所示的物理模型，给出了外磁场 $\overrightarrow{B_0}$ 与坐标 $z$ 方向的涡旋速度 $\overrightarrow{\omega_{fz}}$ 两者同时对固相微粒作用而产生式（4.93）所表达的粘性系数。以及式（5.25c）、式（5.25d）与式（5.23b）～式（5.23d）所表达的 Kelvin 力和附加力。但在铁磁流体的流动中，一般都会存在 $x$，$y$，$z$ 三个方的涡旋速度分量，即 $\overrightarrow{\omega_{fx}}$，$\overrightarrow{\omega_{fy}}$，$\overrightarrow{\omega_{fz}}$，而不独是 $\overrightarrow{\omega_{fz}}$，并且 $\overrightarrow{\omega_{fx}}$、$\overrightarrow{\omega_{fy}}$ 同样地将与外磁场一起对固相微粒产生作用，从而引发相应的粘性系数、Kelvin 力和附加力。所以，在铁磁流体混合流的运动方程中，不同方向有不同的粘性系数，而且任何一个方向的 Kelvin 力和附加力都将由三个部分组成，它们分别是 $\overrightarrow{\omega_{fx}}$、$\overrightarrow{\omega_{fy}}$ 和 $\overrightarrow{\omega_{fz}}$ 所引起的结果。

将式（4.93）写成通用的形式

$$\frac{\eta_{Hj}}{\eta_c} = \frac{\dfrac{(1-\phi_\delta)(1+\beta_e)}{1+\beta_e \cos^2(H,\omega_{fj})} + \phi_\delta(1.5+2C_\delta)}{\left(1-\dfrac{\rho_\delta}{\rho_f}\phi_\delta\right)\dfrac{1+\beta_e}{1+\beta_e \cos^2(H,\omega_{fj})} + \dfrac{\rho_\delta}{\rho_f}\phi_\delta}\left(1-\frac{\omega_{Hj}}{\omega_{fj}}\right) + (1+0.5\phi_\delta+2C_\delta\phi_\delta)\frac{\omega_{Hj}}{\omega_{fj}} \tag{5.30}$$

式中，角度 $(H,\omega_{fj})$ 是外磁场矢量 $\overrightarrow{H}$ 与涡旋速度矢量 $\overrightarrow{\omega_{fj}}$ 之间的夹角，下标 $j = x$，$y$，$z$。

式（5.25c）、式（5.25d）的通用形式是

$$(\overrightarrow{f_{kx}})_j = \vec{i}M_0\left\{[1-(\omega_{fH})_j^2 R_m^2 \sin(H,\omega_{fj})]\sin(H,\omega_{fj})\frac{\partial B_0}{\partial x} + \right.$$
$$\left.(\omega_{fH})_j R_m \sin(H,\omega_{fj})\frac{\partial B_0}{\partial y} + \cos(H,\omega_{fj})\frac{\partial B_0}{\partial z}\right\}\sin(H,\omega_{fj}) \tag{5.31a}$$

$$(\overrightarrow{f_{ky}})_j = 0 \tag{5.31b}$$

$$(\overrightarrow{f_{kz}})_j = \vec{k}M_0\left\{[1-(\omega_{fH})_j^2 R_m^2 \sin(H,\omega_{fj})]\sin(H,\omega_{fj})\frac{\partial B_0}{\partial x} + \right.$$
$$\left.(\omega_{fH})_j R_m \sin(H,\omega_{fj})\frac{\partial B_0}{\partial y} + \cos(H,\omega_{fj})\frac{\partial B_0}{\partial z}\right\}\cos(H,\omega_{fj}) \tag{5.31c}$$

式（5.23b）～式（5.23d）也写成通用的形式，即

$$(\overrightarrow{f_{Lx}})_j = -\vec{i}\frac{1}{2}\left\{\sin(H,\omega_{fj})\frac{\partial}{\partial y}[(\omega_{fH})_j R_m M_0 B_0] + \right.$$
$$\left.\cos(H,\omega_{fj})\frac{\partial}{\partial z}[(\omega_{fH})_j^2 R_m^2 M_0 B_0]\right\}\sin(H,\omega_{fj}) \tag{5.32a}$$

$$(\overrightarrow{f_{Ly}})_j = \vec{j}\frac{1}{2}\left\{\cos(H,\omega_{fj})\frac{\partial}{\partial z}[(\omega_{fH})_j R_m M_0 B_0] + \right.$$
$$\left.\sin(H,\omega_{fj})\frac{\partial}{\partial x}[(\omega_{fH})_j R_m M_0 B_0]\right\}\sin(H,\omega_{fj}) \tag{5.32b}$$

$$(\overrightarrow{f_{Lz}})_j = \vec{k}\frac{1}{2}\left\{\frac{\partial}{\partial x}[(\omega_{fH})_j^2 R_m^2 M_0 B_0] - \frac{\partial}{\partial y}[(\omega_{fH})_j R_m M_0 B_0]\right\}\sin(H,\omega_{fj})\cos(H,\omega_{fj}) \tag{5.32c}$$

在式（5.31a）～式（5.32c）中，左方括号内的下标 $x$，$y$，$z$，均表示力的方向，而 $j$ 表示涡旋矢量的方向，$j=x$，$y$，$z$。

### 5.7.3 在直角坐标系中，铁磁流体磁松弛混合流的运动方程

将方程式（5.29）分解在 $x$，$y$，$z$ 三个方向上，就得

$$\rho_f\left(\frac{\partial u}{\partial t} + u\frac{\partial u}{\partial x} + v\frac{\partial u}{\partial y} + w\frac{\partial u}{\partial z}\right) = \rho_f g_x - \frac{\partial p}{\partial x} + \frac{\partial}{\partial x}\left[\eta_{Hx}\left(\frac{\partial u}{\partial x} + \frac{\partial u}{\partial x}\right)\right] + \frac{\partial}{\partial y}\left[\eta_{Hx}\left(\frac{\partial v}{\partial x} + \frac{\partial u}{\partial y}\right)\right] +$$
$$\frac{\partial}{\partial z}\left[\eta_{Hx}\left(\frac{\partial w}{\partial x} + \frac{\partial u}{\partial z}\right)\right] + \sum(f_{kx})_j + \sum(f_{Lx})_j \tag{5.33a}$$

$$\rho_f\left(\frac{\partial v}{\partial t} + u\frac{\partial v}{\partial x} + v\frac{\partial v}{\partial y} + w\frac{\partial v}{\partial z}\right) = \rho_f g_y - \frac{\partial p}{\partial y} + \frac{\partial}{\partial y}\left[\eta_{Hy}\left(\frac{\partial v}{\partial y} + \frac{\partial v}{\partial y}\right)\right] + \frac{\partial}{\partial z}\left[\eta_{Hy}\left(\frac{\partial w}{\partial y} + \frac{\partial v}{\partial z}\right)\right] +$$
$$\frac{\partial}{\partial x}\left[\eta_{Hy}\left(\frac{\partial u}{\partial z} + \frac{\partial w}{\partial x}\right)\right] + \sum(f_{ky})_j + \sum(f_{Ly})_j \tag{5.33b}$$

$$\rho_f\left(\frac{\partial w}{\partial t}+u\frac{\partial w}{\partial x}+v\frac{\partial w}{\partial y}+w\frac{\partial w}{\partial z}\right)=\rho_f g_z-\frac{\partial p}{\partial z}+\frac{\partial}{\partial z}\left[\eta_{Hz}\left(\frac{\partial w}{\partial z}+\frac{\partial w}{\partial z}\right)\right]+\frac{\partial}{\partial x}\left[\eta_{Hz}\left(\frac{\partial u}{\partial z}+\frac{\partial w}{\partial x}\right)\right]+$$
$$\frac{\partial}{\partial y}\left[\eta_{Hz}\left(\frac{\partial v}{\partial z}+\frac{\partial w}{\partial y}\right)\right]+\sum(f_{kz})_j+\sum(f_{Lz})_j \tag{5.33c}$$

参照图 4-1（b），可以给出式（5.33a）～式（5.33c）中的粘性系数、磁场力和附加力。

1.方程式（5.33a）～式（5.33c）中的粘性系数

由 $\omega_{Hx}=0$，$(\omega_{fH})_x=\omega_{fx}$，角度 $(H,\omega_{fx})=\pi/2-\theta_0$，于是由式（5.30）得

$$\frac{\eta_{Hx}}{\eta_c}=\frac{\dfrac{(1-\phi_\delta)(1+\beta_e)}{1+\beta_e\sin^2\theta_o}+\phi_\delta(1.5+2C_\delta)}{\left(1-\dfrac{\rho_\delta}{\rho_f}\phi_\delta\right)\dfrac{1+\beta_e}{1+\beta_e\sin^2\theta_o}+\dfrac{\rho_\delta}{\rho_f}\phi_\delta} \tag{5.34a}$$

由 $\omega_{Hy}=0$，$(\omega_{fH})_y=\omega_{fy}$，角度 $(H,\omega_{fy})=\pi/2$，则有

$$\frac{\eta_{Hy}}{\eta_c}=\frac{(1-\phi_\delta)(1+\beta_e)+\phi_\delta(1.5+2C_\delta)}{\left(1-\dfrac{\rho_\delta}{\rho_f}\phi_\delta\right)(1+\beta_e)+\dfrac{\rho_\delta}{\rho_f}\phi_\delta} \tag{5.34b}$$

由 $\omega_{Hz}=0$，$(\omega_{fH})_z=\omega_{fz}-\omega_{Hz}$，角度 $(H,\omega_{fz})=\theta_0$，于是

$$\frac{\eta_{Hz}}{\eta_c}=\frac{\dfrac{(1-\phi_\delta)(1+\beta_e)}{1+\beta_e\cos^2\theta_0}+\phi_\delta(1.5+2C_\delta)}{\left(1-\dfrac{\rho_\delta}{\rho_f}\phi_\delta\right)\dfrac{1+\beta_e}{1+\beta_e\cos^2\theta_0}+\dfrac{\rho_\delta}{\rho_f}\phi_\delta}\left(1-\frac{\omega_{Hz}}{\omega_{fz}}\right)+(1+1.5\phi_\delta+2C_\delta\phi_\delta)\frac{\omega_{Hz}}{\omega_{fz}} \tag{5.34c}$$

2.式（5.33a）～式（5.33c）中的磁场力（Kelvin 力）

由式（5.31a）代入以 $\omega_{Hx}=0$，$(H,\omega_{fx})=\pi/2-\theta_0$；$\omega_{Hy}=0$，$(H,\omega_{fy})=\pi/2$；和 $\omega_{Hz}=\omega_H$，$(H,\omega_{fz})=\theta_0$，

则有

$$\left.\begin{aligned}(\vec{f}_{kx})_x&=\vec{i}M_0\left[(1-\omega_{fx}^2R_m^2\cos\theta_0)\cos\theta_0\frac{\partial B_0}{\partial x}+\omega_{fx}R_m\cos\theta_0\frac{\partial B_0}{\partial y}+\sin\theta_0\frac{\partial B_0}{\partial z}\right]\cos\theta_0\\[6pt](\vec{f}_{kx})_y&=\vec{i}M_0\left[(1-\omega_{fy}^2R_m^2)\frac{\partial B_0}{\partial x}+\omega_{fy}R_m\frac{\partial B_0}{\partial y}\right]\\[6pt](\vec{f}_{kx})_z&=\vec{i}M_0\left\{[1-(\omega_{fH})_z^2R_m^2\sin\theta_0]\sin\theta_0\frac{\partial B_0}{\partial x}+(\omega_{fH})_zR_m\sin\theta_0\frac{\partial B_0}{\partial y}+\cos\theta_0\frac{\partial B_0}{\partial z}\right\}\sin\theta_0\end{aligned}\right\} \tag{5.35a}$$

式（5.31b）给出

$$(\vec{f}_{ky})_x=(\vec{f}_{ky})_y=(\vec{f}_{ky})_z=0 \tag{5.35b}$$

由式（5.31c）有

$$\left.\begin{array}{l}(\vec{f}_{kz})_x = \vec{k}M_0\left[(1-\omega_{fx}^2 R_m^2 \cos\theta_0)\cos\theta_0\dfrac{\partial B_0}{\partial x} + \omega_{fx}R_m\cos\theta_0\dfrac{\partial B_0}{\partial y} + \sin\theta_0\dfrac{\partial B_0}{\partial z}\right]\sin\theta_0 \\[2mm] (\vec{f}_{kz})_y = 0 \\[2mm] (\vec{f}_{kz})_z = \vec{k}M_0\left\{[1-(\omega_{fH})_z^2 R_m^2\sin\theta_0]\sin\theta_0\dfrac{\partial B_0}{\partial x} + (\omega_{fH})_z R_m\sin\theta_0\dfrac{\partial B_0}{\partial y} + \cos\theta_0\dfrac{\partial B_0}{\partial z}\right\}\cos\theta_0\end{array}\right\}$$ (5.35c)

3.式（5.33a）～式（5.33c）中的附加力
由式（5.32a），有

$$\left.\begin{array}{l}(\vec{f}_{Lx})_x = -\vec{i}\dfrac{1}{2}\left[\cos\theta_0\dfrac{\partial}{\partial y}(\omega_{fx}R_mM_0B_0) + \sin\theta_0\dfrac{\partial}{\partial z}(\omega_{fx}^2 R_m^2 M_0B_0)\right]\cos\theta_0 \\[2mm] (\vec{f}_{Lx})_y = -\vec{i}\dfrac{1}{2}\dfrac{\partial}{\partial y}(\omega_{fy}R_mM_0B_0) \\[2mm] (\vec{f}_{Lx})_z = -\vec{i}\dfrac{1}{2}\left\{\sin\theta_0\dfrac{\partial}{\partial y}[(\omega_{fH})_z R_mM_0B_0] + \cos\theta_0\dfrac{\partial}{\partial z}[(\omega_{fH})_z^2 R_m^2 M_0B_0]\right\}\sin\theta_0\end{array}\right\}$$ (5.36a)

由式（5.32b），有

$$\left.\begin{array}{l}(\vec{f}_{Ly})_x = \vec{j}\dfrac{1}{2}\left[\sin\theta_0\dfrac{\partial}{\partial z}(\omega_{fx}R_mM_0B_0) + \cos\theta_o\dfrac{\partial}{\partial x}(\omega_{fx}R_mM_0B_0)\right]\cos\theta_0 \\[2mm] (\vec{f}_{Ly})_y = \vec{j}\dfrac{1}{2}\dfrac{\partial}{\partial x}(\omega_{fy}R_mM_0B_0) \\[2mm] (\vec{f}_{Ly})_z = \vec{j}\dfrac{1}{2}\left\{\cos\theta_0\dfrac{\partial}{\partial z}[(\omega_{fH})_z R_mM_0B_0] + \sin\theta_0\dfrac{\partial}{\partial x}[(\omega_{fH})_z R_mM_0B_0]\right\}\sin\theta_0\end{array}\right\}$$ (5.36b)

由式（5.32c），有

$$\left.\begin{array}{l}(\vec{f}_{Lz})_x = \vec{k}\dfrac{1}{2}\left[\dfrac{\partial}{\partial x}(\omega_{fx}^2 R_m^2 M_0B_0) - \dfrac{\partial}{\partial y}(\omega_{fx}R_mM_0B_0)\right]\sin\theta_0\cos\theta_0 \\[2mm] (\vec{f}_{Lz})_y = 0 \\[2mm] (\vec{f}_{Lz})_z = \vec{k}\dfrac{1}{2}\left\{\dfrac{\partial}{\partial x}[(\omega_{fH})_z^2 R_m^2 M_0B_0] - \dfrac{\partial}{\partial y}[(\omega_{fH})_z R_mM_0B_0]\sin\theta_0\cos\theta_0\right\}\end{array}\right\}$$ (5.36c)

　　将式（5.34a）、式（5.35a）、式（5.36a）代入方程式（5.33a）之右方。同样，将式（5.34b）、式（5.35b）、式（5.36b）代入方程式（5.33b）之右方；式（5.34c）、式（5.35c）、式（5.36c）代入方程式（5.33c）之右方，就完整地得到铁磁流体混合流的磁松弛运动方程式。此外，为避开过于繁杂的涡量方程，$\omega_{fx}$、$\omega_{fy}$、$\omega_{fz}$可用 $\omega_f = (1/2)\nabla\times\overrightarrow{U_f}$ 给出的

$$\omega_{fx} = \frac{1}{2}\left(\frac{\partial w}{\partial y} - \frac{\partial v}{\partial z}\right)$$ (5.37a)

$$\omega_{fy} = \frac{1}{2}\left(\frac{\partial u}{\partial z} - \frac{\partial w}{\partial x}\right)$$ (5.37b)

$$\omega_{fz} = \frac{1}{2}\left(\frac{\partial v}{\partial x} - \frac{\partial u}{\partial y}\right)$$ (5.37c)

代入运动方程中，而 $(\omega_{fH})_z = \omega_{fz} - \omega_{Hz}$ 中的 $\omega_{Hz} = \omega_H$ 是已知的常数。同时，注意到在图 4-1（a）和（b）所示的模型中，已经设定外磁场是对称于坐标轴 $z$ 的平行磁场。所以在静止坐标系 $Ox'y'z$ 中，它绕 $z$ 轴旋转形成一个顶角不变的正圆锥。而在旋转坐标系 $Oxyz$ 中，外磁场 $H_0$ 与平衡磁化强度 $M_0$ 均位于坐标平面 $Oxz$ 之上。所以此时二者均系二维函数，即 $H_0 = H_0(x, y)$，$M_0 = M_0(x, z)$。于是，在式（5.31a）～式（5.31c），以及式（5.35a）～式（5.35c）中，应取 $\partial B_0/\partial y = 0$；在式（5.32a）～式（5.32c）以及式（5.36a）、式（5.36c）中应取 $\dfrac{\partial}{\partial y}[(\omega_{fH})_f R_m M_0 B_0] = R_m M_0 B_0 \dfrac{\partial}{\partial y}(\omega_{fH})_j$。

## 5.8 铁磁流体磁松弛混合流的能量守恒方程

能量方程的基本形式就是热力学第一定律，即

$$dE_f = \delta Q_f + \delta W_f$$

此式在物理上的意思是，系统与外部的换热 $\delta Q_f$，及外部对系统所作之功 $\delta W_f$，将转化成系统内能的增加 $dE_f$。因为热和功都和热力学过程相关，并非热力学参数。使用 $\delta Q_f$ 与 $\delta W_f$ 是为避免误解它们在数学上是全微分。$E_f$ 是总能，包括微观内能和宏观机械，对铁磁流体而言，微观上的内能除热内能之外还有磁内能，宏观机械能中只考虑动能而略去势能。于是有

$$E_f = \int_{V_f} \left( e_f + \frac{1}{2} \rho_f U_f^2 \right) dV_f$$

$\delta Q$ 是在 $dt$ 时间内外界通过铁磁流体表面积 $\overrightarrow{S_f}$ 加入到铁磁流体的热量。设通过单位表面积在单位时间内传递的热量即热流是 $\overrightarrow{q_H}$，则有

$$\delta Q_f = \left( -\int_{S_f} \overrightarrow{q_H} \cdot d\overrightarrow{S_f} \right) dt$$

上式积分号前的负号表示热流矢量 $\overrightarrow{q_H}$ 的方向与表面 $\overrightarrow{S_f}$ 之外向法线方向相反。热流 $\overrightarrow{q_H}$ 遵守 Fourier 定律，即

$$\overrightarrow{q_H} = -K_{Hf} \nabla T$$

$K_{Hf}$ 是铁磁流体的导热系数。在没有实验数据的情况下，可以按式（3.58c）作近似估算。上式右方的负号表示热流和温度梯度方向相反。

$\delta W_f$ 是外界作用于铁磁流体的总功，包括表面力 $\overrightarrow{f_s}$ 所作之功 $\delta W_s$，彻体力 $\overrightarrow{f_b}$（包括一般的机械力和 Kelvin 力、附加力等磁场力）所作之功 $\delta W_b$，以及铁磁流体的磁化功 $\delta W_m$。即

$$\delta W_f = \delta W_b + \delta W_s + \delta W_M$$

上式可进一步写成

$$\delta W_f = \left[ \int_{V_f} \overrightarrow{f_b} \cdot \overrightarrow{U_f} dV_f + \int_{S_f} \tau \cdot \overrightarrow{U_f} \cdot d\overrightarrow{S_f} \right] dt + \int_{V_f} \delta w_m \, dV_f$$

上式右方中括号内都是力和速度乘积的积分。所以均是功率，乘以时间 $dt$ 以后才是微元功，$w_m$ 是单位体积铁磁流体的磁化功。

将以上的 $E_f$、$\delta Q_f$ 和 $\delta W_f$ 的式子代入热力学第一定律式内，就有

$$d\int_{V_f} \left( e_f + \frac{1}{2}\rho_f U_f^2 \right) dV_f = \left( -\int_{S_f} \overrightarrow{q_H} \cdot d\overrightarrow{S_f} \right) dt + \left( \int_{V_f} \overrightarrow{f_b} \cdot \overrightarrow{U_f} dV_f + \int_{S_f} \tau \cdot \overrightarrow{U} \cdot d\overrightarrow{S_f} \right) dt + \int_{V_f} \delta w_m dV_f$$

两边同除以 $dt$，而后使用散度定理将上式中的面积分改变为体积分，就得

$$\int_{V_f} \frac{d}{dt}\left( e_f + \frac{1}{2}\rho_f U_f^2 \right) dV_f = -\int_{V_f} \nabla \cdot \overrightarrow{q_H} dV_f + \int_{V_f} \overrightarrow{f_b} \cdot \overrightarrow{U_f} dV_f + \int_{V_f} \nabla \cdot (\tau \cdot \overrightarrow{U_f}) dV_f + \int_{V_f} \frac{\delta w_m}{dt} dV_f$$

因为积分区间 $V_f$ 在流场中是任取的，积分号内的函数经过用 $\delta$ 函数平均以后，均可以视为连续函数，并且流体是不可压缩的，故弃掉积分号之后得

$$\frac{de_f}{dt} + \rho_f \overrightarrow{U_f} \cdot \frac{d\overrightarrow{U_f}}{dt} = -\nabla \cdot \overrightarrow{q_H} + \overrightarrow{f_b} \cdot \overrightarrow{U_f} + \nabla \cdot (\tau \cdot \overrightarrow{U_f}) + \frac{\delta w_m}{dt}$$

$$= -\nabla \cdot \overrightarrow{q_H} + \overrightarrow{f_b} \cdot \overrightarrow{U_f} + (\nabla \cdot \tau) \cdot \overrightarrow{U_f} + (\tau \cdot \nabla) \cdot \overrightarrow{U_f} + \frac{\delta w_m}{dt}$$

式（5.28b）两边点乘以 $\overrightarrow{U_f}$，并且用 $\overrightarrow{f_s} = \nabla \cdot \tau$ 代入，得

$$\rho_f \overrightarrow{U_f} \cdot \frac{d\overrightarrow{U_f}}{dt} = \overrightarrow{f_b} \cdot \overrightarrow{U_f} + (\nabla \cdot \tau) \cdot \overrightarrow{U_f}$$

将此式与上一式相减，就有

$$\frac{de_f}{dt} = -\nabla \cdot \overrightarrow{q_H} + (\tau \cdot \nabla) \cdot \overrightarrow{U_f} + \frac{\delta w_m}{dt} \tag{5.38}$$

以下给出能量方程式（5.38）中的各项。

① $\dfrac{de_f}{dt}$

由式（1.55）给出

$$\frac{de_f}{dt} = \left( c_{vf} \frac{dT_f}{dt} + \mu_0 T_f \frac{\partial M}{\partial T_f} \frac{dH}{dt} \right) + \mu_0 H \left( \frac{\partial M}{\partial T_f} \frac{dT_f}{dt} + \frac{\partial M}{\partial H} \frac{dH}{dt} \right)$$

式中，$c_{vf}$ 是铁磁流体的温度比热容，$\mu_0 T_f (\partial M / \partial T_f)$ 是铁磁流体的磁比热容。右方第二个括号内两项之和是磁化功率转化成的微观磁内能。

② $-\nabla \cdot \overrightarrow{q_H}$

由 Fourier 导热定律 $\overrightarrow{q_H} = -K_{Hf}\nabla T_f$，得

$$-\nabla \cdot \overrightarrow{q_H} = \nabla \cdot (K_{Hf}\nabla T_f)$$

③ $(\tau \cdot \nabla) \cdot \overrightarrow{U_f}$

运用张量

$$(\tau \cdot \nabla) \cdot \overrightarrow{U_f} = \left[\left(\tau_{ij}\overrightarrow{e_i^0}\,\overrightarrow{e_j^0}\right) \cdot \left(\overrightarrow{e_k^0}\frac{\partial}{\partial x_k}\right)\right] \cdot \overrightarrow{e_l^0}u_l = \left(\tau_{ij}\overrightarrow{e_i^0}\delta_{jk}\frac{\partial}{\partial x_k}\right) \cdot \overrightarrow{e_l^0}u_l = \tau_{ij}\overrightarrow{e_i^0}\frac{\partial}{\partial x_j} \cdot \overrightarrow{e_l^0}u_l = \tau_{ij}\frac{\partial}{\partial x_j}\delta_{il}u_l = \tau_{ij}\frac{\partial u_i}{\partial x_j}$$

式中 $\tau_{ij}$ 用式（2.47）代入，则有

$$(\tau \cdot \nabla) \cdot \overrightarrow{U_f} = \left[-p\delta_{ij} + \eta_H\left(\frac{\partial u_i}{\partial x_j} + \frac{\partial u_j}{\partial x_i}\right) - \frac{2}{3}\eta_H(\nabla \cdot \overrightarrow{U_f})\delta_{ij}\right]\frac{\partial u_i}{\partial x_j}$$

由式（2.62）

$$\Phi = \eta_H\left(\frac{\partial u_i}{\partial x_j} + \frac{\partial u_j}{\partial x_i}\right)\frac{\partial u_i}{\partial x_j} - \frac{2}{3}\eta_H\left(\nabla \cdot \overrightarrow{U_j}\right)^2$$

于是有

$$(\tau \cdot \nabla) \cdot \overrightarrow{U_f} = -p\nabla \cdot \overrightarrow{U_f} + \Phi$$

④ $\dfrac{\delta w_m}{dt}$

由式（1.48）得

$$\frac{\delta w_m}{dt} = \mu_0 H\left(\frac{\partial M}{\partial T_f}\frac{dT_f}{dt} + \frac{\partial M}{\partial H}\frac{dH}{dt}\right)$$

将上面的①至④各项结果代入方程（5.38）中，磁内能变化率项与磁化功率项相消，而得

$$c_{vf}\frac{dT_f}{dt} + \mu_0 T_f\frac{\partial M}{\partial T_f}\frac{dH}{dt} = \nabla \cdot (K_{Hf}\nabla T_f) - p\nabla \cdot \overrightarrow{U_f} + \Phi \tag{5.39}$$

式（5.39）左方第二项，即磁热容项内的 $M$ 是铁磁流体磁化强度矢量 $\overrightarrow{M}$ 的模，即

$$\left(\frac{M}{M_0}\right)^2 = \left(\frac{M_x}{M_0}\right)^2 + \left(\frac{M_y}{M_0}\right)^2 + \left(\frac{M_z}{M_0}\right)^2 \tag{5.40}$$

由式（5.20a）与式（5.20b）有

$$\frac{M_y}{M_0} = (\omega_{fH})_z R_m \sin\theta_0, \qquad \frac{M_x}{M_0} = [1 - (\omega_{fH})_z^2 R_m^2]\sin\theta_0$$

注意到 $M_z/M_0 = \cos\theta_0$，于是得到

$$\left(\frac{M}{M_0}\right)^2 = \left[1 - (\omega_{fH})_z^2 R_m^2\right]^2 \sin^2\theta_0 + (\omega_{fH})_z^2 R_m^2 \sin^2\theta_0 + \cos^2\theta_0 = \tag{5.41a}$$
$$1 - (\omega_{fH})_z^2 R_m^2 \sin^2\theta_0 + (\omega_{fH})_z^4 R_m^4 \sin^2\theta_0$$

略去右方含 $R_m$ 的四次方项，于是得

$$M = M_0\sqrt{1 - (\omega_{fH})_z^2 R_m^2 \sin^2\theta_0} \approx M_0\left[1 - \frac{1}{2}(\omega_{fH})_z^2 R_m^2 \sin^2\theta_0\right] \tag{5.41b}$$

用式（5.49）与式（4.77a）改写上式就有

$$\frac{M}{M_0} = 1 - \frac{1}{\left[1 + \left(1 - \phi_\delta\dfrac{\rho_\delta}{\rho_f}\sin^2\theta_0\right)0.5\alpha L(\alpha)\dfrac{t_{r\delta}}{t_{B\delta}}\right]^2} \tag{5.42}$$

上面的式子给出松弛的磁化强度 $\overline{M}$ 与平衡磁化强度 $\overline{M_0}$ 的数值关系，而两矢量间的相位角可以近似估计为

$$\alpha_m \approx \arccos\left(\frac{M}{M_0}\right) = \arccos[1 - 0.5(\omega_{fH})_z^2 R_m^2 \sin^2\theta_0] \tag{5.43}$$

将上式（5.42）代入能量方程（5.39）左方第二项中，即得出 $\partial M/\partial T_f$。注意到 $M_0$ 是可以使用 Langevin 函数计算的平衡磁化强度，它是和温度相关的，再考虑到磁松弛参数 $R_m$ 也是与温度相关的，所以 $M$ 是温度 $T_f$ 的函数。

在方程（5.39）右方第一项，即热传导项，其中 $K_{Hf}$ 是铁磁流体的导热系数，它不仅取决于固相和液相两种成分的材料导热性能，而且和固相在铁磁流体中的存在方式密切相关，其中包括固相的体积分量，固相微粒的尺寸及微粒浓度分布。$K_{Hf}$ 和温度的关系，不仅因为固、液两相的导热系数与温度相关，而且固相微粒的热运动随温度升高而加剧，提高了能量传递的速度。这种动态的导热效应，在式（3.58c）中没有反映，它只适用于两相"凝固"地混合在一起的情况。

5.9 铁磁流体混合流的质量守恒方程

在采用平均密度 $\rho_f$ 的情况下，铁磁流体的质量守恒方程与普通流体没有两样，可以直接引用式（2.42a）～式（2.42c），即

$$\frac{\partial\rho_f}{\partial t} + \nabla\cdot(\rho_f\overline{U_f}) = 0$$

对于可压缩流的稳定状态，有

$$\nabla\cdot(\rho_f\overline{U_f}) = 0$$

在混合流分析中，不能考虑两相间的运动滞后，故铁磁流体被认为是均匀的，即不可压缩的，$\rho_f = \text{const}$，故有

$$\nabla \cdot \overrightarrow{U_f} = 0$$

从上面的 5.7 节到 5.9 节所得出的均是铁磁流体混合流动力学方程组的通式和磁松弛的状况。以下给出磁平衡流和磁冻结流两种特殊状况。

### 5.10 铁磁流体的磁平衡混合流动力学方程
#### 5.10.1 概述

无论是内禀性或非内禀性铁磁流体，只要具磁松弛时间 $t_N$ 或 $t_{B\delta}$ 远远短于任何形式的宏观运动的特征时间，则不管外磁场是否旋转或流动有无涡旋，总可以认为固相微粒的统计平均磁矩与外磁场保持共线。这种状况就是磁平衡流，它具有以下几个特点：

① 磁化强度

$$\overrightarrow{M} = \overrightarrow{M_0}(x, y, z; t)$$

$\overrightarrow{M_0}$ 是平衡磁化强度，可以使用 Langevin 公式估算。

② 磁松弛综合参数 $\beta_e$ 和 $R_m$

由于 $t_N \to 0$ 或 $t_{B\delta} \to 0$，则式（4.77）与式（5.19）

$$\beta_e = M_0 H \frac{\mu_0 t_{r\delta}}{6\phi_\delta \eta_\delta} \to 0, \qquad R_m = \frac{t_{r\delta}}{1 + \left(1 - \phi_\delta \frac{\rho_\delta}{\rho_f} \sin^2 \theta_0\right) \beta_e} \to 0$$

于是，外磁场对铁磁流体的粘度作不出贡献，即 $\eta_m \to 0$ 而有

$$\eta_H \approx \eta_v = \eta_c \left(1 + 0.5\phi_\delta + 2C_\delta \phi_\delta\right) \tag{5.42}$$

③ 由于磁化强度 $\overrightarrow{M}$ 和外磁场 $\overrightarrow{H}$ 平行，Kelvin 力 $\overrightarrow{f_k}$ 与铁磁流体静止时相同，即

$$\overrightarrow{f_k} = M \nabla B_0 = M_0 \nabla B_0 \tag{5.43}$$

④ 不存在磁力矩，同时也没有附加力，即

$$\overrightarrow{L_m} = \overrightarrow{M} \times \overrightarrow{B_0} = 0, \qquad \overrightarrow{f_L} = \frac{1}{2} \nabla \times \overrightarrow{L_m} = 0$$

⑤ 磁力矩既不存在，作为反力矩与之平衡的粘性力矩也趋于零，于是有

$$\overrightarrow{\omega_P} = \overrightarrow{\omega_C} = \overrightarrow{\omega_f}$$

#### 5.10.2 铁磁流体磁平衡混合流的动力学方程组

1.动量守恒方程和能量守恒方程

将以上①～⑤的结果应用于动量守恒方程式（5.29）和能量守恒方程式（5.39），就得到

$$\rho_f \frac{\partial \overrightarrow{U_f}}{\partial t} + \rho_f \overrightarrow{U_f} \cdot \nabla \overrightarrow{U_f} = \rho_f \overrightarrow{g} - \nabla p + \frac{\partial}{\partial x_j}\left[\eta_v\left(\frac{\partial u_i}{\partial x_j} + \frac{\partial u_j}{\partial x_i}\right)\right] - \frac{2}{3}\nabla(\eta_v \cdot \overrightarrow{U_f}) + M_0 \nabla B_0 \quad (5.44)$$

$$c_{vf}\frac{dT_f}{dt} + \mu_0 T_f \frac{\partial M_0}{\partial T_f}\frac{dH}{dt} = \nabla \cdot (K_{Hf}\nabla T_f) - p\nabla \cdot \overrightarrow{U_f} + \Phi \quad (5.45)$$

上两方程中的 $M_0$ 是

$$M_0 = \phi_0 M_P L(\alpha) = \phi_p M_p\left(\coth\alpha - \frac{1}{\alpha}\right)$$

质量守恒方程仍然是

$$\nabla \cdot \overrightarrow{U_f} = 0$$

2.涡量方程

由于涡量的定义是

$$\overrightarrow{\Omega_f} = \nabla \times \overrightarrow{U_f} = 2\overrightarrow{\omega_f}$$

所以可由动量守恒方程取旋度而直接得到涡量方程。为此先将方程式（5.44）作一些简化：略去重力，取 $\rho_f \overrightarrow{g} = 0$；取 $\eta_v$ 在全流场中的平均值，于是 $\eta_v$ 就被认为是常数；铁磁流体是均匀的、不可压缩的，从而有 $\nabla \cdot \overrightarrow{U_f} = 0$，则

$$\frac{\partial}{\partial x_j}\left[\eta_v\left(\frac{\partial u_i}{\partial x_j} + \frac{\partial u_j}{\partial x_i}\right)\right] = \eta_v\frac{\partial^2 u_i}{\partial x_j^2} + \eta_v\frac{\partial^2 u_j}{\partial x_i \partial x_j} = \eta_v\nabla^2\overrightarrow{U_f} + \eta_v\nabla\nabla\cdot\overrightarrow{U_f} = \eta_v\nabla^2\overrightarrow{U_f}$$

于是，方程式（5.44）的常粘度和不可压缩流的形式为

$$\rho_f\frac{\partial \overrightarrow{U_f}}{\partial t} + \rho_f\overrightarrow{U_f}\cdot\nabla\overrightarrow{U_f} = -\nabla p + \eta_v\nabla^2\overrightarrow{U_f} + \mu_0 M_0\nabla H \quad (5.46)$$

上式等号两边取旋度

$$\rho_f\frac{\partial}{\partial t}(\nabla\times\overrightarrow{U_f}) + \rho_f\nabla\times(\overrightarrow{U_f}\cdot\nabla\overrightarrow{U_f}) = -\nabla\times\nabla p + \eta_v\nabla\times(\nabla^2\overrightarrow{U_f}) + \mu_0\nabla\times(M_0\nabla H) \quad (5.47a)$$

利用矢量恒等式，给出上式中的各项：

① $\rho_f\dfrac{\partial}{\partial t}(\nabla\times\overrightarrow{U_f}) = 2\rho_f\dfrac{\partial\overrightarrow{\omega_f}}{\partial t}$

② $\rho_f\nabla\times(\overrightarrow{U_f}\cdot\nabla\overrightarrow{U_f})$

由 $\nabla U_f^2 = \nabla(\overrightarrow{U_f}\cdot\overrightarrow{U_f}) = 2\overrightarrow{U_f}\cdot\nabla\overrightarrow{U_f} + 2\overrightarrow{U_f}\times(\nabla\times\overrightarrow{U_f}) = 2U_f\cdot\nabla\overrightarrow{U_f} + 4\overrightarrow{U_f}\times\overrightarrow{\omega_f}$，于是有

$$\overline{U_f} \cdot \nabla \overline{U_f} = \frac{1}{2} \nabla U_f^2 - 2\overline{U_f} \times \overline{\omega_f}$$

将上式两边取旋度

$$\nabla \times (\overline{U_f} \cdot \nabla \overline{U_f}) = \frac{1}{2} \nabla \times (\nabla U_f^2) - 2\nabla \times (\overline{U_f} \times \overline{\omega_f}) = -2\nabla \times (\overline{U_f} \times \overline{\omega_f}) =$$
$$-2(\overline{U_f} \nabla \cdot \overline{\omega_f} + \overline{\omega_f} \nabla \cdot \overline{U_f} + \overline{\omega_f} \cdot \nabla \overline{U_f} - \overline{U_f} \cdot \nabla \overline{\omega_f}) =$$
$$2\overline{U_f} \cdot \nabla \overline{\omega_f} - 2\overline{\omega_f} \cdot \nabla \overline{U_f}$$

③ $\nabla \times (\nabla p) = 0$

④ $\nabla \times (\nabla^2 \overline{U_f})$

由 $\nabla \times (\nabla \times \overline{U_f}) = \nabla(\nabla \cdot \overline{U_f}) - \nabla^2 \overline{U_f} = -\nabla^2 \overline{U_f}$ ，于是

$$\nabla \times (\nabla^2 \overline{U_f}) = -\nabla \times [\nabla \times (\nabla \times \overline{U_f})] = -\nabla[\nabla \cdot (\nabla \times \overline{U_f})] + \nabla^2(\nabla \times \overline{U_f}) = 2\nabla^2 \omega_f$$

⑤ $\nabla \times (M_0 \nabla H)$

$$\nabla \times (M_0 \nabla H) = M_0 \nabla \times (\nabla H) + (\nabla M_0) \times (\nabla H) = (\nabla M_0) \times (\nabla H)$$

将以上①～⑤的结果代入式（5.47a），就得

$$\rho_f \left( \frac{\partial \overline{\omega_f}}{\partial t} + \overline{U_f} \cdot \nabla \overline{\omega_f} - \overline{\omega_f} \cdot \nabla \overline{U_f} \right) = \eta_v \nabla^2 \omega_f + \frac{1}{2} \mu_0 (\nabla M_0) \times (\nabla H) \tag{5.47b}$$

式（5.47b）即铁磁流体磁平衡混合流的涡量方程。它与方程（2.55a）相比只是在右方多了一个有关磁性的项。

注意到在能量守恒方程（5.45）左方含有 $\partial M_0/\partial T_f$，在涡量方程（5.47b）右方含有 $\nabla M_0$，如果没有 $M_0$ 的完整的实验数据，则只能用 Langevin 方程计算，于是 $\partial M_0/\partial T_f$ 和 $\nabla M_0$ 可以进一步写成

$$\frac{\partial M_0}{\partial T_f} = \phi_p M_p \frac{\partial}{\partial T_f} \left( \coth \alpha - \frac{1}{\alpha} \right) = \phi_p M_p \left( \frac{\alpha}{\sinh^2 \alpha} - \frac{1}{\alpha} \right) \frac{1}{T_f} \tag{5.48a}$$

和

$$\nabla M_0 = \phi_p M_p \nabla \left( \coth \alpha - \frac{1}{\alpha} \right) = \phi_p M_p \left( \frac{\alpha}{\sinh^2 \alpha} - \frac{1}{\alpha} \right) \left( \frac{1}{T_f} \nabla T_f - \frac{1}{H} \nabla H \right) \tag{5.48b}$$

将式（5.48b）代入方程式（5.47b），由于 $(\nabla H) \times (\nabla H) = 0$，而有

$$(\nabla M_0) \times (\nabla H) = \phi_p M_p \left( \frac{\alpha}{\sinh^2 \alpha} - \frac{1}{\alpha} \right) \frac{1}{T_f} (\nabla T_f) \times (\nabla H) \tag{5.48c}$$

## 5.11 铁磁流体磁冻结混合流的动力学方程
### 5.11.1 概述

所谓磁冻结流是指在一段流程中，铁磁流体的磁化状态不发生改变或改变很小的流动。对于大尺

寸的固相微粒，无论其 Neel 松弛，还是 Brown 松弛时间，都显著地增长，尤其是表面附着的分散剂链分子较长时，对 Brown 松弛过程影响格外地大。虽然流动中存在磁力矩和涡粘力矩的作用，由于磁松弛不够快而流动速度高，以致铁磁流体的磁化强度矢量来不及改变或改变很小，就流出所论的流场区段。

铁磁流体磁松弛的几个主要参数是

①Neel 松弛时间

$$t_N = \frac{1}{f_0} \exp\left(\frac{K_1 V_{p1}}{k_0 T_f}\right)$$

②Brown 松弛时间

$$t_{B\delta} = \frac{3\eta_\delta (V_{p1})_\delta}{k_0 T_f}$$

③Langevin 数

$$\alpha = \frac{\mu_0 M_p V_{p1} H}{k_0 T_f}$$

④磁松弛综合参数

$$\beta_e = 0.5 \alpha L(\alpha) \frac{t_{r\delta}}{t_{B\delta}}$$

由以上四个重要参数可以看到影响磁化过程的基本因素是：①固相微粒的尺寸。就是 $V_{p1}$，通常铁磁流体的内禀性或非内禀性依其尺寸来划分。$V_{p1}$ 增大不仅使 $t_N$ 和 $t_{B\delta}$ 增长，而且使 $\alpha$ 和 $\beta_e$ 增大。②温度 $T_f$。它对四个参数都有直接影响，尤其对 $t_{B\delta}$ 的影响是双重的，因为基载液的粘度 $\eta_c$ 随 $T_f$ 升高而降低，$T_f$ 降低 $\eta_c$ 升高。③外磁场强度 $H$。外磁场强度对 $t_N$ 和 $t_{B\delta}$ 没有直接的关联，它只包含在 $\alpha$ 和 $\beta_e$ 之中。

注意到在 $t_N$ 的式子中，$K_1 V_{p1}$ 是一个微粒的体积各向异性能，它实际上是一种能垒[1]，微粒内的原子旋转振动能量，必须超过这个能垒才有可能实现转向。在 $t_{B\delta}$ 中，分子 $\eta_\delta (V_{p1})_\delta$ 是一个微粒受到基载液作用的旋转阻力功。在 $t_N$ 和 $t_{B\delta}$ 分母中的 $k_0 T_f$ 是热运动的动能，这种动能是磁松弛的驱动能量，实际中的 $k_0 T_f$ 比 $K_1 V_{p1}$ 和 $\eta_\delta (V_{p1})_\delta$ 大许多个数量级，所以磁松弛的进程很快，表现在 $t_N$ 和 $t_{B\delta}$ 只有 $10^{-9} \sim 10^{-6}$ 秒的量级[3]。在工程实用中，$T_f$ 只能在铁磁流体的 Curie 温度以下，或者不能高于基载液的沸点，也不能低于基载液的冰点，实际变化范围不超过一个数量级。而 $V_{p1}$ 的上限受铁磁流体需保持胶体稳定性的制约。只有外磁场强度不受限制。在 $\alpha$ 中分子是使铁磁流体磁化的能量。$k_0 T_f$ 是使铁磁流体磁松弛或退

磁的能量。所以 $\alpha$ 值越大，铁磁流体的磁化程度越深。对于非内禀性铁磁流体有 $t_{r\delta}/t_{B\delta} \approx 1$，则 $\alpha$ 越大，$\beta_e$ 也越大，表明保持磁化的因素强过磁松弛的因素，所以说 $\beta_e \gg 1$，就可能出现或接近磁冻结的现象。

既然 $T_f$ 和 $V_{p1}$ 的变化范围有限制，故 $(\omega_{PH})_z t_{B\delta}$ 的值仍然维持为小量，从而由式（4.79）简化为式（4.81）依然近似成立。由 $t_{r\delta}/t_{B\delta} \approx 1$ 以及 $\beta_e \gg 1$，则式（4.84）成为

$$\frac{M_y}{M_0} \approx \frac{(\omega_{CH})_z t_{B\delta}}{0.5\alpha L(\alpha)} \sin\theta_0 \tag{5.49}$$

将式（5.49）代入式（4.83）的右方，得

$$\eta_m = -\frac{3}{2}\eta_c\phi_\delta C_\delta \frac{0.5\alpha L(\alpha)}{\omega_{f\epsilon}t_{B\delta}}\left[\frac{(\omega_{CH})_z t_{B\delta}}{0.5\alpha L(\alpha)}\sin\theta_0\right]\sin\theta_0$$

于是有

$$\eta_m = -\frac{3}{2}\eta_c\phi_\delta C_\delta \frac{(\omega_{CH})_z}{\omega_{f\epsilon}}\sin\theta_0 \tag{5.50}$$

合并式（4.89）与式（4.91）得出

$$\frac{(\omega_{CH})_z}{\omega_{f\epsilon}} = \frac{\dfrac{1+\beta_e}{1+\beta_e\cos^2\theta_0}\left(1-\dfrac{\omega_{Hz}}{\omega_{f\epsilon}}\right)}{\phi_\delta\dfrac{\rho_\delta}{\rho_f}+\left(1-\phi_\delta\dfrac{\rho_\delta}{\rho_f}\right)\dfrac{1+\beta_e}{1+\beta_e\cos^2\theta_0}} = \frac{1-\dfrac{\omega_{Hz}}{\omega_{f\epsilon}}}{1-\dfrac{\beta_e}{1+\beta_e}\phi_\delta\dfrac{\rho_\delta}{\rho_f}\sin^2\theta_0} \tag{5.51}$$

利用式（5.51）将铁磁流体磁松弛混合流的粘性系数式（4.87）与式（4.88）改写成

$$\eta_H = \eta_c\left[(1+0.5\phi_\delta+2C_\delta\phi_\delta)+\frac{3}{2}C_\delta\phi_\delta\frac{\sin^2\theta_0}{1+\dfrac{1}{\beta_e}-\phi_\delta\dfrac{\rho_\delta}{\rho_f}\sin^2\theta_0}\left(1-\frac{\omega_{Hz}}{\omega_{f\epsilon}}\right)\right] \tag{5.52a}$$

和

$$\eta_H = \eta_c\left[\frac{1}{1-2.5\phi_\delta+1.55\phi_\delta^2}+\frac{3}{2}C_\delta\phi_\delta\frac{\sin^2\theta_0}{1+\dfrac{1}{\beta_e}-\phi_\delta\dfrac{\rho_\delta}{\rho_f}\sin^2\theta_0}\left(1-\frac{\omega_{Hz}}{\omega_{f\epsilon}}\right)\right] \tag{5.52b}$$

对于磁冻结流，取 $\beta_e \gg 1$，于是有

$$\eta_H = \eta_c \left[ (1 + 0.5\phi_\delta + 2C_\delta\phi_\delta) + \frac{3}{2}C_\delta\phi_\delta \frac{\sin^2\theta_0}{1 - \phi_\delta \frac{\rho_\delta}{\rho_f}\sin^2\theta_0} \left( 1 - \frac{\omega_{Hz}}{\omega_{fz}} \right) \right] \tag{5.53a}$$

和

$$\eta_H = \eta_c \left[ \frac{1}{1 - 2.5\phi_\delta + 1.55\phi_\delta^2} + \frac{3}{2}C_\delta\phi_\delta \frac{\sin^2\theta_0}{1 - \phi_\delta \frac{\rho_\delta}{\rho_f}\sin^2\theta_0} \left( 1 - \frac{\omega_{Hz}}{\omega_{fz}} \right) \right] \tag{5.53b}$$

对于磁平衡流，$\beta_e \to 0$，于是式（5.52a）、式（5.52b）右方第二项分母中的 $1/\beta_e \to \infty$，从而只剩下第一项，即 $\eta_v$。式（5.52a）和式（5.52b）是磁松弛流粘性系数的一般式。

### 5.11.2 铁磁流体磁冻结混合流的动量守恒方程

在磁冻结流中，铁磁流体的磁化强度是一常矢，设其为 $\overrightarrow{M_i}$。外磁场 $\overrightarrow{B_0}$ 一般是位置和时间的函数，即 $\overrightarrow{B_0} = \overrightarrow{B_0}(x, y, z; t)$。于是磁冻结流中的 Kelvin 力、磁力矩以及附加力分别是

$$\overrightarrow{f_k} = \overrightarrow{M_i} \cdot \nabla \overrightarrow{B_0}(x, y, z; t), \qquad \overrightarrow{L_m} = \overrightarrow{M_i} \times \overrightarrow{B_0}(x, y, z; t), \qquad \overrightarrow{f_L} = \frac{1}{2}\nabla \times (\overrightarrow{M_i} \times \overrightarrow{B_0})$$

展开 $\overrightarrow{f_L}$ 的右方，就有

$$\overrightarrow{f_L} = \frac{1}{2}[\overrightarrow{M_i}(\nabla \cdot \overrightarrow{B_0}) - \overrightarrow{B_0}(\nabla \cdot \overrightarrow{M_i}) + \overrightarrow{B_0} \cdot \nabla \overrightarrow{M_i} - \overrightarrow{M_i} \cdot \nabla \overrightarrow{B_0}]$$

由于 $\overrightarrow{M_i}$ 是常矢，$\overrightarrow{B_0}$ 的散度是零，从而有

$$\overrightarrow{f_L} = -\frac{1}{2}\overrightarrow{M_i} \cdot \nabla \overrightarrow{B_0}$$

将此式代入动量方程（5.29）中，就得磁冻结混合流的动量方程为

$$\rho_f \frac{\partial \overrightarrow{U_f}}{\partial t} + \rho_f \overrightarrow{U_f} \cdot \nabla \overrightarrow{U_f} = \rho_f \overrightarrow{g} - \nabla p + \frac{\partial}{\partial x_j}\left[ \eta_H \left( \frac{\partial u_i}{\partial x_j} + \frac{\partial u_j}{\partial x_i} \right) \right] - \frac{1}{3}\nabla(\eta_H \nabla \cdot \overrightarrow{U_f}) + \frac{1}{2}\overrightarrow{M_i} \cdot \nabla \overrightarrow{B_0} \tag{5.54}$$

方程（5.54）中的粘性系数 $\eta_H$，由式（5.53a）或式（5.53b）两者任选其一。

### 5.11.3 铁磁流体磁冻结混合流的其余动力学方程

1. 涡量方程

考虑一种实用的常见情况，外磁场是静止不旋转的，即 $\omega_{Hz} = 0$，于是式（5.53a）与式（5.53b）成为

$$\eta_H = \eta_c \left[ (1 + 0.5\phi_\delta + 2C_\delta\phi_\delta) + \frac{3}{2}C_\delta\phi_\delta \frac{\sin^2\theta_\delta}{1 - \phi_\delta \frac{\rho_\delta}{\rho_f}\sin^2\theta_0} \right] \tag{5.55a}$$

$$\eta_H = \eta_c \left[ \frac{1}{1 - 2.5\phi_\delta + 1.55\phi_\delta^2} + \frac{3}{2}C_\delta\phi_\delta \frac{\sin^2\theta_0}{1 - \phi_\delta \frac{\rho_\delta}{\rho_f}\sin^2\theta_0} \right] \tag{5.55b}$$

涡量方程可以从动量方程式（5.54）两边取旋度得出。在方程式（5.54）右方磁力项是

$$\frac{1}{2}\overrightarrow{M_i} \cdot \nabla \overrightarrow{B_0}$$

按矢量恒等式

$$\nabla(\overrightarrow{M_i} \cdot \overrightarrow{B_0}) = \overrightarrow{M_i} \cdot \nabla\overrightarrow{B_0} + \overrightarrow{B_0} \cdot \nabla\overrightarrow{M_i} + \overrightarrow{M_i} \times (\nabla \times \overrightarrow{B_0}) + \overrightarrow{B_0} \times (\nabla \times \overrightarrow{M_i})$$

注意到 $\overrightarrow{M_i}$ 时常矢，故 $\overrightarrow{B_0} \cdot \nabla\overrightarrow{M_i} = 0$，$\nabla \times \overrightarrow{M_i} = 0$ 以及没有传导电流穿过磁场，则 Ampere 定理给出

$\nabla \times \overrightarrow{B_0} = 0$，于是有

$$\overrightarrow{M_i} \cdot \nabla\overrightarrow{B_0} = \nabla(\overrightarrow{M_i} \cdot \overrightarrow{B_0})$$

将上式两边取旋度，由 $\overrightarrow{M_i} \cdot \overrightarrow{B_0}$ 是一标量，而 $\nabla \times \nabla\varphi = 0$ 对任意标量函数成立，故

$$\nabla \times (\overrightarrow{M_i} \cdot \nabla\overrightarrow{B_0}) = \nabla \times \nabla(\overrightarrow{M_i} \cdot \overrightarrow{B_0}) = 0$$

此结果表明,在磁冻结流中,磁场力对涡旋度运动没有贡献。于是铁磁流体磁冻结流的涡量方程与式（2.54）或式（2.55）没有形式上的区别,只是密度用 $\rho_f$，粘性系数用式（5.55a）或式（5.55b）所给出的 $\eta_H$。

2.能量守恒方程

能量守恒方程的一般式是方程式（5.39）。将此方程用于磁冻结流时,注意到磁化强度 $M_i$ 是常数,于是方程左方第二项为零,即

$$\mu_0 T_f \frac{\partial M_i}{\partial T_f} \frac{dH}{dt} = 0$$

所以在磁冻结流中,无论外磁场的情况如何,其能量守恒方程与普通两相流之混合流没有区别,即

$$c_{sf} \frac{dT_f}{dt} = \nabla \cdot (K_{Hf}\nabla T_f) - p\nabla \cdot \overrightarrow{U_f} + \Phi \tag{5.56}$$

3.质量守恒方程

可以直接引用式（2.42a）～式（2.42c）。将式（2.42a）写成

$$\frac{\partial \rho_f}{\partial t} + \nabla \cdot (\rho_f \overrightarrow{U_f}) = 0$$

以式（3.26）及式（3.30）代入上式，有

$$\frac{\partial}{\partial t}[\phi_p \rho_{NP} + (1-\phi_p)\rho_{NC}] + \nabla \cdot \left\{[\phi_p \rho_{NP} + (1-\phi_p)\rho_{NC}]\overrightarrow{U_f}\right\} = 0 \qquad (5.57a)$$

于是得到

$$\frac{\partial \phi_p}{\partial t} + \nabla \cdot \left[\left(\phi_p + \frac{\rho_{NC}}{\rho_{NP} - \rho_{NC}}\right)\overrightarrow{U_f}\right] = 0 \qquad (5.57b)$$

无论是磁松弛混合流、磁平衡混合流或是磁冻结混合流，它们的质量守恒方程都是式（5.57b）。对于与时间无关的稳定流动，则有

$$\nabla \cdot \left[\left(\phi_p + \frac{\rho_{NC}}{\rho_{NP} - \rho_{NC}}\right)\overrightarrow{U_f}\right] = 0 \qquad (5.57c)$$

由上面的式（5.57b）可见，虽然铁磁流体的组合成分，都是不可压缩的物质，即 $\rho_{NP} = const$，$\rho_{NC} = const$，但因它们的配比在流场中不均匀，而使得铁磁流体出现了"可压缩"的性质。

但是，在混合流的理论中，铁磁流体是以整体运动为基础的，根本得不到固相和液相运动速度的差别，从而也得不出两相配比在流动中的变化。所谓整体流动就包含了两相运动的一致性，这就是 $\phi_p$ 必须保持为常数。式（5.57b）就成为

$$\nabla \cdot \overrightarrow{U_f} = 0$$

这是混合流实用的质量守恒方程。

## 5.12 铁磁流体方程组的定解条件

### 5.12.1 概述

要确定铁磁流体运动的参数，即速度、压力、温度，就必须知道其磁化强度。所以，描述铁磁流体运动的方程式，除动量守恒方程、能量守恒方程和质量守恒方程之外，还需要磁化方程和磁场方程。

磁化方程就是式（4.64），而磁场方程就是 Gauss 散度定理和 Ampere 旋度定理。

从数学上来看，描述铁磁流体运动的方程和磁场方程，都是泛定的微分方程。它们的解中必定会出现一些积分常数，积分常数的数目取决于待求函数的数目和方程的阶数。在积分常数未具体确定之前，所有的解是普适的或泛定的。

确定积分常数的定解条件，相应于铁磁流体的方程组，包括磁学边界条件、力学边界条件和热学边界条件。对于和时间相关的非稳定过程，还相应地需要初始条件。

定解条件是按照具体的实际情况，抽象为物理上的理想过程，其中不乏简化和近似之处。

### 5.12.2 磁学条件

对外磁场具有响应能力并可接受外磁场的作用，是铁磁流体的基本特性，它的存在价值也在于此。为铁磁流体设置合适的磁场环境，以改善铁磁流体的物性或控制铁磁流体的运动。铁磁流体在进入外磁场之前是不显现磁性的，当外磁场对铁磁流体的固相微粒的热运动产生约束性的规整作用，使铁磁流体在相应的程度上产生磁性，所以外磁场也常叫做磁化场。外磁场 $\overrightarrow{B_0}$ 不因铁磁流体的存在和运动而

有任何改变。但与外磁场相对应的铁磁流体的内磁场 $\vec{B}$，却因铁磁流体的磁化，而在外磁场的基础上产生一个增量 $\vec{B'}$，即

$$\vec{B} = \vec{B_0} + \vec{B'} = \vec{B_0} + \mu_0 \vec{M} = \mu_0(\vec{H} + \vec{M}) \tag{5.58a}$$

式中的 $\vec{B'}$ 一般称之为附加磁感应强度，它实质上就是磁化强度。

在铁磁流体与外磁场相对静止时，$\vec{B}$、$\vec{H}$、$\vec{M}$ 三矢共线，从而有

$$\vec{B} = \mu_0\left(1 + \frac{M}{H}\right)\vec{H} = \mu_0(1 + \chi_m)\vec{H} = \mu\vec{H} \tag{5.58b}$$

式中 $\chi_m$ 称为铁磁流体的磁化率，与真空磁导率 $\mu_0$ 相对应，$\mu$ 称为铁磁流体的磁导率。

描述磁场的方程就是 Gauss 散度定理和 Ampere 旋度定理，即

$$\nabla \cdot \vec{B} = 0 \tag{5.59a}$$

$$\nabla \times \vec{H} = \vec{i_c} \tag{5.59b}$$

式（5.59a）表明，无论有无磁介质存在和有无传导电流穿过磁场，磁感强度场是一种无源场；式（5.59b）右方的 $\vec{i_c}$ 是穿过磁场的传导电流密度。若无传导电流穿过磁场，即 $\vec{i_c} = 0$，则磁场强度场是具势的，它必存在势函数。

在第三章已经阐明过，微分方程（5.59a）和（5.59b）的边界条件，就是他们本身在零厚度的边界面上的应用，即

$$B_{1n} = B_{2n} \tag{5.60a}$$

$$H_{1\tau} = H_{2\tau} \tag{5.60b}$$

当铁磁流体流动时，式（5.58b）不能使用。因为涡粘力矩使铁磁流体的磁化强度矢量 $\vec{M}$ 与外磁场不共线，而且其模也有改变。这时就出现磁化方程，即方程式（4.64），它描述磁松弛过程。

就磁松弛而言，其进程非常之快，从一种被破坏的平衡状态松弛到新的平衡，其时间就是 Neel 扩散时间或 Brown 扩散时间 $t_B$，或两者综合的 $t_r$。在一般常规的时间尺度上，它们都是极短暂的。在流场中，外磁场强度和涡旋强度均可能是时间和位置的函数，所以铁磁流体的微团在运动中，其内部的磁松弛随时随地都发生。在稳定状态下，铁磁流体的磁化强度形成一个不变的分布图。当铁磁流体微团进入某区域时，便具有该区域的磁化强度，当其移至另外一个区域时，则磁化强度变为新区域的磁化强度。

磁平衡流和磁冻结流是铁磁流体运动的两种特殊的极端情况。磁平衡流的指标性参数就是平衡磁化强度 $\vec{M_0}$，其模可以按当地的外磁场强度和温度代入到 Langevin 方程中算出，而其方向则与外磁场 $\vec{H}$ 共线。

　　磁冻结流的指标性参数，是铁磁流体开始进入磁冻结状态时的磁化强度 $\overrightarrow{M_i}$ ，所以 $\overrightarrow{M_i}$ 是初始条件，

并且在冻结流中 $\overrightarrow{M_i}$ 保持为常矢。

　　无论是磁松弛流，还是磁平衡流，磁冻结流，其最重要的磁学定解条件就是由式（5.59a）、式（5.59b）与式（5.60a）、式（5.60b）所给出的外磁场在整个流场中的分布状态。

### 5.12.3 力学条件

　　1.初始条件

　　设初始的时间 $t = t_0$ ，则有

$$\overline{U_f}(r,t_0) = \overline{U_{f0}}(r) \tag{5.61a}$$

$$p(r,t_0) = p_0(r) \tag{5.61b}$$

$$\phi_p(r,t_0) = \phi_{p0}(r) \tag{5.61c}$$

上面三式的右方均为给定的已知量。

　　2.速度边界条件

　　①无穷远处的边界条件

　　所谓的无穷远就是扰动衰减为零的距离，在这里速度、压力、固相体积分量均是已知的，即

$$\overline{U_f}(\infty,t) = \overline{U_{f\infty}}(t) \tag{5.62a}$$

$$p(\infty,t) = p_\infty(t) \tag{5.62b}$$

$$\phi_p(\infty,t) = \phi_{p\infty}(t) \tag{5.62c}$$

　　②接触界面条件

　　在接触界面，运动速度具有连续性（但不一定可微）。设界面坐标为 $r$ ，则有

$$\overline{U_f}(r-0,t) = \overline{U_f}(r+0,t) \tag{5.63a}$$

式中坐标"$-0$"表示在曲率中心一边的界面内侧，而"$+0$"表示界面之外侧。将式（5.63a）在界面的法向和切向分解，就有

$$U_{fn}(r-0,t) = U_{fn}(r+0,t) \tag{5.63b}$$

$$U_{f\tau}(r-0,t) = U_{f\tau}(r+0,t) \tag{5.63c}$$

式（5.63b）表示两种介质在界面上互不分离和互不渗透；式（5.63c）表示在界面上两种介质之间没有相对滑动。若 $U_{f\tau}(r-0,t) \neq U_{f\tau}(r+0,t)$ ，则必存在 $\Delta U_{f\tau}$ ，但界面厚度 $\Delta r \to 0$ ，由牛顿定律知道此时的剪切应力 $\tau$ 为

$$\tau \propto \lim_{\Delta r \to 0} \frac{\Delta U_{f\tau}}{\Delta r} \to \infty$$

任何流动都不可能形成无穷大的剪切应力以致产生相对滑动。

式（5.63b）与式（5.63c）合起来也称为表面粘附条件。

3.界面应力平衡条件

①界面的薄膜内应力——表面张力

如图 5-4（a）所示，界面是一曲面。设界面上任一点 $O$ 处的单位外向法线为 $\overrightarrow{n^0}$，围绕点 $O$ 划取一微元控制面 $\delta S$。有两张垂直相交的平面，其交线就是 $O$ 点处的外向法线。该两平面与界面曲面相交而形成两条在 $O$ 点互相垂直的曲线，即图 5-4（b）中的 $COD$ 和 $FOE$。曲线 $COE$ 在 $O$ 点的曲率 $C_1 = 1/R_1$，$EOF$ 在 $O$ 点的曲率 $C_2 = 1/R_2$，此处 $R_1$ 和 $R_2$ 均为曲率半径。取 $O$ 点曲率的算术平均值为 $C$，即

$$C = \frac{1}{2}(C_1 + C_2) \tag{5.64a}$$

或写成

$$\frac{1}{R} = \frac{1}{2}\left(\frac{1}{R_1} + \frac{1}{R_2}\right) \tag{5.64b}$$

式中 $R$ 可称为平均曲率半径。

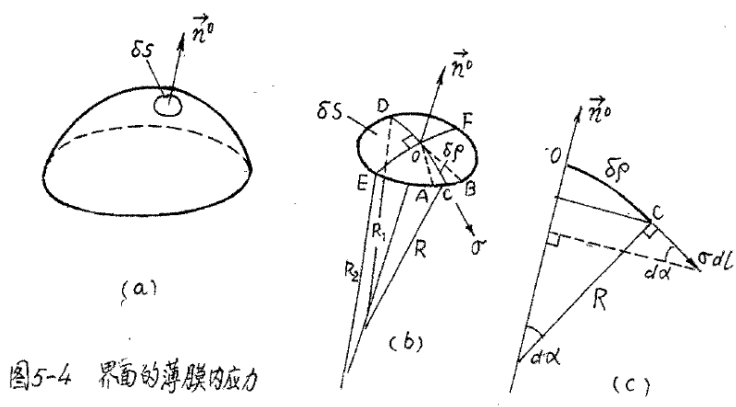

图5-4 界面的薄膜内应力

设界面薄膜的极限张力为 $\sigma$。由于界面没有厚度，故 $\sigma$ 是单位长度上的应力，其单位是 $N/m$。当然，$\sigma$ 与界面两边流体的性质相关。在微元控制面 $\delta S$ 的周界上任取一微段 $AB = dl$，如图 5-4（b）所示。在 $dl$ 上的内力 $f_c$ 是

$$df_c = \sigma dl$$

内力 $f_c$ 在 $\overrightarrow{n^0}$ 方向的分量是

$$df_{cn} = (\sigma dl)\sin(d\alpha) \approx \sigma dl\, d\alpha$$

由图 5-4（c）可以看到

$$d\alpha \approx \frac{\delta\rho}{R}$$

于是有

$$df_{cn} \approx \sigma dl \frac{\delta\rho}{R}$$

在图 5-4（b）上给出 $dl\delta\rho$ 是 $\Delta AOB$ 面积的两倍，即 $dl\delta\rho = 2dS$，于是

$$df_{cn} \approx \sigma(2dS)\frac{1}{R}$$

定义单位界面面积上的张力 $f_{cn}$ 为一当量压力 $p_c$，于是上式两边通除以 $dS$，并将式（5.64b）代入，就得

$$p_c = \frac{df_{cn}}{dS} = \sigma\frac{2}{R} = \sigma\left(\frac{1}{R_1} + \frac{1}{R_2}\right) \tag{5.65}$$

当量压力 $p_c$ 沿 $\overrightarrow{n^0}$ 的负方向，$p_c$ 的存在表示薄膜内应力在法线方向不平衡。因为 $\sigma$ 是常数，只有 $R_1$ 和 $R_2$ 改变才能改变 $p_c$。

只要界面薄膜是各向同性的，即 $\sigma$ 是常数，则界面的薄膜内应力的合力，只有法向分量而无切向分量，即在切线方向薄膜内应力是平衡的。

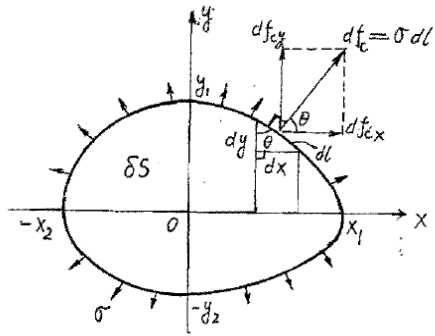

图5-5 界面的切向应力

在图 5-5 中，将微元控制面 $\delta S$ 单独取出。若 $\delta S$ 紧缩得非常微小，则可以认为 $\delta S$ 与 $O$ 点处垂直于 $\overrightarrow{n^0}$ 的切平面重合。设界面薄膜是各向同性的，则 $\delta S$ 的周边上作用的薄膜应力 $\sigma$ 是常量。在周边的微元长度 $dl$ 上的内力是 $df_c = \sigma dl$。其在平面坐标系 $Oxy$ 中，分量是

$$df_{cx} = (\sigma dl)\cos\theta, \qquad df_{cy} = (\sigma dl)\sin\theta$$

由几何关系 $(dl)\cos\theta = dy$，$(dl)\sin\theta = dx$，于是有

$$df_{cx} = \sigma dy, \qquad df_{cy} = \sigma dx$$

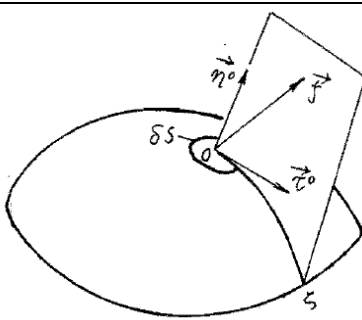

图5-6 界面表面上的 $On\tau$ 坐标系

由图 5-5 不难看出，在 $x$ 轴上方的 $y$ 正值区，$y$ 方向的合成力是 $+\sigma(x_1 + x_2)$，而在 $x$ 轴下方的 $y$ 负值区，$y$ 方向的合成力是 $-\sigma(x_1 + x_2)$。所以 $\delta S$ 全周边合力 $\delta f_c$ 在 $y$ 方向的分量 $\delta f_{cy} = 0$，同样也有 $\delta f_{cx} = 0$。于是可知边界曲面的薄膜内应力在切向是平衡的。

②界面表面上的外力

在界面曲面上，过 $O$ 点的外力 $\vec{f}$ 与 $O$ 点的外向法线 $\overline{n^0}$ 决定一平面，该平面与界面曲面的交线是一条平面曲线 $O\varsigma$，如图 5-6 所示。曲线 $O\varsigma$ 在 $O$ 点的切线就是 $\overline{\tau^0}$。由式（4.78a），作用于单位体积铁磁流体的力是

$$\vec{f} = \overline{f_g} + \overline{f_p} + \overline{f_\eta} + \overline{f_k} + \overline{f_L}$$

式中重力 $\overline{f_g}$ 通常可以忽略。

（1）$\overline{f_p} + \overline{f_\eta}$：由式（2.49）有

$$\overline{f_p} + \overline{f_\eta} = -\nabla \cdot (p\delta_{ij}) + \nabla \cdot (\tau_{ij}) - \frac{2}{3}\nabla \cdot [\eta(\nabla \cdot \vec{U})\delta_{ij}] \tag{5.66a}$$

式中

$$\tau_{ij} = \eta\left(\frac{\partial u_i}{\partial x_j} + \frac{\partial u_j}{\partial x_i}\right)$$

在无厚度的边界曲面表面上的平面坐标系 $On\tau$ 中，有 $i = n$，$j = n$，$\tau$，于是上式给出

$$\tau_{nn} = \eta\frac{\partial u_n}{\partial n}, \qquad \tau_{n\tau} = \eta\frac{\partial u_\tau}{\partial n} \tag{5.66b}$$

（2）$\overline{f_k}$

$$\overline{f_k} = \vec{M} \cdot \nabla\vec{B} = \mu_0\vec{M} \cdot \nabla\vec{H} = (\vec{B} - \mu_0\vec{H}) \cdot \nabla\vec{H}$$

此式右方两项：$\vec{B} \cdot \nabla\vec{H} = \nabla \cdot (\vec{B}\vec{H}) - \vec{H}\nabla \cdot \vec{B} = \nabla \cdot (\vec{B}\vec{H})$，

$$\mu_0 \overline{H} \cdot \nabla \overline{H} = \frac{1}{2} \mu_0 \nabla (\overline{H} \cdot \overline{H}) - \mu_0 \overline{H} \times (\nabla \times \overline{H}) = \frac{1}{2} \mu_0 \nabla (H^2)$$

于是有

$$\overrightarrow{f_k} = \nabla \cdot (\overline{B}\,\overline{H}) - \frac{1}{2} \mu_0 \nabla (H^2) = \nabla \cdot (\overline{B}\,\overline{H}) - \frac{1}{2} \mu_0 \nabla \cdot (H^2 \delta_{ij}) \tag{5.66c}$$

（3） $\overrightarrow{f_L}$

$$\overrightarrow{f_L} = \frac{1}{2} \nabla \times (\overline{M} \times \overline{B_0}) = \frac{1}{2} \nabla \times (\mu_0 \overline{M} \times \overline{H}) = \frac{1}{2} \nabla \times [(\overline{B} - \mu_0 \overline{H}) \times \overline{H}] = \frac{1}{2} \nabla \times (\overline{B} \times \overline{H})$$

上式已使用了关系 $\overline{H} \times \overline{H} = 0$ ，由矢量恒等式进一步可得

$$\overrightarrow{f_L} = \frac{1}{2} \nabla \times (\overline{B} \times \overline{H}) = \frac{1}{2} [\overline{B}(\nabla \cdot \overline{H}) - \overline{H}(\nabla \cdot \overline{B}) + \overline{H} \cdot \nabla \overline{B} - \overline{B} \cdot \nabla \overline{H}]$$

于是得出

$$\overrightarrow{f_L} = \frac{1}{2} \nabla \cdot (\overline{H}\,\overline{B}) - \frac{1}{2} \nabla \cdot (\overline{B}\,\overline{H}) \tag{5.66d}$$

$\overrightarrow{f_k}$ 和 $\overrightarrow{f_L}$ 本质上是铁磁流体的彻体力。但是，使用分子电流环模型描述固相微粒的磁性能时，则如第一章 1.4 节和第五章 5.3 节所阐明的那样， $\overrightarrow{f_k}$ 和 $\overrightarrow{f_L}$ 均能在形式上转化成表面力。于是可以认为作用于铁磁流体的力均是表面力，从而有

$$\overrightarrow{f} = \nabla \cdot \tau_\Sigma = \overrightarrow{f_p} + \overrightarrow{f_\eta} + \overrightarrow{f_k} + \overrightarrow{f_L} \tag{5.67}$$

将式（5.66a）、式（5.66c）及式（5.66d）代入上式，得

$$\nabla \cdot \tau_\Sigma = -\nabla \cdot (p \delta_{ij}) + \nabla \cdot \tau_{ij} - \nabla \cdot \left(\frac{1}{2} \mu_0 H^2 \delta_{ij}\right) + \frac{1}{2} \nabla \cdot (\overline{H}\,\overline{B}) + \frac{1}{2} \nabla \cdot (\overline{B}\,\overline{H})$$

上式两边给出

$$\tau_\Sigma = -p \delta_{ij} + \tau_{ij} - \frac{1}{2} \mu_0 H^2 \delta_{ij} + \frac{1}{2} (\overline{H}\,\overline{B}) + \frac{1}{2} (\overline{B}\,\overline{H}) \tag{5.68}$$

将式（5.68）放在界面表面上写成

$$\overline{n^0} (\tau_\Sigma)_{nj} \overrightarrow{e_j^0} = -p \delta_{nj} + \overline{n^0} \tau_{nj} \overrightarrow{e_j^0} - \frac{1}{2} \mu_0 H^2 \delta_{nj} + \frac{1}{2} (\overline{H}\,\overline{B}) + \frac{1}{2} (\overline{B}\,\overline{H})$$

上式两边点乘以 $\overline{n^0}$ ，并且注意到 $\overline{n^0} \cdot \overline{n^0} = 1$ ， $\overline{n^0} \cdot \delta_{ij} = \overline{n^0}$ ， $\overline{n^0} \cdot (\overline{H}\,\overline{B}) = H_n \overline{B}$ 就有

$$(\tau_\Sigma)_{nj} \overrightarrow{e_j^0} = -\overline{n^0} p + \tau_{nj} \overrightarrow{e_j^0} - \frac{1}{2} \overline{n^0} \mu_0 H^2 + \frac{1}{2} H_n \overline{B} + \frac{1}{2} B_n \overline{H}$$

因为在界面表面上 $j = n, \tau$ ，故上式可以写成

$$\overline{n^0} (\tau_\Sigma)_{nn} + \overline{\tau^0} (\tau_\Sigma)_{n\tau} = -\overline{n^0} p + \overline{n^0} \tau_{nn} + \overline{\tau^0} \tau_{n\tau} - \frac{1}{2} \overline{n^0} \mu_0 H^2 + \frac{1}{2} H_n \overline{B} + \frac{1}{2} B_n \overline{H} \tag{5.69a}$$

改写式（5.69a）右方的各项：

（1） $\tau_{nn}$ 和 $\tau_{n\tau}$

$$\tau_{nn} = \eta \frac{\partial u_n}{\partial n}, \qquad \tau_{n\tau} = \eta \frac{\partial u_\tau}{\partial n}$$

（2）$H^2$

$$H^2 = H_n^2 + H_\tau^2 = \left(\frac{B_n}{\mu_0} - M_n\right)^2 + H_\tau^2 = \left(\frac{B_n}{\mu_0}\right)^2 - \frac{2}{\mu_0} B_n M_n + M_n^2 + H_\tau^2 =$$

$$\left(\frac{B_n}{\mu_0}\right)^2 - \frac{2}{\mu_0} B_n \left(\frac{B_n}{\mu_0} - H_n\right) + M_n^2 + H_\tau^2 =$$

$$-\left(\frac{B_n}{\mu_0}\right)^2 + \frac{2}{\mu_0} B_n H_n + M_n^2 + H_\tau^2$$

（3）$\frac{1}{2} B_n \overline{H}$ 和 $\frac{1}{2} H_n \overline{B}$

$$B_n \overline{H} = B_n(\overline{n^0} H_n + \overline{\tau^0} H_\tau) = \overline{n^0} B_n H_n + \overline{\tau^0} B_n H_\tau$$
$$H_n \overline{B} = H_n(\overline{n^0} B_n + \overline{\tau^0} B_\tau) = \overline{n^0} H_n B_n + \overline{\tau^0} H_n B_\tau = \overline{n^0} H_n B_n + \overline{\tau^0} H_n(\mu H_\tau) = \overline{n^0} H_n B_n + \overline{\tau^0} B_n H_\tau$$

将以上（1）、（2）、（3）的结果代入式（5.69a）之右方，就有

$$\overline{n^0}(\tau_\Sigma)_{nn} + \overline{\tau^0}(\tau_\Sigma)_{n\tau} = -\overline{n^0} p + \overline{n^0} \eta \frac{\partial u_n}{\partial n} + \overline{\tau^0} \eta \frac{\partial u_\tau}{\partial n} + \overline{n^0} \frac{1}{\mu_0} B_n^2 - \overline{n^0} \frac{1}{2} \mu_0 M_n^2 -$$
$$\overline{n^0} \frac{1}{2} \mu_0 H_\tau^2 + \overline{\tau^0} B_n H_\tau \tag{5.69b}$$

将方程式（5.69b）按方向拆开，即得 $\overline{n^0}$ 和 $\overline{\tau^0}$ 两方向的表面应力

$$(\tau_\Sigma)_{nn} = -p + \eta \frac{\partial u_n}{\partial n} - \frac{1}{2} \mu_0 M_n^2 + \frac{1}{\mu_0} B_n^2 - \frac{1}{2} \mu_0 H_\tau^2 \tag{5.70a}$$

$$(\tau_\Sigma)_{n\tau} = \eta \frac{\partial u_\tau}{\partial n} + B_n H_\tau \tag{5.70b}$$

③边界表面上外力与薄膜内力的平衡

约定：与边界曲面之曲率中心同一边的表面为内表面，以下标"1"标志，则内表面上的外力是 $(\tau_{\Sigma 1})_n$ 和 $(\tau_{\Sigma 1})_\tau$；外表面以下标"2"标志，则其上外力是 $(\tau_{\Sigma 2})_n$ 和 $(\tau_{\Sigma 2})_\tau$。在单位面积上力平衡方程式 $\sum \tau = 0$，注意到 $(\tau_{\Sigma 2})_n$ 和 $p_c$ 沿 $\overline{n^0}$ 的负方向，则有

$$(\tau_{\Sigma 1})_n - (\tau_{\Sigma 2})_n = -p_c$$

用式（5.70a）代入，并且注意 $B_{1n} = B_{2n}$，$H_{1\tau} = H_{2\tau}$，于是有

$$p_1 - \eta_1 \frac{\partial u_{1n}}{\partial n_1} + \frac{1}{2} \mu_0 M_{1n}^2 = p_2 - \eta_2 \frac{\partial u_{2n}}{\partial n_2} + \frac{1}{2} \mu_0 M_{2n}^2 + \sigma\left(\frac{1}{R_1} + \frac{1}{R_2}\right) \tag{5.71a}$$

沿表面任何切向的薄膜应力之合力都是零，故有

$$(\tau_{\Sigma 1})_\tau + (\tau_{\Sigma 2})_\tau = 0$$

将式（5.70b）代入，即得

$$\eta_1 \frac{\partial u_{1\tau}}{\partial n_1} = -\eta_2 \frac{\partial u_{2\tau}}{\partial n_2} \tag{5.71b}$$

注意法线 $\overrightarrow{n_1}$ 和 $\overrightarrow{n_2}$ 的方向相反。并且应当考虑粘附条件，紧贴界面内外表面有 $u_{1n} = u_{2n}$ 以及 $u_{1\tau} = u_{2\tau}$。虽然 $u_n$ 和 $u_\tau$ 跨界面连续，但并不可微，因为 $\eta_1$ 和 $\eta_2$ 是间断的。

　　4.边界面的几何形状

　　界面曲面曲面形状，主要取决于力学因素：速度和力。

　　对于铁磁流体与固体壁相接触，则边界面的几何形状完全决定于固体壁面。这时最要紧的力学边界条件是速度的粘附条件。而力的平衡是自然形成的。固体壁面相对于流体而言。它是不变形的刚体，无论有多大的流体力作用于边界上，固体壁面都不会变形而以相等的反作用与之平衡。这样的平衡不能提供任何定解作用。

　　对于铁磁流体与互不相溶的别的流体接触形成的界面。除粘附条件外，力的平衡条件最为关键。已经知道界面薄膜内力的切向分量是自行平衡的，所以界面内外两侧外力的切向分量必须平衡，才能保持界面的形状稳定。由于界面薄膜内力的法向分量表现为表面张力，所以能承受一定的两侧外力之差。尽管 $\sigma$ 是不变的常量，当外力法向分量之差增大，界面的曲率半径减小而表面张力增大，以达到平衡而后止。但表面张力终究有限，通常它相对于外侧的外力很小，有时为了简单而将其忽略。但是忽略表面张力和表面张力等于零是两回事。忽略表面张力，界面形状仍然是曲面，而表面张力等于零，只能是曲率半径无限大，即界面是平面。

　　只有粘附条件中的切向和法向速度处处为零的状态下，界面才是固定不移的。

## 5.12.4 热学条件

　　1.概述

　　在铁磁流体与互不相溶的流体接触的界面上，同时存在传热与传质两类问题，传热问题遵守 Fourier 定律，传质问题遵守 Fick 定律。这两个定律，在数学上具有相似的形式，在物理上也有深刻的联系。在分子运动论中，热量是分子运动的能量。传热就分子运动动能的交换，而传质是分子自身的相互交换。当温度较高，也就是分子动能较高处的分子，进入温度较低（亦即运动动能较低）的区域，既是传质同时也是传热。

　　当铁磁流体接触的是不相溶的流体，传热只是界面上两种流体分子之间通过碰撞进行的动量或能量的交换。但即使不相溶，也需要考虑界面的稳定性问题。[2,1]这种表面不稳定是由于流体界面上浅表波动运动所引起，而参与其中的力包括流体压力、重力、表面张力和磁力。通常重力和磁力在某种状态下，使浅表波动运动失去稳定性，而表面张力是抑制波动的因素。界面一旦不稳定，两种不相溶的流体就形成宏观的迷宫式的交错分布。并且继续向深层扩张。在流动情况下，迅速造成两种流体的掺混。界面不稳定性是铁磁流体不能长期密封液体物质的主要原因，尤其是在动密封中，失效很快。铁磁流体界面不稳定性是一个专门问题。在文献[2]和[1]中有详细的阐释。

　　另外一个问题，是铁磁流体内的固体微粒会不会因热运动而跨过界面，进入接界的流体之中？在图 5-7 中，给出固相微粒跨过界面逸出铁磁流体的示意图。

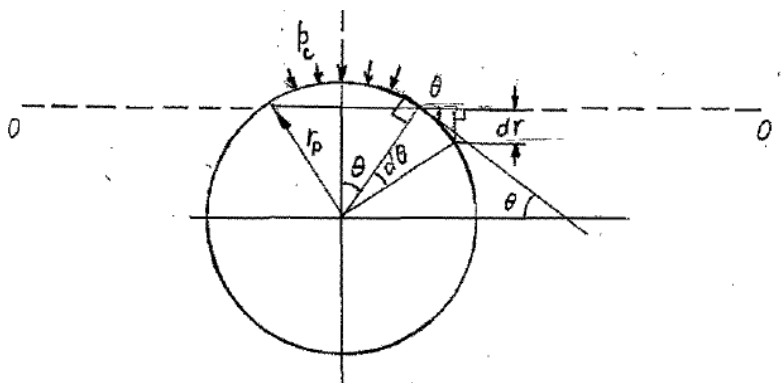

图 5-7  铁磁流体固相微粒跨过界面示意

虽然界面层仅有几个分子尺度的厚度，但是由于固相微粒尺寸的细小而在微粒表面上的表面张力大小十分可观。由式（5.65）中的曲率半径 $R_1$ 和 $R_2$ 均等于微粒半径 $r_p$，而有

$$p_c = \sigma \frac{2}{r_p}$$

球形固相微粒逸出铁磁流体要克服表面张力 $p_c$ 所造成的阻力功是

$$dW_c = p_c \pi (r_p \sin\theta)^2 dr = p_c \pi (r_p \sin\theta)^2 (r_p d\theta) \sin\theta = p_c \pi r_p^3 \sin^3\theta\, d\theta$$

微粒完全逸出所要克服的阻力功是

$$W_c = \int_o^\pi p_c \pi r_p^3 \sin^3\theta\, d\theta = \frac{4}{3}\pi r_p^3 p_c$$

设 $\sigma$ 的量级是 $10^{-2}\, N/m$ [2]，$r_p$ 的量级是 $10^{-8}$ m，则有

$$p_c = 10^{-2} \frac{2}{10^{-8}} = 2 \times 10^6\, N/m^2$$

将 $p_c$ 值代入 $W_c$ 的式子中就得一个固相微粒逸出铁磁流体所需克服的阻力功是

$$W_c = 8.4 \times 10^{-18}\, N \cdot m$$

固相微粒逸出铁磁流体的驱动因素是热运动动能和磁势能。固相微粒在常温即 $T = 293K$ 下的热运动动能是

$$k_0 T = 1.38 \times 10^{-23} \times 293 = 4.04 \times 10^{-21}\, N \cdot m$$

可见热运动动能远小于阻力功，完全不足以使固相微粒逸出铁磁流体，而固相微粒在外磁场中的磁势能必须比热运动能小，以避免铁磁流体内部固相微粒的偏析。所以固相微粒不可能跨过界面而进入相邻流体中。

于是铁磁流体热学边界条件只是热量的交换问题。

2.热学条件

①初始条件

$$T(r,t_0) = 已知 \tag{5.72}$$

②界面导热热流的连续性

由于界面没有厚度，不能存贮热量，所以传入和传出界面的热流必定相等，即

$$K_{H1}\frac{\partial T(r-0,t)}{\partial n} = K_{H2}\frac{\partial T(r+0,t)}{\partial n} \tag{5.73}$$

式中 $n$ 是界面的外向法线，$K_{H1}$ 和 $K_{H2}$ 是界面两边流体的导热系数

③界面温度的连续性

紧贴界面两边的温度不能不相等，若存在温差 $\Delta T$，则热流量为

$$q_H \propto \frac{\Delta T}{\Delta r}$$

由于界面层的厚度 $\Delta r$ 可以视为零，则 $q_H \to \infty$，这样巨大的热流能即时消除温差 $\Delta T$，所以有

$$T(r-0,t) = T(r+0,t) \tag{5.74}$$

虽然跨界面温度具有连续性质，但在界面上温度并不可微，因为式（5.73）中 $K_{H1}$ 和 $K_{H2}$ 不相等，更主要的是 $T$ 在界面上偏导数不相等，而形成一个不光滑的尖点。

④界面绝热条件

一般界面绝热现象发生在铁磁流体和固体壁面相接触的情况。当固壁材料的导热系数极低时，就可以认为这种界面是绝热的。即

$$\left.\frac{\partial T(r,t)}{\partial n}\right|_w = 0 \tag{5.75}$$

式中下标"$w$"表示壁面。

⑤混合边界条件

当铁磁流体沿固壁表面高速流动时，铁磁流体的传导导热效应就居于次要地位，主要的是涡旋对流或湍流对流的传热。这种传热的热流具有下述形式

$$q_H = h(T_f - T_w)$$

式中 $T_f$ 是铁磁流体的特征温度，$T_w$ 是固壁表面温度。$h$ 称为对流传热系数。在稳定状态下，由流体传入固壁的热流必由固壁材料的热传导散走，即

$$\left.-K_H\frac{\partial T}{\partial n}\right|_w = h(T_f - T_w) \tag{5.76}$$

式中 $K_H$ 是固壁材料的导热系数。$h$ 的一般关系式为

$$h = \frac{K_{Hf} \cdot \mathrm{Nu}}{l} \tag{5.77}$$

式中 $l$ 是特征长度，例如圆管内流，则取 $l$ 等于管道直径 $d$，$K_{Hf}$ 是铁磁流体的导热系数。Nu 称为 Nusselt 数，是一种量纲为 1 的准数，通常由半径验公式给出，例如 Drake 实验式

$$\mathrm{Nu} = 2 + 0.459 \cdot \mathrm{Re}^{0.55} \cdot \mathrm{Pr}^{0.33}$$

式中，Re 是 Reynolds 数，即

$$\text{Re} = \frac{\rho_f U_f l}{\eta_H}$$

Pr 是 Prandtl 数，即

$$\text{Pr} = \frac{\eta_H c_v}{K_{Hf}}$$

式中，铁磁流体的温度比热容 $c_v$ 包括温热效应和磁热效应，即式（1.49），$c_v = c_v(T,H)$。$\eta_H$ 是铁磁流体在外磁场中的粘度。

## 5.13 铁磁流体混合流的 Bernoulli 方程

Bernoulli 方程不仅是水力学中最重要的方程，而且也是铁磁流体应用中最普遍使用的方程。Bernoulli 方程只沿流线成立。由于流线具有不可跨越的性质。所以使用一束流线作为母线而形成流管，则在流管横截面上取流体运动的平均参数，Bernoulli 方程同样适用于流管，不言而喻，Bernoulli 方程是一维流动的方程。在内流中，流管就可取为管道本身，因为管道的壁面同样具有不可穿越的性质。

在普通流体中，Bernoulli 方程是一种机械能守恒的方程，但在铁磁流体中，它还包括磁能项。下面由普适的动量守恒方程（5.29）导出 Bernoulli 方程，设

①铁磁流体是不可压缩的

$$\rho_f = \text{const}, \qquad \nabla \cdot \overrightarrow{U_f} = 0$$

②流动是势流，存在速度势 $\varphi_v$

$$\overrightarrow{U_f} = -\nabla \varphi_v, \qquad \overrightarrow{\omega_f} = \frac{1}{2}\nabla \times \overrightarrow{U_f} = \frac{1}{2}\nabla \times (-\nabla \varphi_v) = 0$$

③外磁场是静止的，即 $\overrightarrow{\omega_H} = 0$，加以 $\overrightarrow{\omega_f} = 0$，则有铁磁流体的磁化强度矢量 $\overrightarrow{M}$ 和外磁场 $\overrightarrow{B_0}$ 平行，即 $\overrightarrow{M} \parallel \overrightarrow{B_0}$，于是有

$$\overrightarrow{f_k} = M\nabla B_0, \qquad \overrightarrow{f_L} = \frac{1}{2}\nabla \times (\overrightarrow{M} \times \overrightarrow{B_0}) = 0$$

在 $\overrightarrow{\omega_H} = 0$ 和 $\overrightarrow{\omega_f} = 0$ 的状态下，铁磁流体的粘性系数与外磁场无关，即

$$\eta_H = \eta_v = \text{const}$$

于是表面粘性应力项成为

$$\frac{\partial}{\partial x_j}\left[\eta_H\left(\frac{\partial u_i}{\partial x_j} + \frac{\partial u_j}{\partial x_i}\right)\right] = \eta_v\left[\frac{\partial^2 u_i}{\partial x_j^2} + \frac{\partial}{\partial x_i}\left(\frac{\partial u_j}{\partial x_j}\right)\right] = \eta_v \nabla^2 \overrightarrow{U_f}$$

将上述①～③的结果代入方程（5.29）中，得出

$$\rho_f \frac{\partial \overrightarrow{U_f}}{\partial t} + \rho_f \overrightarrow{U_f} \cdot \nabla \overrightarrow{U_f} = \rho_f \overrightarrow{g} - \nabla p + \eta_v \nabla^2 \overrightarrow{U_f} + M\nabla B_0 \tag{5.78a}$$

方程（5.78a）是三维方程。通过对坐标的积分，得出其一维的形式。积分也就相当于取平均值。改写方程（5.78a）的有关项：

① $\rho_f \dfrac{\partial \overline{U_f}}{\partial t}$

$$\rho_f \frac{\partial \overline{U_f}}{\partial t} = \rho_f \frac{\partial}{\partial t}(-\nabla \varphi_v) = \nabla \left( -\rho_f \frac{\partial \varphi_v}{\partial t} \right)$$

② $\rho_f \overline{U_f} \cdot \nabla \overline{U_f}$

$$\rho_f \overline{U_f} \cdot \nabla \overline{U_f} = \nabla \left( \frac{1}{2} \rho_f U_f^2 \right) - \rho_f \overline{U_f} \times (\nabla \times \overline{U_f}) = \nabla \left( \frac{1}{2} \rho_f U_f^2 \right)$$

③ $\rho_f \vec{g}$

设坐标系 $Oxyz$ 的原点 $O$ 是重力的参考基点。则 $\rho_f \vec{g}$ 的方向是指向 $O$ 点的。若 $\rho_f$ 与 $O$ 点的距离为 $h$，矢径 $\vec{h}$ 的单位矢为 $\vec{h^0}$，则有

$$\rho_f \vec{g} = -\rho_f g \vec{h^0} = -\rho_f g [\vec{i} \cos(x,h) + \vec{j} \cos(y,h) + \vec{k} \cos(z,h)]$$
$$= -\rho_f g \left( \vec{i} \frac{\partial h}{\partial x} + \vec{j} \frac{\partial h}{\partial y} + \vec{k} \frac{\partial h}{\partial z} \right) = -\left( \vec{i} \frac{\partial}{\partial x} + \vec{j} \frac{\partial}{\partial y} + \vec{k} \frac{\partial}{\partial z} \right)(\rho_f gh) = -\nabla(\rho_f gh)$$

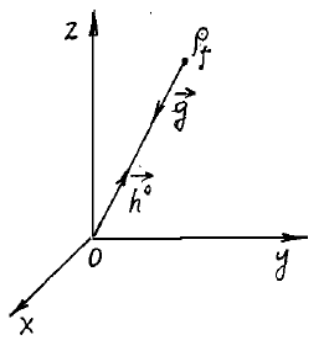

图5-8 重力坐标系

上式已考虑到 $\rho_f$ 和 $g$ 为常数。

④ $\eta_v \nabla^2 \overline{U_f}$

$$\eta_v \nabla^2 \overline{U_f} = \eta_v [\nabla(\nabla \cdot \overline{U_f}) - \nabla \times (\nabla \times \overline{U_f})] = 0$$

⑤ $M \nabla B_0$

铁磁流体的磁化强度 $M$ 是外磁场和温度的函数，即 $M = M(B_0, T)$ 同时有 $B_0 = B_0(x, y, z; t)$，$T = T(x, y, z; t)$ 由 Leibniz 对定积分上限取导的公式

$$\frac{\partial}{\partial x}\int_0^{B_0(x,y,z,t)} M(B_0,T)\,dB_0 = \int_0^{B_0}\frac{\partial M}{\partial x}\,dB_0 + M\frac{\partial B_0}{\partial x}$$

由 $\frac{\partial M}{\partial x} = \frac{\partial M}{\partial B_0}\frac{\partial B_0}{\partial x} + \frac{\partial M}{\partial T}\frac{\partial T}{\partial x}$，在 Leibniz 公式中，被积函数（此处是 $M$）只对参变量（此处是 $T$）取导，

不对积分变量（此处是 $B_0$）取导。因此有 $\frac{\partial M}{\partial x} = \frac{\partial M}{\partial T}\frac{\partial T}{\partial x}$。从而得

$$\frac{\partial}{\partial x}\int_0^{B_0} M\,dB_0 = \int_0^{B_0}\frac{\partial M}{\partial T}\frac{\partial T}{\partial x}\,dB_0 + M\frac{\partial B_0}{\partial x}$$

同样可有

$$\frac{\partial}{\partial y}\int_0^{B_0} M\,dB_0 = \int_0^{B_0}\frac{\partial M}{\partial T}\frac{\partial T}{\partial y}\,dB_0 + M\frac{\partial B_0}{\partial y}$$

$$\frac{\partial}{\partial z}\int_0^{B_0} M\,dB_0 = \int_0^{B_0}\frac{\partial M}{\partial T}\frac{\partial T}{\partial z}\,dB_0 + M\frac{\partial B_0}{\partial z}$$

以上三式合并写成

$$\nabla\int_0^{B_0} M\,dB_0 = \int_0^{B_0}\frac{\partial M}{\partial T}\nabla T\,dB_0 + M\nabla B_0$$

通常铁磁流体温度远低于它的 Curie 温度，所以 $T\approx\text{const}$，而对于等温流动，则 $T=\text{const}$，故有 $\nabla T = 0$，于是得到

$$M\nabla B_0 = \nabla\int_0^{B_0} M\,dB_0$$

将上述①～⑤的结果代入方程（5.78a）中，就有

$$\nabla\left(-\rho_f\frac{\partial\varphi_v}{\partial t} + \frac{1}{2}\rho_f U_f^2 + \rho_f gh + p - \int_0^{B_0} M\,dB_0\right) = 0$$

上式左方括号前的算子 $\nabla$ 表示对坐标的微分。若对坐标积分则 $\nabla$ 被去除。但括号内的项大多与时间相关，所以对坐标积分的积分常数是一个时间函数 $f(t)$，即

$$-\rho_f\frac{\partial\varphi_v}{\partial t} + \frac{1}{2}\rho_f U_f^2 + \rho_f gh + p - \int_0^{B_0} M\,dB_0 = f(t) \tag{5.78b}$$

对于稳定状态，则 $\partial\varphi_v/\partial t = 0$，$f(t) = c$，于是有

$$p + \frac{1}{2}\rho_f U_f^2 + \rho_f gh - \int_0^{B_0} M\,dB_0 = c \tag{5.78c}$$

# 第六章　铁磁流体分相流

6.1 概述

在铁磁流体混合流理论中，将铁磁流体的物性参数，如粘性系数、密度、导热系数、比热容等，按照固相和液相两者的体积分量或质量分量取平均值。就认为铁磁流体是具有这些平均物性的连续的"单相"流体。于是，就能导出牛顿流体的动力学方程组。

混合流理论的问题是：铁磁流体的混合物性参数，无论是按体积平均还是按质量平均，都只能在求解动力学方程组之前确定。然而，这些混合的物性参数，随流动中液两相所占成分的比例改变而变化。混合流的动力学方程组即使求解得到解答之后，也不可能知道铁磁流体成分的变化。这就是说混合流动力学方程的一个假设前提条件就是流动中铁磁流体的成分保持均匀。

铁磁流体最根本的特点是能够受外磁场的控制。这种超距控制也不过是通过磁力和磁力矩来实现。磁力和磁力矩只作用于固相微粒上，通过固相微粒与基载液体接触面上的粘性作用才能传递到液相上。界面粘性作用的机制，就是两者间平动速度和转动速度的滞后而产生粘性阻力和粘性阻力矩，它们对应于磁力与磁力矩，并且在稳态下与之平衡。所以没有运动的滞后，就没有两相间的力和力矩的传递。但是，有运动的滞后，就不可能保持成分的均匀不变。混合流理论通过参数的平均方法，认为磁力和磁力矩直接作用于整个铁磁流体，既回避了两相间的作用，同时又采纳成分均匀的假设。

所谓分相流，就将两相拆开来处理。这样每一相自然是单相流，从而所使用的物性参数就是各相的材料性能。两相间力和力矩的作用，两相间的传热等，在混合流中是内部作用，不影响混合流整体的运动过程。但对于分相流的每相而言，它们都是外部作用，对每相的流动都发生影响。这样，就能显示每个相在流动中的行为细节。

在铁磁流体中，固相和液相都是不连续的。拆开以后变成的单相流动不连续问题更为显著。对于液相成为内部有大量空洞的流体，这些空洞是固相移去之后留下的。而对于固相则是一个个分散的微粒形成的微粒流。

利用流体力学的方法分析铁磁流体分相以后的固相微粒流，是有其合理性的。固相微粒的数密度是每立方厘米 $10^{18}$ 个微粒。在标准状况（温度 273K，压力 1 标准大气压）下，每摩尔的气体体积为 22.4 升，其中含有 $6.02 \times 10^{23}$ 个气体分子（此即 Avogadro 数），故标准状况下气体分子的数密度是每立方厘米 $2.69 \times 10^{19}$ 个分子。而固相微粒的尺寸比气体分子约大 $1 \sim 2$ 个量级，所以将铁磁流体的固相微粒流视为气体的概念是可以成立的。当然，这是以固相微粒遵循气体分子运动论为基础的概念。

无论是带有巨大数量的空洞的液相流或单体分散的固相微粒流，都必须处理成为连续的流体，其最重要的工具就是 Dirac 的 $\delta$ 函数平均方法。在铁磁流体的分相流中分散的空洞或微粒，它们的物性和力学行为与周围情况迥然不同，在界面上突变，完全符合极端狭窄的矩形脉冲函数的规律。若将整个流场划分成很多的小邻域，每个小邻域内包含一个空洞或微粒，而后用 $\delta$ 函数法将空洞或微粒的特性在其邻域上取平均，实际上就是将其"稀薄化"。这样每个邻域内的物理量不改变，但却"稀薄"地充满整个邻域，所有的邻域鳞次栉比地拼接在一起，就形成连续的流场，从而连续介质的力学方法可以应用于其上。

6.2 铁磁流体分相流的动量守恒方程

式（5.28a）给出单位体积铁磁流体混合流的牛顿第二定律式为

$$\rho_f \frac{d\overrightarrow{U_f}}{dt} = \overrightarrow{f_g} + \overrightarrow{f_p} + \overrightarrow{f_\eta} + \overrightarrow{f_k} + \overrightarrow{f_L} = \overrightarrow{f_s} + \overrightarrow{f_b} \tag{6.1a}$$

式中 $\overrightarrow{f_s}$ 是表面力，$\overrightarrow{f_b}$ 是彻体力，即

$$\overrightarrow{f_s} = \overrightarrow{f_p} + \overrightarrow{f_\eta}, \qquad \overrightarrow{f_b} = \overrightarrow{f_g} + \overrightarrow{f_k} + \overrightarrow{f_L} \tag{6.1b}$$

以下逐项分解方程（6.1a）

1.惯性力

设在质量为 $dm_f$ 的铁磁流体中含有固相质量为 $dm_p$ 和液相质量为 $dm_c$，则 $dm_f$ 的惯性力必为 $dm_p$ 与

$dm_c$ 的惯性力之和，即

$$(dm_f)\overrightarrow{a_f} = (dm_p)\overrightarrow{a_p} + (dm_c)\overrightarrow{a_c}$$

上式可写成

$$\rho_f(\vec{r},t)\frac{d\overrightarrow{U_f}(\vec{r},t)}{dt}dV_f = \rho_{NP}\frac{d\overrightarrow{U_p}(\vec{r},t)}{dt}dV_p + \rho_{NC}\frac{d\overrightarrow{U_c}(\vec{r},t)}{dt}dV_c \tag{6.2a}$$

上式两边同乘以 $\delta(\vec{R}-\vec{r})$ 并在邻域 $V_f$ 内取平均就有

$$\int_{V_f}\rho_f(\vec{R},t)\frac{d\overrightarrow{U_f}(\vec{R},t)}{dt}\delta(\vec{R}-\vec{r})dV_f = \int_{V_p}\rho_{NP}\frac{d\overrightarrow{U_p}(\vec{R},t)}{dt}\delta(\vec{R}-\vec{r})dV_p + $$
$$\int_{V_c}\rho_{NC}\frac{d\overrightarrow{U_c}(\vec{R},t)}{dt}\delta(\vec{R}-\vec{r})dV_c \tag{6.2b}$$

用式（3.8）的关系代入，则上式成为

$$\int_{V_f}\rho_f(\vec{R},t)\frac{d\overrightarrow{U_f}(\vec{R},t)}{dt}\delta(\vec{R}-\vec{r})dV_f = \int_{V_f}\rho_{NP}\frac{d\overrightarrow{U_p}(\vec{R},t)}{dt}\delta(\vec{R}-\vec{r})\phi_p(\vec{R},t)dV_f + $$
$$\int_{V_f}\rho_{NC}\frac{d\overrightarrow{U_c}(\vec{R},t)}{dt}\delta(\vec{R}-\vec{r})\phi_c(\vec{R},t)dV_c$$

从而得出平均值的关系

$$\rho_f(\vec{r},t)\frac{d\overrightarrow{U_f}(\vec{r},t)}{dt} = \phi_p(\vec{r},t)\rho_{NP}\frac{d\overrightarrow{U_p}(\vec{r},t)}{dt} + \phi_c(\vec{r},t)\rho_{NC}\frac{d\overrightarrow{U_c}(\vec{r},t)}{dt} \tag{6.2c}$$

由全导数可分解成当地导数与迁移导数之和，即式（2.20）

$$\frac{dq}{dt} = \frac{\partial q}{\partial t} + \vec{U}\cdot\nabla q$$

将此式用于式（6.2c）之两边，就有

$$\rho_f(\vec{r},t)\frac{\partial \overrightarrow{U_f}(\vec{r},t)}{\partial t} + \rho_f(\vec{r},t)\overrightarrow{U_f}(\vec{r},t)\cdot\nabla\overrightarrow{U_f}(\vec{r},t) = $$
$$\phi_p(\vec{r},t)\rho_{NP}\frac{\partial \overrightarrow{U_p}(\vec{r},t)}{\partial t} + \phi_p(\vec{r},t)\rho_{NP}\overrightarrow{U_p}(\vec{r},t)\cdot\nabla\overrightarrow{U_p}(\vec{r},t) + \tag{6.2d}$$
$$\phi_c(\vec{r},t)\rho_{NC}\frac{\partial \overrightarrow{U_c}(\vec{r},t)}{\partial t} + \phi_c(\vec{r},t)\rho_{NC}\overrightarrow{U_c}(\vec{r},t)\cdot\nabla\overrightarrow{U_c}(\vec{r},t)$$

若上式右方使用分密度，则由 $\rho_p(\vec{r},t)=\phi_p(\vec{r},t)\rho_{NP}$, $\rho_c(\vec{r},t)=\phi_c(\vec{r},t)\rho_{NC}$ 得

$$\rho_f(\vec{r},t)\frac{\partial \overrightarrow{U_f}(\vec{r},t)}{\partial t}+\rho_f(\vec{r},t)\overrightarrow{U_f}(\vec{r},t)\cdot\nabla\overrightarrow{U_f}(\vec{r},t)=$$
$$\rho_p(\vec{r},t)\frac{\partial \overrightarrow{U_p}(\vec{r},t)}{\partial t}+\rho_p(\vec{r},t)\overrightarrow{U_p}(\vec{r},t)\cdot\nabla\overrightarrow{U_p}(\vec{r},t)+ \qquad (6.2e)$$
$$\rho_c(\vec{r},t)\frac{\partial \overrightarrow{U_c}(\vec{r},t)}{\partial t}+\rho_c(\vec{r},t)\overrightarrow{U_c}(\vec{r},t)\cdot\nabla\overrightarrow{U_c}(\vec{r},t)$$

2.重力 $\vec{f}_g$

重力加速度 $\vec{g}$ 通常均可认为是物理恒量，于是有

$$(dm_f)\vec{g}=(dm_p)\vec{g}+(dm_c)\vec{g}$$

或写成

$$\rho_f(\vec{r},t)(dV_f)\vec{g}=\rho_{NP}(dV_p)\vec{g}+\rho_{NC}(dV_c)\vec{g}$$

上式两边乘以 $\delta(\vec{R}-\vec{r})$，注意到 $\vec{g}$ 是常矢，则取平均就有

$$\vec{g}\int_{V_f}\rho_f(\vec{R},t)\delta(\vec{R}-\vec{r})dV_f=\vec{g}\int_{V_p}\rho_{NP}\delta(\vec{R}-\vec{r})dV_p+\vec{g}\int_{V_C}\rho_{NC}\delta(\vec{R}-\vec{r})dV_c=$$
$$\vec{g}\int_{V_f}\rho_{NP}\phi_p(\vec{R},t)\delta(\vec{R}-\vec{r})dV_f+\vec{g}\int_{V_f}\rho_{NC}\phi_c(\vec{R},t)\delta(\vec{R}-\vec{r})dV_f$$

于是得到

$$\vec{f}_g=\rho_f(\vec{r},t)\vec{g}=\phi_p(\vec{r},t)\rho_{NP}\vec{g}+\phi_c(\vec{r},t)\rho_{NC}\vec{g}=\rho_p(\vec{r},t)\vec{g}+\rho_c(\vec{r},t)\vec{g} \qquad (6.3)$$

3.Kelvin 力 $\vec{f}_k$

$\vec{f}_k$ 是外磁场作用于单位体积铁磁流体上的磁场力。由于磁场力只产生在固相微粒体上，对于不导磁的基载液体没有作用。所以对铁磁流体混合物而言的 Kelvin 力本质上就是一种平均的概念。

外磁场 $\overrightarrow{B_0}$ 对一个固相微粒产生的磁力 $\overrightarrow{f_{k1}}$ 是

$$\overrightarrow{f_{k1}}=\overrightarrow{m_{p1}}\cdot\nabla\overrightarrow{B_0}$$

式中 $m_{p1}$ 是一个固相微粒的磁矩，对于悬浮于基载液体中的大量固相微粒，由于热运动的影响，其磁矩的统计平均值，即有效磁矩 $(m_p)_e$ 是

$$(m_p)_e=m_{p1}L(\alpha)$$

若单位体积固相物质的极限磁化强度是 $M_p$，则 $m_{p1} = M_p V_{p1}$，从而有

$$(M_p)_e = M_p L(\alpha)$$

由于磁化强度是指单位体积物质而言的,所以体积为 $dV_f$ 的铁磁流体的磁矩必等于其所含的固相和液相的磁矩之和，即

$$\overrightarrow{M_f} dV_f = \overrightarrow{M_p} L(\alpha) dV_p + \overrightarrow{M_c} dV_c \tag{6.4a}$$

上式两边乘以 $\delta(\vec{R} - \vec{r})$，而后在邻域 $V_f$ 内取平均，就有

$$\int_{V_f} \overrightarrow{M_f}(\vec{R},t)\delta(\vec{R}-\vec{r})dV_f = \int_{V_p} \overrightarrow{M_p} L(\alpha)\delta(\vec{R}-\vec{r})dV_p + \int_{V_c} \overrightarrow{M_c}\delta(\vec{R}-\vec{r})dV_c$$

结果得磁化强度平均值的关系为

$$\overrightarrow{M_f}(\vec{r},t) = \phi_p(\vec{r},t)\overrightarrow{M_p} L(\alpha) + \phi_c(\vec{r},t)\overrightarrow{M_c} \tag{6.4b}$$

上式两边与外磁场梯度 $\nabla \overrightarrow{B_0}$ 作数性积，即得 Kelvin 力的关系

$$\overrightarrow{f_k}(\vec{r},t) = M_f(\vec{r},t) \cdot \nabla \overrightarrow{B_0}(\vec{r},t) = \phi_p(\vec{r},t)\overrightarrow{M_p} L(\alpha) \cdot \nabla \overrightarrow{B_0}(\vec{r},t) + \phi_c(\vec{r},t)\overrightarrow{M_c} \cdot \nabla \overrightarrow{B_0}(\vec{r},t) \tag{6.4c}$$

对于现有的铁磁流体，$\overrightarrow{M_c} = 0$。

4.彻体磁力矩引起的附加力 $\overrightarrow{f_L}$

单位体积铁磁流体所受到的磁力矩 $\overrightarrow{L_m}(\vec{r},t)$ 就是式（1.33）所给出的

$$\overrightarrow{L_m}(\vec{r},t) = \overrightarrow{M}(\vec{r},t) \times \overrightarrow{B_0}(\vec{r},t)$$

$\overrightarrow{L_m}(\vec{r},t)$ 既为单位体积铁磁流体所受到的磁力矩，故体积为 $dV_f$ 的铁磁流体受到的磁力矩必定等于它所包含的固相与液相所受的磁力矩之和，即

$$\overrightarrow{M}(\vec{r},t) \times \overrightarrow{B_0}(\vec{r},t) dV_f = \overrightarrow{M_p} L(\alpha) \times \overrightarrow{B_0}(\vec{r},t) dV_p + \overrightarrow{M_c} \times \overrightarrow{B_0}(\vec{r},t) dV_c$$

在邻域 $V_f$ 内取平均，就有

$$\int_{V_f} \overrightarrow{M}(\vec{R},t) \times \overrightarrow{B_0}(\vec{R},t)\delta(\vec{R}-\vec{r}) dV_f = \int_{V_p} \overrightarrow{M_p} L(\alpha) \times \overrightarrow{B_0}(\vec{R},t)\delta(\vec{R}-\vec{r})dV_p +$$
$$\int_{V_c} \overrightarrow{M_c} \times \overrightarrow{B_0}(\vec{R},t)\delta(\vec{R}-\vec{r})dV_c$$

用式（3.8）的关系代入上式右方就得到磁力矩的平均值关系

$$\overrightarrow{M}(\vec{r},t) \times \overrightarrow{B_0}(\vec{r},t) = \phi_p(\vec{r},t)\overrightarrow{M_p} L(\alpha) \times \overrightarrow{B_0}(\vec{r},t) + \phi_c(\vec{r},t)\overrightarrow{M_c} \times \overrightarrow{B_0}(\vec{r},t) \tag{6.5a}$$

对于不导磁的基载液，$\overrightarrow{M_c} = 0$，于是

$$\overrightarrow{L_m} = \overrightarrow{M}(\vec{r},t) \times \overrightarrow{B_0}(\vec{r},t) = \phi_p(\vec{r},t)\overrightarrow{M_p}L(\alpha) \times \overrightarrow{B_0}(\vec{r},t) \tag{6.5b}$$

单位体积的铁磁流体内的彻体力矩 $\overrightarrow{L_b}$ 和它衍生的附加力 $\overrightarrow{f_L}$ 的关系，就是由式（5.11c）所给出的

$$\overrightarrow{f_L}(\vec{r},t) = \frac{1}{2}\nabla \times \overrightarrow{L_b}(\vec{r},t)$$

在铁磁流体内，无论是磁力矩 $\overrightarrow{L_m}$，还是粘性力矩 $\overrightarrow{L_\tau}$，都是作用于固相微粒体之上，而固相微粒遍布于铁磁流体内部，所以 $\overrightarrow{L_m}$ 和 $\overrightarrow{L_\tau}$ 都是彻体力矩。在略去固相微粒体的旋转惯性和铁磁流体内部的质量扩散效应的情况下，式（5.8）给出

$$\overrightarrow{M} \times \overrightarrow{B_0} = 6\phi_\delta\eta_\delta(\overrightarrow{\omega_P} - \overrightarrow{\omega_C})$$

式中 $\overrightarrow{M}, \overrightarrow{B_0}, \phi_\delta, \eta_\delta, \overrightarrow{\omega_P}, \overrightarrow{\omega_C}$ 都是坐标 $\vec{r}$ 和时间 $t$ 的函数。于是有附加力为

$$\overrightarrow{f_L} = \frac{1}{2}\nabla \times (\overrightarrow{M} \times \overrightarrow{B_0}) = 3\nabla \times [\phi_\delta\eta_\delta(\overrightarrow{\omega_P} - \overrightarrow{\omega_C})] \tag{6.6}$$

5.表面力 $\overrightarrow{f_s}$

由式（2.18），单位体积铁磁流体的表面力 $\overrightarrow{f_s}$ 是

$$\overrightarrow{f_s} = \nabla \cdot \tau$$

注意到 $\overrightarrow{f_s}$ 是单位体积铁磁流体内的表面力。所以体积为 $dV_f$ 的铁磁流体之 $\overrightarrow{f_s}$ 必定等于包含在 $dV_f$ 内的体积为 $dV_p$ 的固相的 $\overrightarrow{f_{s,NP}}$ 与体积为 $dV_c$ 的液相的 $\overrightarrow{f_{s,NC}}$ 之和，即

$$\overrightarrow{f_s}(\vec{r},t)\,dV_f = \overrightarrow{f_{s,NP}}(\vec{r},t)\,dV_p + \overrightarrow{f_{s,NC}}(\vec{r},t)\,dV_c$$

在邻域 $V_f$ 内取平均，则有

$$\int_{V_f} \overrightarrow{f_s}(\vec{R},t)\delta(\vec{R}-\vec{r})dV_f = \int_{V_p} \overrightarrow{f_{s,NP}}(\vec{R},t)\delta(\vec{R}-\vec{r})dV_p + \int_{V_c} \overrightarrow{f_{s,NC}}(\vec{R},t)\delta(\vec{R}-\vec{r})dV_c$$

以式（2.18）代入，则得

$$\int_{V_f} (\nabla \cdot \tau)\delta(\vec{R}-\vec{r})dV_f = \int_{V_p} (\nabla \cdot \tau_{NP})\delta(\vec{R}-\vec{r})dV_p + \int_{V_c} (\nabla \cdot \tau_{NC})\delta(\vec{R}-\vec{r})dV_c \tag{6.7a}$$

式中，$\tau_{NP}$ 和 $\tau_{NC}$ 分别是固相和液相的表面应力，它们在微元体积 $dV_p$ 和 $dV_c$ 中都不连续，所以 $\nabla \cdot \tau_{NP}$ 与 $\nabla \cdot \tau_{NC}$ 只有其形式而无数学上的定义。使用式（3.38b）将式（6.7a）右边的两个散度的平均值转换成平均值的散度。平均值是连续的。所以它的散度有确实的数学定义，同时将属于内力或内力矩的两相之

间的相互作用分离出来。于是，式（6.7a）成为式（3.40a）的形式，即

$$\nabla \cdot \tau(\vec{r},t) = \int_{V_f} \nabla \cdot \tau(\vec{R},t) \delta(\vec{R}-\vec{r}) dV_f =$$

$$\nabla \cdot \int_{V_p} \tau_{NP}(\vec{R},t)\delta(\vec{R}-\vec{r}) dV_p + \nabla \cdot \int_{V_c} \tau_{NC}(\vec{R},t)\delta(\vec{R}-\vec{r}) dV_c +$$

$$\int_{S_{p0}} d\overrightarrow{S_{p0}} \cdot \tau_{NP}(\vec{R},t)\delta(\vec{R}-\vec{r}) + \int_{S_{c0}} d\overrightarrow{S_{c0}} \cdot \tau_{NC}(\vec{R},t)\delta(\vec{R}-\vec{r}) + \tag{6.7b}$$

$$\sum_N \int_{S_{p1}} d\overrightarrow{S_{p1}} \cdot \tau_{NP}(\vec{R},t)\delta(\vec{R}-\vec{r}) - \sum_N \int_{S_{p1}} d\overrightarrow{S_{p1}} \cdot \tau_{NC}(\vec{R},t)\delta(\vec{R}-\vec{r})$$

式中 $\overrightarrow{S_{p0}}$ 与 $\overrightarrow{S_{c0}}$ 是邻域 $V_f$ 外表面上固相和液相各占的面积，$\overrightarrow{S_{p1}}$ 是单个固相微粒的表面积。$N$ 是在铁磁流体体积 $V_f$ 内固相微粒数目。设 $\overrightarrow{S_{p0}}$ 和 $\overrightarrow{S_{c0}}$ 的单位法向矢量为 $\vec{n^0}$，则由式（3.6d）有

$$\vec{n^0} \cdot \tau_{NP}(\vec{R},t)\delta(\vec{R}-\vec{r_s}) = \frac{1}{2}\left[\vec{n^0_{s-0}} \cdot \tau_{NP}(\vec{r_s}-0,t) + \vec{n^0_{s+0}} \cdot \tau_{NP}(\vec{r_s}+0,t)\right]\delta(\vec{R}-\vec{r_s})$$

无论 $V_f$ 内外的物质是否相同，在表面 $S_{p0}$ 之内外侧，即区间 $[r_s-0, r_s+0]$ 内应力只能有一个，即

$\tau_{NP}(\vec{r_s}-0,t) = \tau_{NP}(\vec{r_s}+0,t)$，并且 $-\vec{n^0_{s+0}} = \vec{n^0_{s-0}}$，于是上式之右方等于零，即

$$\int_{S_{p0}} d\overrightarrow{S_{p0}} \cdot \overrightarrow{\tau_{NP}}(\vec{R},t)\delta(\vec{R}-\vec{r_s}) = 0$$

同样也有

$$\int_{S_{c0}} d\overrightarrow{S_{c0}} \cdot \overrightarrow{\tau_{NC}}(\vec{R},t)\delta(\vec{R}-\vec{r_s}) = 0$$

于是式（6.7b）成为

$$\vec{f_s}(\vec{r},t) = \nabla \cdot \tau(\vec{r},t) = \int_{V_f} \nabla \cdot \tau(\vec{R},t)\delta(\vec{R}-\vec{r}) dV_f =$$

$$\nabla \cdot \int_{V_p} \tau_{NP}(\vec{R},t)\delta(\vec{R}-\vec{r}) dV_p + \sum_N \int_{S_{p1}} d\overrightarrow{S_{p1}} \cdot \tau_{NP}(\vec{R},t)\delta(\vec{R}-\vec{r}) + \tag{6.7c}$$

$$\nabla \cdot \int_{V_c} \tau_{NC}(\vec{R},t)\delta(\vec{R}-\vec{r}) dV_c - \sum_N \int_{S_{p1}} d\overrightarrow{S_{p1}} \cdot \tau_{NC}(\vec{R},t)\delta(\vec{R}-\vec{r})$$

将以上所得到的式（6.2e）、式（6.3）、式（6.4c）、式（6.6）与式（6.7c）代入到原始形式的动量守恒方程（6.1a）之中，就得

$$\phi_p \rho_{NP} \frac{\partial \overrightarrow{U_p}}{\partial t} + \phi_p \rho_{NP} \overrightarrow{U_p} \cdot \nabla \overrightarrow{U_p} + \phi_c \rho_{NC} \frac{\partial \overrightarrow{U_c}}{\partial t} + \phi_c \rho_{NC} \overrightarrow{U_c} \cdot \nabla \overrightarrow{U_c} =$$

$$\phi_p \rho_{NP} \vec{g} + \phi_c \rho_{NC} \vec{g} + \phi_p \overrightarrow{M_p} L(\alpha) \cdot \nabla \overrightarrow{B_0} + \frac{1}{2}\nabla \times \left[\phi_p \overrightarrow{M_p} L(\alpha) \times \overrightarrow{B_0}\right] +$$

$$\nabla \cdot \int_{V_p} \tau_{NP}(\vec{R},t)\delta(\vec{R}-\vec{r}) dV_p + \sum_N \int_{S_{p1}} d\overrightarrow{S_{p1}} \cdot \tau_{NP}(\vec{R},t)\delta(\vec{R}-\vec{r}) + \tag{6.8a}$$

$$\nabla \cdot \int_{V_c} \tau_{NC}(\vec{R},t)\delta(\vec{R}-\vec{r}) dV_c - \sum_N \int_{S_{p1}} d\overrightarrow{S_{p1}} \cdot \tau_{NC}(\vec{R},t)\delta(\vec{R}-\vec{r})$$

将上述方程式（6.8a）按固液两相拆开，就得到固相和液相的平均动量方程为

$$\phi_p \rho_{NP} \frac{\partial \overrightarrow{U_p}}{\partial t} + \phi_p \rho_{NP} \overrightarrow{U_p} \cdot \nabla \overrightarrow{U_p} = \phi_p \rho_{NP} \vec{g} + \phi_p \overrightarrow{M_p} L(\alpha) \cdot \nabla \overrightarrow{B_0} + \frac{1}{2} \nabla \times [\phi_p \overrightarrow{M_p} L(\alpha) \times \overrightarrow{B_0}] +$$
$$\nabla \cdot \int_{V_p} \tau_{NP} \delta(\vec{R} - \vec{r}) dV_p + \sum_N \int_{S_{p1}} d\overrightarrow{S_{p1}} \cdot \tau_{NP}(\vec{R} - \vec{r}) \tag{6.8b}$$

$$\phi_c \rho_{NC} \frac{\partial \overrightarrow{U_c}}{\partial t} + \phi_c \rho_{NC} \overrightarrow{U_c} \cdot \nabla \overrightarrow{U_c} = \phi_c \rho_{NC} \vec{g} + \nabla \cdot \int_{V_c} \tau_{NC} \delta(\vec{R} - \vec{r}) dV_c + \sum_N \int_{S_{p1}} d\overrightarrow{S_{p1}} \cdot \tau_{NC}(\vec{R} - \vec{r}) \tag{6.8c}$$

在一般情况下，方程（6.8b）和方程（6.8c）中的 $\phi_p$、$\phi_c$、$\overrightarrow{U_p}$、$\overrightarrow{U_c}$、$\overrightarrow{B_0}$、$\alpha$、$\tau_{NP}$ 和 $\tau_{NC}$ 等均系坐标 $\vec{r}$ 和时间 $t$ 的函数。关于表面粘性应力 $\tau_{NP}$ 平均值的散度和它的面积分是

1. $\nabla \cdot \int_{V_p} \tau_{NP} \delta(\vec{R} - \vec{r}) dV_p$

将表面粘性应力的关系式（2.47）代入积分内，就有

$$\nabla \cdot \int_{V_p} \tau_{NP}(\vec{R}, t) \delta(\vec{R} - \vec{r}) dV_p = \nabla \cdot \int_{V_f} \tau_{NP}(\vec{R}, t) \delta(\vec{R} - \vec{r}) \phi_p(\vec{R}, t) dV_f =$$
$$\nabla \cdot \left\{ \phi_p(\vec{r}, t) \left[ -p(\vec{r}, t)\delta_{ij} + \eta_{NP} \left( \frac{\partial u_i}{\partial x_j} + \frac{\partial u_j}{\partial x_i} \right)_p - \frac{2}{3} \eta_{NP} \nabla \cdot \overrightarrow{U_p}(\vec{r}, t)\delta_{ij} \right] \right\}$$

于是得

$$\nabla \cdot \int_{V_p} \tau_{NP}(\vec{R}, t) \delta(\vec{R} - \vec{r}) dV_p = -\nabla(\phi_p p) + \frac{\partial}{\partial x_j} \left[ \phi_p \eta_{NP} \left( \frac{\partial u_i}{\partial x_j} + \frac{\partial u_j}{\partial x_i} \right)_p \right] - \nabla \left( \frac{2}{3} \phi_p \eta_{NP} \nabla \cdot \overrightarrow{U_p} \right) \tag{6.9a}$$

同样，对于液相也有

$$\nabla \cdot \int_{V_c} \tau_{NC}(\vec{R}, t) \delta(\vec{R} - \vec{r}) dV_c = -\nabla(\phi_c p) + \frac{\partial}{\partial x_j} \left[ \phi_c \eta_{NC} \left( \frac{\partial u_i}{\partial x_j} + \frac{\partial u_j}{\partial x_i} \right)_c \right] - \nabla \left( \frac{2}{3} \phi_c \eta_c \nabla \cdot \overrightarrow{U_c} \right) \tag{6.9b}$$

2. $\sum_N \int_{S_{p1}} d\overrightarrow{S_{p1}} \cdot \tau_{NP} \delta(\vec{R} - \vec{r})$

对于铁磁流体的固相微粒，总是假设成圆球形状。对尺寸为纳米级的固相微粒，无论它与液相的相对速度如何，脱离不了低 Re 数运动的范畴。

在第二章中，已经得出在无界流中，圆球作平动运动的表面应力之分离变量形式的解是

$$\tau_{NP} = -p\delta_{ij} + \tau'_{rr} \cos\theta - \tau_{r\theta} \sin\theta$$

上述应力在圆球形固相微粒表面上的积分是

$$\sum_N \int_{S_{p1}} d\overrightarrow{S_{p1}} \cdot \tau_{NP} \delta(\vec{R} - \vec{r}) =$$
$$-\sum_N \int_{S_{p1}} d\overrightarrow{S_{p1}} \cdot (p\delta_{ij}) \delta(\vec{R} - \vec{r}) + \sum_N \int_{S_{p1}} d\overrightarrow{S_{p1}} \cdot (\tau'_{rr} \cos\theta - \tau'_{r\theta} \sin\theta) \delta(\vec{R} - \vec{r}) \tag{A}$$

上式右方第一个积分，利用散度定理得

$$-\sum_N \int_{S_{p1}} d\overrightarrow{S_{p1}} \cdot (p\delta_{ij})\delta(\vec{R}-\vec{r}) = -\int_{V_p} \nabla_R \cdot [p\delta_{ij}\delta(\vec{R}-\vec{r})]dV_p =$$

$$-\int_{V_p} \delta(\vec{R}-\vec{r})\nabla_R \cdot (p\delta_{ij})\,dV_p - \int_{V_p} p\delta_{ij} \cdot \nabla_R \delta(\vec{R}-\vec{r})\,dV_p =$$

$$-\int_{V_p} \delta(\vec{R}-\vec{r})\nabla_R \cdot (p\delta_{ij})\,dV_p + \int_{V_p} p\delta_{ij} \cdot \nabla_r \delta(\vec{R}-\vec{r})\,dV_p =$$

$$-\int_{V_p} \delta(\vec{R}-\vec{r})(\nabla_R p)dV_p + \nabla_r \cdot \int_{V_p} \delta(\vec{R}-\vec{r})p\delta_{ij}\,dV_p$$

于是得

$$-\sum_N \int_{S_{p1}} dS_{p1} \cdot (p\delta_{ij})\delta(\vec{R}-\vec{r}) = -\phi_p \nabla p + \nabla(\phi_p p) \qquad (6.10a)$$

式（A）右方第二个积分求和式，即

$$\sum_N \int_{S_{p1}} d\overrightarrow{S_{p1}} \cdot (\tau'_{rr}\cos\theta - \tau'_{r\theta}\sin\theta)\delta(\vec{R}-\vec{r})$$

注意到在参照坐标系 $O'\vec{r}$ 中，任何一个固相微粒只是一个点，其代表位置即微粒的中心 $O$ 点（参见图 2-9 和图 2-10），而积分中的被积函数都以微粒中心 $O$ 为原点的球坐标系得出，其积分变量是 $dS_{p1} = (2\pi r_p \sin\theta)r_p\,d\theta$，实际积分变量与 $O'\vec{r}$ 坐标系无关，从而可以将 $\delta(\vec{R}-\vec{r})$ 提到积分号的外面，即

$$\sum_{N'} \int_{S_{p1}} d\overrightarrow{S_{p1}} \cdot (\tau'_{rr}\cos\theta - \tau'_{r\theta}\sin\theta)\delta(\vec{R}-\vec{r}) = \sum_N \delta(\vec{R}-\vec{r})\int_{S_{p1}} d\overrightarrow{S_{p1}} \cdot (\tau'_{rr}\cos\theta - \tau'_{r\theta}\sin\theta) \qquad (B)$$

式（B）的右方是在微粒表面上的积分，故在式（2.87b）与式（2.88b）中取 $r = r_p$，得

$$\tau'_{rr} = -\frac{3}{2r_p}\eta_c(\overrightarrow{U_p}-\overrightarrow{U_c}), \qquad \tau'_{r\theta} = \frac{3}{2r_p}\eta_c(\overrightarrow{U_p}-\overrightarrow{U_c})$$

积分结果已在式（2.89c）中给出，即微粒运动的阻力是

$$\overrightarrow{F_{cp}} = -6\pi\eta_c r_p(\overrightarrow{U_p}-\overrightarrow{U_c}) \qquad (C)$$

将式（C）代入式（B）的右方，即得

$$\sum \delta(\vec{R}-\vec{r})6\pi\eta_c r_p(\overrightarrow{U_p}-\overrightarrow{U_c})$$

将上式求和号内同时乘除以 $V_{p1} = (4/3)\pi r_p^3$，就得

$$\sum_N \delta(\vec{R}-\vec{r})\frac{9}{2}\eta_c \frac{1}{r_p^2}(\overrightarrow{U_c}-\overrightarrow{U_p})V_{p1}$$

$V_{p1}$ 相对于 $V_f$ 内所含的 $V_p$ 是极其微小的，可以将它视为 $dV_p$，而 $N$ 是很大的数，所以求和号可以写成极限状况，即积分

$$\sum_N \delta(\vec{R}-\vec{r})\frac{9}{2}\eta_c \frac{1}{r_p^2}(\overrightarrow{U_c}-\overrightarrow{U_p})V_{p1} = \int_{V_p} \delta(\vec{R}-\vec{r})\frac{9}{2}\eta_c \frac{1}{r_p^2}(\overrightarrow{U_c}-\overrightarrow{U_p})\,dV_p = \frac{9}{2}\phi_p\eta_c \frac{1}{r_p^2}(\overrightarrow{U_c}-\overrightarrow{U_p}) \qquad (D)$$

将式（D）代入式（B），就有

$$\sum_N \int_{S_{p1}} d\vec{S_{p1}} (\tau'_{rr} \cos\theta - \tau'_{r\theta} \sin\theta) \delta(\vec{R} - \vec{r}) = \frac{9}{2} \phi_p \eta_c \frac{1}{r_p^2} (\overrightarrow{U_c} - \overrightarrow{U_p}) \tag{6.10b}$$

再将式（6.10a）与式（6.10b）代式（A）的右方，就得到

$$\sum_N \int_{S_{p1}} d\vec{S_{p1}} \cdot \tau_{NP} \delta(\vec{R} - \vec{r}) = -\phi_p \nabla p + \nabla(\phi_p p) + \frac{9}{2} \phi_p \eta_c \frac{1}{r_p^2} (\overrightarrow{U_c} - \overrightarrow{U_p}) \tag{6.11a}$$

同样，对于液相也有

$$-\sum_N \int_{S_{p1}} d\vec{S_{p1}} \cdot \tau_{NC} \delta(\vec{R} - \vec{r}) = \sum_N \int_{S_{c1}} d\vec{S_{c1}} \cdot \tau_{NC} \delta(\vec{R} - \vec{r})$$
$$= -\phi_c \nabla p + \nabla(\phi_c p) + \frac{9}{2} \phi_p \eta_c \frac{1}{r_p^2} (\overrightarrow{U_c} - \overrightarrow{U_p}) \tag{6.11b}$$

在方程（6.11b）中已经考虑到 $\tau_{NC}$ 与 $\tau_{NP}$ 大小相等而方向相反。并且 $d\vec{S_{c1}} = -d\vec{S_{p1}}$。

将式（6.9a）、式（6.11a）与式（6.9b）、式（6.11b）分别代入方程式（6.8b）与式（6.8c）中，就得到以磁平衡流为例的铁磁流体分相流平均动量守恒方程。

1.固相微粒流的平均动量方程

$$\phi_p \rho_{NP} \frac{\partial \overrightarrow{U_p}}{\partial t} + \phi_p \rho_{NP} \overrightarrow{U_p} \cdot \nabla \overrightarrow{U_p} = \phi_p \rho_{NP} \vec{g} + \phi_p \overrightarrow{M_p} L(\alpha) \cdot \nabla \overrightarrow{B_0} + \frac{1}{2} \nabla \times [\phi_p \overrightarrow{M_p} L(\alpha) \times \overrightarrow{B_0}] -$$
$$\phi_p \nabla p + \frac{\partial}{\partial x_j} \left[ \phi_p \eta_{NP} \left( \frac{\partial u_i}{\partial x_j} + \frac{\partial u_j}{\partial x_i} \right)_p \right] - \nabla \left( \frac{2}{3} \phi_p \eta_{NP} \nabla \cdot \overrightarrow{U_p} \right) + \frac{9}{2} \phi_p \eta_c \frac{1}{r_p^2} (\overrightarrow{U_c} - \overrightarrow{U_p}) \tag{6.12a}$$

2.液相流的平均动量方程

$$\phi_c \rho_{NC} \frac{\partial \overrightarrow{U_c}}{\partial t} + \phi_c \rho_{NC} \overrightarrow{U_c} \cdot \nabla \overrightarrow{U_c} = \phi_c \rho_{NC} \vec{g} - \phi_c \nabla p + \frac{\partial}{\partial x_j} \left[ \phi_c \eta_c \left( \frac{\partial u_i}{\partial x_j} + \frac{\partial u_j}{\partial x_i} \right)_c \right] -$$
$$\nabla \left( \frac{2}{3} \phi_c \eta_c \nabla \cdot \overrightarrow{U_c} \right) + \frac{9}{2} \phi_p \eta_c \frac{1}{r_p^2} (\overrightarrow{U_c} - \overrightarrow{U_p}) \tag{6.12b}$$

注意，式（6.12a）与式（6.12b）是在体积为 $V_f$ 内的固相和液相的平均动量方程。若将式（6.12a）两边通除以 $\phi_p$，及式（6.12b）两边通除以 $\phi_c$，就得到单位体积的固相微粒流和单位体积液相流的平均动量方程。两者均以磁平衡流为例。

在满足式（5.17）的前提条件下，则方程（6.12a）中的 Kelvin 力与粘性力相抵清，而方程（6.12b）右方末项换成磁性的 Kelvin 力。于是单位体积的平均动量方程为：

$$\rho_{NP} \frac{\partial \overrightarrow{U_p}}{\partial t} + \rho_{NP} \overrightarrow{U_p} \cdot \nabla \overrightarrow{U_p} = \rho_{NP} \vec{g} + \frac{1}{2\phi_p} \nabla \times [\phi_p \overrightarrow{M_p} L(\alpha) \times \overrightarrow{B_0}] - \nabla p +$$
$$\frac{1}{\phi_p} \frac{\partial}{\partial x_j} \left[ \phi_p \eta_{NP} \left( \frac{\partial u_i}{\partial x_j} + \frac{\partial u_j}{\partial x_i} \right)_p \right] - \frac{2}{3\phi_p} \nabla(\phi_p \eta_{NP} \nabla \cdot \overrightarrow{U_p}) \tag{6.13a}$$

$$\rho_{NC}\frac{\partial \overrightarrow{U_c}}{\partial t} + \rho_{NC}\overrightarrow{U_c}\cdot\nabla\overrightarrow{U_c} = \rho_{NC}\overrightarrow{g} - \nabla p + \frac{1}{\phi_c}\frac{\partial}{\partial x_j}\left[\phi_c\eta_c\left(\frac{\partial u_i}{\partial x_j} + \frac{\partial u_j}{\partial x_i}\right)_c\right] -$$

$$-\frac{2}{3\phi_c}\nabla(\phi_c\eta_c\nabla\cdot\overrightarrow{U_c}) - \frac{\phi_p}{\phi_c}\overrightarrow{M_p}L(\alpha)\cdot\nabla\overrightarrow{B_0}$$

(6.13b)

在固相微粒流的平均动量方程中，即在式（6.12a）与式（6.13a）中，存在粘性系数 $\eta_{NP}$，它是微粒流的粘度。固相微粒流可以与气体相比拟。气体的粘性是由于气体分子在热运动中频繁碰撞产生动量交换的结果。固相微粒在铁磁流体中同样存在热运动，故而固相微粒流的粘性能够按照分子运动论的方法处理。分子运动论的输运理论所给出的气体粘性系数的公式是

$$\eta = \frac{2\alpha}{3\pi d^2}\sqrt{\frac{mk_0T}{\pi}}$$

式中，$\alpha$ 是一常数因子，其值为 1.46，$d$ 是气体分子的有效直径，$m$ 是气体分子的质量，$k_0T$ 即热运动动能。于是利用上式可以估计固相微粒流与空气的粘性之比。由

$$\eta_{NP} = \left(\frac{d}{d_p}\right)^2\sqrt{\frac{m_pT_p}{mT}}\eta_{air}$$

设固相微粒流的材料是 $Fe_3O_4$，微粒直径 $d_p = 8\times10^{-7}$ cm，$\rho_{NP} = 5.24\,g/cm^3$，$m_p$ 是一个微粒的质量，

$m_p = \rho_{NP}(\pi d_p^3/6) = 1.4\times10^{-18}$ g，空气分子的有效直径 $d = 3.72\times10^{-8}$ cm，空气的分子量 $M = 28.97\,g/mol$，

Avogadro 数 $N_0 = 6.02\times10^{23}\,mol^{-1}$，于是空气分子的质量 $m = M/N_0 = 4.81\times10^{-23}\,g$，在相同的温度下，即

$T_p = T$，得出

$$\eta_{NP} = 0.368\eta_{air}$$

由此可见，固相微粒流的粘度只有空气的1/3略强。计算中还没有考虑分散剂链子分子使固相微粒直径增大的作用。基于这样的比较，可以认为固相微粒流如同空气那样，是能够当作理想流体处理的，从而在动量方程中略去含 $\eta_{NP}$ 的粘性项。

对于磁松弛流或磁冻结流，则将方程（6.12a）中的磁力项换成相应磁况下的关系式。

6.3 铁磁流体分相流的平均质量守恒方程
6.3.1 概述

组成铁磁流体的基载液和固相微粒都是不可压缩的物质，所以铁磁流体也是不可压缩的。

若将一块体积为 $V_f$ 的不均匀的铁磁流体，划分成许多体积相等的小块 $\Delta V_f$，则每个小块内的组成

是 $\Delta V_f = \Delta V_p + \Delta V_c$，虽然 $\Delta V_f$ 都相同，但各个 $\Delta V_f$ 内所含的 $\Delta V_p$ 与 $\Delta V_c$ 两者的配比不一样。也就是说，

$\Delta V_f$ 是常数，而其中所含的 $\Delta V_p$ 和 $\Delta V_c$ 却可以是变数。

但是，固相和液相两者的物质密度很不相同，所以 $\Delta V_p$ 和 $\Delta V_c$ 配比不同，导致 $\Delta V_f$ 内铁磁流体的密度变化。这就是说，铁磁流体的体积是不可压缩的。但其密度却是可变的。在这一点上，铁磁流体与普通液固两相流有相同的共性。

6.3.2 铁磁流体分相流的平均质量守恒方程

由铁磁流体混合流的质量守恒方程（5.33a）

$$\frac{\partial \rho_f(\vec{r},t)}{\partial t} + \nabla \cdot [\rho_f(\vec{r},t)\overrightarrow{U_f}(\vec{r},t)] = 0$$

将上式在邻域 $V_f$ 内取平均，则有

$$\int_{V_f} \frac{\partial \rho_f(\vec{R},t)}{\partial t} \delta(\vec{R}-\vec{r})\, dV_f + \int_{V_f} \nabla \cdot [\rho_f(\vec{R},t)\overrightarrow{U_f}(\vec{R},t)] \delta(\vec{R}-\vec{r})\, dV_f = 0 \qquad (E)$$

1. $\int_{V_f} \frac{\partial \rho_f(\vec{R},t)}{\partial t} \delta(\vec{R}-\vec{r})\, dV_f$

由 $dm_f = dm_p + dm_c$，得 $\rho_f(\vec{r},t)\, dV_f = \rho_{NP} dV_p + \rho_{NC} dV_c$，注意到 $\partial/\partial t$ 不对 $V_f$ 作用，故

$$\int_{V_f} \frac{\partial \rho_f(\vec{R},t)}{\partial t} \delta(\vec{R}-\vec{r})\, dV_f = \frac{\partial}{\partial t} \int_{V_f} \rho_f(\vec{R},t)\delta(\vec{R}-\vec{r})\, dV_f \qquad (F)$$

由质量平均式

$$\int_{V_f} \rho_f(\vec{R},t)\, \delta(\vec{R}-\vec{r})\, dV_f = \int_{V_p} \rho_{NP}\delta(\vec{R}-\vec{r})dV_p + \int_{V_c} \rho_{NC}\delta(\vec{R}-\vec{r})dV_c$$

两边对时间 $t$ 取导数

$$\frac{\partial}{\partial t} \int_{V_f} \rho_f(\vec{R},t)\delta(\vec{R}-\vec{r})dV_f = \frac{\partial}{\partial t} \int_{V_p} \rho_{NP}\delta(\vec{R}-\vec{r})dV_p + \frac{\partial}{\partial t} \int_{V_c} \rho_{NC}\delta(\vec{R}-\vec{r})dV_c$$

利用式（F）与式（3.8）就有

$$\int_{V_f} \frac{\partial \rho_f(\vec{R},t)}{\partial t} \delta(\vec{R}-\vec{r})dV_f = \frac{\partial}{\partial t} \int_{V_f} \rho_{NP}\delta(\vec{R}-\vec{r})\phi_p(\vec{R},t)dV_f + \frac{\partial}{\partial t} \int_{V_f} \rho_{NC}\delta(\vec{R}-\vec{r})\phi_c(\vec{R},t)dV_f$$

注意 $\rho_{NP}$ 和 $\rho_{NC}$ 均是常数，于是

$$\int_{V_f} \frac{\partial \rho_f(\vec{R},t)}{\partial t} \delta(\vec{R}-\vec{r})dV_f = \rho_{NP}\frac{\partial \phi_p(\vec{r},t)}{\partial t} + \rho_{NC}\frac{\partial \phi_c(\vec{r},t)}{\partial t} \qquad (6.14a)$$

2. $\int_{V_f} \nabla \cdot [\rho_f(\vec{R},t)\overrightarrow{U_f}(\vec{R},t)]\delta(\vec{R}-\vec{r})dV_f$

将动量守恒式 $(dm_f)\overrightarrow{U_f} = (dm_p)\overrightarrow{U_p} + (dm_c)\overrightarrow{U_c}$ 写成

$$(\rho_f \, dV_f)\overrightarrow{U_f} = (\rho_{NP} \, dV_p)\overrightarrow{U_p} + (\rho_{NC} \, dV_c)\overrightarrow{U_c}$$

上式两边取散度，而后在邻域 $V_f$ 内平均，就得

$$\int_{V_f} [\nabla_R \cdot (\rho_f \overrightarrow{U_f})]\delta(\vec{R}-\vec{r})dV_f =$$
$$\int_{V_p} [\nabla_R \cdot (\rho_{NP}\overrightarrow{U_p})]\delta(\vec{R}-\vec{r})dV_p + \int_{V_c} [\nabla_R \cdot (\rho_{NC}\overrightarrow{U_c})]\delta(\vec{R}-\vec{r})dV_c \qquad \text{(G)}$$

利用式（3.38a）与式（3.38b）得出

$$\int_{V_f} [\nabla_R \cdot (\rho_{NP}\overrightarrow{U_p})]\delta(\vec{R}-\vec{r})dV_p = \nabla_r \cdot \int_{V_p} \rho_{NP}\overrightarrow{U_p}\,\delta(\vec{R}-\vec{r})dV_p + \int_{S_{p0}} d\overrightarrow{S_{p0}} \cdot \rho_{NP}\overrightarrow{U_p}\,\delta(\vec{R}-\vec{r}) +$$
$$\sum_N \int_{S_{p1}} d\overrightarrow{S_{p1}} \cdot \rho_{NP}\overrightarrow{U_p}\,\delta(\vec{R}-\vec{r})$$

同样，可有

$$\int_{V_f} [\nabla_R \cdot \rho_{NC}\overrightarrow{U_c}]\delta(\vec{R}-\vec{r})\,dV_c = \nabla_r \cdot \int_{V_c} \rho_{NC}\overrightarrow{U_c}\,\delta(\vec{R}-\vec{r})\,dV_c + \int_{S_{c0}} d\overrightarrow{S_{c0}} \cdot \rho_{NC}\overrightarrow{U_c}\,\delta(\vec{R}-\vec{r}) +$$
$$\sum_N \int_{S_{c1}} d\overrightarrow{S_{c1}} \cdot \rho_{NC}\overrightarrow{U_c}\,\delta(\vec{R}-\vec{r})$$

将以上两式代入式（G）之右方，就得

$$\int_{V_f} [\nabla_R \cdot (\rho_f \overrightarrow{U})]\delta(\vec{R}-\vec{r})dV_f = \nabla_r \cdot \int_{V_p} \rho_{NP}\overrightarrow{U_p}\,\delta(\vec{R}-\vec{r})dV_p +$$
$$\int_{S_{p0}} d\overrightarrow{S_{p0}} \cdot \rho_{NP}\overrightarrow{U_p}\,\delta(\vec{R}-\vec{r}) + \sum_N \int_{S_{p1}} d\overrightarrow{S_{p1}} \cdot \rho_{NP}\overrightarrow{U_p}\,\delta(\vec{R}-\vec{r}) +$$
$$\nabla_r \cdot \int_{V_c} \rho_{NC}\overrightarrow{U_c}\,\delta(\vec{R}-\vec{r})\,dV_c + \qquad \text{(6.14b)}$$
$$\int_{S_{c0}} d\overrightarrow{S_{c0}} \cdot \rho_{NC}\overrightarrow{U_c}\,\delta(\vec{R}-\vec{r}) + \sum_N \int_{S_{c1}} d\overrightarrow{S_{c1}} \cdot \rho_{NC}\overrightarrow{U_c}\,\delta(\vec{R}-\vec{r})$$

将式（6.14a）和（6.14b）代入式（E）之左边，而后按固相和液相拆开，并且注意 $\rho_{NP} = \text{const}$，$\rho_{NC} = \text{const}$
于是得

$$\frac{\partial \phi_p}{\partial t} + \nabla \cdot \int_{V_p} \overrightarrow{U_p}\,\delta(\vec{R}-\vec{r})\,dV_p + \int_{S_{p0}} d\overrightarrow{S_{p0}} \cdot \overrightarrow{U_p}\,\delta(\vec{R}-\vec{r}) + \sum_N \int_{S_{p1}} d\overrightarrow{S_{p1}} \cdot \overrightarrow{U_p}\,\delta(\vec{R}-\vec{r}) = 0 \qquad \text{(6.15a)}$$

$$\frac{\partial \phi_c}{\partial t} + \nabla \cdot \int_{V_c} \overrightarrow{U_c}\,\delta(\vec{R}-\vec{r})\,dV_c + \int_{S_{c0}} d\overrightarrow{S_{c0}} \cdot \overrightarrow{U_c}\,\delta(\vec{R}-\vec{r}) + \sum_N \int_{S_{c1}} d\overrightarrow{S_{c1}} \cdot \overrightarrow{U_c}\,\delta(\vec{R}-\vec{r}) = 0 \qquad \text{(6.15b)}$$

在方程（6.15a）和方程（6.15b）的左方第三项积分，它是在邻域 $V_f$ 外表面上的积分。设 $V_f$ 外表面的坐标是 $r_s$，同时以 $U$ 代表 $U_p$ 和 $U_c$，由式（3.6d）有

$$\overrightarrow{n_s^0} \cdot \overrightarrow{U}(\vec{R},t)\delta(\vec{R}-\vec{r_s}) = \frac{1}{2}\left[\overrightarrow{n_{s-0}^0} \cdot \overrightarrow{U}(\vec{r_s}-0,t) + \overrightarrow{n_{s+0}^0} \cdot \overrightarrow{U}(\vec{r_s}+0,t)\right]\delta(\vec{R}-\vec{r_s}) \qquad \text{(H)}$$

式中 $\overrightarrow{n_s^0}$ 是 $V_f$ 外表面的外向法线单位矢。$\overrightarrow{n_s^0} = \overrightarrow{n_{s-0}^0} = -\overrightarrow{n_{s+0}^0}$ 若 $V_f$ 的表面是可以穿透的，则因区间 $[\overrightarrow{r_{s-0}}, \overrightarrow{r_{s+0}}]$ 的厚度是零，不能贮存任何物理量；若它是不能穿透的，则不会发生断开和错动。所以，在任何情况下都有

$$\overrightarrow{U}(\overrightarrow{r_s} - 0, t) = \overrightarrow{U}(\overrightarrow{r_s} + 0, t)$$

于是式（H）给出

$$\overrightarrow{n_s^0} \cdot \overrightarrow{U}(\overrightarrow{R}, t)\delta(\overrightarrow{R} - \overrightarrow{r_s}) = 0$$

在方程（6.15a）和方程（6.15b）的左方第三求和号内的面积分，它是在单个微粒表面上的积分，微粒是固体，在平动运动中，表面上各点的速度都是相同的 $\overrightarrow{U_p}$。由图 6-1 可见，若微粒表面的法向单位矢为 $\overrightarrow{n_p^0}$，则 $d\overrightarrow{S_{p1}} \cdot \overrightarrow{U_p} = dS_p \overrightarrow{n_p^0} \cdot \overrightarrow{U_p} = (2\pi r_p \sin\theta)(r_p d\theta)(U_p \cos\theta)$，所以积分变量是和 $\overrightarrow{R}$、$\overrightarrow{r}$ 无关的 $\theta$，因此

$$\int_{S_{p1}} d\overrightarrow{S_{p1}} \cdot \overrightarrow{U_p}(\overrightarrow{R} - \overrightarrow{r}) = \delta(\overrightarrow{R} - \overrightarrow{r})U_p \int_0^\pi 2\pi r_p^2 \sin\theta \cos\theta \, d\theta = 0$$

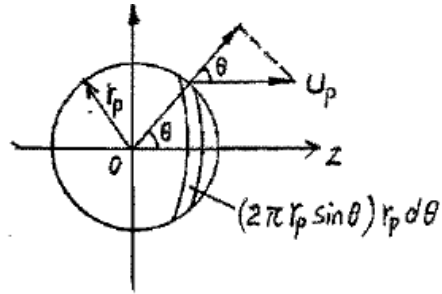

图 6-1 固相微粒的平动运动

方程（6.15a）与方程（6.15b）中第三项和第四项面积分均等于零，从而有

$$\frac{\partial \phi_p(\overrightarrow{r}, t)}{\partial t} + \nabla \cdot [\phi_p(\overrightarrow{r}, t)\overrightarrow{U_p}(\overrightarrow{r}, t)] = 0 \tag{6.16a}$$

$$\frac{\partial \phi_c(\overrightarrow{r}, t)}{\partial t} + \nabla \cdot [\phi_c(\overrightarrow{r}, t)\overrightarrow{U_c}(\overrightarrow{r}, t)] = 0 \tag{6.16b}$$

方程（6.16a）与方程（6.16b）分别是铁磁流体分相流的固相微粒流和液相流的质量守恒方程。

## 6.4 铁磁流体分相流的能量守恒方程
### 6.4.1 概述

铁磁流体能量方程的核心问题是温度。温度在铁磁流体中的重要性有二：改变铁磁流体的磁化强度和铁磁流体的粘度。这两者是相互关联的。温度使两者发生变化的重要原因是固相微粒在铁磁流体内热运动能量正比于温度而变化。

铁磁流体分相流有几个特别的问题需要说明。

①磁松弛。分相流同样存在三种磁况下的流动。所谓磁况即指磁松弛、磁平衡和磁冻结。虽然液相自身没有导磁性质，但是，受固液两相之间存在力学和热学上的相互作用的影响，故而在三种磁况下，液相流的运动状态各不相同。三种磁况的力学效应已经表现在动量守恒方程中。就是磁力所作的

功转化为流动的动能。这种功能的转化在磁和动能之间，不关系到热，所以在能量方程中不出现磁力的功项。

但是，由第一章知道铁磁流体的比热容由两种热容组成，即温度比热容和磁比热容。所以无论铁磁流体混合流还是分相流，它们的内能包括热内能和磁化功转换来的磁内能。热内能和磁内能是互相关联且此消彼涨的。它们在铁磁流体内能中所占比例的分配，直接影响流动的温度。

②热松弛。在混合流中，固相和液相之间的热交换是属于内部热流问题，不改变流动整体的平均热状态。但在分相流中，固相和液相之间的换热，对于每相流动都是与外界的换热，从而成为影响分相流热能状态的重要因素之一。两相之间保持热交换的分相流可称之为热松弛流。

但是，在多数情况下，由于铁磁流体的固相微粒的粒度实在很小，它与周围液相换热的相对面积极其巨大，在第 1-1 节中已经估算过，每毫升体积的铁磁流体中固相微粒与周围液相的接触面积可以达到 45 平方米之谱。所以固相微粒与周围液相介质实现热平衡只是极短促的一瞬间。只有在极高流速下，铁磁流体运动的特征时间使固液两相来不及热平衡。但这样的高速流动，在铁磁流体中实属罕见，甚至在实际中难于实现。铁磁流体固相微粒流和液相流之间处于热平衡，可以说是一种常态。

③机械能耗散。在混合流中，两相间运动的滞后，以及它引起的摩擦均不能显现。这是混合流的缺陷。对于分相流，两相间的平动运动和转动运动的滞后，在两相界面上产生的摩擦力和摩擦力矩所作之摩擦功，将不可逆地转化为摩擦热，它与单相流动的动能，即机械能。这种机械能转化为摩擦热能，按热力学第二定律是自发的不可逆过程，它与单相流动的内摩擦本质上是一样的。因此，也属于流体运动中的机械能耗散。由于铁磁流体分相流比普通的单相流多出两相间的摩擦，所以其机械能耗散也较高。

6.4.2 铁磁流体分相流的能量守恒方程的一般式

铁磁流体混合流能量方程的一般形式是方程（5.38），即

$$\frac{de_f}{dt} = -\nabla \cdot \overrightarrow{q_H} + (\tau \cdot \nabla) \cdot \overrightarrow{U_f} + \frac{\partial w_m}{\partial t}$$

上式或写成

$$\frac{de_f}{dt} + \overrightarrow{U_f} \cdot (\nabla \cdot \tau) = -\nabla \cdot \overrightarrow{q_H} + \nabla \cdot (\tau \cdot \overrightarrow{U_f}) + \frac{\partial w_m}{\partial t}$$

两边取平均，就有

$$\int_{V_f} \frac{de_f}{dt} \delta(\vec{R} - \vec{r}) dV_f + \int_{V_f} \overrightarrow{U_f} \cdot (\nabla \cdot \tau) \delta(\vec{R} - \vec{r}) dV_f =$$
$$-\int_{V_f} \nabla \cdot \overrightarrow{q_H} \delta(\vec{R} - \vec{r}) dV_f + \int_{V_f} \nabla \cdot (\tau \cdot \overrightarrow{U}) \delta(\vec{R} - \vec{r}) dV_f + \int_{V_f} \frac{\partial w_m}{\partial t} \delta(\vec{R} - \vec{r}) dV_f$$

(6.17)

以下逐项给出方程（6.17）的平均结果。

①内能变化率的平均值

$$\int_{V_f} \frac{de_f}{dt} \delta(\vec{R} - \vec{r}) dV_f = \int_{V_p} \frac{de_p}{dt} \delta(\vec{R} - \vec{r}) dV_p + \int_{V_c} \frac{de_c}{dt} \delta(\vec{R} - \vec{r}) dV_c$$

于是有：

$$\int_{V_f} \frac{de_f}{dt} \delta(\vec{R} - \vec{r}) dV_f = \phi_p \frac{de_p}{dt} + \phi_c \frac{de_c}{dt}$$

内能 $e_p$ 和 $e_c$ 以式（1.55）代入，就得

$$\int_{V_f} \frac{de_f}{dt} \delta(\vec{R}-\vec{r}) dV_f = \phi_p \left[ \left( c_s + \mu_0 H \frac{\partial M_{pe}}{\partial T_p} \right) \frac{dT_p}{dt} + \left( \mu_0 T_p \frac{\partial M_{pe}}{\partial T_p} + \mu_0 H \frac{\partial M_{pe}}{\partial H} \right) \frac{dH}{dt} \right] +$$
$$\phi_c \left[ \left( c_c + \mu_0 H \frac{\partial M_c}{\partial T_c} \right) \frac{dT_c}{dt} + \left( \mu_0 T_c \frac{\partial M_c}{\partial T_c} + \mu_0 H \frac{\partial M_c}{\partial H} \right) \frac{dH}{dt} \right] \quad \text{(I)}$$

式中，$M_{pe}$ 是磁平衡状态下，固相微粒群的磁化强度统计平均值，即 $M_{pe} = M_p L(\alpha)$，$M_c$ 是基载液体

的磁化强度，在现有的铁磁流体中，$M_c = 0$。

②表面力的流动功率

注意到 $\nabla \cdot \tau_f$ 是流体的单位体积表面力，所以 $\overline{U_f} \cdot (\nabla \cdot \tau_f)$ 就是单位体积表面力的流动功率，从而有

$$\overline{U_f} \cdot (\nabla \cdot \tau_f) dV_f = \overline{U_p} \cdot (\nabla \cdot \tau_p) dV_p + \overline{U_c} \cdot (\nabla \cdot \tau_c) dV_c$$

在邻域 $V_f$ 内取平均，给出

$$\int_{V_f} \overline{U_f} \cdot (\nabla \cdot \tau_f) \delta(\vec{R}-\vec{r}) \, dV_f = \int_{V_p} \overline{U_p} \cdot (\nabla \cdot \tau_p) \delta(\vec{R}-\vec{r}) \, dV_p + \int_{V_c} \overline{U_c} \cdot (\nabla \cdot \tau_c) \delta(\vec{R}-\vec{r}) \, dV_c =$$
$$\int_{V_f} \overline{U_p} \cdot (\nabla \cdot \tau_p) \delta(\vec{R}-\vec{r}) \phi_p \, dV_f + \int_{V_f} \overline{U_c} \cdot (\nabla \cdot \tau_c) \delta(\vec{R}-\vec{r}) \phi_c \, dV_c$$

于是就有

$$\int_{V_f} \overline{U_f} \cdot (\nabla \cdot \tau_f) \delta(\vec{R}-\vec{r}) dV_f = \phi_p \overline{U_p} \cdot (\nabla \cdot \tau_p) + \phi_c \overline{U_c} \cdot (\nabla \cdot \tau_c) \quad \text{(J)}$$

式（J）中，$\phi_p$、$\phi_c$、$\tau_p$、$\tau_c$、$\overline{U_p}$、$\overline{U_c}$ 均系坐标 $\vec{r}$ 和时间 $t$ 的函数。

③热流散度的平均值

矢量热流是在单位时间内通过单位面积的热量，所以有

$$d\overline{S_f} \cdot \overline{q_{Hf}}(\vec{r},t) = d\overline{S_p} \cdot \overline{q_{Hp}}(\vec{r},t) + d\overline{S_c} \cdot \overline{q_{Hc}}(\vec{r},t)$$

两边积分之

$$\int_{S_f} d\overline{S_f} \cdot \overline{q_{Hf}}(\vec{r},t) = \int_{S_p} d\overline{S_p} \cdot \overline{q_{Hp}}(\vec{r},t) + \int_{S_c} d\overline{S_c} \cdot \overline{q_{Hc}}(\vec{r},t)$$

使用散度定理，上式可以转换成体积积分，即

$$\int_{V_f} \nabla \cdot \overline{q_{Hf}}(\vec{r},t) dV_f = \int_{V_p} \nabla \cdot \overline{q_{Hp}}(\vec{r},t) dV_p + \int_{V_c} \nabla \cdot \overline{q_{Hc}}(\vec{r},t) dV_c$$

由此式显然可见，热流的散度是以单位体积计量的物理量。将上式右方改写

$$\int_{V_f} \nabla \cdot \overline{q_{Hf}}(\vec{r},t) dV_f = \int_{V_f} \phi_p(\vec{r},t) \nabla \cdot \overline{q_{Hp}}(\vec{r},t) dV_f + \int_{V_f} \phi_c(\vec{r},t) \nabla \cdot \overline{q_{Hc}}(\vec{r},t) dV_f$$

被积函数的关系是

$$\nabla \cdot \overline{q_{Hf}}(\vec{r},t) = \phi_p(\vec{r},t) \nabla \cdot \overline{q_{Hp}}(\vec{r},t) + \phi_c(\vec{r},t) \nabla \cdot \overline{q_{Hc}}(\vec{r},t)$$

两边通乘以 $\delta(\vec{R}-\vec{r})$，而后在邻域 $V_f$ 内积分，遂有

$$\int_{V_f} \delta(\vec{R}-\vec{r})\nabla_R \cdot \overrightarrow{q_{Hf}}(\vec{R},t)dV_f =$$

$$\int_{V_f} \phi_p(\vec{R},t)\delta(\vec{R}-\vec{r})\nabla_R \cdot \overrightarrow{q_{Hp}}(\vec{R},t)dV_f + \int_{V_f} \phi_c(\vec{R},t)\delta(\vec{R}-\vec{r})\nabla_R \cdot \overrightarrow{q_{Hc}}(\vec{R},t)dV_f =$$

$$\int_{V_p} \delta(\vec{R}-\vec{r})\nabla_R \cdot \overrightarrow{q_{Hp}}(\vec{R},t)dV_p + \int_{V_c} \delta(\vec{R}-\vec{r})\nabla_R \cdot \overrightarrow{q_{Hc}}(\vec{R},t)dV_c$$

以上式右方第一项为例，按照式（3.40a），有

$$\int_{V_p} \delta(\vec{R}-\vec{r})\nabla_R \cdot \overrightarrow{q_{Hp}}dV_p =$$

$$\int_{V_p} \nabla_R \cdot [\overrightarrow{q_{Hp}}\delta(\vec{R}-\vec{r})]dV_p - \int_{V_p} \overrightarrow{q_{Hp}} \cdot \nabla_R \delta(\vec{R}-\vec{r})dV_p =$$

$$\int_{S_p} d\overrightarrow{S_p} \cdot \overrightarrow{q_{Hp}}\delta(\vec{R}-\vec{r}) + \int_{V_p} \overrightarrow{q_{Hp}} \cdot \nabla_r \delta(\vec{R}-\vec{r})dV_p =$$

$$\int_{S_{p0}} d\overrightarrow{S_{p0}} \cdot \overrightarrow{q_{Hp}}\delta(\vec{R}-\vec{r}) + \sum_N \int_{S_{p1}} d\overrightarrow{S_{p1}} \cdot \overrightarrow{q_{Hcp}}\delta(\vec{R}-\vec{r}) + \nabla_r \cdot \int_{V_p} \overrightarrow{q_{Hp}}\delta(\vec{R}-\vec{r})dV_p$$

下面依次讨论上式右方的各项。

a.第一项是在邻域 $V_f$ 的外表面上，热流对固相微粒流所占有的总面积 $S_{p0}$ 的积分。设 $V_f$ 外表面的坐标是 $\vec{r_s}$，则由式（3.6d）有

$$\overrightarrow{n_s^0} \cdot \overrightarrow{q_{Hp}}(\vec{r_s},t)\delta(\vec{R}-\vec{r_s}) = \frac{1}{2}\left[\overrightarrow{n_{s-0}^0} \cdot \overrightarrow{q_{Hp}}(\vec{r_s}-0,t) + \overrightarrow{n_{s+0}^0} \cdot \overrightarrow{q_{Hp}}(\vec{r_s}+0,t)\right]\delta(\vec{R}-\vec{r_s}) =$$

$$\frac{1}{2}\left[-\overrightarrow{n_s^0} \cdot \overrightarrow{q_{Hp}}(\vec{r_s}-0,t) + \overrightarrow{n_s^0} \cdot \overrightarrow{q_{Hp}}(\vec{r_s}+0,t)\right]\delta(\vec{R}-\vec{r_s})$$

由于在边界表面的内外侧区间 $[r_s-0, r_s+0]$ 没有厚度，不能积贮热量，故必有 $q_{Hp}(r_s-0,t)=q_{Hp}(r_s+0,t)$，从而

$$n_s^0 \cdot q_{Hp}(r_s,t)\delta(R-r_s)=0$$

此即

$$\int_{S_{p0}} d\overrightarrow{S_{p0}} \cdot \overrightarrow{q_{Hp}}\delta(\vec{R}-\vec{r})=0$$

b.第二项是在邻域 $V_f$ 的内部，在单个微粒表面 $s_{p1}$ 上固液两相间的换热。$N$ 是 $V_f$ 内所含有的微粒数，$\sum$ 表示微粒与液相换热之总和。$\overrightarrow{q_{Hpc}}$ 和 $\overrightarrow{q_{Hcp}}$ 分别表示固相向液相的传热流和液相向固相传热的热流，它们都是微粒中心坐标 $\vec{r}$ 和时间 $t$ 的函数。

c.第三项是 $\overrightarrow{q_{Hp}}$ 在邻域 $V_f$ 内的平均值的散度，即

$$\nabla_r \cdot \int_{V_f} \delta(\vec{R}-\vec{r})\overrightarrow{q_{Hp}}(\vec{R},t)dV_p = \nabla_r \cdot \int_{V_f} \delta(\vec{R}-\vec{r})\overrightarrow{q_{Hp}}(\vec{R},t)\phi_p(\vec{R},t)dV_f = \nabla \cdot [\phi_p(\vec{r},t)\overrightarrow{q_{Hp}}(\vec{r},t)]$$

于是得到

$$\int_{V_p} \delta(\vec{R}-\vec{r})\nabla_R \cdot \overrightarrow{q_{Hp}}dV_p = \sum_N \int_{S_{p1}} d\overrightarrow{S_{p1}} \cdot \overrightarrow{q_{Hcp}}\delta(\vec{R}-\vec{r}) + \nabla_r \cdot [\phi_p(\vec{r},t)\overrightarrow{q_{Hp}}(\vec{r},t)]$$

对于 $\nabla \cdot \overrightarrow{q_{Hc}}$ 的平均值与上式有类似的形式。所以铁磁流体的 $\nabla \cdot \overrightarrow{q_{Hf}}$ 的平均值是

$$\int_{V_f} \nabla_R \cdot \overrightarrow{q_{Hf}}(\vec{R},t)\delta(\vec{R}-\vec{r})dV_f = \sum_N \int_{S_{p1}} d\overrightarrow{S_{p1}} \cdot \overrightarrow{q_{Hcp}}\delta(\vec{R}-\vec{r}) + \nabla_r \cdot [\phi_p(\vec{r},t)\overrightarrow{q_{Hp}}(\vec{r},t)] +$$
$$\sum_N \int_{S_{c1}} d\overrightarrow{S_{c1}} \cdot \overrightarrow{q_{Hpc}}\delta(\vec{R}-\vec{r}) + \nabla_r \cdot [\phi_c(\vec{r},t)\overrightarrow{q_{Hc}}(\vec{r},t)]$$

(K)

④表面应力功率的散度之平均值

$\tau \cdot \overrightarrow{U}$ 是单位面上表面力所作的功率，所以有

$$d\overrightarrow{S_f} \cdot (\tau_f \cdot \overrightarrow{U_f}) = d\overrightarrow{S_p} \cdot (\tau_p \cdot \overrightarrow{U_p}) + d\overrightarrow{S_c} \cdot (\tau_c \cdot \overrightarrow{U_c})$$

积分之有

$$\int_{S_f} d\overrightarrow{S_f} \cdot (\tau_f \cdot \overrightarrow{U_f}) = \int_{S_p} d\overrightarrow{S_p} \cdot (\tau_p \cdot \overrightarrow{U_p}) + \int_{S_c} d\overrightarrow{S_c} \cdot (\tau_c \cdot \overrightarrow{U_c})$$

利用散度定理将其转化为体积分的形式

$$\int_{V_f} \nabla \cdot (\tau_f \cdot \overrightarrow{U_f})dV_f = \int_{V_p} \nabla \cdot (\tau_p \cdot \overrightarrow{U_p})dV_p + \int_{V_c} \nabla \cdot (\tau_c \cdot \overrightarrow{U_c})dV_c$$
$$= \int_{V_f} \nabla \cdot (\tau_p \cdot \overrightarrow{U_p})\phi_p dV_f + \int_{V_f} \nabla \cdot (\tau_c \cdot \overrightarrow{U_c})\phi_c dV_f$$

于是有被积函数关系为

$$\nabla \cdot (\tau_f \cdot \overrightarrow{U_f}) = \nabla \cdot (\tau_p \cdot \overrightarrow{U_p})\phi_p + \nabla \cdot (\tau_c \cdot \overrightarrow{U_c})\phi_c$$

两边乘以 $\delta(\vec{R}-\vec{r})$ 并在邻域 $V_f$ 上积分以取平均，得

$$\int_{V_f} \delta(\vec{R}-\vec{r})\nabla \cdot (\tau_f \cdot \overrightarrow{U_f})dV_f = \int_{V_f} \delta(\vec{R}-\vec{r})\nabla \cdot (\tau_p \cdot \overrightarrow{U_p})\phi_p dV_f + \int_{V_f} \delta(\vec{R}-\vec{r})\nabla \cdot (\tau_c \cdot \overrightarrow{U_c})\phi_c dV_f =$$
$$\int_{V_p} \delta(\vec{R}-\vec{r})\nabla \cdot (\tau_p \cdot \overrightarrow{U_p})dV_p + \int_{V_c} \delta(\vec{R}-\vec{r})\nabla \cdot (\tau_c \cdot \overrightarrow{U_c})dV_c$$

以上式右方第一项为例，按照式（3.40a）有

$$\int_{V_p} \delta(\vec{R}-\vec{r})\nabla_R \cdot (\tau_p \cdot \overrightarrow{U_p})dV_p = \int_{V_p} \nabla_R \cdot [(\tau_p \cdot \overrightarrow{U_p})\delta(\vec{R}-\vec{r})]dV_p - \int_{V_p} (\tau_p \cdot \overrightarrow{U_p}) \cdot \nabla_R \delta(\vec{R}-\vec{r})dV_p =$$
$$\int_{S_p} d\overrightarrow{S_p} \cdot (\tau_p \cdot \overrightarrow{U_p})\delta(\vec{R}-\vec{r}) + \int_{V_p} (\tau_p \cdot \overrightarrow{U_p}) \cdot \nabla_r \delta(\vec{R}-\vec{r})dV_p =$$
$$\int_{S_{p0}} d\overrightarrow{S_{p0}} \cdot (\tau_p \cdot \overrightarrow{U_p})\delta(\vec{R}-\vec{r}) + \sum_N \int_{S_{p1}} d\overrightarrow{S_{p1}} \cdot (\tau_p \cdot \overrightarrow{U_p})\delta(\vec{R}-\vec{r}) + \nabla_r \cdot \int_{V_p} (\tau_p \cdot \overrightarrow{U_p})\delta(\vec{R}-\vec{r})dV_p$$

上式右方第一项是在 $V_f$ 外表面上的固相所占面积 $S_{p0}$ 积分，由式（3.6d）有

$$\overrightarrow{n_s^0} dS_{p0} \cdot [\tau_p(\vec{r_s},t) \cdot \overrightarrow{U_p}(\vec{r_s},t)]\delta(\vec{R}-\vec{r_s}) =$$
$$\frac{1}{2}dS_{p0}[\overrightarrow{n_{s-0}^0} \cdot \tau_p(\vec{r_s}-0,t) \cdot \overrightarrow{U_p}(\vec{r_s}-0,t) + \overrightarrow{n_{s+0}^0}\tau_p(\vec{r_s}+0,t) \cdot \overrightarrow{U_p}(\vec{r_s}+0,t)]\delta(\vec{R}-\vec{r_s}) =$$
$$\frac{1}{2}d\overrightarrow{S_{p0}} \cdot [\tau_p(\vec{r_s}-0,t) \cdot \overrightarrow{U_p}(\vec{r_s}-0,t) - \tau_p(\vec{r_s}+0,t) \cdot \overrightarrow{U_p}(\vec{r_s}+0,t)]\delta(\vec{R}-\vec{r_s})$$

在区间 $[\vec{r_s}-0, \vec{r_s}+0]$ 内必有 $\tau_p(\vec{r_s}-0,t) = \tau_p(\vec{r_s}+0,t)$ 和 $U_p(\vec{r_s}-0,t) = U_p(\vec{r_s}+0,t)$，从而

$$\int_{S_{p0}} d\overrightarrow{S_{p0}} \cdot (\tau_p \cdot \overrightarrow{U_p}) \delta(\vec{R} - \vec{r}) = 0$$

于是得到

$$\int_{V_p} \delta(\vec{R} - \vec{r}) \nabla \cdot (\tau_p \cdot \overrightarrow{U_p}) dV_p = \sum_N \int_{S_{p1}} d\overrightarrow{S_{p1}} \cdot (\tau_p \cdot \overrightarrow{U_p}) \delta(\vec{R} - \vec{r}) + \nabla \cdot (\phi_p \tau_p \cdot \overrightarrow{U_p})$$

同样有

$$\int_{V_c} \delta(\vec{R} - \vec{r}) \nabla \cdot (\tau_c \cdot \overrightarrow{U_c}) dV_c = \sum_N \int_{S_{c1}} d\overrightarrow{S_{c1}} \cdot (\tau_c \cdot \overrightarrow{U_c}) \delta(\vec{R} - \vec{r}) + \nabla \cdot (\phi_c \tau_c \cdot \overrightarrow{U_c})$$

于是铁磁流体表面应力的流动功率的散度平均值是

$$\int_{V_f} \delta(\vec{R} - \vec{r}) \nabla \cdot (\tau_f \cdot \overrightarrow{U_f}) dV_f = \sum_N \int_{S_{p1}} d\overrightarrow{S_{p1}} \cdot (\tau_p \cdot \overrightarrow{U_p}) \delta(\vec{R} - \vec{r}) + \nabla \cdot (\phi_p \tau_p \cdot \overrightarrow{U_p}) + \\ \sum_N \int_{S_{c1}} d\overrightarrow{S_{c1}} \cdot (\tau_c \cdot \overrightarrow{U_c}) \delta(\vec{R} - \vec{r}) + \nabla \cdot (\phi_c \tau_c \cdot \overrightarrow{U_c})$$

(L)

⑤磁化功率

将式（1.57）两边除以 $dt$，就得磁化功之变率为

$$\frac{\delta w_m}{dt} = \mu_0 H \frac{\partial M}{\partial T} \frac{dT}{dt} + \mu_0 H \frac{\partial M}{\partial T} \frac{dH}{dt}$$

磁化功是单位体积的量，故有

$$\delta w_{mf}(\vec{r}, t) dV_f = \delta w_{mp}(\vec{r}, t) dV_p + \delta w_{mc}(\vec{r}, t) dV_c$$

两边除以 $dt$ 而后在邻域 $V_f$ 内取平均，就得

$$\int_{V_f} \delta(\vec{R} - \vec{r}) \frac{\delta w_{mf}}{dt} dV_f = \int_{V_p} \delta(\vec{R} - \vec{r}) \frac{\delta w_{mp}}{dt} dV_p + \int_{V_c} \delta(\vec{R} - \vec{r}) \frac{\delta w_{mc}}{dt} dV_c$$

对于磁平衡流，则固相微粒流之磁化强度为 $M_{pe}$ （即 $M_p L(\alpha)$），将式（1.57）代入上式得

$$\int_{V_f} \delta(\vec{R} - \vec{r}) \frac{\delta w_{mf}}{dt} dV_f = \int_{V_p} \delta(\vec{R} - \vec{r}) \mu_0 H \left( \frac{\partial M_{pe}}{\partial T_p} \frac{dT_p}{dt} + \frac{\partial M_{pe}}{\partial H} \frac{dH}{dt} \right) dV_p + \\ \int_{V_c} \delta(\vec{R} - \vec{r}) \mu_0 H \left( \frac{\partial M_c}{\partial T_c} \frac{dT_c}{dt} + \frac{\partial M_c}{\partial H} \frac{dH}{dt} \right) dV_c$$

从而得磁化功率之平均值

$$\int_{V_f} \delta(\vec{R} - \vec{r}) \frac{\delta w_{mf}}{dt} dV_f = \phi_p \mu_0 H \left( \frac{\partial M_{pe}}{\partial T_p} \frac{dT_p}{dt} + \frac{\partial M_{pe}}{\partial H} \frac{dH}{dt} \right) + \phi_p \mu_0 H \left( \frac{\partial M_c}{\partial T_c} \frac{dT_c}{dt} + \frac{\partial M_c}{\partial H} \frac{dH}{dt} \right) \quad (M)$$

将以上得出的式（I）、式（J）、式（K）、式（L）及式（M）代入铁磁流体磁平衡流的平均能量方程中，就得

$$\phi_p \left[ \left( c_s + \mu_0 H \frac{\partial M_{pe}}{\partial T_p} \right) \frac{dT_p}{dt} + \left( \mu_0 T_p \frac{\partial M_{pe}}{\partial T_p} + \mu_0 H \frac{\partial M_{pe}}{\partial H} \right) \frac{dH}{dt} \right] +$$

$$\phi_c \left[ \left( c_c + \mu_0 H \frac{\partial M_c}{\partial T_c} \right) \frac{dT_c}{dt} + \left( \mu_0 T_c \frac{\partial M_c}{\partial T_c} + \mu_0 H \frac{\partial M_c}{\partial H} \right) \frac{dH}{dt} \right] + \phi_p \overrightarrow{U_p} \cdot (\nabla \cdot \tau_p) + \phi_c \overrightarrow{U_c} \cdot (\nabla \cdot \tau_c) =$$

$$-\sum_N \int_{S_{p1}} d\overrightarrow{S_{p1}} \cdot \overrightarrow{q_{Hcp}} \delta(\vec{R} - \vec{r}) - \nabla \cdot (\phi_p \overrightarrow{q_{Hp}}) - \sum_N \int_{S_{c1}} d\overrightarrow{S_{c1}} \cdot \overrightarrow{q_{Hpc}} \delta(\vec{R} - \vec{r}) - \nabla \cdot (\phi_c \overrightarrow{q_{Hc}}) + \tag{6.18}$$

$$\sum_N \int_{S_{p1}} d\overrightarrow{S_{p1}} \cdot (\tau_p \cdot \overrightarrow{U_p}) \delta(\vec{R} - \vec{r}) + \nabla \cdot (\phi_p \tau_p \cdot \overrightarrow{U_p}) + \sum_N \int_{S_{p1}} d\overrightarrow{S_{p1}} \cdot (\tau_c \cdot \overrightarrow{U_c}) \delta(\vec{R} - \vec{r}) + \nabla \cdot (\phi_c \tau_c \cdot \overrightarrow{U_c}) +$$

$$\phi_p \mu_0 H \left( \frac{\partial M_{pe}}{\partial T_p} \frac{dT_p}{dt} + \frac{\partial M_{pe}}{\partial H} \frac{dH}{dt} \right) + \phi_c \mu_0 H \left( \frac{\partial M_c}{\partial T_c} \frac{dT_c}{dt} + \frac{\partial M_c}{\partial H} \frac{dH}{dt} \right)$$

注意到矢量恒等式 $\nabla \cdot (\phi \tau \cdot \vec{U}) - \phi \vec{U} \cdot (\nabla \cdot \tau) = (\tau \cdot \nabla) \cdot (\phi \vec{U})$，并且在方程（6.18）两边消去同类项，就得铁磁流体分相流的平均能量方程之一般形式为

$$\phi_p \left( c_s \frac{dT_p}{dt} + \mu_0 T_p \frac{\partial M_{pe}}{\partial T_p} \frac{dH}{dt} \right) = -\sum_N \int_{S_{p1}} d\overrightarrow{S_{p1}} \cdot \overrightarrow{q_{Hcp}} \delta(\vec{R} - \vec{r}) - \nabla \cdot (\phi_p \overrightarrow{q_{Hp}}) +$$

$$\sum_N \int_{S_{p1}} d\overrightarrow{S_{p1}} \cdot (\tau_p \cdot \overrightarrow{U_p}) \delta(\vec{R} - \vec{r}) + (\tau_p \cdot \nabla) \cdot (\phi_p \overrightarrow{U_p}) \tag{6.19a}$$

$$\phi_c \left( c_c \frac{dT_c}{dt} + \mu_0 T_c \frac{\partial M_c}{\partial T_c} \frac{dH}{dt} \right) = -\sum_N \int_{S_{c1}} d\overrightarrow{S_{c1}} \cdot \overrightarrow{q_{Hpc}} \delta(\vec{R} - \vec{r}) - \nabla \cdot (\phi_c \overrightarrow{q_{Hc}}) +$$

$$\sum_N \int_{S_{c1}} d\overrightarrow{S_{c1}} \cdot (\tau_c \cdot \overrightarrow{U_c}) \delta(\vec{R} - \vec{r}) + (\tau_c \cdot \nabla) \cdot (\phi_c \overrightarrow{U_c}) \tag{6.19b}$$

上两方程中 $c_s$ 和 $c_c$ 分别是固相微粒和液相材料的比热容。

### 6.4.3 两相间的热交换

两相间的热交换是在固相微粒表面上进行的。当固相微粒与周围的液相有相对运动时，两者之间的换热遵从牛顿对流换热定律，即

$$q_H = h(T_p - T_c)$$

式中 $h$ 称为对流传热系数，它是当温度差为 1K 时，在单位面积上和单位时间内所传输的热量。$h$ 在层流和湍流中是很不相同的。由于固相微粒的尺寸极其微小，在液相中的运动是低 Re 数的运动，所以此时的 $h$ 是层流的传热系数。低 Re 数的层流传热实际上是热传导。它应服从 Fourier 定律。

设以一个微粒中心为原点的球坐标系，在微粒表面上取一微元面积 $d\overrightarrow{S_{p1}} = \overrightarrow{r_p^0}(r_p d\theta)(r_p d\varphi)$，则按牛顿对流传热定律有

$$d\overrightarrow{q_{Hcp}} = -\overrightarrow{r_p^0} h(T_c - T_p)(r_p d\theta)(r_p d\varphi)$$

在液相内划出一个包围该微粒的同心圆球面。其半径 $r_\infty \gg r_p$，该球面上的温度是 $T_c$，微粒尺寸极小，它在流场中就是一个点，所以表面上的温度就是 $T_p$，对于所划出的圆球面来说，即使 $r_\infty$ 是 $r_p$ 的几千倍，在流场中的常规尺度看来，这个圆球仍然是一个点，所以在圆球上的温度处处都是 $T_c$。这样一来，就

自然形成一种一维的径向导热的模式。若在圆球和微粒表面之间任意取一个同心圆，其半径为 $r$，则在这个圆球面上，通过微元面积 $\overrightarrow{r_p^0}(rd\theta)(rd\varphi)$ 的传导热流按照 Fourier 定律是

$$d\overrightarrow{q_{Hcp}} = -K_{Hc}\frac{dT}{dr}\overrightarrow{r_p^0}(rd\theta)(rd\varphi)$$

注意到导热是在以微粒中心为顶点，球片 $(rd\theta)(rd\varphi)$ 为底的四棱锥内进行，因为除径向外，没有其它方向的换热，所以通过四棱锥任何横截面的热流都相等，于是有

$$-K_{Hc}\frac{dT}{dr}\overrightarrow{r_p^0}(rd\theta)(rd\varphi) = -\overrightarrow{r_p^0}h(T_c - T_p)(r_p d\theta)(r_p d\varphi)$$

分离变量，得

$$K_{Hc}dT = h(T_c - T_p)r_p^2\frac{dr}{r^2}$$

沿 $\overrightarrow{r_p^0}$ 正向积分，就有

$$K_{Hc}T\Big|_{T_p}^{T_c} = h(T_c - T_p)r_p^2\left(-\frac{1}{r}\right)\Big|_{r_p}^{r_\infty}$$

注意到 $r_\infty \gg r_p$，故相对于 $1/r_p$ 可以略去 $1/r_\infty$，于是得

$$K_{Hc}(T_c - T_p) = h(T_c - T_p)\,r_p$$

上式给出低 Re 数下对流传热系数是

$$h = \frac{K_{Hc}}{r_p} \tag{6.20a}$$

$K_{Hc}$ 是液相物质的导热系数，将其取为常数，故 $h$ 也是常数，但 $T_c$ 和 $T_p$ 在流场内各处不相同，它是 $T_c = T_c(\vec{r},t)$ 和 $T_p = T_p(\vec{r},t)$，也就是说 $\overrightarrow{q_{Hcp}} = \overrightarrow{q_{Hcp}}(\vec{r},t)$。

在传热学中，一个常用的无量纲准数 Nu（即 Nusselt 数），其定义是

$$\mathrm{Nu} = \frac{hd}{K_{Hc}}$$

对于小圆球状的微粒，特征尺寸 $d$ 取为 $2r_p$，于是由式（6.20a）就得低 Re 数下对流传热的 Nu

$$\mathrm{Nu} = 2 \tag{6.20b}$$

在方程（6.19a）右方第一项求和号内积分函数

$$d\overrightarrow{S_{p1}} \cdot \overrightarrow{q_{Hcp}} = dq_{Hcp} = -h(T_c - T_p)dS_{p1}$$

于是有

$$-\sum_N \int_{S_{p1}} d\overrightarrow{S_{p1}} \cdot \overrightarrow{q_{Hcp}}\delta(\vec{R}-\vec{r}) = \sum_N \delta(\vec{R}-\vec{r})h(T_c-T_p)\int_{S_{p1}} dS_{p1} = \sum_N \delta(\vec{R}-\vec{r})h(T_c-T_p)\frac{S_{p1}}{V_{p1}}V_{p1}$$

$N$ 值十分巨大，$V_{p1}$ 很微小以致可以看作微元体积 $dV_p$，则求和式可以写成积分形式，即有

$$-\sum_N \int_{S_{p1}} d\overrightarrow{S_{p1}} \cdot \overrightarrow{q_{Hcp}} \delta(\vec{R}-\vec{r}) = \int_{V_p} \delta(\vec{R}-\vec{r}) h(T_c - T_p) \frac{S_{p1}}{V_{p1}} dV_p$$

注意到 $S_{p1} = 4\pi r_p^2$，$V_{p1} = (4/3)\pi r_p^3$，同时用式（6.20a）置换 $h$

$$-\sum_N \int_{S_{p1}} d\overrightarrow{S_{p1}} \cdot \overrightarrow{q_{Hcp}} \delta(\vec{R}-\vec{r}) = \phi_p \frac{3K_{Hc}}{r_p^2}(T_c - T_p) \tag{N}$$

$K_{Hc} = K_{Hc}(\vec{r}, t)$，上面 $K_{Hc}$ 取为常值只是在 $(4/3)\pi r_\infty^3$ 的范围之内，它是极小的区间，在流场中几乎可以视为一个点。同样，对于 $\overrightarrow{q_{Hpc}}$ 的情况，也可得到

$$-\sum_N \int_{S_{c1}} d\overrightarrow{S_{c1}} \cdot \overrightarrow{q_{Hpc}} \delta(R-r) = \phi_p \frac{3K_{Hc}}{r_p^2}(T_p - T_c) \tag{O}$$

### 6.4.4 两相间的摩擦功率

1.相对平动运动的摩擦功率

相对平动运动在微粒表面形成的摩擦应力是

$$\tau_{rr} + \tau_{r\theta} = \overrightarrow{r_p^0} \tau_{rr} \overrightarrow{r_p^0} + \overrightarrow{r_p^0} \tau_{r\theta} \overrightarrow{\theta^0}$$

将上式代入方程（6.19a）右方第三项求和号内的积分之中，就有

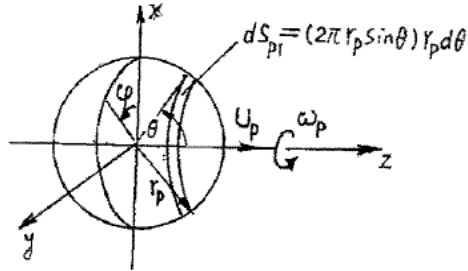

图 6-2　圆球形微粒在球坐标系内的运动

$$\sum_N \int_{S_{p1}} d\overrightarrow{S_{p1}} \cdot [(\overrightarrow{r_p^0} \tau_{rr} \overrightarrow{r_p^0} + \overrightarrow{r_p^0} \tau_{r\theta} \overrightarrow{\theta^0}) \cdot \vec{k} U_p] \delta(\vec{R}-\vec{r}) = \sum_N \int_{S_{p1}} dS_{p1}(\tau_{rr}\cos\theta - \tau_{r\theta}\sin\theta)U_p \delta(\vec{R}-\vec{r})$$

由式（2.87a）、式（2.87b）及式（2.88a）、式（2.88b），当 $r = r_p$ 时，即在微粒表面上，应力是

$$\tau_{rr} = -p_0 - \frac{3}{2}\eta_c \frac{1}{r_p}(U_p - U_c)\cos\theta, \qquad \tau_{r\theta} = \frac{3}{2}\eta_c \frac{1}{r_p}(U_p - U_c)\sin\theta$$

于是有

$$\sum_N \int_{S_{p1}} d\overrightarrow{S_{p1}} \cdot (\tau_{rr} + \tau_{r\theta}) \cdot \overrightarrow{U_p} \delta(\vec{R} - \vec{r}) = \sum_N \delta(\vec{R} - \vec{r}) U_p (2\pi r_p^2) \int_0^{\pi} \left[ -p_0 \cos\theta - \frac{3}{2} \eta_c \frac{1}{r_p} (U_p - U_c) \right] \sin\theta \, d\theta =$$

$$\sum_N \delta(\vec{R} - \vec{r}) U_p (2\pi r_p^2) \left[ -3\eta_c \frac{1}{r_p} (U_p - U_c) \right] =$$

$$\sum_N \delta(\vec{R} - \vec{r}) \frac{9}{2} \eta_c \frac{1}{r_p^2} (U_c - U_p) U_p V_{p1}$$

将求和写成积分的形式，就得

$$\sum_N \int_{S_{p1}} d\overrightarrow{S_{p1}} \cdot (\tau_{rr} + \tau_{r\theta}) \cdot \overrightarrow{U_p} \delta(\vec{R} - \vec{r}) = \int_{V_p} \delta(\vec{R} - \vec{r}) \frac{9}{2} \frac{1}{r_p^2} \eta_c (U_c - U_p) U_p \, dV_P =$$

$$= \frac{9}{2} \frac{1}{r_p^2} \eta_c (U_c - U_p) U_p \phi_p \qquad\qquad\qquad\text{(P)}$$

式中 $U_c = U_c(\vec{r}, t)$，$U_p = U_p(\vec{r}, t)$，$\phi_p = \phi_p(\vec{r}, t)$。

对于液相表面 $S_{p1}$ 作用的应力 $\tau_{rr}$ 是 $\overrightarrow{r_c^0} \tau_{rr} \overrightarrow{r_c^0}$，$\tau_{r\theta}$ 是 $\overrightarrow{r_c^0} \tau_{r\theta} \overrightarrow{\theta_c^0}$，其中 $\overrightarrow{r_c^0}$、$\overrightarrow{\theta_c^0}$ 与 $\overrightarrow{r_p^0}$、$\overrightarrow{\theta^0}$ 的方向相反，而 $\tau_{rr}$ 和 $\tau_{r\theta}$ 的数值大小不变，这当然是作用与反作用相等的表现。即

$$\tau_{rr} = -p_0 - \frac{3}{2} \eta_c \frac{1}{r_p} (U_p - U_c) \cos\theta, \qquad\qquad \tau_{r\theta} = \frac{3}{2} \eta_c \frac{1}{r_p} (U_p - U_c) \sin\theta$$

面积与面积微元 $\overrightarrow{S_{c1}}$、$d\overrightarrow{S_{c1}}$ 大小相等于 $\overrightarrow{S_{p1}}$、$d\overrightarrow{S_{p1}}$，而方向相反。并且，按粘附条件有 $\overrightarrow{U_{c1}} = \overrightarrow{U_p}$，于是

$$\sum_N \int_{S_{c1}} d\overrightarrow{S_{c1}} \cdot (\overrightarrow{\tau_{rr} + \tau_{r\theta}})_c \cdot \overrightarrow{U_{c1}} \delta(\vec{R} - \vec{r}) = \sum_N \int_{S_{c1}} dS_{c1} \overrightarrow{r_c^0} \cdot (\overrightarrow{r_c^0} \tau_{rr} \overrightarrow{r_c^0} + \overrightarrow{r_c^0} \tau_{r\theta} \overrightarrow{\theta_c^0}) \cdot \vec{k} U_{c1} \delta(\vec{R} - \vec{r}) =$$

$$\sum_N \int_{S_{c1}} dS_{c1} (-\tau_{rr} \cos\theta + \tau_{r\theta} \sin\theta) U_{c1} \delta(\vec{R} - \vec{r}) =$$

$$-\sum_N \int_{S_{p1}} dS_{p1} (\tau_{rr} \cos\theta - \tau_{r\theta} \sin\theta) U_p \delta(\vec{R} - \vec{r})$$

于是得到

$$\sum_N \int_{S_{c1}} d\overrightarrow{S_{c1}} \cdot (\tau_{rr} + \tau_{r\theta})_c \cdot \overrightarrow{U_{c1}} \delta(\vec{R} - \vec{r}) = \frac{9}{2} \frac{1}{r_p^2} \eta_c (U_p - U_c) U_p \phi_p \qquad\qquad\text{(Q)}$$

式（Q）右方的 $U_p$ 和 $U_c$ 均系邻域 $V_f$ 内的平均速度。

2.相对转动运动的摩擦功率

液固两相相对转动而在固相微粒表面上形成的应力是 $\tau_{r\varphi}$，显然 $\tau_{r\varphi}$ 是沿圆球形微粒平行圆圆周的应力环流。旋转的固相微粒的表面速度是 $\overrightarrow{\omega_p} \times \overrightarrow{r_p} = \vec{k} \omega_p \times \overrightarrow{r_p^0} r_p$，于是有微粒绕轴旋转的表面速度 $\overrightarrow{\varphi^0} \omega_p r_p \sin\theta$ 可以看出表面速度也是沿平行圆圆周的。故旋转功率为

$$\sum_N \int_{S_{p1}} d\overrightarrow{S_{p1}} \cdot [\tau_{r\varphi} \cdot (\overrightarrow{\omega_p} \times \vec{r}_p)]\delta(\vec{R} - \vec{r}) = \sum_N \int_{S_{p1}} \overrightarrow{r_p^0} dS_{p1} \cdot \left[(\overrightarrow{r_p^0}\tau_{r\varphi}\overrightarrow{\varphi^0}) \cdot (\overrightarrow{\varphi^0}\omega_p r_p \sin\theta)\right]\delta(\vec{R} - \vec{r}) =$$
$$\sum_N \int_{S_{p1}} dS_{p1}(\tau_{r\varphi} r_p \omega_p \sin\theta)\delta(\vec{R} - \vec{r})$$

由式（2.91a）与式（2.91b），在微粒的表面上，即取 $r = r_p$，应力 $\tau_{r\varphi}$ 之值为

$$\tau_{r\varphi} = -3\eta_c(\omega_P - \omega_C)\sin\theta$$

以及 $dS_{p1} = (2\pi r_p \sin\theta)r_p d\theta$，则有

$$\sum_N \int_{S_{p1}} dS_{p1}(\tau_{r\varphi} r_p \omega_p \sin\theta)\delta(\vec{R} - \vec{r}) =$$
$$\sum_N \int_{S_{p1}} \delta(\vec{R} - \vec{r})[-3\eta_c(\omega_P - \omega_C)\sin\theta](r_p \omega_p \sin\theta)(2\pi r_p \sin\theta)r_p d\theta =$$
$$\sum_N \delta(\vec{R} - \vec{r})6\pi r_p^3\eta_c(\omega_C - \omega_P)\omega_p \int_o^\pi \sin^3\theta d\theta = \qquad (R)$$
$$\sum_N \delta(\vec{R} - \vec{r})6(4\pi r_p^3/3)\eta_c(\omega_C - \omega_P)\omega_p = \sum_N \delta(\vec{R} - \vec{r})6V_{p1}\eta_c(\omega_C - \omega_P)\omega_p =$$
$$\int_{V_p} \delta(\vec{R} - \vec{r})6\eta_c(\omega_C - \omega_P)\omega_p dV_p = 6\phi_p\eta_c(\omega_C - \omega_P)\omega_p$$

若考虑 $\phi_p$ 的影响，则将 $\eta_c$ 换为 $\eta_\phi$，若同时考虑分散剂的影响可使用 $\eta_\delta$ 代替 $\eta_c$。

对于与微粒接触的液相，$\overrightarrow{S_{c1}} = -\overrightarrow{S_{p1}}$，$d\overrightarrow{S_{c1}} = -d\overrightarrow{S_{p1}}$，表面应力 $(\tau_{r\varphi})_c$ 与 $\tau_{r\varphi}$ 大小相等而方向相反，它所作用的液相表面的外向法线是 $\vec{r}_c$，且 $\overrightarrow{r_c^0} = -\overrightarrow{r_p^0}$，故 $(\tau_{r\varphi})_c = \overrightarrow{r_c^0}\tau_{r\varphi}(-\overrightarrow{\varphi^0})$，表面速度由粘附条件知道仍然是 $\overrightarrow{\omega_p} \times \vec{r}_p$，于是有

$$\sum_N \int_{S_{c1}} d\overrightarrow{S_{c1}} \cdot [(\tau_{r\varphi})_c \cdot (\overrightarrow{\omega_p} \times \vec{r}_p)]\delta(\vec{R} - \vec{r}) =$$
、 $$\sum_N \int_{S_{c1}} \overrightarrow{r_c^0} dS_{c1} \cdot [\overrightarrow{r_c^0}\tau_{r\varphi}(-\overrightarrow{\varphi^0}) \cdot (\overrightarrow{\omega_p} \times \vec{r}_p)]\delta(\vec{R} - \vec{r}) = \qquad (S)$$
$$\sum_N \int_{S_{p1}} dS_{p1}(-\tau_{r\varphi})\omega_P r_p(\sin\theta)\delta(\vec{R} - \vec{r}) = 6\phi_p\eta_c(\omega_P - \omega_c)\omega_p$$

6.4.5 在固相微粒流及液相流各自内部的流动功率和摩擦功率

方程（6.19a）和方程（6.19b）右方的末项 $(\tau_p \cdot \nabla) \cdot (\phi_p \overline{U_p})$ 与 $(\tau_c \cdot \nabla) \cdot (\phi_c \overline{U_c})$ 都已经是邻域 $V_f$ 内的平均值。由

$$(\tau_p \cdot \nabla) \cdot (\phi_p \overline{U_p}) = \left[(\vec{i}\,\tau_{pij}\,\vec{j}) \cdot \left(\vec{i}\frac{\partial}{\partial x_i} + \vec{j}\frac{\partial}{\partial x_j} + \vec{k}\frac{\partial}{\partial x_k}\right)\right] \cdot \phi_p(\vec{i}u_{pi} + \vec{j}u_{pj} + \vec{k}u_{pk}) =$$
$$\left(\vec{i}\,\tau_{pij}\frac{\partial}{\partial x_j}\right) \cdot \phi_p(\vec{i}u_{pi} + \vec{j}u_{pj} + \vec{k}u_{pk}) = \tau_{pij}\frac{\partial}{\partial x_j}(\phi_p u_{pi})$$

由式（2.47）有

$$\tau_{pij} = -p\delta_{ij} + \eta_{NP}\left(\frac{\partial u_{pi}}{\partial x_j} + \frac{\partial u_{pj}}{\partial x_i}\right) - \frac{2}{3}\eta_{NP}(\nabla \cdot \overrightarrow{U_p})\delta_{ij}$$

于是有

$$
\begin{aligned}
&\tau_{pij}\frac{\partial}{\partial x_j}(\phi_p u_{pi}) = \\
&-p\delta_{ij}\frac{\partial}{\partial x_j}(\phi_p u_{pi}) + \eta_{NP}\left(\frac{\partial u_{pi}}{\partial x_j} + \frac{\partial u_{pj}}{\partial x_i}\right)\frac{\partial(\phi_p u_{pi})}{\partial x_j} - \frac{2}{3}\eta_{NP}(\nabla \cdot \overrightarrow{U_p})\delta_{ij}\frac{\partial(\phi_p u_{pi})}{\partial x_j} = \\
&-p\frac{\partial}{\partial x_i}(\phi_p u_{pi}) + \eta_{NP}\left(\frac{\partial u_{pi}}{\partial x_j} + \frac{\partial u_{pj}}{\partial x_i}\right)\frac{\partial}{\partial x_j}(\phi_p u_{pi}) - \frac{2}{3}\eta_{NP}(\nabla \cdot \overrightarrow{U_p})\frac{\partial}{\partial x_i}(\phi_p u_{pi}) = \\
&-p\nabla \cdot (\phi_p \overrightarrow{U_p}) + \eta_{NP}\left(\frac{\partial u_{pi}}{\partial x_j} + \frac{\partial u_{pj}}{\partial x_i}\right)\frac{\partial}{\partial x_j}(\phi_p u_{pi}) - \frac{2}{3}\eta_{NP}(\nabla \cdot \overrightarrow{U_p})\nabla \cdot (\phi_p \overrightarrow{U_p})
\end{aligned}
\tag{T}
$$

同样可有

$$(\tau_c \cdot \nabla) \cdot (\phi_c \overrightarrow{U_c}) = -p\nabla \cdot (\phi_c \overrightarrow{U_c}) + \eta_c\left(\frac{\partial u_{ci}}{\partial x_j} + \frac{\partial u_{cj}}{\partial x_i}\right)\frac{\partial}{\partial x_j}(\phi_c u_{ci}) - \frac{2}{3}\eta_c(\nabla \cdot \overrightarrow{U_c})\nabla \cdot (\phi_c \overrightarrow{U_c}) \tag{U}$$

### 6.4.6 铁磁流体磁平衡分相流的能量守恒方程

将以上得出的式（N）～式（U）按固、液相分别代入能量方程（6.19a）、（6.19b）中，得

$$
\begin{aligned}
\phi_p\left(c_s\frac{dT_p}{dt} + \mu_0 T_p\frac{\partial M_{pe}}{\partial T_p}\frac{dH}{dt}\right) = &\phi_p\frac{3K_{Hc}}{r_p^2}(T_c - T_p) - \nabla \cdot (\phi_p K_{Hp}\nabla T_p) + \\
&\frac{9}{2}\phi_p\eta_c\frac{1}{r_p^2}(U_c - U_p)U_p + 6\phi_p\eta_c(\omega_C - \omega_p)\omega_p - p\nabla \cdot (\phi_p \overrightarrow{U_p}) + \\
&\eta_{NP}\left(\frac{\partial u_{pi}}{\partial x_j} + \frac{\partial u_{pj}}{\partial x_i}\right)\frac{\partial}{\partial x_j}(\phi_p u_{pi}) - \frac{2}{3}\eta_{NP}(\nabla \cdot \overrightarrow{U_p})\nabla \cdot (\phi_p \overrightarrow{U_p})
\end{aligned}
\tag{6.21a}
$$

和

$$
\begin{aligned}
\phi_c\left(c_c\frac{dT_c}{dt} + \mu_0 T_c\frac{\partial M_c}{\partial T_c}\frac{dH}{dt}\right) = &\phi_p\frac{3K_{Hc}}{r_p^2}(T_p - T_c) - \nabla \cdot (\phi_c K_{Hc}\nabla T_c) + \\
&\frac{9}{2}\phi_p\eta_c\frac{1}{r_p^2}(U_p - U_c)U_p + 6\phi_p\eta_c(\omega_p - \omega_C)\omega_p - p\nabla \cdot (\phi_c \overrightarrow{U_c}) + \\
&\eta_c\left(\frac{\partial u_{ci}}{\partial x_j} + \frac{\partial u_{cj}}{\partial x_i}\right)\frac{\partial}{\partial x_j}(\phi_c u_{ci}) - \frac{2}{3}\eta_c(\nabla \cdot \overrightarrow{U_c})\nabla \cdot (\phi_c \overrightarrow{U_c})
\end{aligned}
\tag{6.21b}
$$

### 6.4.7 机械能耗散函数

在铁磁流体混合流理论中，不考虑两相运动的滞后，也就不能反映出两相之间的摩擦。在分相流理论中，认为每个相都单独地占满场空间，而另外一相在物质上并不存在，代之以相间的力、力矩、热交换等作用。这些力和力矩的发生源于运动的滞后和基载液体的粘性。显然它们是粘性摩擦力和摩擦力矩。这类摩擦的结果都引起机械能转化成摩擦热。在热力学上就是自发的不可逆过程，因而造成

机械不能恢复的损失。在能量方程中表示这种损失的就是耗散函数 $\Phi_p$ 和 $\Phi_c$：

$$\Phi_p = \frac{9}{2}\eta_c \frac{1}{r_p^2}(U_c - U_p)U_p + 6\eta_c(\omega_c - \omega_p)\omega_p + \frac{1}{\phi_p}\eta_{NP}\left(\frac{\partial u_{pi}}{\partial x_j} + \frac{\partial u_{pj}}{\partial x_i}\right)\frac{\partial}{\partial x_j}(\phi_p u_{pi}) -$$
$$\frac{1}{\phi_p}\frac{2}{3}\eta_{NP}(\nabla \cdot \overrightarrow{U_p})\nabla \cdot (\phi_p \overrightarrow{U_p}) \tag{6.22a}$$

$$\Phi_c = \frac{9}{2}\frac{\phi_p}{\phi_c}\eta_c \frac{1}{r_p^2}(U_p - U_c)U_p + 6\frac{\phi_p}{\phi_c}\eta_c(\omega_p - \omega_c)\omega_p + \frac{1}{\phi_c}\eta_c\left(\frac{\partial u_{ci}}{\partial x_j} + \frac{\partial u_{cj}}{\partial x_i}\right)\frac{\partial}{\partial x_j}(\phi_c u_{ci}) -$$
$$\frac{2}{3}\frac{1}{\phi_c}\eta_c(\nabla \cdot \overrightarrow{U_c})\nabla \cdot (\phi_c \overrightarrow{U_c}) \tag{6.22b}$$

上式中 $\Phi_p$ 是单位体积固相微粒流的耗散函数，$\Phi_c$ 是单位体积液相流的耗散函数。将能量方程式（6.21a）与式（6.21b）分别除以 $\phi_p$ 与 $\phi_c$，就得单位体积的能量方程：

$$c_s\frac{dT_p}{dt} + \mu_0\frac{\partial M_{pe}}{\partial T_p}\frac{dH}{dt} = \frac{3K_{Hc}}{r_p^2}(T_c - T_p) - \frac{1}{\phi_p}\nabla \cdot (\phi_p K_{Hp}\nabla T_p) - \frac{1}{\phi_p}p\nabla \cdot (\phi_p \overrightarrow{U_p}) + \Phi_p \tag{6.23a}$$

$$c_c\frac{dT_c}{dt} + \mu_0\frac{\partial M_c}{\partial T_c}\frac{dH}{dt} = \frac{\phi_p}{\phi_c}\frac{3K_{Hc}}{r_p^2}(T_p - T_c) - \frac{1}{\phi_c}\nabla \cdot (\phi_c K_{Hc}\nabla T_c) - \frac{1}{\phi_c}p\nabla \cdot (\phi_c \overrightarrow{U_c}) + \Phi_c \tag{6.23b}$$

上两方程中 $c_s$ 和 $c_c$ 分别是固相微粒流与液相物质的体积比热容，$K_{Hp}$ 是固相微粒流的热传导系数，它不同于固相物质的热传导系数，$K_{Hc}$ 就是液相物质的热传导系数。在现有的铁磁流体中，基载液体都是不导磁的，即 $M_c = 0$，其它参数 $T_p$、$T_c$、$M_{pe}$、$H$、$\phi_p$、$\phi_c$、$U_p$、$U_c$ 以及 $\Phi_p$、$\Phi_c$ 等均是坐标和时间的函数。

### 6.4.8 固相微粒流的几个物性参数由初级输运理论得到的近似结果

#### 1.概述

固相微粒流由分散的大量固相微粒流体组成。单个固相微粒尺寸大致为最小气体分子的 100 倍左右，大的气体分子的尺寸可以达到小气体分子的几千至万倍以上。所以铁磁流体中的固相微粒的常规尺寸相当于中等的气体分子，它们具有气体分子的热运动是不言而喻的。在标准状况下所有气体每毫升含有的分子数目是一常数，为 $2.688 \times 10^{19}$（此即 Loschmidt 数），而在铁磁流内，通常每毫升含有的固相微粒数目的量级为 $10^{16} \sim 10^{17}$ 个。所以，将固相微粒流当作气体那样的分子流来分析，有一定的合理性。

固相微粒之间是有距离的，在这种分散状态下，其输运过程与连续的大块固相物质不一样，它的输运过程主要依靠微粒的热运动来实现。粘性系数是热运动中微粒间动量交换的体现，而热传导就是微粒热运动能变换的结果。温度越高，热运动越旺盛，微粒间互相碰撞的几率越大，因而无论是微粒流的粘性系数，还是微粒流的热传导系数都随温度升而相应地增大。

#### 2.固相微粒流的粘性系数 $\eta_{NP}$

按照分子运动论，气体的粘性系数是[11]

$$\eta = \frac{\sqrt{2} m_0 U_e}{6\pi d^2}$$

式中，$m_0$ 是单个气体分子的质量，$U_e$ 是其热运动速度之平均值，$d$ 是气体分子的直径。若取 $U_e$ 是统计平均速度，则有[2,12,13]

$$U_e = \sqrt{\frac{8k_0 T}{\pi m_0}}$$

将上式代入 $\eta$ 的关系式中，得

$$\eta = \frac{2\alpha}{3\pi d^2} \sqrt{\frac{m_0 k_0 T}{\pi}} \tag{6.24}$$

式中 $\alpha$ 是一个常系数，$\alpha = 1.46$[12]。在本章 6.2 节末，曾给出一个 $Fe_3O_4$ 微粒流的粘性系 $\eta_{NP}$ 与空气的粘性系数 $\eta_{air}$ 之间的关系为

$$\eta_{NP} = 0.368 \eta_{air}$$

利用已知的空气粘度系数之实验值[9]，计算微粒流的粘性系数列于表 6-1 中。

表 6-1 $Fe_3O_4$ 固相微粒流的粘性系数

| 温度 $T$ （℃） | 0 | 20 | 40 | 60 | 80 | 100 |
|---|---|---|---|---|---|---|
| 粘度 $\eta_{NP}$（微泊） | 62.9 | 66.6 | 69.9 | 73.6 | 76.4 | 80.2 |

（注：1 微泊 $= 10^{-6}$ 泊 $= 10^{-7}$ $N \cdot s/m^2$ ）

表 6-1 中所列数据可以作为定性的参考。

3．固相微粒流的体积比热容是 $c_s$

含有大量微粒的固相微粒群，其单个微粒的热运动能是[12,13]

$$\frac{1}{2} m_{p1} (U_p^2)_e$$

式中 $m_{p1}$ 是一个微粒的质量，$m_{p1} = \rho_{NP} V_{p1}$，$(U_p^2)_e$ 是微粒流速度平方的统计平均值，由分子运动论给出

$$(U_p^2)_e = \frac{3k_0 T}{m_{p1}}$$

此式给出热运动动能为

$$\frac{1}{2} m_{p1} (U_p^2)_e = \frac{3}{2} k_0 T$$

于是单位质量的固相微粒的运动动能是

$$\frac{3}{2}\frac{k_0 T}{m_{p1}} = \frac{3}{2}\frac{k_0 T}{\rho_{NP}V_{p1}}$$

单位质量的固相微粒群的总内能 $de'_p$ 是其热内能 $c'_{NP}dT$ 与其热运动动能 $\frac{3}{2}\frac{k_0}{\rho_{NP}V_{p1}}dT$ 之和，即

$$de'_p = \left(c'_{NP} + \frac{3}{2}\frac{k_0}{\rho_{NP}V_{p1}}\right)dT$$

式中 $c'_{NP}$ 是固相微粒材料的质量比热容，而固相微粒群的质量比热容 $c'_s$ 是

$$c'_s = c'_{NP} + \frac{3}{2}\frac{k_0}{\rho_{NP}V_{p1}} \tag{6.25a}$$

由体积比热容 $c_v$ 与质量比热容 $c'_s$ 间的关系，$c_v = \rho c'_s$，得固相微粒群的体积比热容为

$$c_s = \rho_{NP}c'_s = c_{NP} + \frac{3}{2}\frac{k_0}{V_{p1}} \tag{6.25b}$$

式中 $c_{NP}$ 是固相微粒材料的体积比热容。注意到 Boltzmann 常数 $k_0$ 的量级是 $10^{-23}$，所以固相微粒的体积 $V_{p1}$ 必须很微小，式（6.25b）右方第二项才有实际意义。对于分子尺寸为 $10^{-10}$ m 量级的小分子气体，右方第一项远小于第二项而可略，所以单原子气体的热容量基本上是其平动热运动的动能。如果固相微粒的尺寸比较大，则右方第二项远小于第一项而可略，或固相微粒的尺寸超过微米级，微粒就不出现热运动，这时的比热容仅体现固相材料内部的热内能。

4.固相微粒的热传导系数 $K_{Hp}$

按照分子运动论，气态物质的热传导过程是热运动分子之间的动能交换，而粘性则是热运动分子之间的动量交换，所以这两者必有关联。设两者具有简单关系，即

$$K_{Hp} = A\eta_{NP}$$

式中 $A$ 是一个系数。用量纲分析可以对 $A$ 进行推断。为了简单明了，此处直接使用单位分析来推断 $A$。热传系数 $K_{Hp}$ 的单位是 $N \cdot m/(m \cdot s \cdot K)$，粘性系数 $\eta_{NP}$ 的单位是 $N \cdot s/m^2 = (kg \cdot m/s^2) \cdot (s/m^2) = kg/(m \cdot s)$。于是 $A$ 的单位是

$$\left(\frac{N \cdot m}{m \cdot s \cdot K}\right) \Big/ \left(\frac{kg}{m \cdot s}\right) = \frac{N \cdot m}{kg \cdot K}$$

显然，等号右方是质量比热容的单位，由此可见 $A$ 即 $c'_s$，于是

$$K_{Hp} = c'_s \eta_{NP} \tag{6.26a}$$

按照分子运动论严格地推导得出关系式正是式（6.26a）。[11]

利用 $c_s = \rho_{NP} c'_s$，动力粘性系数 $\eta_{NP}$ 与运动粘性系数 $\nu_{NP}$ 的关系，即 $\nu_{NP} = \eta_{NP}/\rho_{NP}$，可以将式（6.26a）

改写成

$$K_{Hp} = \frac{c_s}{\rho_{NP}} \eta_{NP} = c_s \nu_{NP} \tag{6.26b}$$

## 6.5 铁磁流分相流的三种磁况

所谓磁况均指磁化强度的状况。即磁松弛、磁平衡和磁冻结。在分析分相流的磁松弛问题上，情况比混合流略为简单一些。当然，引起磁松弛的涡粘力矩和外磁场的旋转仍然存在，其物理模型就是图 4-1（a）和（b）。

1.磁松弛的磁化强度

这里的磁强度 $M$ 是指固相微粒流的磁化强度。由式（5.18b）给出

$$\frac{M_y}{M_0} = \frac{(\omega_{PH})_z}{1 + \beta_e \cos^2 \theta_0} t_{r\delta} \sin \theta_0$$

将上式代入式（4.68b）中，得

$$\frac{M_x}{M_0} = \frac{\sin \theta_0}{1 + \beta_e \cos^2 \theta_0}$$

由图 4-1（a）可见

$$\frac{M_z}{M_0} = \cos \theta_0$$

于是得到磁松弛的 $M$

$$M = \sqrt{M_x^2 + M_y^2 + M_z^2} = M_0 \left\{ \left( \frac{\sin \theta_0}{1 + \beta_e \cos^2 \theta_0} \right)^2 [1 + (\omega_{PH})_z^2 t_{r\delta}^2] + \cos^2 \theta_0 \right\}^{1/2} \tag{6.27a}$$

$M$ 和外磁场之间的夹角 $\alpha_m$ 是

$$\alpha_m \approx \arccos \left( \frac{M}{M_0} \right) = \arccos \left\{ \left( \frac{\sin \theta_0}{1 + \beta_e \cos^2 \theta_0} \right)^2 [1 + (\omega_{PH})_z^2 t_{r\delta}^2] + \cos^2 \theta_0 \right\}^{1/2} \tag{6.27b}$$

式中（6.27a）与式（6.27b）右方的 $(\omega_{PH})_z^2 t_{r\delta}^2$ 通常是二阶小量，相对 1 可以略去。在上述两式中，综合磁松弛时间 $t_{r\delta}$ 见式（4.51b），磁松弛参数 $\beta_e$ 见式（4.77a）与式（4.77b），基载液体的粘性系数 $\eta_\delta$ 见式（4.32）。此外，两式中的 $M_0$ 即平衡磁化强度 $M_{pe}$

$$\overline{M_0} = \overline{M_{pe}} = \overline{M_p} L(\alpha) = \frac{\overline{m_{p1}}}{V_{p1}} \left( \coth \alpha - \frac{1}{\alpha} \right) \tag{6.28}$$

式中 $\alpha$ 见式（1.16）。$\overrightarrow{m_{p1}}$ 是固相微粒的磁矩。

2.磁平衡的磁化强度

磁平衡的基本条件是磁松弛时间 $t_{r\delta}$ 相对转动的特征时间是高阶小，这时磁松弛参数 $\beta_e$ 趋近于零，

这时式（6.27a）给出 $M = M_0$，式（3.27b）给出 $\alpha_m = 0$。此即在磁平衡状态下，固相微粒流磁化强度

由 Langevin 方程决定，并且磁化强度的方程保持与外磁场一致。

3.磁冻结状况

此时的固相微粒流的磁化强度，在流动过程中保持动始值不变，即

$$\overline{M} = \overline{M_i} = 常矢 \tag{6.29}$$